\documentclass[11pt,urlcolor=blue, linkcolor=blue]{article} 
\usepackage{cite}

\usepackage{float}

\usepackage{multicol}
\usepackage{tablefootnote}
\usepackage{amsmath, amsthm, amssymb,slashed,mathtools}
\usepackage{ifpdf}
\ifpdf
  \usepackage[pdftex]{graphicx}
  \usepackage{epstopdf}
\else
  \usepackage[dvips]{graphicx}
\fi
\parskip=\baselineskip

\usepackage[usenames, dvipsnames]{color}

\allowdisplaybreaks[1]

\numberwithin{equation}{section}

\usepackage{booktabs}
\newtheorem{thm}{Theorem}
\newtheorem{lem}[thm]{Lemma}

\theoremstyle{definition}

\definecolor{mygray}{gray}{0.6}

\usepackage{upgreek}
\usepackage{bbm}

\newenvironment{myfont}[2][]{\csname#2\endcsname[#1]}{}

\newcommand\hcup[1]{\underset{{\scriptscriptstyle #1}}{\cup}}

\usepackage{slashed}
\usepackage[makeroom]{cancel}
\usepackage[normalem]{ulem}
\usepackage{soul}
\newcommand{\stkout}[1]{\ifmmode\text{\sout{\ensuremath{#1}}}\else\sout{#1}\fi}

\usepackage{sseq}
\usepackage[all,cmtip]{xy}
\usepackage{tikz-cd}
\usepackage{tikz}
\usetikzlibrary{matrix}

\newcommand{\bea}{\begin{eqnarray}}
\newcommand{\eea}{\end{eqnarray}}
\def\be{\begin{equation}}
\def\ee{\end{equation}}

\newcommand{\e}{\hspace{1pt}\mathrm{e}}

\newcommand{\ii}{\hspace{1pt}\mathrm{i}\hspace{1pt}}

\def\RP{{\mathbb{RP}}}
\def\CP{{\mathbb{CP}}}

\newcommand{\nn}{\nonumber}

\usepackage{color}
\usepackage[colorlinks,citecolor=blue]{hyperref}
\definecolor{red}{rgb}{1,0,0}
\definecolor{blue}{rgb}{0,0,1}
\definecolor{dblue}{rgb}{0,0,0.4}
\definecolor{green}{rgb}{0,1,0}
\definecolor{black}{rgb}{0,0,0}
\definecolor{white}{rgb}{1,1,1}

\definecolor{brn}{rgb}{.8,.4,.0}
\definecolor{redo}{rgb}{1,.5,.0}
\definecolor{ddgrn}{rgb}{0,0.4,0}
\definecolor{dgrn}{rgb}{0,0.55,0}
\definecolor{dbl}{rgb}{0,0,0.5}

\usepackage[bbgreekl]{mathbbol}
\usepackage{amscd}

\newcommand{\Z}{\mathbb{Z}}
\newcommand{\C}{\mathbb{C}}
\newcommand{\R}{\mathbb{R}}

\newcommand{\dd}{\hspace{1pt}\mathrm{d}}

\newcommand{\Ref}[1]{Ref.~\cite{#1}}
\newcommand{\Eq}[1]{(\ref{#1})} 
\newcommand{\eq}[1]{(\ref{#1})}

\newcommand{\Tr}{{\rm Tr}}

\newcommand{\prt}{\partial}

\newcommand{\bpm}{\begin{pmatrix}}
\newcommand{\epm}{\end{pmatrix}}
\newcommand{\bmm}{\begin{matrix}}
\newcommand{\emm}{\end{matrix}}

\newcommand{\cG}{ {\cal G} }

\newcommand{\vphi}{\varphi}

\def\Z{{\mathbb{Z}}}

\def\R{{\mathbb{R}}}
\def\C{{\mathbb{C}}}

\def\Tr{{\mathrm{Tr}}}

\def \To{\longrightarrow}
\def \Hom{\operatorname{Hom}}

\def \H{\operatorname{H}}

\def \Z{\mathbb{Z}}
\def \Pin{\mathrm{Pin}}
\def \A{\mathcal{A}}
\def \RP{\mathbb{RP}}
\def \CP{\mathbb{CP}}

\def\Ext{\operatorname{Ext}}

\newcommand{\Sec}[1]{Sec.~\ref{#1}}

\usetikzlibrary{decorations.markings}
\usetikzlibrary{positioning}
\usetikzlibrary{shadings}

\newcommand{\SO}{{\rm SO}}
\newcommand{\Spin}{{\rm Spin}}
\newcommand{\U}{{\rm U}}
\newcommand{\SU}{{\rm SU}}

\renewcommand{\O}{{\rm O}}

\newcommand{\rS}{{\rm S}}

\newcommand{\rF}{{\rm F}}

\def\Sq{\mathrm{Sq}}

\def\B{\mathrm{B}}

\def\TP{\mathrm{TP}}

\usepackage{datetime}
\usepackage{enumitem} 
\usepackage{moreenum}

\newcommand{\sharpfootnote}[1]{%
\let\oldthefootnote=\thefootnote%
\stepcounter{mpfootnote}%
\addtocounter{footnote}{-1}%
\renewcommand{\thefootnote}{{W$^+\sharp$}}
\footnote{#1}%
\let\thefootnote=\oldthefootnote%
}

\newcommand{\naturalfootnote}[1]{%
\let\oldthefootnote=\thefootnote%
\stepcounter{mpfootnote}%
\addtocounter{footnote}{-1}%
\renewcommand{\thefootnote}{{W$^-\natural$}}
\footnote{#1}%
\let\thefootnote=\oldthefootnote%
}

\newcommand{\flatfootnote}[1]{%
\let\oldthefootnote=\thefootnote%
\stepcounter{mpfootnote}%
\addtocounter{footnote}{-1}%
\renewcommand{\thefootnote}{{W$^-\flat$}}
\footnote{#1}%
\let\thefootnote=\oldthefootnote%
}

\usepackage{comment}
\def\bZ{{\mathbf{Z}}}

\usepackage{upgreek}
\newcommand{\Fig}[1]{Fig.~\ref{#1}}

\begin{document}
\begin{titlepage}
\begin{flushright}
\end{flushright}
\begin{center}

{\bf\LARGE{ 
Beyond Standard Models and Grand Unifications: \\[8mm]
Anomalies, 
Topological Terms, and \\[10mm]
Dynamical Constraints via Cobordisms
}}

\vskip0.5cm 
\Large{Zheyan Wan$^1$\flatfootnote{e-mail: {\tt wanzheyan@mail.tsinghua.edu.cn}} 
and Juven Wang$^{2,3}$\sharpfootnote{e-mail: {\tt jw@cmsa.fas.harvard.edu} (Corresponding Author) \href{http://sns.ias.edu/~juven/}{http://sns.ias.edu/$\sim$juven/} 
} 
\\[2.75mm]  
} 
\vskip.5cm
{\small{\textit{$^1${Yau Mathematical Sciences Center, Tsinghua University, Beijing 100084, China}\\}}
}
 \vskip.2cm
 {\small{\textit{$^2${Center of Mathematical Sciences and Applications, Harvard University,  Cambridge, MA 02138, USA}}\\}}
 \vskip.2cm
 {\small{\textit{$^3${School of Natural Sciences, Institute for Advanced Study, Einstein Drive, Princeton, NJ 08540, USA} \\}}
}

\end{center}

\vskip1.5cm
\baselineskip 16pt
\begin{abstract}

We classify and characterize 
all invertible anomalies and all allowed topological terms related to
various Standard Models (SM), 
Grand Unified Theories (GUT), and 
Beyond Standard Model (BSM) physics.
By all anomalies, we mean the inclusion of 
(1) perturbative/local anomalies captured by perturbative Feynman diagram loop calculations, classified by $\mathbb{Z}$ free classes,
and (2) non-perturbative/global anomalies, classified by finite group $\mathbb{Z}_N$ torsion classes.
Our work built from [arXiv:1812.11967] fuses the math tools of Adams spectral sequence, Thom-Madsen-Tillmann spectra,  and Freed-Hopkins theorem.
For example, we compute bordism groups $\Omega^{G}_{d}$ and their invertible topological field theory invariants, 
which characterize $d$d topological terms and $(d-1)$d anomalies, 
protected by the following symmetry group $G$: 
$\Spin\times \frac{\SU(3)\times \SU(2)\times \U(1)}{\mathbb{Z}_q}$ for SM with   
$q=1,2,3,6$; 
$\frac{\Spin \times \Spin(n)}{\Z_2^F}$  
or $\Spin \times \Spin(n)$  
for SO(10) or SO(18) GUT as $n=10, 18$;
$\Spin \times \SU(n)$ 
for Georgi-Glashow SU(5) GUT
as $n=5$;
$\frac{\Spin\times \frac{\SU(4)\times(\SU(2)\times \SU(2))}{\mathbb{Z}_{q'}}}{\Z_2^F}$
for Pati-Salam GUT as $q'=1,2$; and others. 
For SM with an extra discrete symmetry, 
we obtain \emph{new} anomaly matching conditions of $\Z_{16}$, $\Z_{4}$, and $\Z_{2}$ classes in 4d 
\emph{beyond} the familiar Witten anomaly.
Our approach offers an alternative view of all anomaly matching conditions  
built from the lower-energy (B)SM or GUT, in contrast to high-energy Quantum Gravity or String Theory Landscape v.s.~Swampland program,
as bottom-up/top-down complements.
Symmetries and anomalies provide constraints of kinematics,
we further suggest constraints of quantum gauge dynamics, and new predictions of possible extended defects/excitations plus hidden BSM non-perturbative topological sectors.\\[6mm]



\flushright
October 2019\\

\end{abstract}

\end{titlepage}

  \pagenumbering{arabic}
    \setcounter{page}{2}
    

\tableofcontents


\section{Introduction}

\subsection{Physics Guide}

The world where we reside, to our present knowledge, 
can be described by quantum theory, gravity theory, and the underlying long-range entanglement.
Quantum field theory (QFT), specifically 
gauge field theory, under the name of \emph{Gauge Principle} following Maxwell, Hilbert, Weyl \cite{Weyl1929ZPhy}, Pauli, and others,
forms a cornerstone of the fundamental physics. 
Yang-Mills (YM) gauge theory \cite{PhysRev96191YM1954}, generalizing the U(1) abelian gauge group to a non-abelian Lie group, 
has been proven theoretically and experimentally essential to 
describe the Standard Model (SM) physics  
\cite{Glashow1961trPartialSymmetriesofWeakInteractions, Salam1964ryElectromagneticWeakInteractions, Weinberg1967tqSMAModelofLeptons}.


The SM of particle physics is a gauge theory 
encoding three of the four known fundamental forces or interactions 
(the electromagnetic, weak, and strong forces, but without gravity) 
in the universe. 
The SM also classifies all experimentally known elementary particles: 
fermions including three generations of quarks and leptons, while 
bosons including the electromagnetic force mediator 
photon $\gamma$, 
the strong force mediator 
gluon g, 
the weak force mediator 
W$^\pm$ and Z$^0$ gauge bosons, and Higgs particle; 
while the graviton has not yet been detected and is not in SM.
Physics experiments have confirmed that at a higher energy of SM, 
the electromagnetic and weak forces are unified into an electroweak interaction sector.  
Grand Unifications and 
Grand Unified Theories (GUT) 
predict that at further higher energy, 
the strong and the electroweak interactions will be unified into an electroweak-nuclear GUT interaction sector.
The GUT interaction is characterized by one larger gauge group 
and its force carrier mediator gauge bosons 
with a single unified 
coupling constant.\footnote{Unifying gravity 
with the GUT interaction gives rise to a Theory of Everything (TOE). 
However, in our present work, 
the gravity only plays the role of the background probed fields instead of dynamical gravity. 
As we will classify and characterize, the background probed gravity also gives new constraints, such as in 
the gravitational anomaly or the mixed gauge-gravitational anomaly. 
We however will comment the implications for dynamical gravity such as in Quantum Gravity in 
\Sec{sec:ConclusionsExplorationsonNon-PerturbativeBSM}.}
Examples of GUT that we will encounter in this article includes 
Georgi-Glashow SU(5) GUT \cite{Georgi1974syUnityofAllElementaryParticleForces},
Fritzsch-Minkowski SO(10) GUT \cite{Fritzsch1974nnMinkowskiUnifiedInteractionsofLeptonsandHadrons}
and Pati-Salam model \cite{Pati1974yyPatiSalamLeptonNumberastheFourthColor} and others. 


In our present work, we aim to 
classify and characterize fully all (invertible) anomalies and all allowed topological terms associated with
various Standard Models (SM), Grand Unified Theories (GUT), and 
Beyond Standard Model physics (BSM) in 4d.\footnote{{We denote 
$d$d means the $d$ spacetime dimensions.
The $d+1$D means the $d$ spatial and 1 time dimensions.
The $\bar{D}$D means the $\bar{D}$ space dimensions.}}
Then we will suggest the dynamical constraints on SM, GUT and BSM via non-perturbative statements based on 
anomalies and topological terms.

By ``anomalies'' of a theory in physics terminology, 
physicists may mean one of the following:
\begin{enumerate}[label=\textcolor{blue}{(\arabic*)}:, ref={(\arabic*)}]
\item \label{a1}
Classical global symmetry is violated in a quantum theory, such that the 
classical global symmetry fails to be a quantum global symmetry, e.g. Adler-Bell-Jackiw anomaly \cite{Adler1969gkABJ,Bell1969tsABJ}.

\item \label{a2} 
Quantum global symmetry is well-defined kinematically. 
However, there is an obstruction known as `{`'t Hooft anomaly \cite{tHooft1979ratanomaly},''}
to gauge the global symmetry, detectable via coupling the charge operator 
(i.e., symmetry generators or symmetry defects, which measures the global symmetry charge of charged objects) 
to background fields.\footnote{Throughout our article, we explicitly or implicitly use the modern language of symmetries and higher symmetries of 
QFTs, introduced in \cite{Gaiotto2014kfa1412.5148}.}
Specifically, we may detect an obstruction to even \emph{weakly gauge}
the symmetry or couple the symmetry to a \emph{non-dynamical background probed field} (sometimes as background gauge field/connection).
`{`'t Hooft anomaly \cite{tHooft1979ratanomaly},''} is sometimes
regarded as a ``background gauged anomaly'' in condensed matter.
Namely, the path integral or partition function $\bZ$ does not sum over background gauge fields. 
We only fix a background gauge field and the $\bZ$ 
only depends on the background gauge connection as a classical field or as a classical coupling constant.

\item \label{a3}
Quantum global symmetry is well-defined kinematically. 
However, once we promote the global symmetry to a dynamical local gauge symmetry of the dynamical gauge theory,
then the gauge theory becomes ill-defined. Some people call this as 
a ``dynamical gauge anomaly'' prohibiting a quantum theory to be well-defined.
Namely, the path integral after summing over dynamical gauge fields becomes ill-defined.
Therefore, the anomaly-free or anomaly-matching conditions are crucial to
avoid the sickness and ill-defineness of quantum gauge theory.

\end{enumerate}

In fact, it is obvious to observe that the anomalies from \ref{a3} are descendants of anomalies from \ref{a2}.\\[-10mm]
\begin{enumerate}[label=\textcolor{blue}{(\greek*)}., ref={(\greek*)}]
\item \label{alp}
Anomalies from \ref{a3} can be related to anomalies from \ref{a2} via the gauging principle.
\item \label{bet}
Anomalies from \ref{a2} can be related to anomalies from \ref{a3} via the ungauging principle.\\[-10mm]
\end{enumerate}
Thus our key idea is that if we know the gauge group of a gauge theory (e.g., SM, GUT or BSM),
we may identify its ungauged global symmetry group as an internal symmetry group, say ${\mathbb{G}_{\text{internal}} }$ via 
ungauging.\footnote{By gauging or ungauging, also depending on the representation of the matter fields that couple to the gauge theory,
we may gain or lose symmetries or higher symmetries \cite{Gaiotto2014kfa1412.5148}.
It will soon become clear, 
for our purpose, 
we only need to firstly focus on 
the ordinary (0-form) internal global symmetries
and their anomalies. See also \Sec{sec:ConclusionsExplorationsonNon-PerturbativeBSM}. 
}

To start, 
we should rewrite the global symmetries of an ungauging theory into the form of
\bea  \label{eq:Gall}
G\equiv ({\frac{{G_{\text{spacetime} }} \ltimes  {\mathbb{G}_{\text{internal}} }}{{N_{\text{shared}}}}}),
\eea
where the ${G_{\text{spacetime} }}$ is the spacetime symmetry,
the ${\mathbb{G}_{\text{internal}} }$ is the internal symmetry,\footnote{ \label{footnote:wj}
Later we denote the probed background spacetime $M$ connection over the spacetime tangent bundle $TM$, e.g. as
$w_j(TM) \equiv w_j$ where $w_j$ is $j$-th Stiefel-Whitney (SW) class.
We also denote the probed background internal-symmetry/gauge connection over the principal bundle $E$, e.g. as
$w_j(E)\equiv w_j(V_{{\mathbb{G}_{\text{internal}} }})$ where $w_j$ is also $j$-th SW class. 
} 
the $\ltimes$ is a semi-direct product from a ``twisted'' 
extension,\footnote{The  ``twisted'' 
extension is due to the symmetry extension from ${\mathbb{G}_{\text{internal}} }$ by ${G_{\text{spacetime} }}$, for a trivial extension 
$\ltimes$ becomes a direct product $\times$.}
and the ${N_{\text{shared}}}$ is the shared common normal subgroup symmetry between ${G_{\text{spacetime} }}$ 
and ${\mathbb{G}_{\text{internal}} }$.

In the later sections of our work, we write down the \emph{ungauged} global symmetry groups $G$ of SMs, GUTs and BSMs.
Then we should determine, 
classify and characterize all of their associated (invertible) anomalies and topological terms. Moreover, based on the 
descendant relations between the anomalies from \ref{a2} and \ref{a3}, and the gauging/ungauging principles
relate \ref{alp} and \ref{bet}, we thus also determine both:\\[-10mm]
\begin{enumerate}[leftmargin=2mm, label=\textcolor{blue}{(\Alph*)}., ref={(\Alph*)}]
\item \label{A} \emph{For the ungauged SM, GUT and BSM theories}: \\
\emph{Invertible 't Hooft anomalies and background probed topological terms} associated to a global symmetry group $G$. This is related to a relation \ref{bet}.
\item \label{B}  \emph{For the gauged SM, GUT and BSM theories}: \\
\emph{Dynamical gauge anomalies and dynamical topological terms} associated to a gauge group $G$, descent via the relation \ref{alp}.
\end{enumerate}

However, by far, there are some pertinent basic questions below that the readers may wonder. 
We should provide the answers to the readers immediately:
\begin{enumerate}[leftmargin=-4.mm, label=\textcolor{blue}{[\Roman*]}., ref={[\Roman*]}]
\item \label{q1}
\emph{What do we mean by all} (\emph{invertible}) \emph{anomalies and all topological terms}? (See a disclaimer in Footnote \ref{ft:disclaimer}.)

By ``all (invertible) anomalies,'' we mean the inclusion of: 
\begin{enumerate}[leftmargin=2mm, label=\textcolor{blue}{(\roman*)}., ref={(\roman*)}]
\item
{\bf Perturbative local anomalies} captured by perturbative Feynman diagram loop calculations, 
classified by the integer group $\mathbb{Z}$ classes, or the so-called  free classes in mathematics. 
Some selective examples from QFT or gravity include:
\begin{enumerate} [label=\textcolor{blue}{(\arabic*)}:, ref={(\arabic*)}]
\item
Perturbative fermionic anomalies from chiral fermions with U(1) symmetry, originated from Adler-Bell-Jackiw (ABJ) anomalies \cite{Adler1969gkABJ,Bell1969tsABJ}
with $\mathbb{Z}$ classes.
\item Perturbative bosonic anomalies from bosonic systems with U(1) symmetry
 with $\mathbb{Z}$ classes.
 \item Perturbative gravitational anomalies \cite{AlvarezGaume1983igWitten1984}.
\end{enumerate}
\item {\bf  Non-perturbative global anomalies}, classified by a product of finite groups such as $\mathbb{Z}_N$, or the so-called torsion classes  in mathematics.
Some selective examples from QFT or gravity include:
\begin{enumerate} [label=\textcolor{blue}{(\arabic*)}:, ref={(\arabic*)}]
\item An SU(2) anomaly of Witten in 4d or in 5d \cite{Witten1982fp} with a $\mathbb{Z}_2$ class, which is a gauge anomaly. 
\item A new SU(2) anomaly  in 4d or in 5d \cite{WangWenWitten2018qoy1810.00844} with another $\mathbb{Z}_2$ class, which is a mixed gauge-gravity anomaly.
\item Some higher 't Hooft anomalies for a pure 4d SU(2) YM theory  
with a second-Chern-class topological term \cite{Gaiotto2017yupZoharTTT, Wan2018zqlWWZ1812.11968, Wan2019oyr1904.00994} (or the so-called SU(2)$_{\theta =\pi}$ YM):
The higher anomaly involves a discrete 0-form time-reversal symmetry and a 1-form center $\Z_2$-symmetry.  
The first anomaly is discovered in \cite{Gaiotto2017yupZoharTTT}; later the anomaly is refined via a mathematical well-defined 5d {bordism invariant as its topological term},\footnote{We shall briefly clarify the physical notations and usages of a co/bordism theory.
More detailed mathematical definitions are organized in \Sec{sec:MP}. More physical interpretations of
co/bordism invariants would be given in 
\Sec{sec:interpretations}.

\begin{itemize}

\item bordism group: We denote it as $\Omega_{d}^{G}$, see \Eq{eq:bordism-def}.
It is the set of equivalence classes of closed $d$-manifolds $M$ with a $G$-structure under the equivalence relation $\sim$. Here $M\sim M'$ if and only if there is a compact $d+1$-manifold $N$ with a $G$-structure such that the boundary of $N$ is the disjoint union of $M$ and $M'$, and the $G$-structures on $M$ and $M'$ are induced from the $G$-structure on $N$, see \Fig{fig:Cobordism}. The disjoint union operation on closed $d$-manifolds induces an abelian group structure on $\Omega_d^G$.
\item cobordism group: We denote it as $\Omega^{d}_{G} \equiv \TP_d(G)$ as the TP (topological phases) in Freed-Hopkins \cite{Freed2016}, see \Eq{eq:sec1-cobordism-group} and \Eq{eq:TP-def}. To be precise, this is not exactly the commonly defined
Pontryagin dual of the torsion subgroup ($\equiv$ {tors}) of the bordism group $\Omega^{G}_{d}$: 
$\mathrm{Hom}(\Omega^{G,\mathrm{tors}}_{d}, \mathrm{U(1)})$, as the homomorphism map to U(1).
The $\Omega^{d}_{G}$ and $\mathrm{Hom}(\Omega^{G,\mathrm{tors}}_{d}, \mathrm{U(1)})$ are only the same for the finite group sectors (the torsion part), they are differed by the integer $\Z$ classes (the free part). Also,
Freed-Hopkins \cite{Freed2016} suggests that the torsion part
$(\Omega^{d}_{G})_{\rm{tors}} \equiv (\TP_d(G)))_{\rm{tors}} \equiv \Omega_{d}^{G,{\rm{tors}}}$
classifies the \emph{deformation} classes of reflection positive invertible $d$-dimensional extended topological field theories with a symmetry group $G(d)$.

Alternatively, there is another kind of cobordism group defined as $\Hom(\Omega_d^G,\U(1))$ in \Ref{2019CMaPh.368.1121Y} by Yonekura. 
This group classifies the \emph{isomorphism} classes of $d$-dimensional unitary invertible topological field theories with the symmetry group $G(d)$. (We may denote $G$
as $G(d)$ because the spacetime symmetry sector can be $d$-dependent.) 

{Here are some physical meanings of the different versions of cobordism groups:
\begin{itemize}[leftmargin=3mm]
\item $\Omega^{d}_{G} \equiv \TP_d(G)$: Classify the $G$-symmetric invertible topological orders in $d$-dimension, with symmetry $G(d)$. 
\item  $\TP_d(\SO)$: Classify the bosonic invertible topological orders (iTO) in $d$-dimension, without any internal symmetry $G$ (but with a spacetime symmetry SO($d$) group). 
\item  $\TP_d(\Spin)$: Classify the fermionic invertible topological orders in $d$-dimension, without any internal symmetry $G$ (but with a spacetime symmetry Spin($d$) group). 
\item  $\TP_d(G)/\TP_d(\SO)$: Classify the $G$-symmetric bosonic SPTs in $d$-dimension, with symmetry $G \supseteq \SO$. 
(Mod out $\TP_d(\SO)$ means that excluding the bosonic iTO protected by no internal symmetry.)
\item  $\TP_d(G)/\TP_d(\Spin)$: Classify the $G$-symmetric fermionic SPTs in $d$-dimension, with symmetry $G \supseteq \Spin$. 
(Mod out $\TP_d(\Spin)$ means that excluding the fermionic iTO protected by no internal symmetry except a fermionic parity symmetry $\Z_2^F$.)
\item $\mathrm{Hom}(\Omega^{G,\mathrm{tors}}_{d}, \mathrm{U(1)})$:
Classify the torsion (the finite subgroup $\Z_n$) classes of the  $G$-symmetric invertible topological orders, but which miss the integer $\Z$ classes.
\item $\Hom(\Omega_d^G,\U(1))$:
Classify the topological terms of $G$-symmetric phases. For example,  $\Hom(\Omega_d^G,\U(1))$ includes the $\mathrm{Hom}(\Omega^{G,\mathrm{tors}}_{d}, \mathrm{U(1)})$, 
which classifies the finite subgroup $\Z_n$ classes as the subclasses of the  $G$-symmetric invertible topological orders.
However, $\Hom(\Omega_d^G,\U(1))$ also maps the $\Z \in \Omega_d^G$ to U(1), which specifies a theta angle $\theta \in \U(1)$, which is in fact the
$\theta$-term  in physics of $G$-symmetric phases.
\end{itemize}
Here are some physical and entanglement meanings of the aforementioned condensed matter phases, based on the definition of deformation classes of the local unitary transformations \cite{Wen2016ddy1610.03911}:
\begin{itemize}[leftmargin=3mm]
\item invertible topological orders (iTO) includes both the invertible short-range and long-range entangled gapped phases in condensed matter.
\item Symmetry-Protected Topological states (SPTs) includes only the invertible short-range entangled gapped phases in condensed matter  \cite{Wen2016ddy1610.03911}.
\end{itemize}
}

\item bordism generators: The generators of bordism group $\Omega_{d}^{G}$ are the manifolds. We also call the
bordism generators as the manifold generators.
\item cobordism generators: The generators of the torsion part of the cobordism group $\Omega^{d}_{G} \equiv \TP_d(G)$ are the reflection positive extended $d$-dimensional
invertible topological quantum field theories (iTQFTs) with symmetry group $G(d)$. 
Its partition function ${\bf Z}(M^d)$ on any closed manifold $M^d$ must have its absolute value $|{\bf Z}(M^d)|=1$,
namely ${\bf Z}(M^d)=\e^{\ii \theta}$ can only be a complex phase which is invertible ${\bf Z}^\dagger(M^d)=\e^{-\ii \theta}$.
The generators of the cobordism group $\Omega^{d}_{G} \equiv \TP_d(G)$ correspond to the $d$-dimensional invertible topological orders (iTOs) with a symmetry group $G(d)$. 

\item {bordism invariants vs cobordism invariants: 
\begin{itemize}
\item Torsion part $\Z_n$ class (bordism invariants $\equiv$ cobordism invariants):
A bordism invariant
is invariant for all manifolds in the same equivalence class of bordism group. Thus, bordism invariants {$\vphi$ (given in our Tables, implicitly $\vphi$
paired with a manifold as $\int_{M} \vphi = \langle M, \vphi \rangle$)} and partition functions of iTQFTs 
{$\bf{Z}$} are related by $\vphi\mapsto\e^{\frac{2\pi\ii k}{n}\vphi}$ (if $\vphi$ is $\Z_n$ valued, and $k \in \Z_n$ specifies the level of $\Z_n$ class).
Conventionally, bordism invariants and cobordism invariants mean exactly the same for the torsion part.
\item Free part $\Z$ class: bordism invariants $\neq$ cobordism invariants.
For bordism invariants of the $\Z$ class of $\Omega_{d}^{G}$, the bordism invariants {$\vphi$ (given in our Tables, implicitly $\vphi$
paired with a manifold as $\int_{M} \vphi = \langle M, \vphi \rangle$)} and partition functions as the theta term
{$\bf{Z}$} are related by $\vphi\mapsto {\bf{Z}}=\e^{\ii \theta \vphi}$ (if $\vphi$ is $\Z$ valued, and the periodic $\theta \in [0, 2 \pi)$ in U(1)).
This bordism invariant of $\Omega_{d}^{G,{\rm{free}}}$ is related to the cobordism invariants of $\Omega^{d-1}_{G,{\rm{free}}}\equiv \TP_{d-1}(G)_{\rm{free}}$.
For cobordism invariants of the $\Z$ class of $\Omega^{d-1}_{G}\equiv \TP_{d-1}(G)$, 
the bordism invariants {$\vphi$} and partition functions as the iTQFTs (as a $\Z$ class of invertible topological order)
{$\bf{Z}$} are related by $\vphi\mapsto {\bf{Z}}=\e^{\ii 2 \pi k \vphi}$ (if $\vphi$ is $\Z$ valued, and $k \in \Z$).
\end{itemize}
}

\item reflection positive extended iTQFTs:
Reflection positivity means stability, namely if we remove the condition of reflection positivity in Freed-Hopkin's theorem, then we should change the Madsen-Tillmann spectrum $MTH$ to $\Sigma^nMTH(n)$ where $MTH$ is the colimit of $\Sigma^nMTH(n)$. 
Reflection positivity is also the manifestation of unitarity.
An extended $n$-dimensional TQFT means a symmetric monoidal functor from the $(\infty,n)$-category of extended cobordisms to another symmetric monoidal $(\infty,n)$-category. 
There is also a conjecture in \cite{Freed2016} that the full $\TP_d(G)$ corresponds to removing ``topological'' in {the previous} Freed-Hopkin's theorem. Here ``topological'' means that the field theory does not depend on any continuously varying background fields, such as a spacetime metric or conformal structure. 

\item deformation classes v.s. isomorphism classes: There is a deformation between two theories if there is a
continuous path of theories connecting them. Two theories are isomorphic if there is a natural monoidal transformation between them.

\item  topological invariants vs.~geometric invariants:
We say that a cobordism invariant is topological if it can be defined purely using topological data, such as a cohomology class. While we say that a cobordism invariant is geometric if it can be defined purely using geometric data like metrics, connections, and curvatures.
These two definitions have no conflict, a cobordism invariant can be both topological and geometric.
\item topological invariants vs.~topological terms vs.~iTQFTs:
In physics, loosely speaking, for co/bordism invariants of a cobordism theory,
people sometimes interchangeably use topological invariants, topological terms, and iTQFTs for the same thing.
The $G$-iTQFTs with a global symmetry $G$ (obtained from the cobordism invariants of cobordism group for manifolds with a $G$-structure) 
describes the low energy physics of the short-range entangle SPTs in condensed matter.
However, in the general context, 
topological invariants and topological terms may not need to be invertible.
\end{itemize}
} 
with additional new anomalies found for Lorentz symmetry-enriched 
four siblings of YM \cite{Wan2018zqlWWZ1812.11968, Wan2019oyr1904.00994}.
\item Global gravitational anomalies \cite{Witten1985xe}.
\end{enumerate}

\end{enumerate}

By ``all topological terms,'' we mean the anomaly-inflow relation \cite{Callan1984saCallanHarvey}:
The $(d-1)$d anomalies can be systematically captured by a one higher dimensional $d$d topological invariants or $d$d topological terms.
Recently \Ref{Witten2019bou1909.08775}
gives a non-perturbative description of anomaly inflow including both local and global anomalies 
based on Dai-Freed theorem \cite{Dai1994kqFreed9405012} and the Atiyah-Patodi-Singer $\eta$-Invariant \cite{Atiyah1975jfAPS}.

Therefore, by determining all $(d-1)$d anomalies, we also determine all $d$d topological terms, and 
vice versa.\footnote{There \label{ft:disclaimer}
are however a 
disclaimer and some caveats: 
\begin{enumerate}[leftmargin=2mm]
\item
By all anomalies and all topological terms, their
classifications and characterizations depend on the category of manifolds that can detect them. The
categories of manifolds can be: TOP (topological manifolds), PL (piecewise linear manifolds), or DIFF (differentiable thus equivalently smooth manifolds), etc. 
These categories are different, and they are related by 
\bea
\rm{TOP} \supseteq \rm{PL}  \supseteq \rm{DIFF}.
\eea
Since the SM, GUT and BSM are given by continuum QFT data, in this work, 
we only focus on the 
DIFF manifolds and their associated all possible anomalies and topological terms.
However, if we refine the data of QFT later in the future to include PL or TOP data from  PL or TOP manifolds,
we may also need to refine the corresponding SM, GUT and BSM. Thus, we will have a new set of 
so-called all anomalies and all topological terms.
The tools we use in either case would be a certain version of cobordism theory suitable for a 
specific category of manifolds. See more in \cite{toappear}.
\item 
By anomalies and topological terms for some SM, GUT and BSM theories in our work, we either mean\\ 
\ref{A}.\emph{Invertible 't Hooft anomalies and background probed topological terms}
 for the \emph{ungauged SM, GUT and BSM}, or\\
\ref{B}.\emph{Dynamical gauge anomalies and dynamical topological terms}
for the \emph{gauged SM, GUT and BSM}.\\
But after gauging $G$-symmetry of \ref{A}, for the \emph{gauged SM, GUT and BSM} in \ref{B},
there could be additional new higher 't Hooft anomalies associated to the higher symmetries (depending on the group representations of the matter fields)
whose charged objects are dynamical extend objects (e.g. 1-lines, 2-surfaces, etc.). 
In this present work, 
we do \emph{not} discuss these additionally gained new higher 't Hooft anomalies after dynamically gauging,
but will leave them for future work \cite{toappear}. Examples of such higher  't Hooft anomalies can be found in 
\cite{Gaiotto2017yupZoharTTT, Cordova2018acb1806.09592DumitrescuClay,
Wan2018djlW2.1812.11955,
Wan2018zqlWWZ1812.11968, Wan2019oyr1904.00994} and References therein.

\item Another possible loop hole is that we do only focus on invertible anomalies captured by invertible topological quantum field theories (iTQFTs), we do not study non-invertible anomalies (e.g. \cite{Ji2019eqoJiWen1905.13279}). 
Different experts and different research fields may regard and define anomalies in different ways.
After all,
\href{http://sns.ias.edu/~juven/}{Laozi (600 B.C.) in Dao De Jing} had long ago educated us that
``\href{http://sns.ias.edu/~juven/}{The Way that can be told of is not an eternal way;}
\href{http://sns.ias.edu/~juven/}{The names that can be named are not eternal names.}
It was from the Nameless that Heaven and Earth sprang.
\end{enumerate}
}

\item \label{q2}
\emph{What tools are we using 
to classify and characterize all anomalies and all topological terms}? 

The short answer is based on the Freed-Hopkin's theorem \cite{Freed2016} and our prior 
work \cite{WanWang2018bns1812.11967, Wan2019sooWWZHAHSII1912.13504, Wan2019oaxWWHAHSIII1912.13514,  WanWangv2}.\footnote{See also a precursor of works for \Ref{WanWang2018bns1812.11967, Wan2019sooWWZHAHSII1912.13504, Wan2019oaxWWHAHSIII1912.13514, WanWangv2} 
by one of the present authors: \Ref{1711.11587GPW}.}

The long answer is follows.
Based on the Freed-Hopkin's theorem \cite{Freed2016} and an extended generalization that we propose \cite{WanWang2018bns1812.11967, WanWangv2},
there exists a one-to-one correspondence between  ``the invertible topological quantum field theories (iTQFTs) with symmetry 
(including higher symmetries or generalized global symmetries  \cite{Gaiotto2014kfa1412.5148})'' 
and ``a cobordism group.''
In condensed matter physics, this means that there is  
a relation from iTQFT to 
``the symmetric invertible topological order (iTO, see a review \cite{Wen2016ddy1610.03911}) with symmetry (including higher symmetries) or
symmetry-protected topological state (SPTs, see \cite{Chen2011pg1106.4772, Senthil1405.4015, Wang1405.7689, Wen2016ddy1610.03911})
that can be regularized on a lattice in its own dimensions.''

More precisely, it is a one-to-one correspondence (isomorphism ``$\cong$'') between the following two well-defined ``mathematical objects'' (which turn out to be abelian groups):
\bea \label{eq:thm}
&&\left\{\begin{array}{ccc}\text{Deformation classes of the reflection positive }\\\text{invertible } d
\text{-dimensional extended}\\
\text{topological field theories (iTQFT) with} \\
\text{symmetry group }
{\frac{{G_{\text{spacetime} }} \ltimes  {\mathbb{G}_{\text{internal}} }}{{N_{\text{shared}}}}}
\end{array}\right\}\nn \\
&&\cong[MT({\frac{{G_{\text{spacetime} }} \ltimes  {\mathbb{G}_{\text{internal}} }}{{N_{\text{shared}}}}}),\Sigma^{d+1}I\Z]_{\text{tors}}.
\eea
We shall explain the notation above:
 $MTG$ is the Madsen-Tillmann-Thom spectrum \cite{MadsenTillmann4} of the group $G$,
$\Sigma$ is the suspension, $I\Z$ is the Anderson dual spectrum, and ${\text{tors}}$ means  the torsion group by taking only the finite group sector.
The right hand side is the torsion subgroup of homotopy classes of maps from a Madsen-Tillmann-Thom spectrum ($MTG$) to
a suspension shift ($\Sigma^{d+1}$) of the Anderson dual to the sphere spectrum ($I\Z$).

In other words, we classify the deformation classes of symmetric iTQFTs and also symmetric invertible topological orders (iTOs), via
this \emph{particular cobordism group} defined as follows
\bea \label{eq:sec1-cobordism-group}
\Omega^{d}_{G} &\equiv&
\Omega^{d}_{({\frac{{G_{\text{spacetime} }} \ltimes  {\mathbb{G}_{\text{internal}} }}{{N_{\text{shared}}}}})} \nn\\
&\equiv&
\TP_d(G)\equiv[MTG,\Sigma^{d+1}I\Z].
\eea
 by classifying the cobordant relations of smooth, differentiable and triangulable
manifolds with a stable $G$-structure, via associating them to the homotopy
groups of Thom-Madsen-Tillmann spectra \cite{thom1954quelques,MadsenTillmann4},
given by a theorem in \Ref{Freed2016}. \Ref{Freed2016} introduced TP which means the abbreviation of ``Topological Phases''
classifying the above symmetric iTQFT,
where our notations follow \cite{Freed2016} and \cite{WanWang2018bns1812.11967}.
(For an introduction of the mathematical background and mathematical notations explained for physicists, the readers can consult the Appendix A of 
\cite{1711.11587GPW} or \cite{WanWang2018bns1812.11967}.)

{Now let us pause for a moment to trace back some recent history of relating these anomalies/topological terms to a cobordism theory.}
The $d$ dimensional  't Hooft anomaly of ordinary 0-form 
global symmetries is known to be captured by a $(d+1)$ dimensional iTQFT. 
In the condensed matter literatures, these  $(d+1)$d iTQFTs describe Symmetry-Protected Topological states (SPTs)
or symmetric invertible topological orders (iTO)\footnote{We abbreviate both Symmetry-Protected Topological \emph{state} and Symmetry-Protected Topological \emph{states} as SPTs.\\
We also abbreviate both Symmetry-Enriched Topologically ordered \emph{state} and Symmetry-Enriched Topologically ordered  \emph{states} as SETs.} \cite{Chen2011pg1106.4772, Senthil1405.4015, Wang1405.7689, Wen2016ddy1610.03911}. 
SPTs and symmetric iTO are interacting systems (interacting systems of bosons and fermions at the lattice scale UV with a local Hilbert space)
beyond the free fermion or K theory classification  \cite{Kitaevperiod} for the
(non-interacting or so-called free) topological insulators/superconductors \cite{2010RMP_HasanKane, 2011_RMP_Qi_Zhang}.
The relations between the SPTs and the response probe field theory partition functions have been \emph{systematically} studied, selectively, in 
 \cite{QiHughesZhang, Wang1405.7689, Witten:2015aba, Witten2016cio1605.02391, 1711.11587GPW} (and References therein),\footnote{For example,
 the interacting versions of 10 Cartan symmetry classes of fermionic superconductors/insulators classifications (e.g. \cite{CWang1401.1142} in 4d or 3+1D) from condensed matter can be
 captured precisely by {bordism invariants} as invertible TQFTs \cite{1711.11587GPW}.} and climaxed 
 to the evidence of cobordism classification of SPTs\cite{Kapustin2014tfa1403.1467, Kapustin1406.7329}.
Recently, the iTQFTs and SPTs are found to be systematically classified
by a powerful cobordism theory of Freed-Hopkins \cite{Freed2016}, following the earlier framework of Thom-Madsen-Tillmann spectra \cite{thom1954quelques,MadsenTillmann4}.

A \emph{new} ingredient in our work \cite{WanWang2018bns1812.11967, WanWangv2} is a generalization of 
the calculations and the cobordism theory of Freed-Hopkins \cite{Freed2016} involving higher symmetries:
Instead of the ordinary group $G$ or ordinary classifying space $\B G$, we consider a generalized cobordism theory studying spacetime manifolds 
endorsed with ${G_{\text{spacetime} }}$ structure, with an additional higher group $\mathbb{G}$ (i.e., generalized as principal-$\mathbb{G}$ bundles) 
and \emph{higher classifying spaces} 
$\B \mathbb{G}$.\footnote{Although
most of our results in this article focus on the ordinary symmetry group, our framework does allow us to consider
possible higher symmetries and higher anomalies \cite{WanWang2018bns1812.11967, WanWangv2}.}

\item \label{q3}
\emph{What do we mean by  classifications and characterizations}?

\begin{itemize}

\item By \emph{classification}, 
we mean that given certain physics theories or phenomena (here, higher-iTQFT and higher quantum anomalies),
 given a spacetime dimensions (here $d+1$d for higher-iTQFT or $d$d for higher quantum anomalies),
and their spacetime ${G_{\text{spacetime} }}$-structure and the internal higher global symmetry ${\mathbb{G}_{\text{internal}} }$,
we compute how \emph{many classes} (a number to count them) there are?
Also, we aim to determine the mathematical structures of classes (i.e. here group structure as for (co)bordism groups:
would the classes be a finite group $\Z_N$ or an infinite group $\Z$ or others, etc.).

\item By \emph{characterization}, we mean that we formulate their mathematical invariants (here, we mean the {bordism invariants}) to
fully describe or capture their mathematical essences and physics properties. Hopefully, one may further compute their physical observables from mathematical invariants.
%


\end{itemize}

\end{enumerate}

Since some of readers are still with us reading this sentence (after we answer the three questions \ref{q1}, \ref{q2} and  \ref{q3}),
we believe that these readers decide to be interested in understanding our results in details.
Here we concern theories of 4d SMs, GUTs and BSMs and their anomalies and topological terms.
Their 4d 't Hooft anomalies captured by 5d iTQFTs. 
These 5d iTQFTs or {bordism invariants} are defined on the $d$d manifolds ($d=5$).
In fact, in our work, we present all
$$
\text{
$(d-1)$d 't Hooft anomalies captured by $d$d iTQFTs, for $d=1,2,3,4,5$, 
}
$$
associated with various SMs, GUTs and BSM (ungauged) symmetries.
The manifold generators for the bordism groups are actually the closed $d$d manifolds.
We should clarify that although there are 't Hooft anomalies for $(d-1)$d QFTs (so ${\mathbb{G}_{\text{internal}} }$ may not be gauge-able on the boundary), 
the SPTs/topological invariants defined in the closed $d$d  
actually have ${\mathbb{G}_{\text{internal}} }$ always gauge-able
in that $d$d.
This is related to the fact that the \emph{bulk} $d$d  SPTs  in condensed matter physics
has an onsite local internal ${\mathbb{G}_{\text{internal}} }$-symmetry (or on-$n$-simplex-symmetry as a generalization for higher-SPTs), thus this 
${\mathbb{G}_{\text{internal}} }$ must be gauge-able.
{By gauging the topological terms, this idea has been used to study the vacua of YM gauge theories coupling to dynamically gauged SPT terms (like the orbifold techniques in string theory, but here we generalize this thinking to any dimension), for example, in \cite{1711.11587GPW} and 
references therein.}
There are other uses and interpretations of our cobordism theory data that we will explain in \Sec{sec:ConclusionsExplorationsonNon-PerturbativeBSM}.



We should emphasize that several recent pursuits are also along the fusions between
the non-perturbative physics of SMs, GUTs, and BSMs via a cobordism theory: 
\begin{itemize}
\item
Garcia-Etxebarria-Montero  \cite{GarciaEtxebarriaMontero2018ajm1808.00009} studies global anomalies of some SMs and GUTs model
via a Dai-Freed theorem and Atiyah-Hirzebruch spectral sequence (AHSS) \cite{Dai1994kqFreed9405012}.


\item
Wang-Wen \cite{WangWen2018cai1809.11171}, 
independently, studies the non-perturbative definitions (e.g. on a lattice) and 
the global anomalies of SO(10) GUTs or SO(18) GUTs via $\Omega_5^{\frac{\Spin(5) \times \Spin(n)}{\Z_2^F}}$ with $n=10,18$
and SU(5) GUTs via $\Omega_5^{\Spin}(\B\SU(5))$.
They ask what are all allowed thus all possible anomalies for a fermionic theory 
with 
${\frac{\Spin(d) \times \Spin(10)}{\Z_2^F}}$-symmetry.\footnote{Before dynamically gauging Spin(10),
{SO(10)-GUT} is one kind of such theory: an Spin(10)-{chiral} gauge theory with {fermions in the half-integer (iso)spin}-representation.
So we may call this ungauged theory as SO(10)-GUT chiral fermion theory.} 
Under the interaction effects,
the answer turns out to be a $\Z_2$ class (or a mod 2 class) global anomaly captured by the 5d iTQFT:
\bea
\label{topinv}
e^{\ii \pi \int_{M^5} w_2(TM)w_3(TM)},
\eea
where $w_j(TM)$ is the $j$-{th}-Stiefel-Whitney
class 
for the tangent bundle of $5$D spacetime $M^5$.  We note that on a $M^5$, we have a $\frac{\Spin(d=5) \times
\Spin(n)}{{\Z_2^F}}$ connection --- a mixed gravity-gauge connection,
rather than the pure gravitational $\Spin(d=5)$ connection, {such that
$w_2(TM)=w_2(V_{\SO(n)})\equiv w_2({\SO(n)})$ and $w_3(TM)=w_3(V_{\SO(n)})\equiv w_3({\SO(n)})$, where
$w_j(V_{\SO(n)}) :=w_j({\SO(n)}) $ is the $j$-{th}-Stiefel-Whitney class for the associated vector bundle of an ${\SO(n)}$
gauge bundle}.  The $M^5$ can be a non-spin manifold. 
This is the same global anomaly known as ``a new SU(2) anomaly''  studied in \Ref{WangWenWitten2018qoy1810.00844}. 
But \Ref{WangWen2018cai1809.11171} and \cite{WangWenWitten2018qoy1810.00844} show explicitly, since 
$$
\Spin(10)\supset \Spin(3)= \SU(2), 
$$ 
if the SO(10) GUT chiral fermion theory is free from ``the new SU(2) anomaly \cite{WangWenWitten2018qoy1810.00844}'' (which indeed is true), 
then the SO(10) GUT chiral fermion theory contains \emph{no} anomaly at all. Thus this SO(10) GUT is all anomaly-free \cite{WangWen2018cai1809.11171, WangWenWitten2018qoy1810.00844}.'' 
This leads to a possible non-perturbative construction of SO(10) GUTs on the lattice proposed in \cite{Wen2013oza1303.1803, Wen2013ppa1305.1045},
rooted in the idea of gapping the mirror chiral fermions of Eichten-Preskill\cite{Eichten1985ftPreskill1986}.
However, we will not pursue this idea of \cite{{WangWen2018cai1809.11171}} nor \cite{Wen2013oza1303.1803, Wen2013ppa1305.1045} 
 (such as the lattice regularization) further in this work, but leave this for a future exploration \cite{JWangmirror}.\footnote{
In addition to \Ref{WangWen2018cai1809.11171, Wen2013oza1303.1803, Wen2013ppa1305.1045},
related ideas about interactions 
 gapping the mirror chiral fermions also occur in 1+1d in \Ref{Wang2013ytaJW1307.7480, BenTov2014eea1412.0154, Wang2018ugfJW1807.05998} based on the 
 anomaly cancellation  constraints, or
 in 3+1d based on arguments about the topological defects in \Ref{You2014oaaYouBenTovXu1402.4151, BenTov2015graZee1505.04312}.}


\item Wan-Wang \cite{WanWang2018bns1812.11967, Wan2019sooWWZHAHSII1912.13504, Wan2019oaxWWHAHSIII1912.13514, WanWangv2} attempts to classify all invertible local or global anomalies and the invertible higher-anomalies,
based on a generalized cobordism group classification of invertible TQFTs and invertible higher-TQFTs in one higher dimensions
via Adams spectral sequence.
\Ref{WanWang2018bns1812.11967} computes the cobordism classification relevant for 
perturbative anomalies of chiral fermions (e.g. originated from Adler-Bell-Jackiw \cite{Adler1969gkABJ,Bell1969tsABJ}) or chiral bosons with U(1) symmetry in any even spacetime dimensions; they also compute the cobordism classification for
the non-perturbative global anomalies such as Witten anomaly  \cite{Witten1982fp}  and the new SU(2) anomaly \cite{WangWenWitten2018qoy1810.00844} in 4d and 5d.
\Ref{WanWang2018bns1812.11967} also obtains 
the cobordism classification relevant for 
higher 't Hooft anomalies for a pure 4d SU(N) YM theory  
with a second-Chern-class $\theta=\pi$ topological term \cite{Gaiotto2017yupZoharTTT, Wan2018zqlWWZ1812.11968, Wan2019oyr1904.00994}.

\item McNamara-Vafa \cite{McNamara2019rupVafa1909.10355} studies
the cobordism classes and the constraints on the Quantum Gravity or String Theory Landscape v.s.~Swampland. 
QFT must satisfy some consistent criteria in order to be part of a consistent theory of Quantum Gravity. 
Those QFT not obeying those criteria are quoted to reside in Swampland.

\item Davighi-Gripaios-Lohitsiri \cite{2019arXiv191011277D}  
studies also
the {global anomalies in various SMs and BSMs}, 
 based on Atiyah-Hirzebruch spectral sequence.

\item Kaidi-Parra-Martinez-Tachikawa \cite{Kaidi2019pzjJulioTachikawa1908.04805,Kaidi2019tyfJulioTachikawa1911.11780}
studies the possible fermionic SPTs (or invertible spin TQFTs) on the worldsheet of the string as the Gliozzi-Scherk-Olive (GSO) projections in the superstring theory.
Their approach, based on the various relevant bordism groups, also shows the relationship to the K-theoretic classification of D-branes.

\item Freed-Hopkins \cite{Freed2019scoFreedHopkins1908.09916} studies a global anomaly cancellation involving the time-reversal symmetry relevant for the 11-dimensional M theory.

\end{itemize}

The outline of our article is the following. We consider the following models/theories, their co/bordism groups, TP groups,
topological terms and anomalies, written in terms of iTQFTs.

\begin{enumerate}
\item
Standard Models (SM) in \Sec{sec:SM} and \Sec{sec:SM4}:

\begin{itemize}
\item 
 \Sec{sec:SM}: 
${\Spin\times \frac{\SU(3)\times \SU(2)\times \U(1)}{\Z_q} }$ model for 
$\Z_q$
from 
 $\Omega_d^{\Spin}(\B\frac{\SU(3)\times \SU(2)\times \U(1)}{\Z_q})$ with $q=1,2,3,6$.
 The Lie algebra of SM is known to be $su(3) \times su(2) \times u(1)$,
 but it is known that the global structure of gauge group allows a quotient group of 
 \bea
 {\SU(3)\times \SU(2)\times \U(1)}
 \eea
as 
\bea
\frac{\SU(3)\times \SU(2)\times \U(1)}{\Z_q},
\eea
which is well-explained, for example, in \cite{Tong2017oea1705.01853}.\footnote{More generally, the
global structure of gauge group as $G$ or $G/\Gamma$ makes the real physical differences in observables (see \cite{AharonyASY2013hdaSeiberg1305.0318}
and \cite{Tong2017oea1705.01853}).}


\item 
 \Sec{sec:SM4}: ${\Spin \times_{\Z_2} \Z_4\times {\frac{\SU(3)\times \SU(2)\times \U(1)}{\Z_q}}}$ model 
from 
 $\Omega_d^{\Spin \times_{\Z_2} \Z_4}(\B\frac{\SU(3)\times \SU(2)\times \U(1)}{\Z_q})$ with $q=1,2,3,6$. 
This is an interesting group suggested by  \cite{GarciaEtxebarriaMontero2018ajm1808.00009, Montero}. 
\end{itemize}

\item Grand Unified Theories (GUT) in \Sec{sec:PS}, \Sec{sec:GUSO} and  \Sec{sec:GUSU}:

\begin{itemize}
\item  \Sec{sec:PS}: Pati-Salam model $\frac{\Spin\times \frac{\SU(4)\times(\SU(2)\times \SU(2))}{\Z_{q'}}}{\Z_2^F}$ with $q'=1,2$.
Here the ${\Z_2^F}$ is the well-known fermion parity symmetry, which acts on any fermionic operator $\Psi$ by 
\bea
{\Z_2^F}: \Psi \to - \Psi,
\eea
such that the operation is $(-)^{N_F}$ where ${N_F}$ is the fermion number.

\item \Sec{sec:GUSO}: {SO(10) and SO(18)} GUT from either $\Omega_d^{\frac{\Spin(d) \times \Spin(n)}{\Z_2^F}}$ and $\Omega_d^{\Spin}(\B\SO(n))$. 

\item \Sec{sec:GUSU}:
{SU(5)} GUT from either $\Omega_d^{\Spin}(\B\SU(5))$ or more generally $\Omega_d^{\Spin}(\B\SU(n))$.

\end{itemize}

\end{enumerate}

We provide an overview how this data can be used to dynamically constrain SMs, GUTs and BSMs in Conclusions in \Sec{sec:ConclusionsExplorationsonNon-PerturbativeBSM}.
After all these physics stories and inputs, 
now let us introduce some mathematics preliminary in \Sec{sec:MP}.



\subsection{Mathematics Preliminary}

\label{sec:MP}

\subsubsection{Definition of bordism groups}

\begin{figure}[h!] 
  \centering
  \includegraphics[width=4.in]{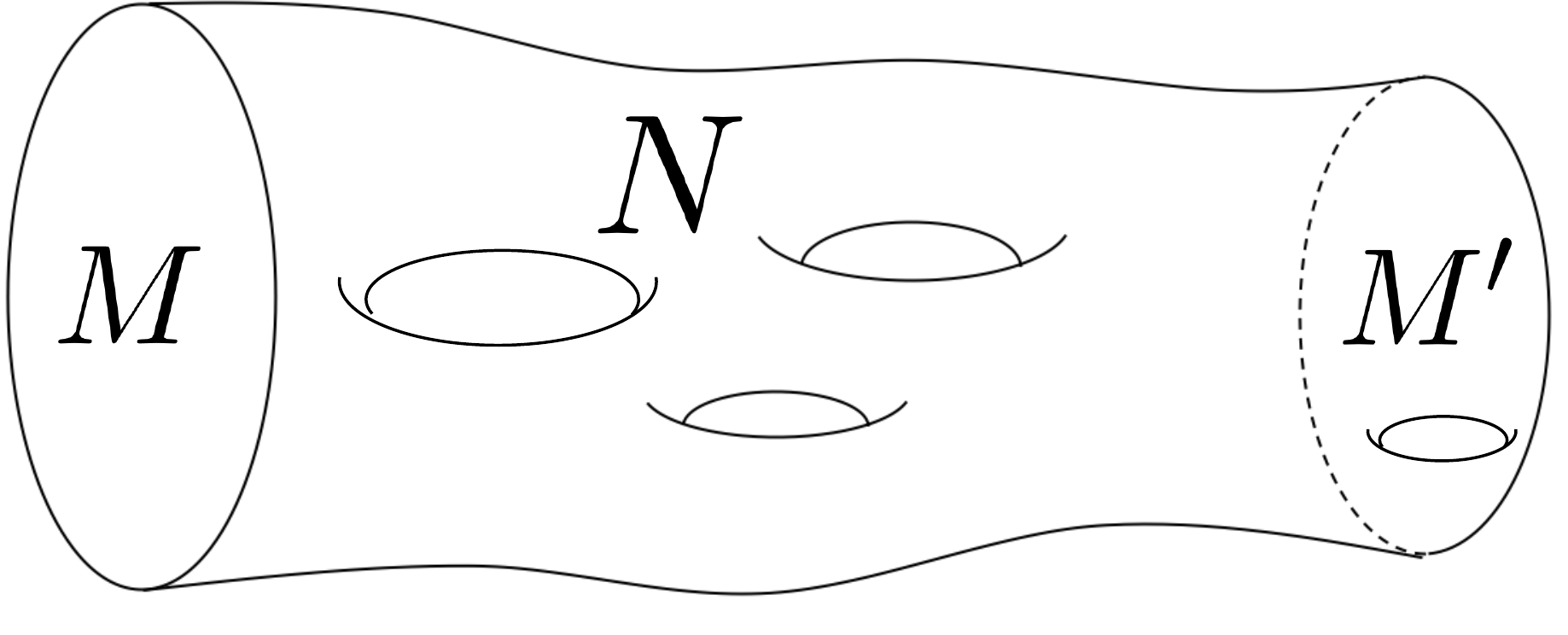}
  \caption{Bordism between manifolds. Here $M$ and $M'$ are two closed $d$-manifolds, $N$ is a compact $d+1$-manifold whose boundary is the disjoint union of $M$ and $M'$, {so $\prt N= M \sqcup M'$.} If there are additional $G$-structures on these manifolds, then the $G$-structure on  $N$ is required to be compatible with the $G$-structures on $M$ and $M'$. 
  If there are additional maps from these manifolds to a fixed topological space, then the maps are also required to be compatible with each other. If these conditions are satisfied, then $M$ and $M'$ are called bordant and $N$ is called a bordism between $M$ and $M'$. This relation is an equivalence relation.}\label{fig:Cobordism}
\end{figure}

{
We assume that the readers are familiar with the basic algebraic topology, such as (co)homology and homotopy. To fill in the gap between this knowledge and the mathematical tools used in this article, we introduce some prerequisites in this subsection. 
}

{
First, we introduce the notion of bordism group. To define the bordism group, we need to define a tangential structure which involves the notion of classifying spaces.}
{
For any Lie group $G$, the $\B G$ is the classifying space of principal $G$-bundles, namely the homotopy classes of maps $X\to \B G$ are in one-to-one correspondence with the isomorphism classes of principal $G$-bundles over $X$ for any topological space $X$. For any abelian group $G$, the iterated classifying space $\B^2G$ is also the Eilenberg-MacLane space $K(G,2)$.
For any abelian group $G$, more generally, we have the $\B^d G=K(G,d)$.
}

{To define tangential structure, we require the orthogonal group $\O(n)$ since a tangential structure involves a fibration over the classifying space of $\O(n)$.  
There is a natural inclusion $\O(n)\to\O(n+1)$. Let $\O$ denote the colimit $\O:=\text{colim}_{n\to\infty}\O(n)$.
The inclusions $\R^q\to\R^{q+1}$ induce the closed inclusions of Grassmannian spaces
$Gr_n(\R^q)\to Gr_n(\R^{q+1})$. 
The colimit $\text{colim}_{q\to\infty}Gr_n(\R^q)$ is the classifying space $\B\O(n)$.  
There are closed inclusions $Gr_n(\R^q)\to Gr_{n+1}(\R^{q+1})$ obtained by sending $W\mapsto\R\oplus W$ where we write $\R^{q+1}=\R\oplus\R^q$. These induce maps $\B\O(n)\to\B\O(n+1)$, and we define
\bea
\B\O:=\text{colim}_{n\to\infty}\B\O(n).
\eea
The $\B\O$ is a classifying space for the infinite orthogonal group $\O$.
}

{
An \emph{$n$-dimensional tangential structure} is a topological space $\B G(n)$ and a fibration $\pi(n):\B G(n)\to\B\O(n)$. A \emph{stable tangential structure} is a topological space $\B G$ and a fibration $\pi:\B G\to \B\O$. It gives rise to an $n$-dimensional tangential structure for each $n\in\Z^{\ge0}$ by letting $\pi(n):\B G(n)\to \B\O(n)$ be the fiber product (also called pullback)
\bea
\xymatrix{
\B G(n)\ar@{-->}[r]\ar@{-->}[d]_{\pi(n)}&\B G\ar[d]^{\pi}\\
\B\O(n)\ar[r]&\B\O.
}
\eea
Here the map $\B\O(n)\to \B\O$ is the inclusion map.
If $M$ is an $m$-dimensional manifold, then a $G(n)$-structure on $M$ is a lift $M\to\B G(n)$ of a classifying map $M\to\B\O(n)$ of $\widetilde{TM}$, where we have stabilized the tangent bundle $TM$ of the $m$-dimensional manifold $M$ to the rank $n$ bundle 
\bea
\widetilde{TM}:=\underline{\R^{n-m}}\oplus TM.
\eea
Here $\underline{\R^{n-m}}$ is the trivial real vector bundle of dimension $n-m$.
A $G$-structure on $M$ is a family of coherent $G(n)$-structures for $n$ sufficiently large.
Here coherent $G(n)$-structures for $n$ sufficiently large means that the composite of the $G(l)$-structure on $M$ and the inclusion map $\B G(l)\to \B G(n)$: $M\to\B G(l)\to \B G(n)$ is exactly the $G(n)$-structure on $M$ for $l<n$ where $l$ and $n$ are sufficiently large.
Notice that an $n$-dimensional tangential structure induces an $l$-dimensional tangential structure for all $l<n$ by taking the fiber product (also called pullback)
\bea
\xymatrix{
\B G(l)\ar@{-->}[r]\ar@{-->}[d]_{\pi(l)}&\B G(n)\ar[d]^{\pi(n)}\\
\B\O(l)\ar[r]&\B\O(n).
}
\eea
See \cite{Freed2013bordism} for more details.
}

{
For any structure group $G$ and any topological space $X$, we define the bordism group as the set of equivalence classes:
\bea\label{eq:bordism-def}
\Omega_d^G(X):=\{(M,f)\mid M\text{ is a closed }d\text{-manifold with a }G\text{-structure, }f:M\to X\text{ is a map}\}/\sim.
\eea
{The $(M,f)$ is a pair of data of the manifold $M$ and map $f$.}
The $\sim$ is an equivalence relation defined on the set of pairs of a manifold $M$ and a map $f$. $(M,f)\sim(M',f')$ if and only if there is a compact $(d+1)$-manifold $N$ with a $G$-structure and a map $F:N\to X$ such that the boundary of $N$ is the disjoint union of $M$ and $M'$ (Figure \ref{fig:Cobordism}), the $G$-structures on $M$ and $M'$ are induced from the $G$-structure on $N$, and $F|_{M}=f$ and $F|_{M'}=f'$.
The disjoint union operation on closed $d$-manifolds induces an abelian group structure on $\Omega_d^G(X)$. Here a $G$-structure on a manifold is a $G$-structure on the \emph{stable} tangent bundle of this manifold. 
In this article, we will focus on the case when $X$ is a classifying space. 
In most cases, $X$ is just a point.
}

{
If $\Omega_d^{G}(X)=G_1\times G_2\times\cdots\times G_r$ where $G_i$ are cyclic groups, then the
group homomorphisms $\vphi_i:\Omega_d^{G}(X)\to G_i$ are called \emph{bordism invariants}, and they form a complete set of bordism invariants if 
$\vphi=(\vphi_1,\vphi_2,\dots,\vphi_r):\Omega_d^{G}(X)\to G_1\times G_2\times\cdots\times G_r$ is a group isomorphism.
So $(M,f)$ and $(M',f')$ are in the same equivalence class of the bordism group $\Omega_d^{G}(X)$ if and only if the values of each bordism invariant $\vphi_i$ on $(M,f)$ and $(M',f')$ are the same for $1\le i\le r$. 
}

{
In this article, we will compute the bordism groups $\Omega_d^{G}$ for $G=\Spin\times G'$ or $\Spin\times_{\Z_2}G'$ for some groups $G'$. So we first clarify these notations here.
$\Spin$ (or $\Spin(d)$) is a nontrivial extension of the group $\SO$ (or $\SO(d)$) by $\Z_2$, namely there are short exact sequences of groups
\bea
1\to\Z_2\to\Spin\to\SO\to1,\\
1\to\Z_2\to\Spin(d)\to\SO(d)\to1,
\eea
where $\SO$ is the colimit of $\SO(d)$, and $\SO(d)$ is the special orthogonal group.
The $\Spin$ is the colimit of $\Spin(d)$.
The $\Spin \times_{\Z_2} G'$ is the quotient group of $\Spin\times G'$ by the diagonal central subgroup $\Z_2$.
}

\subsubsection{Spectra}
{
For a pointed topological space $X$, the $\Sigma$ denotes a suspension $\Sigma X=S^1\wedge X=(S^1\times X)/(S^1\vee X)$ where $\wedge$ and $\vee$ are smash product and wedge sum (a one point union) of pointed topological spaces respectively. 
}

{Now we recall some basic notions regarding spectrum, see \cite{Freed2013bordism,WanWang2018bns1812.11967} for more details.
}
{
A \emph{prespectrum} $T_{\bullet}$ is a sequence $\{T_n\}_{n\in\Z^{\ge0}}$ of pointed spaces and the maps $s_n:\Sigma T_n\to T_{n+1}$.
}
{
An \emph{$\Omega$-prespectrum} is a prespectrum $T_{\bullet}$ such that the adjoints $t_q: T_n\to\Omega T_{n+1}$ of the structure maps are weak homotopy equivalences.
Here $\Omega$ means the loop space, its meaning is different from the bordism notation $\Omega$.
}
{
A \emph{spectrum} is a prespectrum $T_{\bullet}$ such that the adjoints $t_n: T_n\to\Omega T_{n+1}$ of the structure maps are homeomorphisms.
}
For example, let $X$ be a pointed space, $T_n=\Sigma^nX$ for $n\ge0$, then $T_{\bullet}$ is a prespectrum.
In particular, if $T_n=S^n$, then $T_{\bullet}$ is a prespectrum.
Let $G$ be an abelian group, $T_n=K(G,n)$ be the Eilenberg-MacLane space, then $T_{\bullet}$ is an $\Omega$-prespectrum. {In general, these examples are not spectra. But we can always construct a spectrum from a prespectrum using spectrification.}

{
Let $T_{\bullet}$ be a prespectrum, define
$(LT)_n$ to be the colimit of 
{ 
$$T_n\xrightarrow{t_n}\Omega T_{n+1}\xrightarrow{\Omega t_{n+1}}\Omega^2T_{n+2}
\xrightarrow{\Omega^2 t_{n+2}}
\dots
\xrightarrow{\Omega^{l-1} t_{n+l-1}}\Omega^l T_{n+l} \xrightarrow{\Omega^{l} t_{n+l}} 
\dots.$$
}
Namely,
$$(LT)_n=\text{colim}_{l\to\infty}\Omega^lT_{n+l},$$
then $(LT)_{\bullet}$ is a spectrum, called the spectrification of $T_{\bullet}$.
}
For example, if $T_n=S^n$, then $(LT)_{\bullet}$ is a spectrum $\mathbb{S}$ (the sphere spectrum).
 Let $G$ be an abelian group, $T_q=K(G,n)$ the Eilenberg-MacLane space, then $(LT)_{\bullet}$ is a spectrum $HG$ (the Eilenberg-MacLane spectrum).

{
Next we define the homotopy groups and cohomology rings of prespectra.
Let $T_{\bullet}$ be a prespectrum, we define the (stable) homotopy group 
$\pi_dT_{\bullet}$ to be the colimit of 
$$\pi_{d+n}T_n\xrightarrow{\pi_{d+n}t_n}\pi_{d+n}\Omega T_{n+1}\xrightarrow{\text{adjunction}}\pi_{d+n+1}T_{n+1}\dots.$$
Namely,
$$\pi_dT_{\bullet}=\text{colim}_{n\to\infty}\pi_{d+n}T_n.$$
}

{If $\rm{M}_{\bullet}$ and $\rm{N}_{\bullet}$ are two prespectra, and $\rm{N}_{\bullet}$ is an $\Omega$-prespectrum, then for any integer $k$, the abelian group of homotopy classes of maps from $\rm{M}_{\bullet}$ to $\rm{N}_{\bullet}$ of degree $-k$ is defined as follows: a map from $\rm{M}_{\bullet}$ to $\rm{N}_{\bullet}$ of degree $-k$ is a sequence of maps $\rm{M}_n\to \rm{N}_{n+k}$ such that the following diagram commutes
\bea
\xymatrix{\Sigma \rm{M}_n\ar[r]\ar[d]&\Sigma \rm{N}_{n+k}\ar[d]\\
\rm{M}_{n+1}\ar[r]&\rm{N}_{n+k+1}}
\eea
where the columns are the structure maps of the prespectra $\rm{M}_{\bullet}$ and $\rm{N}_{\bullet}$. Two maps of prespectra of degree $-k$: $f,g:\rm{M}_{\bullet}\to \rm{N}_{\bullet+k}$ are homotopic, denoted $f\simeq g$, if there is a map $H:\rm{M}_{\bullet}\wedge I_+\to \rm{N}_{\bullet+k}$ which restricts to $f\vee g$ along the inclusion $\rm{M}_{\bullet}\vee \rm{M}_{\bullet}\xrightarrow{i_0\vee i_1}\rm{M}_{\bullet}\wedge I_+\xrightarrow{H}\rm{N}_{\bullet+k}$ where the interval $I=[0,1]$, $I_+$ is the disjoint union of $I$ and a base point. 
Here $i_0$ and $i_1$ are the inclusion maps at $0 \in I$ and $1 \in I$ respectively.
Note that $\Sigma(\rm{M}_{\bullet}\times I)\ne \Sigma \rm{M}_{\bullet}\times I$, so we use $\rm{M}\wedge I_+$ instead of $\rm{M}_{\bullet}\times I$. 
The $[\rm{M}_{\bullet}, \rm{N}_{\bullet}]_{-k}$ is an abelian group since $\rm{N}_{\bullet}$ is an $\Omega$-prespectrum. If in addition, $\rm{N}_{\bullet}$ is a ring spectrum, then the abelian groups $[\rm{M}_{\bullet}, \rm{N}_{\bullet}]_{-k}$ form a graded ring $[\rm{M}_{\bullet}, \rm{N}_{\bullet}]_{-*}$. In particular, $[\rm{M}_{\bullet}, \rm{N}_{\bullet}]:=[\rm{M}_{\bullet}, \rm{N}_{\bullet}]_0$.
For example,
$\pi_d\rm{M}_{\bullet}=[\mathbb{S}, \rm{M}_{\bullet}]_d$.
Let $R$ be a ring, then the Eilenberg-MacLane spectrum $HR$ is a ring spectrum.
The cohomology ring of a prespectrum $\rm{M}_{\bullet}$ with coefficients in $R$ is defined to be $\H^*(\rm{M}_{\bullet},R):=[\rm{M}_{\bullet},HR]_{-*}$.
}

{
Let $V\to Y$ be a real vector bundle, and fix a Euclidean metric. The \emph{Thom space} $\text{Thom}(Y;V)$ is the quotient $D(V)/S(V)$ where $D(V)$ is the unit disk bundle and $S(V)$ is the unit sphere bundle. Namely, $D(V)=\{v\in V\mid |v|\le1\}$ and $S(V)=\{v\in V\mid |v|=1\}$ where $|\cdot|$ denotes the Euclidean norm.
}

{Thom spaces satisfy
\bea\label{thomsum}
\text{Thom}(X\times Y;V\times W)&=&\text{Thom}(X;V)\wedge \text{Thom}(Y;W),\notag\\
\text{Thom}(X;V\oplus \underline{\R^n})&=&\Sigma^n\text{Thom}(X;V),\notag\\
\text{Thom}(X;\underline{\R^n})&=&\Sigma^nX_+,
\eea
where $V\to X$ and $W\to Y$ are real vector bundles, $\underline{\R^n}$ is the trivial real vector bundle of dimension $n$, {and} $X_+$ is the disjoint union of $X$ and a point.
}

{
Let $G$ be a group with a group homomorphism $\rho: G\to\O$.
{Let} $MG(n)=\text{Thom}(\B G(n);V_n)$, where $V_n$ is the induced vector bundle (of dimension $n$) by the map $\B G(n)\to\B\O(n)$.
{Let} $T_n=MG(n)$, then $T_{\bullet}$ is a prespectrum by the property of Thom spaces. The \emph{Thom spectrum} $MG$ is the spectrification of $T_{\bullet}$.
In other words, $MG=\text{Thom}(\B G;V)$ where $V$ is the colimit of $V_n-n$.
}

{
Let $G$ be a group with a group homomorphism $\rho: G\to\O$.
The \emph{Madsen-Tillmann spectrum}
$MTG$ is the colimit of $\Sigma^nMTG(n)$, where the virtual Thom spectrum
$MTG(n)=\text{Thom}(\B G(n);-V_n)$
is the spectrification of the prespectrum whose $(n+q)$-th space is 
$\text{Thom}(\B G(n,n+q);Q_q)$ where  $\B G(n,n+q)$ is the pullback 
\bea
\xymatrix{
\B G(n,n+q)\ar@{-->}[r]\ar@{-->}[d]&\B G(n)\ar[d]\\
Gr_n(\R^{n+q})\ar[r]&\B\O(n)
}
\eea
and there is a direct sum $\underline{\R^{n+q}}=V_n\oplus Q_q$ of vector bundles over $Gr_n(\R^{n+q})$ and, by pullback, over $\B G(n,n+q)$ where $\underline{\R^{n+q}}$ is the trivial real vector bundle of dimension $n+q$.
In other words, $MTG=\text{Thom}(\B G;-V)$ where $V$ is the colimit of $V_n-n$.
}

\subsubsection{Adams spectral sequence}

Adams spectral sequence is a mathematical tool to compute the homotopy groups of spectra \cite{Adams1958}. In particular, the homotopy group of the Madsen-Tillmann spectrum $MTG$ \cite{MadsenTillmann4}
is the bordism group $\Omega_d^G$. We use Adams spectral sequence to compute several bordism groups related to Standard Models (SM), Grand Unified Theories (GUT) and beyond. We also compute the group $\TP_d(G)$ classifying the topological phases (i.e., the topological terms in QFT or the topological phases of quantum matter) 
based on the computation of bordism groups and a short exact sequence.
{See \cite{Freed2016, 1711.11587GPW, WanWang2018bns1812.11967} for primers.
We will call the group $\TP_d(G)$ the cobordism group. The relation between $\Omega_d^G$ and $\TP_d(G)$ is like that between homology group and cohomology group, as we will see later. }

{The Adams spectral sequence shows \cite{Adams1958}:
\bea\label{eq:ExtA_p}
\Ext_{\A_p}^{s,t}(\H^*(Y,\Z_p),\Z_p)\Rightarrow(\pi_{t-s}(Y))_p^{\wedge},
\eea
where the Ext denotes the extension functor. Note that this extension functor has 2 upper indices. It is different from but similar to the usual extension functor. The index $s$ refers to the degree of resolution, and the index $t$ is the internal degree of the graded module. Here $\A_p=[H\Z_p, H\Z_p]_{-*}$ is the mod $p$ Steenrod algebra which consists of mod $p$ cohomology operations \cite{Steenrod1962}, and $Y$ is any spectrum. In particular, the mod 2 Steenrod algebra $\A_2$ is generated by the Steenrod squares $\Sq^i:\H^*(-,\Z_2)\to\H^{*+i}(-,\Z_2)$. By the Yoneda lemma, the $\H^*(Y,\Z_p)=[Y,H\Z_p]_{-*}$ is automatically an $\A_p$-module whose internal degree $t$ is given by the $*$.
The $(\pi_{t-s}(Y))_p^{\wedge}$ is the $p$-completion of the $(t-s)$-th homotopy group of the spectrum $Y$.
For any finitely generated abelian group $\cG$, the abelian group $\cG_p^{\wedge}=\lim_{n\to\infty}\cG/p^n\cG$ is the $p$-completion of $\cG$.
We note that, 
if $\cG$ is a finite abelian group, then the $\cG_p^{\wedge}$ is the Sylow $p$-subgroup of $\cG$;
 if $\cG$ is the infinite cyclic group $\Z$, then the $\cG_p^{\wedge}$ is the ring of $p$-adic integers. Here the $\cG$ is meant to be substituted by a homotopy group $\pi_{t-s}(Y)$ in \eqref{eq:ExtA_p}. 
Here are some explanations and inputs:
\begin{enumerate}[leftmargin=2mm, label=\textcolor{blue}{(\arabic*)}., ref={(\arabic*)}] 
\item
Here the double-arrow 
``$\Rightarrow$'' means ``convergent to.'' 
The $E_2$ page are groups $\Ext^{s,t}$ 
with double indices $(s,t)$, we reindex the bidegree by $(t-s,s)$. 
Inductively, there are differentials $d_r$ in $E_r$ page which are arrows from $(t-s,s)$ to $(t-s-1,s+r)$ 
(that is, $\Ext^{s,t}\to \Ext^{s+r,t+r-1}$). Take Ker$d_r$/Im$d_r$ at each $(t-s,s)$, then we get the $E_{r+1}$ page. 
Finally $E_r$ page equals $E_{r+1}$ page (there are no differentials) for $r \geq N$, we call this $E_N$ page as the $E_{\infty}$ page, we can read the result $\pi_d$ at $t-s=d$.
See further details discussed in \Ref{WanWang2018bns1812.11967}'s Sec.~2.3.
\begin{figure}[H]
\begin{center}
\begin{tikzpicture}[scale=0.5]
\node at (6,-1) {$t-s$};
\node at (-1,6) {$s$};
\draw[->] (-0.5,-0.5) -- (-0.5,6);
\draw[->] (-0.5,-0.5) -- (6,-0.5);
\node[right] at (3,1) {$E_r^{s,t}$};
\node[above] at (2,3) {$E_r^{s+r,t+r-1}$};
\draw[->] (3,1) -- (2,3);
\node at (3,2) {$d_r$};
\end{tikzpicture}
\end{center}
\caption{The $E_r$ page of Adams spectral sequence}
\label{fig:E_r}
\end{figure}
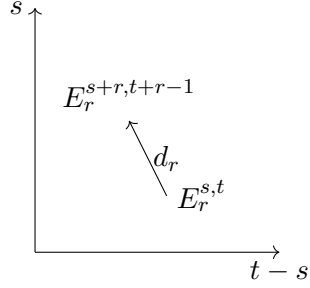
\item
In Adams spectral sequence, we consider $\Ext_R^{s,t}(L,\Z_p)$. In this article, we will consider the algebra $R=\A_p$ or $\A_2(1)$ for $p=2$, and $L$ is an $R$-module. The $\A_2(1)$ is the subalgebra of $\A_2$ generated by the Steenrod squares $\Sq^1$ and $\Sq^2$. 
Ext groups are  defined by firstly taking a projective $R$-resolution $P_{\bullet}$ of $L$, then secondly computing the (co)homology group of the (co)chain complex $\Hom(P_{\bullet},\Z_p)$.
Here a projective $R$-resolution $P_{\bullet}$ is an exact sequence of $R$-modules $\cdots\to P_s\to P_{s-1}\to \cdots\to P_0\to L$ where $P_s$ is a projective $R$-module for $s\ge0$. An $R$-module is projective if and only if it is a direct summand of a free $R$-module.
\end{enumerate}
}

For $Y=MTG$, where $MTG$ is the Madsen-Tillmann spectrum of a group $G$, the Adams spectral sequence \eqref{eq:ExtA_p} shows:
\bea\label{eq:ExtA_pMTG}
\Ext_{\A_p}^{s,t}(\H^*(MTG,\Z_p),\Z_p)\Rightarrow\pi_{t-s}(MTG)_p^{\wedge}=(\Omega_{d=t-s}^G)_p^{\wedge}.
\eea
The last equality is by the generalized Pontryagin-Thom isomorphism, we have an equality between the $d$-th bordism group of $G$ given by $\Omega_d^G$ and the $d$-th homotopy group of $MTG$ given by $\pi_d(MTG)$, namely
\bea\label{eq:PonThom}
\Omega_d^G=\pi_d(MTG).
\eea
The $T$ in $MTG$ means that the $G$-structures are on the {stable} tangent bundles instead of {stable} normal bundles. For Spin, the Madsen-Tillmann spectrum $MT\Spin=M\Spin$ is equivalent to the Thom spectrum.

We also compute the cobordism group of topological phases (TP) defined in \cite{Freed2016} as
\bea\label{eq:TP-def}
\TP_d(G):=[MTG,\Sigma^{d+1}I\Z].
\eea
Here $I\Z$ is the Anderson dual spectrum.
{By \cite{Freed2016}, the torsion part of} $\TP_d(G)$ classifies deformation classes of reflection positive invertible $d$-dimensional extended topological field theories with a symmetry group $G(d)$. {Here $G(d)$ means the $d$-dimensional spacetime version of the group $G$.}
The $\TP_d(G)$ and the bordism group $\Omega_d^G$ are related by a short exact sequence
\bea\label{eq:TPexact}
0\to\Ext^1(\Omega_d^G,\Z)\to\TP_d(G)\to\Hom(\Omega_{d+1}^G,\Z)\to0.
\eea
{
This short exact sequence is very similar to the universal coefficient theorem relating homology group and cohomology group. It is split, since $\Ext^1(\Omega_d^G,\Z)$ is always torsion, $\Hom(\Omega_{d+1}^G,\Z)$ is always free, and $\Ext^1(\Z,\Z_n)=0$. So we can directly derive the group $\TP_d(G)$ from the data of $\Omega_d^G$ and $\Omega_{d+1}^G$. 
}

We will use the \eqref{eq:ExtA_pMTG} and \eqref{eq:PonThom} to compute the $d$-th bordism group of $G$ given by $\Omega_d^G$. Then we will use \eqref{eq:TPexact} to compute the $d$-th cobordism group of topological phases of $G$ given by $\TP_d(G)$.

If $G=\Spin\times G'$, then $\B G=\B\Spin\times\B G'$. By definition, the Madsen-Tillmann spectrum $MTG=\text{Thom}(\B G;-V)$ where $V$ is the induced virtual bundle of dimension 0 by the map $\B G\to\B\O$.
By the properties of Thom space \eqref{thomsum},
we have 
\bea
MT(\Spin\times G')=M\Spin\wedge(\B G')_+.
\eea
The $\wedge$ is the smash product.\footnote{Smash product 
between a spectrum $M_{\bullet}$ and a topological space $X$ is a spectrum whose $n$-th space is $M_n\wedge X$ which is the ordinary smash product of topological spaces.}
The $(\B G')_+$ is the disjoint union of the classifying space $\B G'$ and a point.\footnote{\label{ft:X_+}For a topological space $X$, it is a standard convention to 
denote that $X_+$ as the disjoint union of $X$ and a point.
Note that the reduced cohomology of $X_+$ is exactly the ordinary cohomology of $X$.}
{
By the generalized Pontryagin-Thom isomorphism, for any structure group $G$ and any topological space $X$, we have
\bea\label{eq:genPonThom}
\Omega_d^{G}(X)=\pi_d(MTG\wedge X_+).
\eea
Since $MT\Spin=M\Spin$, we have
\bea\label{eq:Spin-BG'}
\Omega_d^{\Spin\times G'}=\pi_d(MT(\Spin\times G'))=\pi_d(M\Spin\wedge(\B G')_+)=\Omega_d^{\Spin}(\B G').
\eea
}

{If $MTG=M\Spin\wedge X$ where $X$ is any {topological space},
by Corollary 5.1.2 of \cite{2018arXiv180107530B} which is based on \cite{ABP1967}, we have
\bea
\Ext_{\A_2}^{s,t}(\H^*(M\Spin\wedge X,\Z_2),\Z_2)=\Ext_{\A_2(1)}^{s,t}(\H^*(X,\Z_2),\Z_2)
\eea
for $t-s<8$.
Here $\A_2(1)$ is the subalgebra of $\A_2$ generated by $\Sq^1$ and $\Sq^2$.
}

So for the dimension $d=t-s<8$, we have
\bea\label{eq:ExtA_2(1)}
\Ext_{\A_2(1)}^{s,t}(\H^*(X,\Z_2),\Z_2)\Rightarrow(\Omega_{t-s}^G)_2^{\wedge}.
\eea

The {reduced cohomology} $\H^*(X,\Z_2)$ is an $\A_2(1)$-module whose internal degree $t$ is given by the $*$.

{
\begin{lem}[Lemma 11 of  \cite{WanWang2018bns1812.11967}]\label{principle}
Given a short exact sequence of $\A_2(1)$-modules
\bea
0\to L_1\to L_2\to L_3\to0,
\eea
then for any $t$, there is a long exact sequence
\bea
&&\cdots\to\Ext_{\A_2(1)}^{s,t}(L_3,\Z_2)\to\Ext_{\A_2(1)}^{s,t}(L_2,\Z_2)\to\Ext_{\A_2(1)}^{s,t}(L_1,\Z_2)\\\notag
&&\stackrel{d_1}{\to}\Ext_{\A_2(1)}^{s+1,t}(L_3,\Z_2)\to\Ext_{\A_2(1)}^{s+1,t}(L_2,\Z_2)\to\cdots
\eea
\end{lem}
}
{We can compute the $E_2$ page of $\A_2(1)$-module based on Lemma \ref{principle}. More precisely, in order to compute 
$\Ext_{\A_2(1)}^{s,t}(L_2,\Z_2)$, we find a short exact sequence of 
$\A_2(1)$-modules 
\bea
0\to L_1\to L_2\to L_3\to 0,
\eea
then we apply Lemma \ref{principle} to compute $\Ext_{\A_2(1)}^{s,t}(L_2,\Z_2)$ by the given data of $\Ext_{\A_2(1)}^{s,t}(L_1,\Z_2)$ and $\Ext_{\A_2(1)}^{s,t}(L_3,\Z_2)$. 
Our strategy is choosing $L_1$ to be the direct sum of suspensions of $\Z_2$ on which $\Sq^1$ and $\Sq^2$ act trivially, then we take $L_3$ to be the quotient of $L_2$ by $L_1$.
{If $\Ext_{\A_2(1)}^{s,t}(L_3,\Z_2)$ is undetermined, then we take $L_3$ to be the new $L_2$ and repeat this procedure.
We can use this procedure again and again until $\Ext_{\A_2(1)}^{s,t}(L_2,\Z_2)$ is determined.}
}

\subsubsection{Characteristic classes}

Throughout the article, {we use the standard notation for characteristic classes: $w_i$ for the Stiefel-Whitney class, $c_i$ for the Chern class, $p_i$ for the Pontryagin class, and 
$e_n$ for the Euler class. Note that the Euler class only appears in the total dimension of the vector bundle. We use the notation $w_i(G)$, $c_i(G)$, $p_i(G)$, and $e_n(G)$ to denote the characteristic classes of the associated vector bundle of the principal $G$ bundle (normally denoted as $w_i(V_{G})$, $c_i(V_{G})$, $p_i(V_{G})$, and $e_n(V_{G})$). For simplicity, we may denote
the Stiefel-Whitney class of the tangent bundle $TM$ as $w_j \equiv w_j(TM)$; 
if we do not specify $w_j$ with which bundle, then we really mean $TM$.}

We will use $\text{CS}_{2n-1}^V$ to denote the Chern-Simons $2n-1$-form for the Chern class (if $V$ is a complex vector bundle) or the Pontryagin class (if $V$ is a real vector bundle). Note that $p_i(V)=(-1)^ic_{2i}(V\otimes\C)$. The relation between the Chern-Simons form and the Chern class is
\bea
c_n(V)=\dd \text{CS}_{2n-1}^V
\eea 
where the $\dd$ is the exterior differential and the $c_n(V)$ is regarded as a closed differential form in de Rham cohomology.

There is also another kind of Chern-Simons form for Euler class {$e_{2n}(V)$} \cite{MULLERHOISSEN1990235}, we denote it by $\text{CS}_{2n-1,e}^V$, it satisfies
\bea
e_{2n}(V)=\dd \text{CS}_{2n-1,e}^V.
\eea

The relations between Pontryagin class, Euler class and Stiefel-Whitney class are
\bea
p_i(V)=w_{2i}(V)^2\mod2
\eea
and
\bea
e_{2n}(V)=w_{2n}(V)\mod2
\eea
where $2n=\dim V$.

By the Hirzebruch signature theorem, the relation between the signature and the first Pontryagin class of a 4-manifold $M$ is
\bea
\sigma=\frac{1}{3}\int_{M}p_1(TM).
\eea


\subsection{Lie algebra to Lie group and the representation theory}

To justify the spacetime symmetry group $G_{\text{spacetime}}$ and internal symmetry group $G_{\text{internal}}$ relevant for Standard Model physics, 
we shall first review the Lie algebra of Standard Models, to the representation theory of matter field contents,
and to the Lie groups of Standard Models.
\begin{enumerate}[leftmargin=-4.mm, label=\textcolor{blue}{[\Roman*]}., ref={[\Roman*]}]
\item The local gauge structure of Standard Model is given by the Lie algebra $u(1) \times su(2) \times su(3)$.
This means that the Lie algebra valued 1-form gauge fields take values in  $u(1) \times su(2) \times su(3)$.
The 1-form gauge fields are the 1-connections of the principals $G_{\text{internal}}$-bundles that we should determine.

\item  Fermions as the spinor fields.
A spinor field is a section of the spinor bundle. For the left-handed Weyl spinor $\Psi_L$,
it is a doublet spin-1/2 representation of spacetime symmetry group $G_{\text{spacetime}}$ (Minkowski $\Spin(3,1)$ or Euclidean $\Spin(4)$), denoted as
\bea
\Psi_L \sim {\bf 2}_L \text{ of } \Spin(3,1), \text{ or } \Psi_L \sim  {\bf 2}_L \text{ of } \Spin(4).
\eea 
On the other hand, the matter fields as Weyl spinors $\Psi_L$ contain:
\begin{itemize}
\item
The left-handed up and down quarks ($u$ and $d$) form a doublet $\begin{pmatrix}
u\\
d
\end{pmatrix}_L$ in ${\bf 2}$ for the SU(2)$_{\text{weak}}$, and they are in ${\bf 3}$ for the SU(3)$_{\text{strong}}$.
\item
The right-handed up and down quarks, each forms a singlet $u_R$ and $d_R$ in ${\bf 1}$ for the SU(2)$_{\text{weak}}$. They are in ${\bf 3}$ for the SU(3)$_{\text{strong}}$.
\item
The left-handed electron and neutrino form a doublet $\begin{pmatrix}
\nu_e\\
e
\end{pmatrix}_L$ in ${\bf 2}$ for the SU(2)$_{\text{weak}}$, and they are in ${\bf 1}$ for the SU(3)$_{\text{strong}}$.
\item
The right-handed electron forms a singlet in ${\bf 1}$ for the SU(2)$_{\text{weak}}$, 
and it is in ${\bf 1}$ for the SU(3)$_{\text{strong}}$.
\end{itemize}
There are two more families of quarks: 
charm and strange quarks ($c$ and $s$),
and top and bottom quarks ($t$ and $b$).
There are also two more families of leptons:
muon and its neutrino ($\mu$ and $\nu_\mu$), 
and 
tauon and its neutrino ($\tau$ and $\nu_\tau$). 
So there are three families (i.e., generations) of quarks and leptons:
\bea
\Bigg(
\begin{pmatrix}
u\\
d
\end{pmatrix}_L \times {\bf 3}_{\text{color}}, \quad\quad \quad u_R \times {\bf 3}_{\text{color}}, \quad\quad\quad\quad d_R \times {\bf 3}_{\text{color}}, 
\quad\quad\quad
\begin{pmatrix}
\nu_e\\
e
\end{pmatrix}_L,\quad\quad  \quad e_R 
\quad  \Bigg),\\
\Bigg(
\begin{pmatrix}
c\\
s
\end{pmatrix}_L \times {\bf 3}_{\text{color}}, \quad\quad \quad c_R \times {\bf 3}_{\text{color}}, \quad\quad\quad\quad s_R \times {\bf 3}_{\text{color}}, 
\quad\quad\quad
\begin{pmatrix}
\nu_\mu\\
\mu
\end{pmatrix}_L,\quad\quad  \quad \mu_R 
\quad  \Bigg),\\
\Bigg(
\begin{pmatrix}
t\\
b
\end{pmatrix}_L \times {\bf 3}_{\text{color}}, \quad\quad \quad t_R \times {\bf 3}_{\text{color}}, \quad\quad\quad\quad b_R \times {\bf 3}_{\text{color}}, 
\quad\quad\quad
\begin{pmatrix}
\nu_\tau\\
\tau
\end{pmatrix}_L,\quad\quad  \quad \tau_R 
\quad  \Bigg).
\eea
In short, for all of them as three families, we can denote them as:
\bea
\Bigg(
\begin{pmatrix}
u\\
d
\end{pmatrix}_L \times {\bf 3}_{\text{color}}, \quad\quad \quad u_R \times {\bf 3}_{\text{color}}, \quad\quad\quad\quad d_R \times {\bf 3}_{\text{color}}, 
\quad\quad\quad
\begin{pmatrix}
\nu_e\\
e
\end{pmatrix}_L,\quad\quad  \quad e_R 
\quad  \Bigg)\times \text{3 families}. 
\eea
In fact, all the following \emph{four} kinds of
\bea
G_{\text{internal}}=\frac{\SU(3)\times \SU(2)\times \U(1)}{\Z_q}
\eea
with $q=1,2,3,6$ are compatible with the above representations of fermion fields.
These Weyl spinors can be written in the following more succinct forms of representations for any of the internal symmetry group $G_{\text{internal}}$ with $q=1,2,3,6$:
\bea \label{eq:rep-3families}
\Bigg( ({\bf 3},{\bf 2}, 1/6)_L,({\bf 3},{\bf 1}, 2/3)_R,({\bf 3},{\bf 1},-1/3)_R,({\bf 1},{\bf 2},-1/2)_L,({\bf 1},{\bf 1},-1)_R  \Bigg)\times \text{3 families} \nn\\
\Rightarrow
\Bigg( ({\bf 3},{\bf 2}, 1/6)_L,(\overline{\bf 3},{\bf 1}, -2/3)_L,(\overline{\bf 3},{\bf 1},1/3)_L, ({\bf 1},{\bf 2},-1/2)_L,({\bf 1},{\bf 1},1)_L  \Bigg)\times \text{3 families}.
\eea
The triplet given above is listed by their (SU(3) representation, SU(2) representation, hypercharge $Y$).\footnote{Note that 
the hypercharge $Y$ is conventionally given by the relation: $Q_{\rm{EM}}=T_3 +Y$ where 
$Q_{\rm{EM}}$ is the unbroken (un-Higgsed) electromagnetic gauge charge
and $T_3= \frac{1}{2}
\begin{pmatrix}
1 & 0\\
0 & -1
\end{pmatrix}$ is a generator of SU(2)$_{\text{weak}}$. However, some other conventions are possible, one convention is
$Q_{\rm{EM}}=T_3 +\frac{1}{2} Y'$, another convention is $Q_{\rm{EM}}=T_3 +\frac{1}{6} \tilde{Y}$  \cite{Tong2017oea1705.01853}.
In the $Q_{\rm{EM}}=T_3 +\frac{1}{6} \tilde{Y}$ version, we have the integer quantized $\tilde{Y} = 6 Y$. We can rewrite \eq{eq:rep-3families} as:
 \bea \label{eq:rep-3families-tildeY}
\Bigg( ({\bf 3},{\bf 2}, \tilde{Y}=1)_L,({\bf 3},{\bf 1}, \tilde{Y}=4)_R,({\bf 3},{\bf 1},\tilde{Y}=-2)_R,({\bf 1},{\bf 2},\tilde{Y}=-3)_L,({\bf 1},{\bf 1},\tilde{Y}=-6)_R  \Bigg)\times \text{3 families} \nn\\
\Rightarrow
\Bigg( ({\bf 3},{\bf 2}, \tilde{Y}=1)_L,(\overline{\bf 3},{\bf 1}, \tilde{Y}=-4)_L,(\overline{\bf 3},{\bf 1},\tilde{Y}=2)_L, ({\bf 1},{\bf 2},\tilde{Y}=-3)_L,({\bf 1},{\bf 1},\tilde{Y}=6)_L  \Bigg)\times \text{3 families}.
\eea
Then the right-handed neutrino stays the same form since $\tilde{Y} = 3 Y'= 6 Y =0$. 
} 
For example,
$({\bf 3},{\bf 2}, 1/6)$ means that ${\bf 3}$ in SU(3), ${\bf 2}$ in SU(2) and 1/6 for hypercharge.
In the second line, we transforms the right-handed Weyl spinor 
$\Psi_R \sim {\bf 2}_R \text{ of } \Spin(3,1)$
to its left-handed $\Psi_L \sim {\bf 2}_L \text{ of } \Spin(3,1)$, while we flip their (SU(3) representation, SU(2) representation, hypercharge)
to its (complex) conjugation representations.\footnote{Note that ${\bf 2}$ and $\overline{\bf 2}$ are the same representation in SU(2), see, e.g., \cite{1711.11587GPW}.} 
If we include the right-handed neutrinos (say ${\nu_{e}}_R$, ${\nu_{\mu}}_R$, and ${\nu_{\tau}}_R$), they are all in the representation 
$({\bf 1},{\bf 1},0)_R$ with no hypercharge.
We can also represent a right-handed neutrino by the left-handed (complex) conjugation version $({\bf 1},{\bf 1},0)_L$.

Also the complex scalar Higgs field $\phi_H$ is in a representation $({\bf 1},{\bf 2}, 1/2)$.\footnote{The Higgs field $\phi_H$ is in
$({\bf 1},{\bf 2}, Y=1/2)
=({\bf 1},{\bf 2}, Y'=1)
=({\bf 1},{\bf 2}, \tilde{Y} =3)$
where $\tilde{Y} =3 Y'= 6 Y$.
}
In the Higgs condensed phase, the conventional Higgs vacuum expectation value (vev) is chosen to be $\langle \phi_H \rangle =\frac{1}{\sqrt 2}
\begin{pmatrix}
0\\
v
\end{pmatrix}$, which vev has  $Q_{\rm{EM}}=0$.

\item If we include the $3 \times 2 + 3 + 3 + 2 +1 =15$ left-handed Weyl spinors from one single family, 
we can combine them as a multiplet of $\overline{\bf 5}$ and {\bf 10} left-handed Weyl spinors of SU(5):
\bea
(\overline{\bf 3},{\bf 1},1/3)_L \oplus ({\bf 1},{\bf 2},-1/2)_L &\sim& \overline{\bf 5} \text{ of } \SU(5),\\
  ({\bf 3},{\bf 2}, 1/6)_L \oplus (\overline{\bf 3},{\bf 1}, -2/3)_L \oplus ({\bf 1},{\bf 1},1)_L  &\sim& {\bf 10} \text{ of } \SU(5).
\eea
Hence we can study the SU(5) GUT with a SU(5) gauge group.

\item If we include the $3 \times 2 + 3 + 3 + 2 +1 =15$ left-handed Weyl spinors from one single family, 
and also a right-handed neutrino, we can combine them as a multiplet of 16 left-handed Weyl spinors:
\bea
\quad  \Psi_L \sim {\bf 16}^+ \text{ of } \Spin(10),
\eea
which sits at the 16-dimensional irreducible spinor representation of Spin(10).
(There are ${\bf 16}^+$ and ${\bf 16}^-$-dimensional irreducible spinor representation 
together form a ${\bf 32}$-dimensional reducible spinor representation of Spin(10).)
Namely, we should study the SO(10) GUT with a Spin(10) gauge group instead of a SO(10) gauge group.
\item We find the following Lie group embedding for the internal symmetry of GUTs and Standard Models:
\bea
\SO(10) 
\supset 
 \SU(5) 
\supset 
\frac{\U(1) \times \SU(2)\times\SU(3)}{\Z_6}.\\
\Spin(10) 
\supset 
 \SU(5) 
\supset 
\frac{\U(1) \times \SU(2)\times\SU(3)}{\Z_6}.
\eea
The other ${\Z_q}$ for $q=1,2,3$ cannot be embedded into Spin(10) nor SU(5).
So from the GUT perspective, it is natural to consider the Standard Model gauge group 
$\frac{\U(1) \times \SU(2)\times\SU(3)}{\Z_6}$.
\item

We find the following group embedding for the spacetime and internal symmetries of GUTs and Standard Models (see also \cite{WanWangv2} for the derivations):
\bea \label{eq:SMembed1}
{\frac{\Spin(d) \times
\Spin(10)}{{\Z_2^F}} 
\supset 
\Spin(d) \times \SU(5) 
\supset 
\Spin(d) \times \frac{\SU(3) \times   \SU(2) \times \U(1)}{\Z_6}}.\\
 \label{eq:SMembed2}
{{\Spin(d) \times
\Spin(10)} 
\supset 
\Spin(d) \times \SU(5) 
\supset 
\Spin(d) \times \frac{\SU(3) \times   \SU(2) \times \U(1)}{\Z_6}}.
\eea
\end{enumerate}
We shall study the cobordism theory of these SM, BSM, and GUT groups
in the following subsections.


\section{Standard Models}\label{sec:SM}

Now we consider the co/bordism classes relevant for Standard Model (SM) physics \cite{Glashow1961trPartialSymmetriesofWeakInteractions, Salam1964ryElectromagneticWeakInteractions, Weinberg1967tqSMAModelofLeptons}.

\subsection{
${\Spin\times {\SU(3)\times \SU(2)\times \U(1)}}$ model
}
We consider $G={\Spin\times {\SU(3)\times \SU(2)\times \U(1)}}$, the Madsen-Tillmann spectrum $MTG$ of the group $G$ is
\bea
MTG=M\Spin\wedge(\B(\SU(3)\times \SU(2)\times \U(1)))_+.
\eea
%
%

 The $(\B(\SU(3)\times \SU(2)\times \U(1)))_+$ is the disjoint union of the classifying space $\B(\SU(3)\times \SU(2)\times \U(1))$ and a point, see footnote \ref{ft:X_+}.

For the dimension $d=t-s<8$, since there is no odd torsion,\footnote{\label{ft:no-odd-torsion}By computation using the mod $p$ Adams spectral sequence for an odd prime $p$, we find there is no odd torsion.}
%
%
for $MTG=M\Spin\wedge(\B(\SU(3)\times \SU(2)\times \U(1)))_+$, for the dimension $d=t-s<8$, by \eqref{eq:ExtA_2(1)}, we have the Adams spectral sequence
\bea
\Ext_{\A_2(1)}^{s,t}(\H^*(\B(\SU(3)\times \SU(2)\times \U(1)),\Z_2),\Z_2)\Rightarrow\Omega_{t-s}^{\Spin\times {\SU(3)\times \SU(2)\times \U(1)}}.
\eea

We have
\bea
\H^*(\B\SU(n),\Z_2)=\Z_2[c_2,\dots,c_n]
\eea
and
\bea
\H^*(\B\U(n),\Z_2)=\Z_2[c_1,\dots,c_n].
\eea

We also have the Wu formula\footnote{There is another Wu formula which will be used later:
\bea
\Sq^i(x_{d-i})=u_ix_{d-i}
\eea
on $d$-manifold $M$ where $x_{d-i}\in\H^{d-i}(M,\Z_2)$ and $u_i$ is the Wu class.}
\bea
\Sq^{2j}(c_i)=\sum_{k=0}^j\binom{i-k-1}{j-k}c_{i+j-k}c_k \text{ for } 0\le j\le i.
\eea


By K\"unneth formula, we have
\bea
\H^*(\B(\SU(3)\times \SU(2)\times \U(1)),\Z_2)=\Z_2[c_2,c_3]\otimes\Z_2[c_2']\otimes\Z_2[c_1''].
\eea
{Here only in this subsection, $c_i$ is the Chern class of SU(3) bundle, and $c_i'$ is the Chern class of SU(2) bundle,
and $c_i''$ is the Chern class of U(1) bundle.}

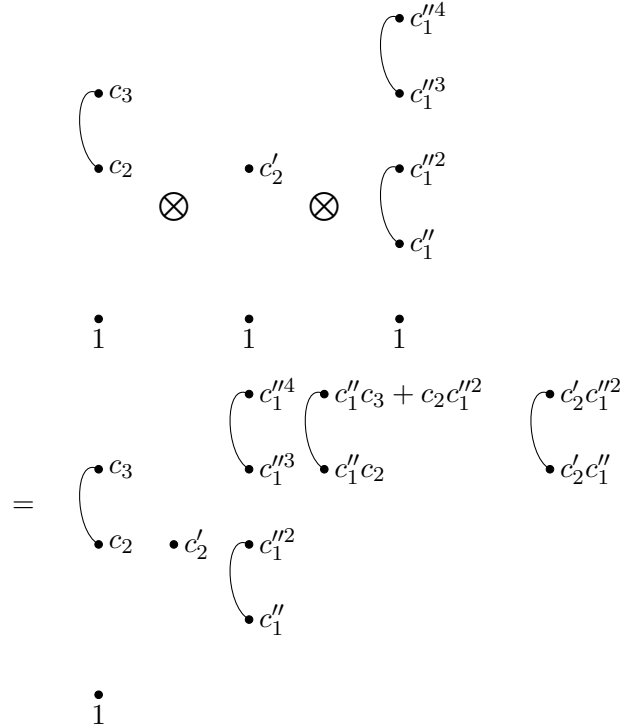
\begin{figure}[H]
\begin{center}
\begin{tikzpicture}[scale=0.5]

\node[below] at (0,0) {1};

\draw[fill] (0,0) circle(.1);
\draw[fill] (0,4) circle(.1);
\draw[fill] (0,6) circle(.1);
\draw (0,4) to [out=150,in=150] (0,6);

\node[right] at (0,4) {$c_2$};

\node[right] at (0,6) {$c_3$};

\node at (2,3) {$\bigotimes$};

\node[below] at (4,0) {1};

\draw[fill] (4,0) circle(.1);
\draw[fill] (4,4) circle(.1);

\node[right] at (4,4) {$c_2'$};

\node at (6,3) {$\bigotimes$};

\node[below] at (8,0) {1};

\draw[fill] (8,0) circle(.1);

\draw[fill] (8,2) circle(.1);

\draw[fill] (8,4) circle(.1);
\draw (8,2) to [out=150,in=150] (8,4);

\draw[fill] (8,6) circle(.1);

\draw[fill] (8,8) circle(.1);

\draw (8,6) to [out=150,in=150] (8,8);

\node[right] at (8,2) {$c_1''$};

\node[right] at (8,4) {$c_1''^2$};

\node[right] at (8,6) {$c_1''^3$};

\node[right] at (8,8) {$c_1''^4$};

\node at (-2,-5) {$=$};

\node[below] at (0,-10) {1};

\draw[fill] (0,-10) circle(.1);
\draw[fill] (0,-6) circle(.1);
\draw[fill] (0,-4) circle(.1);
\draw (0,-6) to [out=150,in=150] (0,-4);

\node[right] at (0,-6) {$c_2$};

\node[right] at (0,-4) {$c_3$};

\draw[fill] (2,-6) circle(.1);

\node[right] at (2,-6) {$c_2'$};

\draw[fill] (4,-8) circle(.1);

\draw[fill] (4,-6) circle(.1);
\draw (4,-8) to [out=150,in=150] (4,-6);

\draw[fill] (4,-4) circle(.1);

\draw[fill] (4,-2) circle(.1);

\draw (4,-4) to [out=150,in=150] (4,-2);

\node[right] at (4,-8) {$c_1''$};

\node[right] at (4,-6) {$c_1''^2$};

\node[right] at (4,-4) {$c_1''^3$};

\node[right] at (4,-2) {$c_1''^4$};

\draw[fill] (6,-4) circle(.1);

\node[right] at (6,-4) {$c_1''c_2$};

\draw[fill] (6,-2) circle(.1);

\node[right] at (6,-2) {$c_1''c_3+c_2c_1''^2$};

\draw (6,-4) to [out=150,in=150] (6,-2);

\draw[fill] (12,-4) circle(.1);

\node[right] at (12,-4) {$c_2'c_1''$};

\draw[fill] (12,-2) circle(.1);

\node[right] at (12,-2) {$c_2'c_1''^2$};

\draw (12,-4) to [out=150,in=150] (12,-2);

\end{tikzpicture}
\end{center}
\caption{The $\A_2(1)$-module structure of $\H^*(\B(\SU(3)\times \SU(2)\times \U(1)),\Z_2)$ below degree 6.}
\label{fig:A_2(1)SU3SU2U1}
\end{figure}

\begin{figure}[H]
\begin{center}
\begin{tikzpicture}
\node at (0,-1) {0};
\node at (1,-1) {1};
\node at (2,-1) {2};
\node at (3,-1) {3};
\node at (4,-1) {4};
\node at (5,-1) {5};
\node at (6,-1) {6};
\node at (7,-1) {$t-s$};
\node at (-1,0) {0};
\node at (-1,1) {1};
\node at (-1,2) {2};
\node at (-1,3) {3};
\node at (-1,4) {4};
\node at (-1,5) {5};
\node at (-1,6) {$s$};

\draw[->] (-0.5,-0.5) -- (-0.5,6);
\draw[->] (-0.5,-0.5) -- (7,-0.5);

\draw (0,0) -- (0,5);

\draw (0,0) -- (2,2);

\draw (2.1,0) -- (2.1,5);

\draw (4,3) -- (4,5);

\draw (3.9,1) -- (3.9,5);

\draw (4.1,0) -- (4.1,5);

\draw (4.1,0) -- (6.1,2);

\draw (4.2,0) -- (4.2,5);

\draw (6,2) -- (6,5);

\draw (6.2,0) -- (6.2,5);

\draw (6.3,0) -- (6.3,5);

\draw (6.4,1) -- (6.4,5);

\draw (6.5,0) -- (6.5,5);

\end{tikzpicture}
\end{center}
\caption{$\Omega_*^{\Spin\times \SU(3)\times \SU(2)\times \U(1)}$.}
\label{fig:E_2SU3SU2U1}
\end{figure}
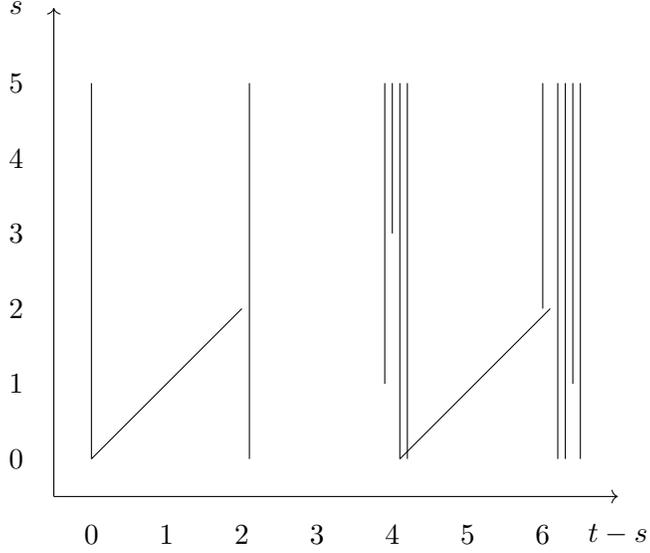

The $\A_2(1)$-module structure of $\H^*(\B(\SU(3)\times \SU(2)\times \U(1)),\Z_2)$ below degree 6
and the $E_2$ page are shown in Figure \ref{fig:A_2(1)SU3SU2U1}, \ref{fig:E_2SU3SU2U1}.
Here we have used the correspondence between $\A_2(1)$-module structure and the $E_2$ page shown in Figure \ref{fig:Z_2} and \ref{fig:Ceta}.

In Adams chart, the horizontal axis labels the integer degree $d=t-s$ and the vertical axis labels the integer degree $s$. The differential $d_r^{s,t}:E_r^{s,t}\to E_r^{s+r,t+r-1}$ is an arrow starting at the bidegree $(t-s,s)$ with direction $(-1,r)$.
$E_{r+1}^{s,t}:=\frac{\text{Ker}d_r^{s,t}}{\text{Im}d_r^{s-r,t-r+1}}$ for $r\ge2$. There exists $N$ such that $E_{N+k}=E_N$ stabilized for $k>0$, we denote the stabilized page $E_{\infty}:=E_N$.

To read the result from the Adams chart in Figure \ref{fig:E_2SU3SU2U1}, we look at the stabilized $E_{\infty}$ page, one dot indicates a finite group $\Z_p$, a vertical finite line segment connecting $n$ dots indicates a finite group $\Z_{p^n}$. But when $n=\infty$, the infinite line connecting infinite dots indicates a $\Z$. Here $p$ is given by the mod $p$ Adams spectral sequence in \eqref{eq:ExtA_pMTG}. Here in Figure \ref{fig:E_2SU3SU2U1}, $p=2$, we can read from the Adams chart $\Omega^{\Spin\times {\SU(3)\times \SU(2)\times \U(1)}}_0=\Z$ (an infinite line), $\Omega^{\Spin\times {\SU(3)\times \SU(2)\times \U(1)}}_1=\Z_2$ (a dot), $\Omega^{\Spin\times {\SU(3)\times \SU(2)\times \U(1)}}_2=\Z\times\Z_2$ (an infinite line and a dot), $\Omega^{\Spin\times {\SU(3)\times \SU(2)\times \U(1)}}_3=0$ (nothing), $\Omega^{\Spin\times {\SU(3)\times \SU(2)\times \U(1)}}_4=\Z^4$ (four infinite lines), $\Omega^{\Spin\times {\SU(3)\times \SU(2)\times \U(1)}}_5=\Z_2$ (a dot), $\Omega^{\Spin\times {\SU(3)\times \SU(2)\times \U(1)}}_6=\Z^5\times\Z_2$ (five infinite lines and a dot). Thus we obtain the bordism group $\Omega^{\Spin\times {\SU(3)\times \SU(2)\times \U(1)}}_d$ shown in Table \ref{table:SU3SU2U1Bordism}.



\begin{table}[H]
\centering
\hspace*{-25mm}
\begin{tabular}{c c c }
\hline
\multicolumn{3}{c}{Bordism group}\\
\hline
$d$ & 
$\Omega^{\Spin\times {\SU(3)\times \SU(2)\times \U(1)}}_d$
& bordism invariants \\
\hline
0& $\Z$  \\
\hline
1& $\Z_2$ & $\tilde\eta$ \\
\hline
2&  $\Z\times\Z_2$ & $c_1(\U(1))$, Arf \\
\hline
3 & 0 \\
\hline
4 & $\Z^4$ & $\frac{\sigma}{16} = \frac{p_1(TM)}{48},\frac{c_1(\U(1))^2}{2},c_2(\SU(2)),c_2(\SU(3))$ \\
\hline
5 & $\Z_2$  & $c_2(\SU(2))\tilde\eta$\\
\hline
6 & $\Z^5\times\Z_2$ &
$\begin{matrix}
\frac{c_1(\U(1))(\sigma-\rF \cdot \rF )}{8}, c_1(\U(1))^3,c_1(\U(1))c_2(\SU(2)),\\
c_1(\U(1))c_2(\SU(3)),\frac{c_3(\SU(3))}{2}, c_2(\SU(2))\text{Arf}
 \end{matrix}$  \\
\hline
\end{tabular}
\caption{Bordism group. 
$\tilde\eta$ is a mod 2 index of 1d Dirac operator.
Arf is a 2d Arf invariant. $c_i(G)$ is the Chern class of the associated vector bundle of the principal $G$-bundle.
$\sigma = \frac{p_1(TM)}{3}$ is the signature of manifold.
{$\rF $ is the characteristic 2-surface \cite{Saveliev} in a 4-manifold $M^4$, it satisfies the condition $\rF\cdot x=x\cdot x\mod2$ for all $x\in\H_2(M^4,\Z)$. 
Here $\cdot$ is the intersection form of $M^4$.
By the Freedman-Kirby theorem, $(\frac{\sigma-\rF\cdot\rF}{8} )(M^4)=\text{Arf}(M^4,\rF)\mod2$.}
Note that $c_1(\U(1))^2=\Sq^2c_1(\U(1))=(w_2+w_1^2)c_1(\U(1))=0\mod2$ on Spin 4-manifolds, so $c_1(\U(1))^2/2 \in \Z$.
Note that $c_3(\SU(3))=\Sq^2c_2(\SU(3))=(w_2+w_1^2)c_2(\SU(3))=0\mod2$ on Spin 6-manifolds, 
so $c_3(\SU(3))/2 \in \Z$.
}
\label{table:SU3SU2U1Bordism}
\end{table}
By \eqref{eq:TPexact}, we obtain the cobordism group $\TP_d(\Spin\times {\SU(3)\times \SU(2)\times \U(1)})$ shown in Table \ref{table:SU3SU2U1TP}.
\begin{table}[H]
\centering
\hspace*{-8mm}
\begin{tabular}{c c c }
\hline
\multicolumn{3}{c}{Cobordism group}\\
\hline
$d$ & 

$\begin{matrix}\TP_d\\
(\Spin\times {\SU(3)\times \SU(2)\times \U(1)})
\end{matrix}$
& topological terms \\
\hline
0& 0 \\
\hline
1& $\Z\times\Z_2$ & $\text{CS}_1^{\U(1)},\tilde\eta$ \\
\hline
2& $\Z_2$ & Arf \\
\hline
3 & $\Z^4$ & $\frac{\text{CS}_3^{TM}}{48}, \frac{1}{2}\text{CS}_1^{\U(1)}c_1(\U(1)), \text{CS}_3^{\SU(2)},\text{CS}_3^{\SU(3)}$ \\
\hline
4 & 0 \\
\hline
5 & $\Z^5\times\Z_2$ & 
$\begin{matrix}
\mu(\text{PD}(c_1(\U(1)))), \text{CS}_1^{\U(1)}c_1(\U(1))^2,
\text{CS}_1^{\U(1)}c_2(\SU(2)) {\sim c_1(\U(1))\text{CS}_3^{\SU(2)}},\\
 \text{CS}_1^{\U(1)}c_2(\SU(3)) {\sim c_1(\U(1))\text{CS}_3^{\SU(3)}}, \frac{1}{2}\text{CS}_5^{\SU(3)}, c_2(\SU(2))\tilde\eta 
 \end{matrix}$\\
\hline
\end{tabular}
\caption{Topological phase classification ($\equiv$ TP) as a cobordism group, following Table \ref{table:SU3SU2U1Bordism}. 
$\tilde\eta$ is a mod 2 index of 1d Dirac operator.
Arf is a 2d Arf invariant.
$c_i(G)$ is the Chern class of the associated vector bundle of the principal $G$-bundle.
$\text{CS}_{2n-1}^{V}$ or $\text{CS}_{2n-1}^{G}$ is the Chern-Simons form of the vector bundle $V$ or the associated vector bundle of the principal $G$-bundle.
The PD is the Poincar\'e dual.
The $TM$ is the spacetime tangent bundle.
The $\mu$ is the 3d Rokhlin invariant.
If $\partial M^4=M^3$, then $\mu(M^3)=(\frac{\sigma-\rF\cdot\rF}{8} )(M^4)$, thus
$\mu(\text{PD}(c_1(\U(1))))$ is related to $\frac{c_1(\U(1))(\sigma-\rF \cdot \rF )}{8}$
in Table \ref{table:SU3SU2U1Bordism}.
{Here the $c_2(\SU(2))\tilde\eta$ in 5d captures the Witten anomaly in 4d.
}
{See Appendix \ref{sec:comment} for comment on the difference in 5d between Table \ref{table:SU3SU2U1TP} and \ref{table:SU3SU2U1Z2TP}.
We use the notation ``$\sim$'' to indicate the two sides are equal in that dimension up to a total derivative term.}
}
\label{table:SU3SU2U1TP}
\end{table}

{In Table \ref{table:SU3SU2U1TP}, note that \footnote{\label{ft:Chern-Simons}Locally
$\dd(\text{CS}_1^{\U(1)}c_2(\SU(2)))=\dd(c_1(\U(1)) \text{CS}_3^{\SU(2)})=c_1(\U(1))c_2(\SU(2))$, so by the Poincar\'e Lemma, $\text{CS}_1^{\U(1)}c_2(\SU(2))$ and $c_1(\U(1)) \text{CS}_3^{\SU(2)}$ differ by an exact form locally. The locally defined Chern-Simons form can be glued together to be defined globally, and globally $\text{CS}_1^{\U(1)}c_2(\SU(2))= c_1(\U(1)) \text{CS}_3^{\SU(2)}$ up to a total derivative term (vanishing on a closed 5-manifold).}
 $\text{CS}_1^{\U(1)}c_2(\SU(2))= c_1(\U(1)) \text{CS}_3^{\SU(2)}$ 
and $\text{CS}_1^{\U(1)}c_2(\SU(3))= c_1(\U(1)) \text{CS}_3^{\SU(3)}$ up to a total derivative term (vanishing on a closed 5-manifold).}

\subsection{
${\Spin\times \frac{\SU(3)\times \SU(2)\times \U(1)}{\Z_2}}$ model
}
We consider $G={\Spin\times \frac{\SU(3)\times \SU(2)\times \U(1)}{\Z_2}}=\Spin\times\SU(3)\times\U(2)$, the Madsen-Tillmann spectrum $MTG$ of the group $G$ is 
\bea
MTG=M\Spin\wedge(\B(\SU(3)\times\U(2)))_+.
\eea
 The $(\B(\SU(3)\times \U(2)))_+$ is the disjoint union of the classifying space $\B(\SU(3)\times \U(2))$ and a point, see footnote \ref{ft:X_+}.

For the dimension $d=t-s<8$, since there is no odd torsion (see footnote \ref{ft:no-odd-torsion}), by \eqref{eq:ExtA_2(1)},  we have the Adams spectral sequence
\bea
\Ext_{\A_2(1)}^{s,t}(\H^*(\B(\SU(3)\times \U(2)),\Z_2),\Z_2)\Rightarrow\Omega_{t-s}^{\Spin\times {\SU(3)\times \U(2)}}.
\eea

By K\"unneth formula, we have
\bea
\H^*(\B(\SU(3)\times \U(2)),\Z_2)=\Z_2[c_2,c_3]\otimes\Z_2[c_1',c_2'].
\eea
{Here only in this subsection, $c_i$ is the Chern class of SU(3) bundle, and $c_i'$ is the Chern class of U(2) bundle.}

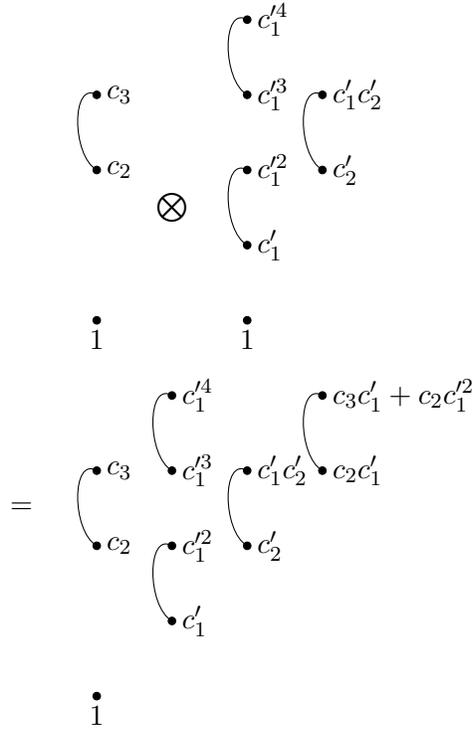
\begin{figure}[H]
\begin{center}
\begin{tikzpicture}[scale=0.5]

\node[below] at (0,0) {1};

\draw[fill] (0,0) circle(.1);
\draw[fill] (0,4) circle(.1);
\draw[fill] (0,6) circle(.1);
\draw (0,4) to [out=150,in=150] (0,6);

\node[right] at (0,4) {$c_2$};

\node[right] at (0,6) {$c_3$};

\node at (2,3) {$\bigotimes$};

\node[below] at (4,0) {1};

\draw[fill] (4,0) circle(.1);

\draw[fill] (4,2) circle(.1);

\draw[fill] (4,4) circle(.1);
\draw (4,2) to [out=150,in=150] (4,4);

\draw[fill] (4,6) circle(.1);

\draw[fill] (4,8) circle(.1);

\draw (4,6) to [out=150,in=150] (4,8);

\node[right] at (4,2) {$c_1'$};

\node[right] at (4,4) {$c_1'^2$};

\node[right] at (4,6) {$c_1'^3$};

\node[right] at (4,8) {$c_1'^4$};

\draw[fill] (6,4) circle(.1);
\draw (6,4) to [out=150,in=150] (6,6);

\draw[fill] (6,6) circle(.1);

\node[right] at (6,4) {$c_2'$};

\node[right] at (6,6) {$c_1'c_2'$};

\node at (-2,-5) {$=$};

\node[below] at (0,-10) {1};

\draw[fill] (0,-10) circle(.1);
\draw[fill] (0,-6) circle(.1);
\draw[fill] (0,-4) circle(.1);
\draw (0,-6) to [out=150,in=150] (0,-4);

\node[right] at (0,-6) {$c_2$};

\node[right] at (0,-4) {$c_3$};

\draw[fill] (2,-8) circle(.1);

\draw[fill] (2,-6) circle(.1);
\draw (2,-8) to [out=150,in=150] (2,-6);

\draw[fill] (2,-4) circle(.1);

\draw[fill] (2,-2) circle(.1);

\draw (2,-4) to [out=150,in=150] (2,-2);

\node[right] at (2,-8) {$c_1'$};

\node[right] at (2,-6) {$c_1'^2$};

\node[right] at (2,-4) {$c_1'^3$};

\node[right] at (2,-2) {$c_1'^4$};

\draw[fill] (4,-6) circle(.1);
\draw (4,-6) to [out=150,in=150] (4,-4);

\draw[fill] (4,-4) circle(.1);

\node[right] at (4,-6) {$c_2'$};

\node[right] at (4,-4) {$c_1'c_2'$};

\draw[fill] (6,-4) circle(.1);

\node[right] at (6,-4) {$c_2c_1'$};

\draw[fill] (6,-2) circle(.1);

\node[right] at (6,-2) {$c_3c_1'+c_2c_1'^2$};

\draw (6,-4) to [out=150,in=150] (6,-2);

\end{tikzpicture}
\end{center}
\caption{The $\A_2(1)$-module structure of $\H^*(\B(\SU(3)\times \U(2)),\Z_2)$ below degree 6.}
\label{fig:A_2(1)SU3U2}
\end{figure}

The $\A_2(1)$-module structure of $\H^*(\B(\SU(3)\times \U(2)),\Z_2)$ below degree 6
and the $E_2$ page are shown in Figure \ref{fig:A_2(1)SU3U2}, \ref{fig:E_2SU3U2}.
Here we have used the correspondence between $\A_2(1)$-module structure and the $E_2$ page shown in Figure \ref{fig:Z_2} and \ref{fig:Ceta}.

\begin{figure}[H]
\begin{center}
\begin{tikzpicture}
\node at (0,-1) {0};
\node at (1,-1) {1};
\node at (2,-1) {2};
\node at (3,-1) {3};
\node at (4,-1) {4};
\node at (5,-1) {5};
\node at (6,-1) {6};
\node at (7,-1) {$t-s$};
\node at (-1,0) {0};
\node at (-1,1) {1};
\node at (-1,2) {2};
\node at (-1,3) {3};
\node at (-1,4) {4};
\node at (-1,5) {5};
\node at (-1,6) {$s$};

\draw[->] (-0.5,-0.5) -- (-0.5,6);
\draw[->] (-0.5,-0.5) -- (7,-0.5);

\draw (0,0) -- (0,5);

\draw (0,0) -- (2,2);

\draw (2.1,0) -- (2.1,5);

\draw (4,3) -- (4,5);

\draw (3.9,0) -- (3.9,5);

\draw (4.1,1) -- (4.1,5);

\draw (4.2,0) -- (4.2,5);

\draw (6,1) -- (6,5);

\draw (6.1,2) -- (6.1,5);

\draw (6.2,1) -- (6.2,5);

\draw (6.3,0) -- (6.3,5);

\draw (6.4,0) -- (6.4,5);

\end{tikzpicture}
\end{center}
\caption{$\Omega_*^{\Spin\times \SU(3)\times \U(2)}$.}
\label{fig:E_2SU3U2}
\end{figure}



Thus we obtain the bordism group $\Omega^{\Spin\times \frac{\SU(3)\times \SU(2)\times \U(1)}{\Z_2}}_d$ shown in Table \ref{table:SU3SU2U1Z2Bordism}.

\begin{table}[H]
\centering
\hspace*{-14mm}
\begin{tabular}{ c c c}
\hline
\multicolumn{3}{c}{Bordism group}\\
\hline
$d$ & 
$\Omega^{\Spin\times \frac{\SU(3)\times \SU(2)\times \U(1)}{\Z_2}}_d$
& bordism invariants \\
\hline
0& $\Z$ \\
\hline
1& $\Z_2$  & $\tilde\eta$\\
\hline
2&  $\Z\times\Z_2$ & $c_1(\U(2))$, Arf \\
\hline
3 & 0\\
\hline
4 &  $\Z^4$ & $\frac{\sigma}{16}  = \frac{p_1(TM)}{48}, \frac{1}{2}c_1(\U(2))^2, c_2(\U(2)), c_2(\SU(3))$ \\
\hline
5 & 0  & \\
\hline
6 & $\Z^5$ &  
$\begin{matrix}
\frac{c_1(\U(2))(\sigma-\rF \cdot \rF )}{8},c_1(\U(2))^3, \frac{c_1(\U(2))c_2(\U(2))}{2}, \\
 c_1(\U(2))c_2(\SU(3)), \frac{c_3(\SU(3))}{2}
\end{matrix}$\\
\hline
\end{tabular}
\caption{{Bordism group. 
$\tilde\eta$ is a mod 2 index of 1d Dirac operator.
Arf is a 2d Arf invariant.
$\sigma  = \frac{p_1(TM)}{3}$ is the signature of manifold.
$c_i(G)$ is the Chern class of the associated vector bundle of the principal $G$-bundle.
$\rF $ is the characteristic 2-surface \cite{Saveliev} in a 4-manifold $M^4$, it satisfies the condition $\rF\cdot x=x\cdot x\mod2$ for all $x\in\H_2(M^4,\Z)$.
Here $\cdot$ is the intersection form of $M^4$.
By the Freedman-Kirby theorem, $(\frac{\sigma-\rF\cdot\rF}{8} )(M^4)=\text{Arf}(M^4,\rF)\mod2$.
Note that $c_1(\U(2))^2=\Sq^2c_1(\U(2))=(w_2+w_1^2)c_1(\U(2))=0\mod2$ on Spin 4-manifolds, $c_3(\SU(3))=\Sq^2c_2(\SU(3))=(w_2+w_1^2)c_2(\SU(3))=0\mod2$ on Spin 6-manifolds.
}}
\label{table:SU3SU2U1Z2Bordism}
\end{table}

By \eqref{eq:TPexact}, we obtain the cobordism group $\TP_d(\Spin\times \frac{\SU(3)\times \SU(2)\times \U(1)}{\Z_2})$ shown in Table \ref{table:SU3SU2U1Z2TP}.

\begin{table}[H]
\centering
\hspace*{-8mm}
\begin{tabular}{ c c c}
\hline
\multicolumn{3}{c}{Cobordism group}\\
\hline
$d$ & 
$\TP_d(\Spin\times \frac{\SU(3)\times \SU(2)\times \U(1)}{\Z_2})$
& topological terms \\
\hline
0& 0 \\
\hline
1& $\Z\times\Z_2$ & $\text{CS}_1^{\U(2)},\tilde\eta$\\
\hline
2& $\Z_2$ & Arf \\
\hline
3 & $\Z^4$ & $\frac{\text{CS}_3^{TM}}{48}, \frac{1}{2}\text{CS}_1^{\U(2)}c_1(\U(2)), \text{CS}_3^{\U(2)},\text{CS}_3^{\SU(3)}$ \\
\hline
4 & 0 \\
\hline
5 & $\Z^5$ &  
$\begin{matrix}\mu(\text{PD}(c_1(\U(2)))), \text{CS}_1^{\U(2)}c_1(\U(2))^2, \frac{1}{2}\text{CS}_1^{\U(2)}c_2(\U(2)) {\sim  c_1(\U(2)) \text{CS}_3^{\U(2)}},\\ 
 \text{CS}_1^{\U(2)}c_2(\SU(3))
{\sim c_1(\U(2)) \text{CS}_3^{\SU(3)}},  \frac{1}{2}\text{CS}_5^{\SU(3)}
\end{matrix}$\\
\hline
\end{tabular}
\caption{Topological phase classification ($\equiv$ TP) as a cobordism group, following Table \ref{table:SU3SU2U1Z2Bordism}. 
$\tilde\eta$ is a mod 2 index of 1d Dirac operator.
Arf is a 2d Arf invariant.
$c_i(G)$ is the Chern class of the associated vector bundle of the principal $G$-bundle.
$\text{CS}_{2n-1}^{V}$ or $\text{CS}_{2n-1}^{G}$ is the Chern-Simons form of the vector bundle $V$ or the associated vector bundle of the principal $G$-bundle.
The PD is the Poincar\'e dual.
The $TM$ is the spacetime tangent bundle.
The $\mu$ is the 3d Rokhlin invariant.
If $\partial M^4=M^3$, then $\mu(M^3)=(\frac{\sigma-\rF\cdot\rF}{8} )(M^4)$, thus
$\mu(\text{PD}(c_1(\U(2))))$ is related to $\frac{c_1(\U(2))(\sigma-\rF \cdot \rF )}{8}$
in Table \ref{table:SU3SU2U1Z2Bordism}.
{See Appendix \ref{sec:comment} for comment on the difference in 5d between Table \ref{table:SU3SU2U1TP} and \ref{table:SU3SU2U1Z2TP}.}
}
\label{table:SU3SU2U1Z2TP}
\end{table}
{In Table \ref{table:SU3SU2U1Z2TP}, note that $\text{CS}_1^{\U(2)}c_2(\U(2))= c_1(\U(2)) \text{CS}_3^{\U(2)}$
and $\text{CS}_1^{\U(2)}c_2(\SU(3))= c_1(\U(2)) \text{CS}_3^{\SU(3)}$ up to a total derivative term (vanishing on a closed 5-manifold). See footnote \ref{ft:Chern-Simons}.}


\subsection{
${\Spin\times \frac{\SU(3)\times \SU(2)\times \U(1)}{\Z_3}}$ model
}\label{sec:SU3SU2U1Z3}
We consider $G={\Spin\times \frac{\SU(3)\times \SU(2)\times \U(1)}{\Z_3}}=\Spin\times\U(3)\times\SU(2)$, the Madsen-Tillmann spectrum $MTG$ of the group $G$ is 
\bea
MTG=M\Spin\wedge(\B(\U(3)\times\SU(2)))_+.
\eea
 The $(\B(\U(3)\times \SU(2)))_+$ is the disjoint union of the classifying space $\B(\U(3)\times \SU(2))$ and a point, see footnote \ref{ft:X_+}.

The localization of $M\Spin$ at the prime 3 is the wedge sum of suspensions of the Brown-Peterson spectrum $BP$ (here $M\Spin_{(3)}=BP\vee\Sigma^8BP\vee\cdots$) and $\H^*(BP,\Z_3)=\A_3/(\beta_{(3,3)})$ where $(\beta_{(3,3)})$ is the two-sided ideal generated by $\beta_{(3,3)}$, and $\beta_{(3,3)}$ is the Bockstein homomorphism associated to the extension $\Z_3\to\Z_9\to\Z_3$. Note that
\bea
\cdots\To\Sigma^2\A_3\oplus\Sigma^6\A_3\oplus\cdots\To\Sigma\A_3\oplus\Sigma^5\A_3\oplus\cdots\To\A_3\To\A_3/(\beta_{(3,3)})
\eea
is {a projective} $\A_3$-resolution of $\A_3/(\beta_{(3,3)})$ (denoted by $P_{\bullet}$) where the differentials $d_1$ are induced by $\beta_{(3,3)}$.

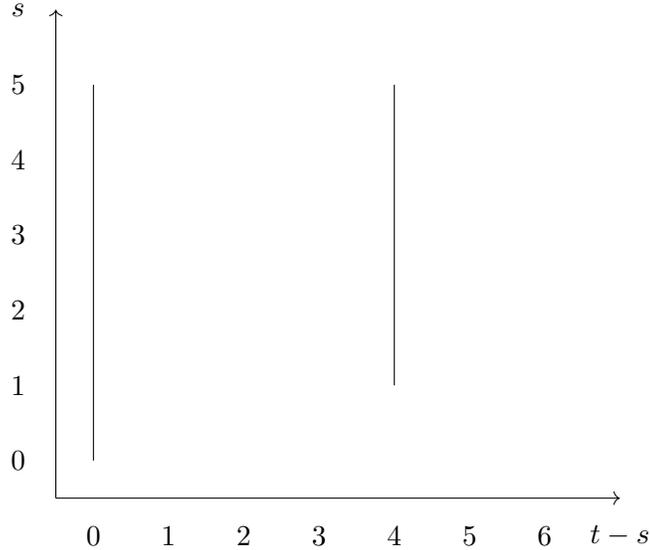
\begin{figure}[H]
\begin{center}
\begin{tikzpicture}
\node at (0,-1) {0};
\node at (1,-1) {1};
\node at (2,-1) {2};
\node at (3,-1) {3};
\node at (4,-1) {4};
\node at (5,-1) {5};
\node at (6,-1) {6};
\node at (7,-1) {$t-s$};
\node at (-1,0) {0};
\node at (-1,1) {1};
\node at (-1,2) {2};
\node at (-1,3) {3};
\node at (-1,4) {4};
\node at (-1,5) {5};

\node at (-1,6) {$s$};

\draw[->] (-0.5,-0.5) -- (-0.5,6);
\draw[->] (-0.5,-0.5) -- (7,-0.5);

\draw (0,0) -- (0,5);

\draw (4,1) -- (4,5);

\end{tikzpicture}
\end{center}
\caption{Adams chart of $\Ext_{\A_3}^{s,t}(\H^*(M\Spin,\Z_3),\Z_3)$.}
\label{fig:A_3}
\end{figure}

The Adams chart of $\Ext_{\A_3}^{s,t}(\H^*(M\Spin,\Z_3),\Z_3)$ is shown in Figure \ref{fig:A_3}. $P_{\bullet}\otimes\H^*(\B(\U(3)\times\SU(2)),\Z_3)$ is a projective $\A_3$-resolution of $\H^*(BP,\Z_3)\otimes\H^*(\B(\U(3)\times\SU(2)),\Z_3)$ (since $P_{\bullet}$ is actually a free $\A_3$-resolution), the differentials $d_1$ are induced by $\beta_{(3,3)}$.

We have the Adams spectral sequence
\bea
\Ext_{\A_3}^{s,t}(\H^*(M\Spin,\Z_3)\otimes\H^*(\B(\U(3)\times\SU(2)),\Z_3),\Z_3)\Rightarrow(\Omega_{t-s}^{\Spin\times \U(3)\times\SU(2)})_3^{\wedge}.
\eea

 By K\"unneth formula, we have
\bea
\H^*(\B(\U(3)\times\SU(2)),\Z_3)=\Z_3[c_1,c_2,c_3]\otimes\Z_3[c_2'].
\eea
{Here only in this subsection, $c_i$ is the Chern class of U(3) bundle, and $c_i'$ is the Chern class of SU(2) bundle.}

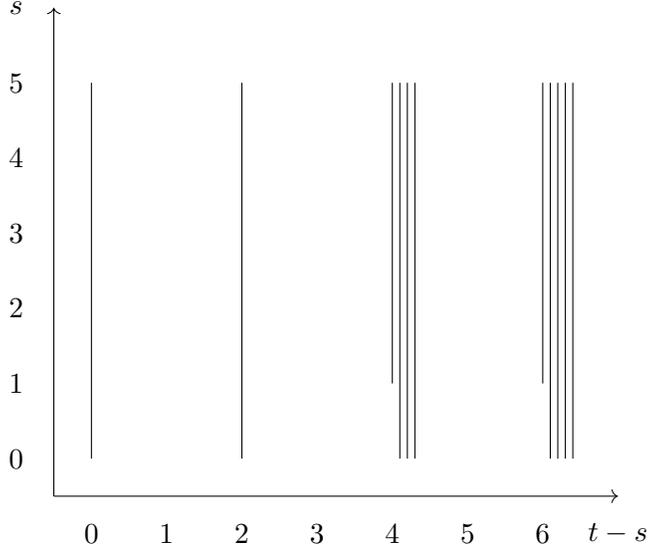
\begin{figure}[H]
\begin{center}
\begin{tikzpicture}
\node at (0,-1) {0};
\node at (1,-1) {1};
\node at (2,-1) {2};
\node at (3,-1) {3};
\node at (4,-1) {4};
\node at (5,-1) {5};
\node at (6,-1) {6};
\node at (7,-1) {$t-s$};
\node at (-1,0) {0};
\node at (-1,1) {1};
\node at (-1,2) {2};
\node at (-1,3) {3};
\node at (-1,4) {4};
\node at (-1,5) {5};

\node at (-1,6) {$s$};

\draw[->] (-0.5,-0.5) -- (-0.5,6);
\draw[->] (-0.5,-0.5) -- (7,-0.5);

\draw (0,0) -- (0,5);

\draw (2,0) -- (2,5);
\draw (4,1) -- (4,5);
\draw (4.1,0) -- (4.1,5);
\draw (4.2,0) -- (4.2,5);
\draw (4.3,0) -- (4.3,5);

\draw (6,1) -- (6,5);

\draw (6.1,0) -- (6.1,5);
\draw (6.2,0) -- (6.2,5);
\draw (6.3,0) -- (6.3,5);
\draw (6.4,0) -- (6.4,5);
\end{tikzpicture}
\end{center}
\caption{Adams chart of $\Ext_{\A_3}^{s,t}(\H^*(M\Spin,\Z_3)\otimes\H^*(\B(\U(3)\times\SU(2)),\Z_3),\Z_3)$.}
\label{fig:A_3U3SU2}
\end{figure}

The Adams chart of $\Ext_{\A_3}^{s,t}(\H^*(M\Spin,\Z_3)\otimes\H^*(\B(\U(3)\times\SU(2)),\Z_3),\Z_3)$ is shown in Figure \ref{fig:A_3U3SU2}. There is no differential since the arrow of the differential $d_r$ is of bidegree $(-1,r)$, while all lines are of interval 2 at degree $t-s$.

So there is actually no 3-torsion in $\Omega_d^{\Spin\times \U(3)\times\SU(2)}$.

For the dimension $d=t-s<8$, since there is no odd torsion (see footnote \ref{ft:no-odd-torsion}), by \eqref{eq:ExtA_2(1)},  we have the Adams spectral sequence
\bea
\Ext_{\A_2(1)}^{s,t}(\H^*(\B(\U(3)\times \SU(2)),\Z_2),\Z_2)\Rightarrow\Omega_{t-s}^{\Spin\times {\U(3)\times \SU(2)}}.
\eea

By K\"unneth formula, we have
\bea
\H^*(\B(\U(3)\times \SU(2)),\Z_2)=\Z_2[c_1,c_2,c_3]\otimes\Z_2[c_2'].
\eea
{Here only in this subsection, $c_i$ is the Chern class of U(3) bundle, and $c_i'$ is the Chern class of SU(2) bundle.}

\begin{figure}[H]
\begin{center}
\begin{tikzpicture}[scale=0.5]

\node[below] at (0,0) {1};

\draw[fill] (0,0) circle(.1);

\draw[fill] (0,2) circle(.1);

\draw[fill] (0,4) circle(.1);
\draw[fill] (0,6) circle(.1);
\draw[fill] (0,8) circle(.1);
\draw (0,2) to [out=150,in=150] (0,4);

\draw (0,6) to [out=150,in=150] (0,8);

\node[right] at (0,2) {$c_1$};

\node[right] at (0,4) {$c_1^2$};

\node[right] at (0,6) {$c_1^3$};

\node[right] at (0,8) {$c_1^4$};

\draw[fill] (2,4) circle(.1);

\draw[fill] (2,6) circle(.1);

\draw (2,4) to [out=150,in=150] (2,6);

\node[right] at (2,4) {$c_2$};

\node[right] at (2,6) {$c_1c_2+c_3$};

\draw[fill] (6,6) circle(.1);

\draw[fill] (6,8) circle(.1);

\draw (6,6) to [out=150,in=150] (6,8);

\node[right] at (6,6) {$c_3$};

\node[right] at (6,8) {$c_1c_3$};

\node at (8,3) {$\bigotimes$};

\draw[fill] (10,0) circle(.1);

\draw[fill] (10,4) circle(.1);

\node[below] at (10,0) {1};

\node[right] at (10,4) {$c_2'$};

\node at (-2,-5) {$=$};

\node[below] at (0,-10) {1};

\draw[fill] (0,-10) circle(.1);

\draw[fill] (0,-8) circle(.1);

\draw[fill] (0,-6) circle(.1);
\draw (0,-8) to [out=150,in=150] (0,-6);

\draw[fill] (0,-4) circle(.1);

\draw[fill] (0,-2) circle(.1);

\draw (0,-4) to [out=150,in=150] (0,-2);

\node[right] at (0,-8) {$c_1$};

\node[right] at (0,-6) {$c_1^2$};

\node[right] at (0,-4) {$c_1^3$};

\node[right] at (0,-2) {$c_1^4$};

\draw[fill] (2,-6) circle(.1);
\draw[fill] (2,-4) circle(.1);
\draw (2,-6) to [out=150,in=150] (2,-4);

\node[right] at (2,-6) {$c_2$};

\node[right] at (2,-4) {$c_1c_2+c_3$};

\draw[fill] (6,-4) circle(.1);

\node[right] at (6,-4) {$c_3$};

\draw[fill] (6,-2) circle(.1);

\node[right] at (6,-2) {$c_1c_3$};

\draw (6,-4) to [out=150,in=150] (6,-2);

\draw[fill] (8,-6) circle(.1);

\node[right] at (8,-6) {$c_2'$};

\draw[fill] (10,-4) circle(.1);

\node[right] at (10,-4) {$c_1c_2'$};

\draw[fill] (10,-2) circle(.1);

\node[right] at (10,-2) {$c_1^2c_2'$};

\draw (10,-4) to [out=150,in=150] (10,-2);

\end{tikzpicture}
\end{center}
\caption{The $\A_2(1)$-module structure of $\H^*(\B(\U(3)\times \SU(2)),\Z_2)$ below degree 6.}
\label{fig:A_2(1)U3SU2}
\end{figure}
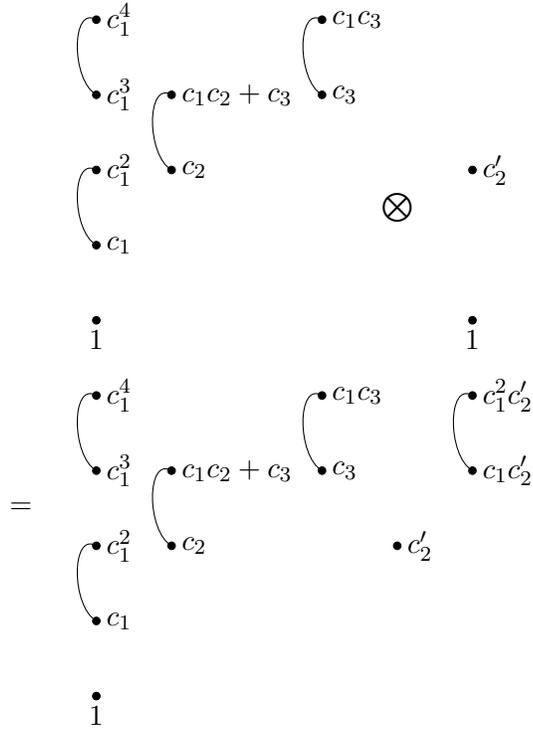

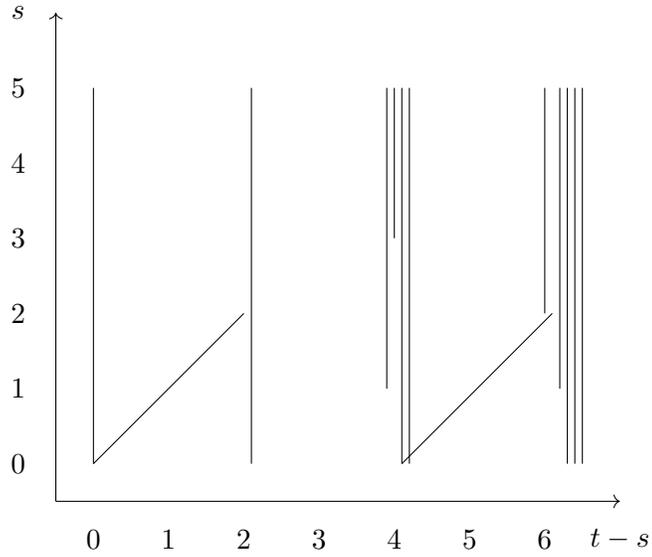
\begin{figure}[H]
\begin{center}
\begin{tikzpicture}
\node at (0,-1) {0};
\node at (1,-1) {1};
\node at (2,-1) {2};
\node at (3,-1) {3};
\node at (4,-1) {4};
\node at (5,-1) {5};
\node at (6,-1) {6};
\node at (7,-1) {$t-s$};
\node at (-1,0) {0};
\node at (-1,1) {1};
\node at (-1,2) {2};
\node at (-1,3) {3};
\node at (-1,4) {4};
\node at (-1,5) {5};
\node at (-1,6) {$s$};

\draw[->] (-0.5,-0.5) -- (-0.5,6);
\draw[->] (-0.5,-0.5) -- (7,-0.5);

\draw (0,0) -- (0,5);

\draw (0,0) -- (2,2);

\draw (2.1,0) -- (2.1,5);

\draw (4,3) -- (4,5);

\draw (3.9,1) -- (3.9,5);

\draw (4.1,0) -- (4.1,5);

\draw (4.1,0) -- (6.1,2);

\draw (4.2,0) -- (4.2,5);

\draw (6,2) -- (6,5);

\draw (6.2,1) -- (6.2,5);

\draw (6.3,0) -- (6.3,5);

\draw (6.4,0) -- (6.4,5);

\draw (6.5,0) -- (6.5,5);

\end{tikzpicture}
\end{center}
\caption{$\Omega_*^{\Spin\times \U(3)\times \SU(2)}$}
\label{fig:E_2U3SU2}
\end{figure}

The $\A_2(1)$-module structure of $\H^*(\B(\U(3)\times \SU(2)),\Z_2)$ below degree 6
and the $E_2$ page are shown in Figure \ref{fig:A_2(1)U3SU2}, \ref{fig:E_2U3SU2}.
Here we have used the correspondence between $\A_2(1)$-module structure and the $E_2$ page shown in Figure \ref{fig:Z_2} and \ref{fig:Ceta}.

Thus we obtain the bordism group $\Omega^{\Spin\times \frac{\SU(3)\times \SU(2)\times \U(1)}{\Z_3}}_d$ shown in Table \ref{table:SU3SU2U1Z3Bordism}.

\begin{table}[H]
\centering
\hspace*{-25mm}
\begin{tabular}{ c c c}
\hline
\multicolumn{3}{c}{Bordism group}\\
\hline
$d$ & 
$\Omega^{\Spin\times \frac{\SU(3)\times \SU(2)\times \U(1)}{\Z_3}}_d$
& bordism invariants \\
\hline
0& $\Z$ \\
\hline
1& $\Z_2$  & $\tilde\eta$\\
\hline
2& $\Z\times\Z_2$  & $c_1(\U(3))$, Arf \\
\hline
3 & 0 \\
\hline
4 & $\Z^4$ & $\frac{\sigma}{16}, \frac{1}{2}c_1(\U(3))^2, c_2(\SU(2)) ,  c_2(\U(3))$ \\
\hline
5 &  $\Z_2$ & $c_2(\SU(2))\tilde\eta$\\
\hline
6 & $\Z^5\times\Z_2$ & 
$\begin{matrix}\frac{c_1(\U(3))(\sigma-\rF \cdot \rF )}{8}, 
c_1(\U(3))^3,
c_1(\U(3))c_2(\SU(2)),\\
\frac{c_1(\U(3))c_2(\U(3))+c_3(\U(3))}{2},
c_3(\U(3)), c_2(\SU(2))\text{Arf}
\end{matrix}$  \\
\hline
\end{tabular}
\caption{{Bordism group. 
$\tilde\eta$ is a mod 2 index of 1d Dirac operator.
Arf is a 2d Arf invariant.
$\sigma$ is the signature of manifold.
$c_i(G)$ is the Chern class of the associated vector bundle of the principal $G$-bundle.
$\rF $ is the characteristic 2-surface \cite{Saveliev} in a 4-manifold $M^4$, it satisfies the condition $\rF\cdot x=x\cdot x\mod2$ for all $x\in\H_2(M^4,\Z)$.
Here $\cdot$ is the intersection form of $M^4$.
By the Freedman-Kirby theorem, $(\frac{\sigma-\rF\cdot\rF}{8} )(M^4)=\text{Arf}(M^4,\rF)\mod2$.
Note that $c_1(\U(3))^2=\Sq^2c_1(\U(3))=(w_2+w_1^2)c_1(\U(3))=0\mod2$ on Spin 4-manifolds, $c_1(\U(3))c_2(\U(3))+c_3(\U(3))=\Sq^2c_2(\U(3))=(w_2+w_1^2)c_2(\U(3))=0\mod2$ on Spin 6-manifolds.
}}
\label{table:SU3SU2U1Z3Bordism}
\end{table}

By \eqref{eq:TPexact}, we obtain the cobordism group $\TP_d(\Spin\times \frac{\SU(3)\times \SU(2)\times \U(1)}{\Z_3})$ shown in Table \ref{table:SU3SU2U1Z3TP}.

\begin{table}[H]
\centering
\hspace*{-8mm}
\begin{tabular}{ c c c}
\hline
\multicolumn{3}{c}{Cobordism group}\\
\hline
$d$ & 
$\TP_d(\Spin\times \frac{\SU(3)\times \SU(2)\times \U(1)}{\Z_3})$
& topological terms \\
\hline
0& 0 \\
\hline
1& $\Z\times\Z_2$ & $\text{CS}_1^{\U(3)},\tilde\eta$\\
\hline
2& $\Z_2$ & Arf\\
\hline
3 & $\Z^4$ & $\frac{\text{CS}_3^{TM}}{48}, \frac{1}{2}\text{CS}_1^{\U(3)}c_1(\U(3)),
\text{CS}_3^{\SU(2)},  \text{CS}_3^{\U(3)}$\\
\hline
4 & 0 \\
\hline
5 & $\Z^5\times\Z_2$ & 
$\begin{matrix}\mu(\text{PD}(c_1(\U(3)))), 
\text{CS}_1^{\U(3)}c_1(\U(3))^2,
\text{CS}_1^{\U(3)}c_2(\SU(2)) {\sim c_1(\U(3)) \text{CS}_3^{\SU(2)}},
 \\
\frac{\text{CS}_1^{\U(3)}c_2(\U(3))+\text{CS}_5^{\U(3)}}{2}
{\sim 
\frac{
 c_1(\U(3)) \text{CS}_3^{\U(3)}
+\text{CS}_5^{\U(3)}}{2}
}
, \text{CS}_5^{\U(3)},  c_2(\SU(2))\tilde\eta
 \end{matrix}$ \\
\hline
\end{tabular}
\caption{Topological phase classification ($\equiv$ TP) as a cobordism group, following Table \ref{table:SU3SU2U1Z3Bordism}. 
$\tilde\eta$ is a mod 2 index of 1d Dirac operator.
Arf is a 2d Arf invariant.
$c_i(G)$ is the Chern class of the associated vector bundle of the principal $G$-bundle.
$\text{CS}_{2n-1}^{V}$ or $\text{CS}_{2n-1}^{G}$ is the Chern-Simons form of the vector bundle $V$ or the associated vector bundle of the principal $G$-bundle.
The PD is the Poincar\'e dual.
The $TM$ is the spacetime tangent bundle.
The $\mu$ is the 3d Rokhlin invariant.
If $\partial M^4=M^3$, then $\mu(M^3)=(\frac{\sigma-\rF\cdot\rF}{8} )(M^4)$, thus
$\mu(\text{PD}(c_1(\U(3))))$ is related to $\frac{c_1(\U(3))(\sigma-\rF \cdot \rF )}{8}$
in Table \ref{table:SU3SU2U1Z3Bordism}.
}
\label{table:SU3SU2U1Z3TP}
\end{table}
{In Table \ref{table:SU3SU2U1Z3TP}, note that $\text{CS}_1^{\U(3)} c_2(\U(3))= c_1(\U(3)) \text{CS}_3^{\U(3)}$ 
and $\text{CS}_1^{\U(3)}c_2(\SU(2))= c_1(\U(3)) \text{CS}_3^{\SU(2)}$ up to a total derivative term (vanishing on a closed 5-manifold). See footnote \ref{ft:Chern-Simons}.}


\subsection{
${\Spin\times \frac{\SU(3)\times \SU(2)\times \U(1)}{\Z_6}}$ model
}
We consider $G={\Spin\times \frac{\SU(3)\times \SU(2)\times \U(1)}{\Z_6}}=\Spin\times \rS(\U(3)\times \U(2))$.
This SM group is particularly interesting because:
\bea
\Spin(10) \supset \SU(5) \supset \frac{\SU(3)\times \SU(2)\times \U(1)}{\Z_6}, \text{ and } \SO(10) \supset \SU(5) \supset \frac{\SU(3)\times \SU(2)\times \U(1)}{\Z_6}.
\eea 
Thus the SO(10) GUT and SO(5) GUT can be Higgs down to $\frac{\SU(3)\times \SU(2)\times \U(1)}{\Z_6}$ SM.
We have
\bea
(\SU(3) \times \SU(2) \times \U(1))/\Z_6 
= (\SU(3) \times \U(2) )/\Z_3 
= (\SU(2) \times \U(3) )/\Z_2.
\eea


Let the hypercharge be:
\bea
Y \equiv \frac{Y'}{2}\equiv \frac{\tilde Y}{6}\equiv\text{diag}(-\frac{1}{3},-\frac{1}{3},-\frac{1}{3},\frac{1}{2},\frac{1}{2}).
\eea
The group $\frac{\SU(3)\times \SU(2)\times \U(1)}{\Z_6}$ is just the subgroup of $\SU(5)$ commuting with the group generated by $Y$.

The subgroup is $\rS(\U(3)\times \U(2))={(A,B)\in \U(3)\times \U(2) | \det A\cdot \det B=1} $
and
$(\SU(3) \times \SU(2) \times \U(1))/\Z_6 = \rS(\U(3)\times \U(2)) \subset \SU(5)$.



Since 
\bea
\H^*(\B(\frac{\SU(3)\times \SU(2)\times \U(1)}{\Z_6}),\Z_3)=\H^*(\B(\frac{\SU(3)\times \SU(2)\times \U(1)}{\Z_3}),\Z_3),
\eea
similarly as the discussion in \Sec{sec:SU3SU2U1Z3}, there is no 3-torsion in $\Omega_d^{\Spin\times \rS({\U(3)\times \U(2)})}$.

The Madsen-Tillmann spectrum $MTG$ of the group $G$ is 
\bea
MTG=M\Spin\wedge(\B(\rS({\U(3)\times \U(2)})))_+.
\eea
 The $(\B(\rS(\U(3)\times \U(2))))_+$ is the disjoint union of the classifying space $\B(\rS(\U(3)\times \U(2)))$ and a point, see footnote \ref{ft:X_+}.

For the dimension $d=t-s<8$, since there is no odd torsion (see footnote \ref{ft:no-odd-torsion}), by \eqref{eq:ExtA_2(1)},  we have the Adams spectral sequence
\bea
\Ext_{\A_2(1)}^{s,t}(\H^*(\B (\rS(\U(3)\times \U(2))),\Z_2),\Z_2)\Rightarrow\Omega_{t-s}^{\Spin\times \rS({\U(3)\times \U(2)})}.
\eea

We have the following commutative diagram with exact columns
\bea \label{eq:U3U2detmap}
\xymatrix{\rS(\U(3)\times \U(2))\ar[d]\ar@{^{(}->}[r]&\SU(5)\ar[d]\\
\U(3)\times\U(2)\ar[d]^{\det}\ar@{^{(}->}[r]&\U(5)\ar[d]^{\det}\\
\U(1)&\U(1).}
\eea

So by K\"unneth formula, we have
\bea
\H^*(\B (\rS(\U(3)\times \U(2))),\Z_2)=\Z_2[c_1,c_2,c_3]\otimes\Z_2[c_1',c_2']/(c_1=c_1').
\eea
{Here only in this subsection, $c_i$ is the Chern class of U(3) bundle, and $c_i'$ is the Chern class of U(2) bundle.}

The $\A_2(1)$-module structure of $\H^*(\B (\rS(\U(3)\times \U(2))),\Z_2)$ below degree 6
and the $E_2$ page are shown in Figure \ref{fig:A_2(1)S(U3U2)}, \ref{fig:E_2S(U3U2)}.
Here we have used the correspondence between $\A_2(1)$-module structure and the $E_2$ page shown in Figure \ref{fig:Z_2} and \ref{fig:Ceta}.

\begin{figure}[H]
\begin{center}
\begin{tikzpicture}[scale=0.5]

\node[below] at (0,0) {1};

\draw[fill] (0,0) circle(.1);

\draw[fill] (0,2) circle(.1);

\draw[fill] (0,4) circle(.1);
\draw[fill] (0,6) circle(.1);
\draw[fill] (0,8) circle(.1);
\draw (0,2) to [out=150,in=150] (0,4);

\draw (0,6) to [out=150,in=150] (0,8);

\node[right] at (0,2) {$c_1$};

\node[right] at (0,4) {$c_1^2$};

\node[right] at (0,6) {$c_1^3$};

\node[right] at (0,8) {$c_1^4$};

\draw[fill] (2,4) circle(.1);

\draw[fill] (2,6) circle(.1);

\draw (2,4) to [out=150,in=150] (2,6);

\node[right] at (2,4) {$c_2$};

\node[right] at (2,6) {$c_1c_2+c_3$};

\draw[fill] (6,6) circle(.1);

\draw[fill] (6,8) circle(.1);

\draw (6,6) to [out=150,in=150] (6,8);

\node[right] at (6,6) {$c_3$};

\node[right] at (6,8) {$c_1c_3$};

\draw[fill] (8,4) circle(.1);
\draw[fill] (8,6) circle(.1);

\node[right] at (8,4) {$c_2'$};

\node[right] at (8,6) {$c_1'c_2'=c_1c_2'$};

\draw (8,4) to [out=150,in=150] (8,6);

\end{tikzpicture}
\end{center}
\caption{The $\A_2(1)$-module structure of $\H^*(\B (\rS(\U(3)\times \U(2))),\Z_2)$ below degree 6.}
\label{fig:A_2(1)S(U3U2)}
\end{figure}


\begin{figure}[H]
\begin{center}
\begin{tikzpicture}
\node at (0,-1) {0};
\node at (1,-1) {1};
\node at (2,-1) {2};
\node at (3,-1) {3};
\node at (4,-1) {4};
\node at (5,-1) {5};
\node at (6,-1) {6};
\node at (7,-1) {$t-s$};
\node at (-1,0) {0};
\node at (-1,1) {1};
\node at (-1,2) {2};
\node at (-1,3) {3};
\node at (-1,4) {4};
\node at (-1,5) {5};
\node at (-1,6) {$s$};

\draw[->] (-0.5,-0.5) -- (-0.5,6);
\draw[->] (-0.5,-0.5) -- (7,-0.5);

\draw (0,0) -- (0,5);

\draw (0,0) -- (2,2);

\draw (2.1,0) -- (2.1,5);

\draw (4,3) -- (4,5);

\draw (4.1,1) -- (4.1,5);

\draw (4.2,0) -- (4.2,5);

\draw (4.3,0) -- (4.3,5);

\draw (6,1) -- (6,5);

\draw (6.1,2) -- (6.1,5);

\draw (6.2,1) -- (6.2,5);

\draw (6.3,0) -- (6.3,5);

\draw (6.4,0) -- (6.4,5);

\end{tikzpicture}
\end{center}
\caption{$\Omega_*^{\Spin\times \rS(\U(3)\times \U(2))}$.}
\label{fig:E_2S(U3U2)}
\end{figure}



Thus we obtain the bordism group $\Omega^{\Spin\times \frac{\SU(3)\times \SU(2)\times \U(1)}{\Z_6}}_d$ shown in Table \ref{table:SU3SU2U1Z6Bordism}.

\begin{table}[H]
\centering
\hspace*{-25mm}
\begin{tabular}{c c c}
\hline
\multicolumn{3}{c}{Bordism group}\\
\hline
$d$ & 
$\Omega^{\Spin\times \frac{\SU(3)\times \SU(2)\times \U(1)}{\Z_6}}_d$
& bordism invariants \\
\hline
0& $\Z$ \\
\hline
1& $\Z_2$  & $\tilde\eta$\\
\hline
2&  $\Z\times\Z_2$ & $c_1(\U(3))$, Arf\\
\hline
3 & 0 \\
\hline
4 & $\Z^4$ & $\frac{\sigma}{16}, 
{ \frac{1}{2}c_1(\U(2))^2=\frac{1}{2}c_1(\U(3))^2},
c_2(\U(2)),  c_2(\U(3))$ \\
\hline
5 & 0  & \\
\hline
6 & $\Z^5$  & 
$\begin{matrix}\frac{c_1(\U(3))(\sigma-\rF \cdot \rF )}{8},c_1(\U(3))^3,
\frac{c_1(\U(3))c_2(\U(2))}{2},
\\
\frac{c_1(\U(3))c_2(\U(3))+c_3(\U(3))}{2},c_3(\U(3))
\end{matrix}$ \\
\hline
\end{tabular}
\caption{{Bordism group. 
$\tilde\eta$ is a mod 2 index of 1d Dirac operator.
Arf is a 2d Arf invariant.
$\sigma$ is the signature of manifold.
$c_i(G)$ is the Chern class of the associated vector bundle of the principal $G$-bundle.
$\rF $ is the characteristic 2-surface \cite{Saveliev} in a 4-manifold $M^4$, it satisfies the condition $\rF\cdot x=x\cdot x\mod2$ for all $x\in\H_2(M^4,\Z)$.
Here $\cdot$ is the intersection form of $M^4$.
By the Freedman-Kirby theorem, $(\frac{\sigma-\rF\cdot\rF}{8} )(M^4)=\text{Arf}(M^4,\rF)\mod2$.
Here $c_1(\U(3))$ is identified with $c_1(\U(2))$.
Note that $c_1(\U(3))^2=\Sq^2c_1(\U(3))=(w_2+w_1^2)c_1(\U(3))=0\mod2$ on Spin 4-manifolds, $c_1(\U(3))c_2(\U(3))+c_3(\U(3))=\Sq^2c_2(\U(3))=(w_2+w_1^2)c_2(\U(3))=0\mod2$ on Spin 6-manifolds, and
$c_1(\U(3))c_2(\U(2))=c_1(\U(2))c_2(\U(2))=\Sq^2(c_2(\U(2)))=(w_2+w_1^2)c_2(\U(2))=0\mod2$ on Spin 6-manifolds.
}}
\label{table:SU3SU2U1Z6Bordism}
\end{table}

By \eqref{eq:TPexact}, we obtain the cobordism group $\TP_d(\Spin\times \frac{\SU(3)\times \SU(2)\times \U(1)}{\Z_6})$ shown in Table \ref{table:SU3SU2U1Z6TP}.

\begin{table}[H]
\centering
\hspace*{-16mm}
\begin{tabular}{ c c c}
\hline
\multicolumn{3}{c}{Cobordism group}\\
\hline
$d$ & 
$\TP_d(\Spin\times \frac{\SU(3)\times \SU(2)\times \U(1)}{\Z_6})$
& topological terms \\
\hline
0& 0\\
\hline
1& $\Z\times\Z_2$ & $\text{CS}_1^{\U(3)},\tilde\eta$ \\
\hline
2& $\Z_2$ & Arf \\
\hline
3 & $\Z^4$ & 
$\frac{\text{CS}_3^{TM}}{48},
{
\frac{1}{2}\text{CS}_1^{\U(2)}c_1(\U(2)) 
=
\frac{1}{2}\text{CS}_1^{\U(3)}c_1(\U(3)) },
\text{CS}_3^{\U(2)}, 
 \text{CS}_3^{\U(3)}$\\
\hline
4 & 0 \\
\hline
5 & $\Z^5$ & 
$\begin{matrix}\mu(\text{PD}(c_1(\U(3)))),
\text{CS}_1^{\U(3)}c_1(\U(3))^2,
\frac{\text{CS}_1^{\U(3)}c_2(\U(2))}{2} {\sim \frac{c_1(\U(3)) \text{CS}_3^{\U(2)}}{2}},
 \\
\frac{\text{CS}_1^{\U(3)}c_2(\U(3))+\text{CS}_5^{\U(3)}}{2}
 {\sim 
\frac{ c_1(\U(3)) \text{CS}_3^{\U(3)}
+\text{CS}_5^{\U(3)}}{2}
 },  
 \text{CS}_5^{\U(3)}
\end{matrix}$ \\
\hline
\end{tabular}
\caption{Topological phase classification ($\equiv$ TP) as a cobordism group, following Table \ref{table:SU3SU2U1Z6Bordism}. 
$\tilde\eta$ is a mod 2 index of 1d Dirac operator.
Arf is a 2d Arf invariant.
$c_i(G)$ is the Chern class of the associated vector bundle of the principal $G$-bundle.
$\text{CS}_{2n-1}^{V}$ or $\text{CS}_{2n-1}^{G}$ is the Chern-Simons form of the vector bundle $V$ or the associated vector bundle of the principal $G$-bundle.
The PD is the Poincar\'e dual.
The $TM$ is the spacetime tangent bundle.
The $\mu$ is the 3d Rokhlin invariant.
If $\partial M^4=M^3$, then $\mu(M^3)=(\frac{\sigma-\rF\cdot\rF}{8} )(M^4)$, thus
$\mu(\text{PD}(c_1(\U(3))))$ is related to $\frac{c_1(\U(3))(\sigma-\rF \cdot \rF )}{8}$
in Table \ref{table:SU3SU2U1Z6Bordism}.
}
\label{table:SU3SU2U1Z6TP}
\end{table}
%
{In Table \ref{table:SU3SU2U1Z6TP}, note that $\text{CS}_1^{\U(3)} c_2(\U(3))= c_1(\U(3)) \text{CS}_3^{\U(3)}$ 
and $\text{CS}_1^{\U(3)}c_2(\U(2))= c_1(\U(3)) \text{CS}_3^{\U(2)}$ up to a total derivative term (vanishing on a closed 5-manifold). See footnote \ref{ft:Chern-Simons}.}

\subsection{Comparison between Adams spectral sequence and Atiyah-Hirzebruch spectral sequence}
\label{sec:Comparison}

\begin{itemize}[leftmargin=2mm]
\item
Our approach  \cite{WanWang2018bns1812.11967, WanWangv2} is based on Adams spectral sequence (ASS), which includes the more refined data, containing both \emph{module} and \emph{group} structure, thus with the benefits of having less differentials.
In addition, as another advantage, 
we can conveniently read and extract the iTQFT (namely, co/bordism invariants) from the Adams chart and 
ASS \cite{WanWang2018bns1812.11967, WanWangv2}.


\item
\Ref{GarciaEtxebarriaMontero2018ajm1808.00009, 2019arXiv191011277D} is based on  Atiyah-Hirzebruch spectral sequence (AHSS), which includes only the \emph{group} structure, but with the disadvantage of having more differentials
and some undetermined extensions.
{It is also not known or difficult, if not impossible, to extract the iTQFT data directly (namely, co/bordism invariants) from the AHSS.
Therefore, 
\Ref{GarciaEtxebarriaMontero2018ajm1808.00009, 2019arXiv191011277D} cannot provide
the explicit iTQFT data from the AHSS calculations.}

{By \eqref{eq:Spin-BG'} and the Atiyah-Hirzebruch spectral sequence 
\bea
\H_p(X,\Omega_q^{\Spin})\Rightarrow\Omega_{p+q}^{\Spin}(X),
\eea
one can compute the bordism groups $\Omega_d^{\Spin}(\B G')=\Omega_d^{\Spin\times G'}$.
However, in general, one can not compute the bordism groups $\Omega_d^{\Spin\times_{\Z_2}G'}$ using the Atiyah-Hirzebruch spectral sequence since by \eqref{eq:PonThom},
$\Omega_d^{\Spin\times_{\Z_2}G'}=\pi_d(MT(\Spin\times_{\Z_2}G'))=\pi_d(M\Spin\wedge X')$ for some topological space $X'$, but $X'$ is not the disjoint union of a topological space $X$ and a point, while by \eqref{eq:genPonThom}, $\Omega_d^{\Spin}(X)=\pi_d(M\Spin\wedge X_+)$ where $X_+$ is the disjoint union of $X$ and a point.
}

{
In contrast, in this article, using Adams spectral sequence, we compute the bordism groups $\Omega_d^{\Spin\times_{\Z_2}G'}$ for several groups $G'$, such as $G'=\SU(4)\times\SU(2)\times\SU(2)$ or $\SU(4)\times_{\Z_2}(\SU(2)\times\SU(2))$ in the Pati-Salam models, and $G'=\Spin(10)$ or $\Spin(18)$ in the Grand Unified Theories.
}
\end{itemize}

Specifically, in \Ref{2019arXiv191011277D}, using Atiyah-Hirzebruch spectral sequence, the authors compute the cobordism groups $\Omega_d^{\Spin}(\B(\frac{\SU(3)\times\SU(2)\times\U(1)}{\Gamma}))$ for $0\le d\le 5$ and $\Gamma=1,\Z_2,\Z_3,\Z_6$, but their result for the $\Gamma=\Z_6$ case is \emph{not} completely determined. \Ref{2019arXiv191011277D} determines the cobordism groups $\Omega_d^{\Spin}(\B(\frac{\SU(3)\times\SU(2)\times\U(1)}{\Z_6}))$ in 2d and 4d as some \emph{undetermined} extensions. Namely, they obtain that
$\Omega_2^{\Spin}(\B(\frac{\SU(3)\times\SU(2)\times\U(1)}{\Z_6}))=e(\Z_3,\Z\times\Z_2)$, and $\Omega_4^{\Spin}(\B(\frac{\SU(3)\times\SU(2)\times\U(1)}{\Z_6}))=e(\Z_3,e(\Z_3,\Z^4))$ where $e(\rm{Q},\rm{N})$ is the group extension of $\rm{Q}$ by $\rm{N}$. So the group $e(\rm{Q},\rm{N})$ fits into the short exact sequence $0\to\rm{N}\to e(\rm{Q},\rm{N})\to\rm{Q}\to0$ but it may not be uniquely determined. 

In contrast, our results are more refined and can \emph{uniquely determine} the extension in this case. 
Our result from Adams   spectral sequence demonstrates that each step of extensions is nontrivial, while the trivial extension yields 3-torsion. Using Adams spectral sequence, we find that there is no 3-torsion for the $\Gamma=\Z_6$ case. 
So we also provide the solutions to the extension problems in \Ref{2019arXiv191011277D}, given by the nontrivial extension $\Z\to\Z\to\Z_3$.
We obtained the precise answer
$$\Omega_2^{\Spin}(\B(\frac{\SU(3)\times\SU(2)\times\U(1)}{\Z_6}))=\Z\times\Z_2$$
and
$$\Omega_4^{\Spin}(\B(\frac{\SU(3)\times\SU(2)\times\U(1)}{\Z_6}))=\Z^4.$$

\section{Standard Models with additional discrete symmetries}\label{sec:SM4}

In \Eq{eq:SMembed1} and \Eq{eq:SMembed2},
we have found the group embedding for the spacetime and internal symmetries of GUTs and SMs, when
the SM groups are $\Spin(d) \times \frac{\SU(3) \times   \SU(2) \times \U(1)}{\Z_q}$ with $q=6$.
Furthermore, 
inspired by the SMs with 
additional discrete symmetries (see some of the earlier 
work \cite{Ibanez1991hvRossPLB, Banks1991xjDine9109045, Csaki1997awMurayama9710105, Dreiner2005rd0512163} and References therein 
\cite{GarciaEtxebarriaMontero2018ajm1808.00009, Hsieh2018ifc1808.02881}) and motivated by a version of Smith homomorphism map between 5d and 4d 
bordism groups \cite{2018arXiv180502772T},
\bea
\Omega_5^{\Spin \times_{\Z_2} \Z_4} = 
\Omega_4^{\Pin^+} =\Z_{16}, 
\eea
we find the following group embedding for the spacetime and internal symmetries for 
GUTs and the SMs with additional discrete symmetries (see also \cite{WanWangv2} for the derivations):
\bea 
 \label{eq:SMembed3}
{\frac{\Spin(d) \times
\Spin(10)}{{\Z_2^F}} 
\supset 
\Spin(d) \times_{\Z_2} \Z_4 \times \SU(5) 
\supset 
\Spin(d) \times_{\Z_2} \Z_4 \times \frac{\SU(3) \times   \SU(2) \times \U(1)}{\Z_6} }.
\eea
Let us discuss the physics role of the group $\Z_4$ in $\Spin(d) \times_{\Z_2} \Z_4$:\\[-8mm]
\begin{itemize}
\item The $\Z_4$ as a gauge symmetry:
The center of Spin(10) is $Z(\Spin(10))=\Z_4$, which is naturally \emph{dynamically gauged} in the Spin(10) gauge group. So 
we have $[\Z_4] \subset [\Spin(10)]$ for SO(10) GUTs,
with the bracket $[\dots]$ indicating the groups as (part of) gauge groups.
\item The $\Z_4$ as a global symmetry:
However, this $\Z_4$ can simply be a
 part of the \emph{internal global symmetry}, for the SU(5) GUT and for the SM with $q=6$, assuming 
 that if we do not descend these models from the gauged $[\Spin(10)]$ of the SO(10) GUT. 
 This $\Z_4$ also contains the $\Z_2^F$ fermion parity (where $(-1)^F$ is also a $\Z_2$ normal subgroup
of the spacetime symmetry $\Spin(d)$).
Thus $\Z_4$ remains ungauged and a \emph{global symmetry} in
$\Spin(d) \times_{\Z_2} \Z_4 \times [\SU(5)]$ 
and $\Spin(d) \times_{\Z_2} \Z_4 \times [\frac{\SU(3) \times   \SU(2) \times \U(1)}{\Z_6}]$, 
even when the bracket $[\dots]$ become gauge groups, 
for the SU(5) GUT and for the SM with $q=6$.\\[-8mm]
\end{itemize}
We shall study the cobordism theory of the SM groups,
${\Spin \times_{\Z_2} \Z_4\times \frac{\SU(3)\times \SU(2)\times \U(1)}{\Z_q}}$, 
with $q=1,2,3,6$ in the following subsections.
For SM with this discrete symmetry, 
we obtain \emph{new} anomaly matching conditions of $\Z_{16}$, $\Z_{4}$ and $\Z_{2}$ classes \emph{beyond} the familiar Witten anomaly.
Depend on whether this $\Z_4$ is a global symmetry or a gauge symmetry, 
we shall interpret some of the 4d anomalies obtained from the 5d cobordism groups below
as 't Hooft anomalies, and some of others as a dynamical gauge anomalies.
%

\subsection{
${\Spin \times_{\Z_2} \Z_4\times {\SU(3)\times \SU(2)\times \U(1)}}$ model
}

Below we consider the co/bordism classes relevant for Standard Models with 
additional discrete symmetries.\footnote{JW is grateful to Miguel Montero \cite{GarciaEtxebarriaMontero2018ajm1808.00009} 
for informing his unpublished note \cite{Montero}. To make comparison, our approach is based on the Adams spectral sequence,
while \cite{GarciaEtxebarriaMontero2018ajm1808.00009, Montero} uses Atiyah-Hirzebruch spectral sequence (AHSS).
Two approaches between ours \cite{WanWang2018bns1812.11967, WanWangv2} and 
Garcia-Etxebarria-Montero \cite{GarciaEtxebarriaMontero2018ajm1808.00009, Montero} are rather different.
See more physics implications for future work, see \cite{Montero, MMVWangYau}.}

We consider $G={\Spin \times_{\Z_2} \Z_4 \times {\SU(3)\times \SU(2)\times \U(1)}}$.

We have a homotopy pullback square
\bea
\xymatrix{
\B(\Spin \times_{\Z_2} \Z_4)\ar[r]\ar[d]&\B\Z_2\ar[d]^{a^2}\\
\B\SO\ar[r]^{w_2}&\B^2\Z_2}
\eea
where $a$ is the generator of $\H^1(\B\Z_2,\Z_2)$.

By \cite{2017arXiv170804264C}, since there is a homotopy pullback square
\bea
\xymatrix{
\B(\Spin \times_{\Z_2} \Z_4)\ar[r]\ar[d]&{*}\times\B\Z_2\ar[d]\\
\B\O\ar[r]^-{(w_1,w_2)}&K(\Z_2,1)\times K(\Z_2,2),}
\eea
which is equivalent to the homotopy pullback square
\bea
\xymatrix{
\B(\Spin \times_{\Z_2} \Z_4)\ar[r]\ar[d]&\B\Spin\ar[d]\\
\B\SO\times\B\Z_2\ar[r]^{(\text{Id},2\xi)}&\B\SO,}
\eea
$MT(\Spin\times_{\Z_2}\Z_4)=M\Spin\wedge(\B\Z_2)^{2\xi}=M\Spin\wedge\Sigma^{-2}\RP_2^{\infty}$ where $2\xi:\B\Z_2\to\B\SO$ is twice the sign representation, the final identification is by \cite{AtiyahThom}.

The Madsen-Tillmann spectrum $MTG$ of the group $G$ is 
\bea
MTG
&=&M\Spin\wedge(\B\Z_2)^{2\xi}\wedge (\B (\SU(3)\times \SU(2)\times\U(1)))_+\nn\\
&=&M\Spin\wedge\Sigma^{-2}\RP_2^{\infty}\wedge (\B (\SU(3)\times \SU(2)\times\U(1)))_+.
\eea
 The $(\B(\SU(3)\times \SU(2)\times \U(1)))_+$ is the disjoint union of the classifying space $\B(\SU(3)\times \SU(2)\times \U(1))$ and a point, see footnote \ref{ft:X_+}.

For the dimension $d=t-s<8$, since there is no odd torsion (see footnote \ref{ft:no-odd-torsion}), by \eqref{eq:ExtA_2(1)}, we have the Adams spectral sequence
\bea
\Ext_{\A_2(1)}^{s,t}(\H^{*+2}(\RP_2^{\infty},\Z_2)\otimes\H^*(\B(\SU(3)\times \SU(2)\times \U(1)),\Z_2),\Z_2)\Rightarrow\Omega_{t-s}^{\Spin \times_{\Z_2} \Z_4\times {\SU(3)\times \SU(2)\times \U(1)}}.
\eea

The $\A_2(1)$-module structure of $\H^{*+2}(\RP_2^{\infty},\Z_2)$ is shown in Figure \ref{fig:A_2(1)RP_2}.

\begin{figure}[H]
\begin{center}
\begin{tikzpicture}[scale=0.5]

\draw[fill] (0,0) circle(.1);
\draw[fill] (0,1) circle(.1);
\draw[fill] (0,2) circle(.1);
\draw (0,0) to [out=30,in=30] (0,2);
\draw (0,1) -- (0,2);
\draw[fill] (0,3) circle(.1);
\draw (0,1) to [out=150,in=150] (0,3);
\draw[fill] (0,4) circle(.1);
\draw (0,3) -- (0,4);
\draw[fill] (0,5) circle(.1);
\draw[fill] (0,6) circle(.1);
\draw (0,4) to [out=150,in=150] (0,6);
\draw (0,5) -- (0,6);
\draw[fill] (0,7) circle(.1);
\draw (0,5) to [out=30,in=30] (0,7);
\draw[fill] (0,8) circle(.1);
\draw (0,7) -- (0,8);

\node at (-0.5,1) {$a$};
\node at (1,2) {$a^2$};

\end{tikzpicture}
\end{center}
\caption{The $\A_2(1)$-module structure of $\H^{*+2}(\RP_2^{\infty},\Z_2)$.}
\label{fig:A_2(1)RP_2}
\end{figure}

The $\A_2(1)$-module structure of $\H^{*+2}(\RP_2^{\infty},\Z_2)\otimes\H^{*+2}(C\upeta,\Z_2)$ is shown in Figure \ref{fig:A_2(1)RP_2Ceta}.
Here $\upeta: S^3\to S^2$ is the Hopf fibration,\footnote{Beware that we use $\eta$ to denote the $\eta$ invariant (e.g. the APS $\eta$ invariant), 
while we use the up-greek font eta $\upeta: S^3\to S^2$ to denote the Hopf fibration.} 
the mapping cone is $C\upeta=S^2\hcup{\upeta}e^4=\CP^2$. The $\A_2(1)$-module structure of $\H^{*+2}(C\upeta,\Z_2)$ has two elements in degree 0 and 2 attached by a $\Sq^2$.

\begin{figure}[H]
\begin{center}
\begin{tikzpicture}[scale=0.5]

\draw[fill] (0,0) circle(.1);
\draw[fill] (0,1) circle(.1);
\draw[fill] (0,2) circle(.1);
\draw (0,0) to [out=30,in=30] (0,2);
\draw (0,1) -- (0,2);
\draw[fill] (0,3) circle(.1);
\draw (0,1) to [out=150,in=150] (0,3);
\draw[fill] (0,4) circle(.1);
\draw (0,3) -- (0,4);
\draw[fill] (0,5) circle(.1);
\draw[fill] (0,6) circle(.1);
\draw (0,4) to [out=150,in=150] (0,6);
\draw (0,5) -- (0,6);
\draw[fill] (0,7) circle(.1);
\draw (0,5) to [out=30,in=30] (0,7);
\draw[fill] (0,8) circle(.1);
\draw (0,7) -- (0,8);

\node at (2,4) {$\bigotimes$};

\draw[fill] (4,0) circle(.1);
\draw[fill] (4,2) circle(.1);
\draw (4,0) to [out=150,in=150] (4,2);

\def\rx{8}
\def\ry{12}

\node at (-2+\rx,-6+\ry) {$=$};

\draw[fill] (0+\rx,-12+\ry) circle(.1);
\draw[fill] (0+\rx,-10+\ry) circle(.1);
\draw (0+\rx,-12+\ry) to [out=150,in=150] (0+\rx,-10+\ry);

\draw[fill] (2+\rx,-11+\ry) circle(.1);
\draw[fill] (2+\rx,-10+\ry) circle(.1);
\draw (2+\rx,-11+\ry) -- (2+\rx,-10+\ry);
\draw[fill] (2+\rx,-9+\ry) circle(.1);
\draw (2+\rx,-11+\ry) to [out=150,in=150] (2+\rx,-9+\ry);
\draw[fill] (3+\rx,-9+\ry) circle(.1);
\draw[fill] (3+\rx,-8+\ry) circle(.1);
\draw (3+\rx,-9+\ry) -- (3+\rx,-8+\ry);
\draw (2+\rx,-10+\ry) to [out=30,in=150] (3+\rx,-8+\ry);
\draw[fill] (2+\rx,-8+\ry) circle(.1);
\draw (2+\rx,-9+\ry) -- (2+\rx,-8+\ry);
\draw[fill] (2+\rx,-7+\ry) circle(.1);
\draw[fill] (2+\rx,-6+\ry) circle(.1);
\draw (2+\rx,-7+\ry) -- (2+\rx,-6+\ry);
\draw (2+\rx,-8+\ry) to [out=150,in=150] (2+\rx,-6+\ry);
\draw[fill] (3+\rx,-7+\ry) circle(.1);
\draw (3+\rx,-9+\ry) to [out=30,in=30] (3+\rx,-7+\ry);
\draw[fill] (3+\rx,-6+\ry) circle(.1);
\draw (3+\rx,-7+\ry) -- (3+\rx,-6+\ry);
\draw[fill] (2+\rx,-5+\ry) circle(.1);
\draw (2+\rx,-7+\ry) to [out=30,in=30] (2+\rx,-5+\ry);
\draw[fill] (3+\rx,-5+\ry) circle(.1);
\draw[fill] (3+\rx,-4+\ry) circle(.1);
\draw (3+\rx,-5+\ry) -- (3+\rx,-4+\ry);
\draw (3+\rx,-6+\ry) to [out=150,in=150] (3+\rx,-4+\ry);
\draw[fill] (2+\rx,-4+\ry) circle(.1);
\draw (2+\rx,-5+\ry) -- (2+\rx,-4+\ry);
\draw[fill] (3+\rx,-3+\ry) circle(.1);
\draw[fill] (3+\rx,-2+\ry) circle(.1);
\draw (3+\rx,-3+\ry) -- (3+\rx,-2+\ry);
\draw (3+\rx,-5+\ry) to [out=30,in=30] (3+\rx,-3+\ry);
\end{tikzpicture}
\end{center}
\caption{The $\A_2(1)$-module structure of $\H^{*+2}(\RP_2^{\infty},\Z_2)\otimes\H^{*+2}(C\upeta,\Z_2)$.}
\label{fig:A_2(1)RP_2Ceta}
\end{figure}

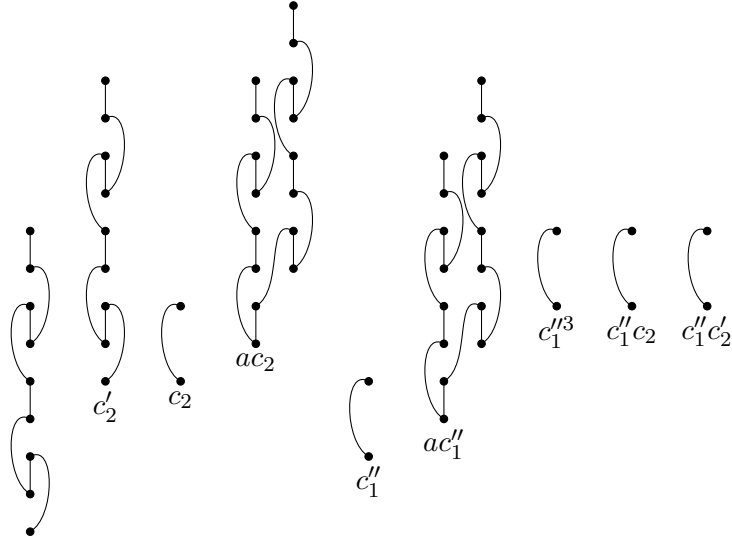
\begin{figure}[H]
\begin{center}
\begin{tikzpicture}[scale=0.5]

\draw[fill] (0,0) circle(.1);
\draw[fill] (0,1) circle(.1);
\draw[fill] (0,2) circle(.1);
\draw (0,0) to [out=30,in=30] (0,2);
\draw (0,1) -- (0,2);
\draw[fill] (0,3) circle(.1);
\draw (0,1) to [out=150,in=150] (0,3);
\draw[fill] (0,4) circle(.1);
\draw (0,3) -- (0,4);
\draw[fill] (0,5) circle(.1);
\draw[fill] (0,6) circle(.1);
\draw (0,4) to [out=150,in=150] (0,6);
\draw (0,5) -- (0,6);
\draw[fill] (0,7) circle(.1);
\draw (0,5) to [out=30,in=30] (0,7);
\draw[fill] (0,8) circle(.1);
\draw (0,7) -- (0,8);

\node[below] at (2,4) {$c_2'$};

\draw[fill] (2,4) circle(.1);
\draw[fill] (2,5) circle(.1);
\draw[fill] (2,6) circle(.1);
\draw (2,4) to [out=30,in=30] (2,6);
\draw (2,5) -- (2,6);
\draw[fill] (2,7) circle(.1);
\draw (2,5) to [out=150,in=150] (2,7);
\draw[fill] (2,8) circle(.1);
\draw (2,7) -- (2,8);
\draw[fill] (2,9) circle(.1);
\draw[fill] (2,10) circle(.1);
\draw (2,8) to [out=150,in=150] (2,10);
\draw (2,9) -- (2,10);
\draw[fill] (2,11) circle(.1);
\draw (2,9) to [out=30,in=30] (2,11);
\draw[fill] (2,12) circle(.1);
\draw (2,11) -- (2,12);

\node[below] at (4,4) {$c_2$};

\draw[fill] (4,4) circle(.1);
\draw[fill] (4,6) circle(.1);
\draw (4,4) to [out=150,in=150] (4,6);

\node[below] at (6,5) {$ac_2$};

\draw[fill] (6,5) circle(.1);
\draw[fill] (6,6) circle(.1);
\draw (6,5) -- (6,6);
\draw[fill] (6,7) circle(.1);
\draw (6,5) to [out=150,in=150] (6,7);
\draw[fill] (7,7) circle(.1);
\draw[fill] (7,8) circle(.1);
\draw (7,7) -- (7,8);
\draw (6,6) to [out=30,in=150] (7,8);
\draw[fill] (6,8) circle(.1);
\draw (6,7) -- (6,8);
\draw[fill] (6,9) circle(.1);
\draw[fill] (6,10) circle(.1);
\draw (6,9) -- (6,10);
\draw (6,8) to [out=150,in=150] (6,10);
\draw[fill] (7,9) circle(.1);
\draw (7,7) to [out=30,in=30] (7,9);
\draw[fill] (7,10) circle(.1);
\draw (7,9) -- (7,10);
\draw[fill] (6,11) circle(.1);
\draw (6,9) to [out=30,in=30] (6,11);
\draw[fill] (7,11) circle(.1);
\draw[fill] (7,12) circle(.1);
\draw (7,11) -- (7,12);
\draw (7,10) to [out=150,in=150] (7,12);
\draw[fill] (6,12) circle(.1);
\draw (6,11) -- (6,12);
\draw[fill] (7,13) circle(.1);
\draw[fill] (7,14) circle(.1);
\draw (7,13) -- (7,14);
\draw (7,11) to [out=30,in=30] (7,13);

\node[below] at (9,2) {$c_1''$};

\draw[fill] (9,2) circle(.1);
\draw[fill] (9,4) circle(.1);
\draw (9,2) to [out=150,in=150] (9,4);

\node[below] at (11,3) {$ac_1''$};

\draw[fill] (11,3) circle(.1);
\draw[fill] (11,4) circle(.1);
\draw (11,3) -- (11,4);
\draw[fill] (11,5) circle(.1);
\draw (11,3) to [out=150,in=150] (11,5);
\draw[fill] (12,5) circle(.1);
\draw[fill] (12,6) circle(.1);
\draw (12,5) -- (12,6);
\draw (11,4) to [out=30,in=150] (12,6);
\draw[fill] (11,6) circle(.1);
\draw (11,5) -- (11,6);
\draw[fill] (11,7) circle(.1);
\draw[fill] (11,8) circle(.1);
\draw (11,7) -- (11,8);
\draw (11,6) to [out=150,in=150] (11,8);
\draw[fill] (12,7) circle(.1);
\draw (12,5) to [out=30,in=30] (12,7);
\draw[fill] (12,8) circle(.1);
\draw (12,7) -- (12,8);
\draw[fill] (11,9) circle(.1);
\draw (11,7) to [out=30,in=30] (11,9);
\draw[fill] (12,9) circle(.1);
\draw[fill] (12,10) circle(.1);
\draw (12,9) -- (12,10);
\draw (12,8) to [out=150,in=150] (12,10);
\draw[fill] (11,10) circle(.1);
\draw (11,9) -- (11,10);
\draw[fill] (12,11) circle(.1);
\draw[fill] (12,12) circle(.1);
\draw (12,11) -- (12,12);
\draw (12,9) to [out=30,in=30] (12,11);

\node[below] at (14,6) {$c_1''^3$};

\draw[fill] (14,6) circle(.1);
\draw[fill] (14,8) circle(.1);
\draw (14,6) to [out=150,in=150] (14,8);

\node[below] at (16,6) {$c_1''c_2$};

\draw[fill] (16,6) circle(.1);
\draw[fill] (16,8) circle(.1);
\draw (16,6) to [out=150,in=150] (16,8);

\node[below] at (18,6) {$c_1''c_2'$};

\draw[fill] (18,6) circle(.1);
\draw[fill] (18,8) circle(.1);
\draw (18,6) to [out=150,in=150] (18,8);

\end{tikzpicture}
\end{center}
\caption{The $\A_2(1)$-module structure of $\H^{*+2}(\RP_2^{\infty},\Z_2)\otimes\H^*(\B(\SU(3)\times\SU(2)\times\U(1)),\Z_2)$ below degree 6. {Here, $c_i$ is the Chern class of SU(3) bundle, and $c_i'$ is the Chern class of SU(2) bundle,
and $c_i''$ is the Chern class of U(1) bundle.}}
\label{fig:A_2(1)RP_2SU3SU2U1}
\end{figure}

Based on Figure \ref{fig:A_2(1)SU3SU2U1} and \ref{fig:A_2(1)RP_2Ceta}, we obtain the $\A_2(1)$-module structure of $\H^{*+2}(\RP_2^{\infty},\Z_2)\otimes\H^*(\B(\SU(3)\times\SU(2)\times\U(1)),\Z_2)$ below degree 6, as shown in Figure \ref{fig:A_2(1)RP_2SU3SU2U1}.

The $E_2$ page is shown in Figure \ref{fig:E_2Z4SU3SU2U1}.
Here we have used the correspondence between $\A_2(1)$-module structure and the $E_2$ page shown in Figure \ref{fig:Ceta}, \ref{fig:RP_2} and \ref{fig:L_1}.

\begin{figure}[H]
\begin{center}
\begin{tikzpicture}
\node at (0,-1) {0};
\node at (1,-1) {1};
\node at (2,-1) {2};
\node at (3,-1) {3};
\node at (4,-1) {4};
\node at (5,-1) {5};
\node at (6,-1) {6};
\node at (7,-1) {$t-s$};
\node at (-1,0) {0};
\node at (-1,1) {1};
\node at (-1,2) {2};
\node at (-1,3) {3};
\node at (-1,4) {4};
\node at (-1,5) {5};
\node at (-1,6) {$s$};

\draw[->] (-0.5,-0.5) -- (-0.5,6);
\draw[->] (-0.5,-0.5) -- (7,-0.5);

\draw (0,0) -- (0,5);

\draw (1,0) -- (1,1);

\draw (2,0) -- (2,5);

\draw[fill] (3,0) circle(0.05);

\draw (4,1) -- (4,5);

\draw (4.1,2) -- (4.1,5);
\draw (4.2,0) -- (4.2,5);

\draw (4.3,0) -- (4.3,5);

\draw (5,0) -- (5,1);
\draw (5.1,0) -- (5.1,3);

\draw (5.2,0) -- (5.2,1);

\draw[fill] (5.3,0) circle(0.05);

\draw (6,1) -- (6,5);

\draw (6.1,2) -- (6.1,5);

\draw (6.2,0) -- (6.2,5);

\draw (6.3,0) -- (6.3,5);

\draw (6.4,0) -- (6.4,5);

\end{tikzpicture}
\end{center}
\caption{$\Omega_*^{\Spin\times_{\Z_2}\Z_4\times \SU(3)\times\SU(2)\times\U(1)}$.}
\label{fig:E_2Z4SU3SU2U1}
\end{figure}

Thus we obtain the bordism group $\Omega^{\Spin \times_{\Z_2} \Z_4 \times {\SU(3)\times \SU(2)\times \U(1)}}_d$ shown in Table \ref{table:Z4SU3SU2U1Bordism}.

\begin{table}[H]
\centering
\hspace*{-25mm}
\begin{tabular}{ c c c}
\hline
\multicolumn{3}{c}{Bordism group}\\
\hline
$d$ & 
$\Omega^{\Spin \times_{\Z_2} \Z_4 \times {\SU(3)\times \SU(2)\times \U(1)}}_d$
& bordism invariants \\
\hline
0&  $\Z$ \\
\hline
1&  $\Z_4$& $\eta'$\\
\hline
2&  $\Z$ & $c_1(\U(1))$ \\
\hline
3 & $\Z_2$ &$ac_1(\U(1))$\\
\hline
4 & $\Z^4$ & $\frac{\sigma-\rF\cdot \rF }{8}, \frac{a^2c_1(\U(1))+c_1(\U(1))^2}{2},c_2(\SU(2)),c_2(\SU(3))$ \\
\hline
5 &  $\Z_2\times\Z_4^2\times\Z_{16}$ & $ac_2(\SU(3)),c_2(\SU(2))\eta',c_1(\U(1))^2\eta',\eta(\text{PD}(a))$ \\
\hline
6 & $\Z^5$  & $\begin{matrix}c_1(\U(1))\frac{\sigma-\rF\cdot \rF }{8},c_1(\U(1))^3,c_1(\U(1))c_2(\SU(2)),\\c_1(\U(1))c_2(\SU(3)),\frac{a^2c_2(\SU(3))+c_3(\SU(3))}{2}\end{matrix}$ \\
\hline
\end{tabular}
\caption{Bordism group. 
$\eta'$ is a $\Z_4$ valued 1d eta invariant which is the extension of $a$ by $\tilde\eta$ where $a$ is the generator of $\H^1(\B\Z_2,\Z_2)$, $\tilde\eta$ is the mod 2 index of 1d Dirac operator.
$\sigma$ is the signature of manifold.
$\eta$ is the $\Z_{16}$ valued 4d eta invariant.
$\rF $ is the characteristic 2-surface \cite{Saveliev} in a 4-manifold $M^4$, it satisfies the condition $\rF\cdot x=x\cdot x\mod2$ for all $x\in\H_2(M^4,\Z)$.
Here $\cdot$ is the intersection form of $M^4$.
By the Freedman-Kirby theorem, $(\frac{\sigma-\rF\cdot\rF}{8} )(M^4)=\text{Arf}(M^4,\rF)\mod2$. 
The PD is the Poincar\'e dual.
Note that
$c_1(\U(1))^2=\Sq^2c_1(\U(1))=(w_2+w_1^2)c_1(\U(1))=a^2c_1(\U(1))\mod2$.
$c_3(\SU(3))=\Sq^2c_2(\SU(3))=(w_2+w_1^2)c_2(\SU(3))=a^2c_2(\SU(3))\mod2$.
On 5-manifolds with $\Spin\times_{\Z_2}\Z_4$ structure, there is a $\Pin^+$ structure on $\text{PD}(a)$ and there is an isomorphism $\Omega_5^{\Spin\times_{\Z_2}\Z_4}\xrightarrow{\cap a}\Omega_4^{\Pin^+}$ \cite{2018arXiv180502772T}, so the bordism invariant of $\Omega_5^{\Spin\times_{\Z_2}\Z_4}=\Z_{16}$ is $\eta(\text{PD}(a))$.
{The Witten anomaly $c_2(\SU(2))\tilde\eta$ appears in $c_2(\SU(2))\eta'$ as the normal subgroup $\Z_2$ of $\Z_4$. While the other three bordism invariants in 5d are beyond Witten anomaly.
}
{See Appendix \ref{sec:comment-2} for comment on the difference in 5d between Table \ref{table:Z4SU3SU2U1Bordism} and \ref{table:Z4SU3SU2U1Z2Bordism}.}
}
\label{table:Z4SU3SU2U1Bordism}
\end{table}
By \eqref{eq:TPexact}, we obtain a cobordism group $\TP_d
({\Spin \times_{\Z_2} \Z_4 \times {\SU(3)\times \SU(2)\times \U(1)}})
$ shown in Table \ref{table:Z4SU3SU2U1TP}.
\begin{table}[H]
\centering
\hspace*{-18mm}
\begin{tabular}{ c c c}
\hline
\multicolumn{3}{c}{Cobordism group}\\
\hline
$d$ & 

$\begin{matrix}\TP_d\\
({\Spin \times_{\Z_2} \Z_4 \times {\SU(3)\times \SU(2)\times \U(1)}})
\end{matrix}$
& topological terms \\
\hline
0& 0 \\
\hline
1& $\Z\times\Z_4$ &  $\text{CS}_1^{\U(1)},\eta'$\\
\hline
2& 0 &  \\
\hline
3 & $\Z^4\times\Z_2$ & $\mu,
\frac{a^2\text{CS}_1^{\U(1)}+c_1(\U(1))\text{CS}_1^{\U(1)}}{2},\text{CS}_3^{\SU(2)},\text{CS}_3^{\SU(3)},ac_1(\U(1))$ \\
\hline
4 & 0 & \\
\hline
5 & $\Z^5\times\Z_2\times\Z_4^2\times\Z_{16}$ & 
$\begin{matrix}\mu(\text{PD}(c_1(\U(1)))),
\text{CS}_1^{\U(1)} c_1(\U(1))^2, 
{\text{CS}_1^{\U(1)}c_2(\SU(3)) \sim c_1(\U(1))\text{CS}_3^{\SU(2)}},\\ 
 {\text{CS}_1^{\U(1)}c_2(\SU(3)) \sim c_1(\U(1))\text{CS}_3^{\SU(3)}}, 
 \frac{a^2\text{CS}_3^{\SU(3)}+\text{CS}_5^{\SU(3)}}{2},
\\
ac_2(\SU(3)),c_2(\SU(2))\eta',c_1(\U(1))^2\eta',\eta(\text{PD}(a))
\end{matrix}$\\
\hline
\end{tabular}
\caption{Topological phase classification ($\equiv$ TP) as a cobordism group, following Table \ref{table:Z4SU3SU2U1Bordism}. 
$\eta'$ is a $\Z_4$ valued 1d eta invariant explained in Table \ref{table:Z4SU3SU2U1Bordism}.
$a$ is the generator of $\H^1(\B\Z_2,\Z_2)$.
The PD is the Poincar\'e dual.
The $\mu$ is the 3d Rokhlin invariant.
If $\partial M^4=M^3$, then $\mu(M^3)=(\frac{\sigma-\rF\cdot\rF}{8} )(M^4)$, thus
$\mu(\text{PD}(c_1(\U(1))))$ is related to $\frac{c_1(\U(1))(\sigma-\rF \cdot \rF )}{8}$
in Table \ref{table:Z4SU3SU2U1Bordism}. The $\eta(\text{PD}(a))$ is explained in Table \ref{table:Z4SU3SU2U1Bordism}.
{The Witten anomaly $c_2(\SU(2))\tilde\eta$ appears in $c_2(\SU(2))\eta'$ as the normal subgroup $\Z_2$ of $\Z_4$. While the other topological terms in 5d are beyond Witten anomaly.}
}
\label{table:Z4SU3SU2U1TP}
\end{table}

{In Table \ref{table:Z4SU3SU2U1TP}, note that $\text{CS}_1^{\U(1)} c_2(\SU(2))= c_1(\U(1)) \text{CS}_3^{\SU(2)}$ 
and $\text{CS}_1^{\U(1)}c_2(\SU(3))= c_1(\U(1)) \text{CS}_3^{\SU(3)}$ up to a total derivative term (vanishing on a closed 5-manifold). See footnote \ref{ft:Chern-Simons}.}

\subsection{
${\Spin \times_{\Z_2} \Z_4\times {\frac{\SU(3)\times \SU(2)\times \U(1)}{\Z_2}}}$ model
}
We consider $G={\Spin \times_{\Z_2} \Z_4 \times {\frac{\SU(3)\times \SU(2)\times \U(1)}{\Z_2}}}=\Spin \times_{\Z_2} \Z_4 \times {\SU(3)\times \U(2)}$, the Madsen-Tillmann spectrum $MTG$ of the group $G$ is 
\bea
MTG
&=&M\Spin\wedge(\B\Z_2)^{2\xi}\wedge (\B (\SU(3)\times \U(2)))_+\nn\\
&=&M\Spin\wedge\Sigma^{-2}\RP_2^{\infty}\wedge (\B (\SU(3)\times \U(2)))_+.
\eea
 The $(\B(\SU(3)\times \U(2)))_+$ is the disjoint union of the classifying space $\B(\SU(3)\times \U(2))$ and a point, see footnote \ref{ft:X_+}.

For the dimension $d=t-s<8$, since there is no odd torsion (see footnote \ref{ft:no-odd-torsion}), by \eqref{eq:ExtA_2(1)}, we have the Adams spectral sequence
\bea
\Ext_{\A_2(1)}^{s,t}(\H^{*+2}(\RP_2^{\infty},\Z_2)\otimes\H^*(\B(\SU(3)\times \U(2)),\Z_2),\Z_2)\Rightarrow\Omega_{t-s}^{\Spin \times_{\Z_2} \Z_4\times {\SU(3)\times \U(2)}}.
\eea

Based on Figure \ref{fig:A_2(1)SU3U2} and \ref{fig:A_2(1)RP_2Ceta}, we obtain the $\A_2(1)$-module structure of $\H^{*+2}(\RP_2^{\infty},\Z_2)\otimes\H^*(\B(\SU(3)\times\U(2)),\Z_2)$ below degree 6, as shown in Figure \ref{fig:A_2(1)RP_2SU3U2}.

\begin{figure}[H]
\begin{center}
\begin{tikzpicture}[scale=0.5]

\draw[fill] (0,0) circle(.1);
\draw[fill] (0,1) circle(.1);
\draw[fill] (0,2) circle(.1);
\draw (0,0) to [out=30,in=30] (0,2);
\draw (0,1) -- (0,2);
\draw[fill] (0,3) circle(.1);
\draw (0,1) to [out=150,in=150] (0,3);
\draw[fill] (0,4) circle(.1);
\draw (0,3) -- (0,4);
\draw[fill] (0,5) circle(.1);
\draw[fill] (0,6) circle(.1);
\draw (0,4) to [out=150,in=150] (0,6);
\draw (0,5) -- (0,6);
\draw[fill] (0,7) circle(.1);
\draw (0,5) to [out=30,in=30] (0,7);
\draw[fill] (0,8) circle(.1);
\draw (0,7) -- (0,8);

\node[below] at (4,4) {$c_2$};

\draw[fill] (4,4) circle(.1);
\draw[fill] (4,6) circle(.1);
\draw (4,4) to [out=150,in=150] (4,6);

\node[below] at (6,5) {$ac_2$};

\draw[fill] (6,5) circle(.1);
\draw[fill] (6,6) circle(.1);
\draw (6,5) -- (6,6);
\draw[fill] (6,7) circle(.1);
\draw (6,5) to [out=150,in=150] (6,7);
\draw[fill] (7,7) circle(.1);
\draw[fill] (7,8) circle(.1);
\draw (7,7) -- (7,8);
\draw (6,6) to [out=30,in=150] (7,8);
\draw[fill] (6,8) circle(.1);
\draw (6,7) -- (6,8);
\draw[fill] (6,9) circle(.1);
\draw[fill] (6,10) circle(.1);
\draw (6,9) -- (6,10);
\draw (6,8) to [out=150,in=150] (6,10);
\draw[fill] (7,9) circle(.1);
\draw (7,7) to [out=30,in=30] (7,9);
\draw[fill] (7,10) circle(.1);
\draw (7,9) -- (7,10);
\draw[fill] (6,11) circle(.1);
\draw (6,9) to [out=30,in=30] (6,11);
\draw[fill] (7,11) circle(.1);
\draw[fill] (7,12) circle(.1);
\draw (7,11) -- (7,12);
\draw (7,10) to [out=150,in=150] (7,12);
\draw[fill] (6,12) circle(.1);
\draw (6,11) -- (6,12);
\draw[fill] (7,13) circle(.1);
\draw[fill] (7,14) circle(.1);
\draw (7,13) -- (7,14);
\draw (7,11) to [out=30,in=30] (7,13);

\node[below] at (9,2) {$c_1'$};

\draw[fill] (9,2) circle(.1);
\draw[fill] (9,4) circle(.1);
\draw (9,2) to [out=150,in=150] (9,4);

\node[below] at (11,3) {$ac_1'$};

\draw[fill] (11,3) circle(.1);
\draw[fill] (11,4) circle(.1);
\draw (11,3) -- (11,4);
\draw[fill] (11,5) circle(.1);
\draw (11,3) to [out=150,in=150] (11,5);
\draw[fill] (12,5) circle(.1);
\draw[fill] (12,6) circle(.1);
\draw (12,5) -- (12,6);
\draw (11,4) to [out=30,in=150] (12,6);
\draw[fill] (11,6) circle(.1);
\draw (11,5) -- (11,6);
\draw[fill] (11,7) circle(.1);
\draw[fill] (11,8) circle(.1);
\draw (11,7) -- (11,8);
\draw (11,6) to [out=150,in=150] (11,8);
\draw[fill] (12,7) circle(.1);
\draw (12,5) to [out=30,in=30] (12,7);
\draw[fill] (12,8) circle(.1);
\draw (12,7) -- (12,8);
\draw[fill] (11,9) circle(.1);
\draw (11,7) to [out=30,in=30] (11,9);
\draw[fill] (12,9) circle(.1);
\draw[fill] (12,10) circle(.1);
\draw (12,9) -- (12,10);
\draw (12,8) to [out=150,in=150] (12,10);
\draw[fill] (11,10) circle(.1);
\draw (11,9) -- (11,10);
\draw[fill] (12,11) circle(.1);
\draw[fill] (12,12) circle(.1);
\draw (12,11) -- (12,12);
\draw (12,9) to [out=30,in=30] (12,11);

\node[below] at (2,6) {$c_1'^3$};

\draw[fill] (2,6) circle(.1);
\draw[fill] (2,8) circle(.1);
\draw (2,6) to [out=150,in=150] (2,8);

\node[below] at (19,6) {$c_1'c_2$};

\draw[fill] (19,6) circle(.1);
\draw[fill] (19,8) circle(.1);
\draw (19,6) to [out=150,in=150] (19,8);

\node[below] at (14,4) {$c_2'$};

\draw[fill] (14,4) circle(.1);
\draw[fill] (14,6) circle(.1);
\draw (14,4) to [out=150,in=150] (14,6);

\node[below] at (16,5) {$ac_2'$};

\draw[fill] (16,5) circle(.1);
\draw[fill] (16,6) circle(.1);
\draw (16,5) -- (16,6);
\draw[fill] (16,7) circle(.1);
\draw (16,5) to [out=150,in=150] (16,7);
\draw[fill] (17,7) circle(.1);
\draw[fill] (17,8) circle(.1);
\draw (17,7) -- (17,8);
\draw (16,6) to [out=30,in=150] (17,8);
\draw[fill] (16,8) circle(.1);
\draw (16,7) -- (16,8);
\draw[fill] (16,9) circle(.1);
\draw[fill] (16,10) circle(.1);
\draw (16,9) -- (16,10);
\draw (16,8) to [out=150,in=150] (16,10);
\draw[fill] (17,9) circle(.1);
\draw (17,7) to [out=30,in=30] (17,9);
\draw[fill] (17,10) circle(.1);
\draw (17,9) -- (17,10);
\draw[fill] (16,11) circle(.1);
\draw (16,9) to [out=30,in=30] (16,11);
\draw[fill] (17,11) circle(.1);
\draw[fill] (17,12) circle(.1);
\draw (17,11) -- (17,12);
\draw (17,10) to [out=150,in=150] (17,12);
\draw[fill] (16,12) circle(.1);
\draw (16,11) -- (16,12);
\draw[fill] (17,13) circle(.1);
\draw[fill] (17,14) circle(.1);
\draw (17,13) -- (17,14);
\draw (17,11) to [out=30,in=30] (17,13);

\end{tikzpicture}
\end{center}
\caption{The $\A_2(1)$-module structure of $\H^{*+2}(\RP_2^{\infty},\Z_2)\otimes\H^*(\B(\SU(3)\times\U(2)),\Z_2)$ below degree 6. {Here, $c_i$ is the Chern class of SU(3) bundle, and $c_i'$ is the Chern class of U(2) bundle.}}
\label{fig:A_2(1)RP_2SU3U2}
\end{figure}

The $E_2$ page is shown in Figure \ref{fig:E_2Z4SU3U2}.
Here we have used the correspondence between $\A_2(1)$-module structure and the $E_2$ page shown in Figure \ref{fig:Ceta}, \ref{fig:RP_2} and \ref{fig:L_1}.

\begin{figure}[H]
\begin{center}
\begin{tikzpicture}
\node at (0,-1) {0};
\node at (1,-1) {1};
\node at (2,-1) {2};
\node at (3,-1) {3};
\node at (4,-1) {4};
\node at (5,-1) {5};
\node at (6,-1) {6};
\node at (7,-1) {$t-s$};
\node at (-1,0) {0};
\node at (-1,1) {1};
\node at (-1,2) {2};
\node at (-1,3) {3};
\node at (-1,4) {4};
\node at (-1,5) {5};
\node at (-1,6) {$s$};

\draw[->] (-0.5,-0.5) -- (-0.5,6);
\draw[->] (-0.5,-0.5) -- (7,-0.5);

\draw (0,0) -- (0,5);

\draw (1,0) -- (1,1);

\draw (2,0) -- (2,5);

\draw[fill] (3,0) circle(0.05);

\draw (4,2) -- (4,5);

\draw (4.1,1) -- (4.1,5);
\draw (4.2,0) -- (4.2,5);

\draw (4.3,0) -- (4.3,5);

\draw (5,0) -- (5,3);
\draw (5.1,0) -- (5.1,1);

\draw[fill] (5.2,0) circle(0.05);

\draw[fill] (5.3,0) circle(0.05);

\draw (6,1) -- (6,5);

\draw (6.1,2) -- (6.1,5);

\draw (6.2,1) -- (6.2,5);

\draw (6.3,0) -- (6.3,5);

\draw (6.4,0) -- (6.4,5);

\end{tikzpicture}
\end{center}
\caption{$\Omega_*^{\Spin\times_{\Z_2}\Z_4\times \SU(3)\times\U(2)}$.}
\label{fig:E_2Z4SU3U2}
\end{figure}

Thus we obtain the bordism group $\Omega^{\Spin \times_{\Z_2} \Z_4 \times {\frac{\SU(3)\times \SU(2)\times \U(1)}{\Z_2}}}_d$ shown in Table \ref{table:Z4SU3SU2U1Z2Bordism}.

\begin{table}[H]
\centering
\hspace*{-15mm}
\begin{tabular}{ c c c}
\hline
\multicolumn{3}{c}{Bordism group}\\
\hline
$d$ & 
$\Omega^{\Spin \times_{\Z_2} \Z_4 \times {\frac{\SU(3)\times \SU(2)\times \U(1)}{\Z_2}}}_d$
& bordism invariants \\
\hline
0&  $\Z$ \\
\hline
1& $\Z_4$  & $\eta'$\\
\hline
2&  $\Z$  &  $c_1(\U(2))$ \\
\hline
3 & $\Z_2$ & $ac_1(\U(2))$\\
\hline
4 & $\Z^4$ & ${\frac{\sigma-\rF\cdot \rF}{8}},
\frac{a^2c_1(\U(2))+c_1(\U(2))^2}{2},c_2(\U(2)),c_2(\SU(3))$ \\
\hline
5 & $\Z_2^2\times\Z_4\times\Z_{16}$  & $ac_2(\SU(3)),ac_2(\U(2)),c_1(\U(2))^2\eta',\eta(\text{PD}(a))$ \\
\hline
6 & $\Z^5$  & $\begin{matrix}c_1(\U(2))\frac{\sigma-\rF\cdot \rF }{8},c_1(\U(2))^3,\frac{a^2c_2(\U(2))+c_1(\U(2))c_2(\U(2))}{2},\\
c_1(\U(2))c_2(\SU(3)),\frac{a^2c_2(\SU(3))+c_3(\SU(3))}{2}\end{matrix}$ \\
\hline
\end{tabular}
\caption{Bordism group. 
$\eta'$ is a $\Z_4$ valued 1d eta invariant explained in Table \ref{table:Z4SU3SU2U1Bordism}.
$\eta$ is the 4d eta invariant.
$\sigma$ is the signature of manifold.
$c_i(G)$ is the Chern class of the associated vector bundle of the principal $G$-bundle.
$a$ is the generator of $\H^1(\B\Z_2,\Z_2)$.
$\rF $ is the characteristic 2-surface \cite{Saveliev} in a 4-manifold $M^4$, it satisfies the condition $\rF\cdot x=x\cdot x\mod2$ for all $x\in\H_2(M^4,\Z)$.
Here $\cdot$ is the intersection form of $M^4$.
By the Freedman-Kirby theorem, $(\frac{\sigma-\rF\cdot\rF}{8} )(M^4)=\text{Arf}(M^4,\rF)\mod2$. 
The PD is the Poincar\'e dual.
Note that 
$c_1(\U(2))^2=\Sq^2c_1(\U(2))=(w_2+w_1^2)c_1(\U(2))=a^2c_1(\U(2))\mod2$.
$c_3(\SU(3))=\Sq^2c_2(\SU(3))=(w_2+w_1^2)c_2(\SU(3))=a^2c_2(\SU(3))\mod2$.
$c_1(\U(2))c_2(\U(2))=\Sq^2c_2(\U(2))=(w_2+w_1^2)c_2(\U(2))=a^2c_2(\U(2))\mod2$. The $\eta(\text{PD}(a))$ is explained in Table \ref{table:Z4SU3SU2U1Bordism}.
{See Appendix \ref{sec:comment-2} for comment on the difference in 5d between Table \ref{table:Z4SU3SU2U1Bordism} and \ref{table:Z4SU3SU2U1Z2Bordism}.}
}
\label{table:Z4SU3SU2U1Z2Bordism}
\end{table}

By \eqref{eq:TPexact}, we obtain the cobordism group $\TP_d
({\Spin \times_{\Z_2} \Z_4 \times {\frac{\SU(3)\times \SU(2)\times \U(1)}{\Z_2}}})
$ shown in Table \ref{table:Z4SU3SU2U1Z2TP}.

\begin{table}[H]
\centering
\hspace*{-19mm}
\begin{tabular}{ c c c}
\hline
\multicolumn{3}{c}{Cobordism group}\\
\hline
$d$ & 

$\begin{matrix}\TP_d\\
({\Spin \times_{\Z_2} \Z_4 \times {\frac{\SU(3)\times \SU(2)\times \U(1)}{\Z_2}}})
\end{matrix}$
& topological terms \\
\hline
0& 0 \\
\hline
1& $\Z\times\Z_4$  & $\text{CS}_1^{\U(2)},\eta'$  \\
\hline
2& 0 &  \\
\hline
3 & $\Z^4\times\Z_2$ &  ${\mu},
\frac{a^2\text{CS}_1^{\U(2)}+c_1(\U(2))\text{CS}_1^{\U(2)}}{2},\text{CS}_3^{\U(2)},\text{CS}_3^{\SU(3)},ac_1(\U(2))$ \\
\hline
4 & 0 & \\
\hline
5 & $\Z^5\times\Z_2^2\times\Z_4\times\Z_{16}$  &
$\begin{matrix}
\mu(\text{PD}(c_1(\U(2)))),
{\text{CS}_1^{\U(2)}c_1(\U(2))^2},
{\frac{a^2\text{CS}_3^{\U(2)}+\text{CS}_1^{\U(2)}c_2(\U(2)) }{2}\sim}
\frac{a^2\text{CS}_3^{\U(2)}+c_1(\U(2))\text{CS}_3^{\U(2)}}{2},\\
{\text{CS}_1^{\U(2)}c_2(\SU(3)) \sim} c_1(\U(2))\text{CS}_3^{\SU(3)},\frac{a^2\text{CS}_3^{\SU(3)}+\text{CS}_5^{\SU(3)}}{2},
\\
ac_2(\SU(3)),ac_2(\U(2)),c_1(\U(2))^2\eta',\eta(\text{PD}(a))
\end{matrix}$ \\
\hline
\end{tabular}
\caption{Topological phase classification ($\equiv$ TP) as a cobordism group, following Table \ref{table:Z4SU3SU2U1Z2Bordism}. 
$\eta'$ is a $\Z_4$ valued 1d eta invariant explained in Table \ref{table:Z4SU3SU2U1Bordism}.
$\eta$ is the 4d eta invariant.
$c_i(G)$ is the Chern class of the associated vector bundle of the principal $G$-bundle.
$\text{CS}_{2n-1}^{G}$ is the Chern-Simons form of the associated vector bundle of the principal $G$-bundle.
$a$ is the generator of $\H^1(\B\Z_2,\Z_2)$.
The PD is the Poincar\'e dual.
The $\mu$ is the 3d Rokhlin invariant.
If $\partial M^4=M^3$, then $\mu(M^3)=(\frac{\sigma-\rF\cdot\rF}{8} )(M^4)$, thus
$\mu(\text{PD}(c_1(\U(2))))$ is related to $\frac{c_1(\U(2))(\sigma-\rF \cdot \rF )}{8}$
in Table \ref{table:Z4SU3SU2U1Z2Bordism}. The $\eta(\text{PD}(a))$ is explained in Table \ref{table:Z4SU3SU2U1Bordism}.
}
\label{table:Z4SU3SU2U1Z2TP}
\end{table}
{In Table \ref{table:Z4SU3SU2U1Z2TP}, note that $\text{CS}_1^{\U(2)} c_2(\U(2))= c_1(\U(2)) \text{CS}_3^{\U(2)}$ 
and $\text{CS}_1^{\U(2)}c_2(\SU(3))= c_1(\U(2)) \text{CS}_3^{\SU(3)}$ up to a total derivative term (vanishing on a closed 5-manifold). See footnote \ref{ft:Chern-Simons}.}

\subsection{
${\Spin \times_{\Z_2} \Z_4\times {\frac{\SU(3)\times \SU(2)\times \U(1)}{\Z_3}}}$ model
}
We consider $G={\Spin \times_{\Z_2} \Z_4 \times {\frac{\SU(3)\times \SU(2)\times \U(1)}{\Z_3}}}=\Spin \times_{\Z_2} \Z_4\times {\U(3)\times \SU(2)}$, the Madsen-Tillmann spectrum $MTG$ of the group $G$ is 
\bea
MTG
&=&M\Spin\wedge(\B\Z_2)^{2\xi}\wedge (\B (\U(3)\times \SU(2)))_+\nn\\
&=&M\Spin\wedge\Sigma^{-2}\RP_2^{\infty}\wedge (\B (\U(3)\times \SU(2)))_+.
\eea
 The $(\B(\U(3)\times \SU(2)))_+$ is the disjoint union of the classifying space $\B(\U(3)\times \SU(2))$ and a point, see footnote \ref{ft:X_+}.

Since 
\bea
\H^*(M\Spin\wedge(\B\Z_2)^{2\xi},\Z_3)=\H^*(M\Spin,\Z_3),
\eea
similarly as the discussion in \Sec{sec:SU3SU2U1Z3}, there is no 3-torsion in $\Omega_d^{\Spin \times_{\Z_2} \Z_4\times {\U(3)\times \SU(2)}}$.

For the dimension $d=t-s<8$, since there is no odd torsion (see footnote \ref{ft:no-odd-torsion}), by \eqref{eq:ExtA_2(1)}, we have the Adams spectral sequence
\bea
\Ext_{\A_2(1)}^{s,t}(\H^{*+2}(\RP_2^{\infty},\Z_2)\otimes\H^*(\B(\U(3)\times \SU(2)),\Z_2),\Z_2)\Rightarrow\Omega_{t-s}^{\Spin \times_{\Z_2} \Z_4\times {\U(3)\times \SU(2)}}.
\eea

Based on Figure \ref{fig:A_2(1)U3SU2} and \ref{fig:A_2(1)RP_2Ceta}, we obtain the $\A_2(1)$-module structure of $\H^{*+2}(\RP_2^{\infty},\Z_2)\otimes\H^*(\B(\U(3)\times\SU(2)),\Z_2)$ below degree 6, as shown in Figure \ref{fig:A_2(1)RP_2U3SU2}.

\begin{figure}[H]
\begin{center}
\begin{tikzpicture}[scale=0.5]

\draw[fill] (0,0) circle(.1);
\draw[fill] (0,1) circle(.1);
\draw[fill] (0,2) circle(.1);
\draw (0,0) to [out=30,in=30] (0,2);
\draw (0,1) -- (0,2);
\draw[fill] (0,3) circle(.1);
\draw (0,1) to [out=150,in=150] (0,3);
\draw[fill] (0,4) circle(.1);
\draw (0,3) -- (0,4);
\draw[fill] (0,5) circle(.1);
\draw[fill] (0,6) circle(.1);
\draw (0,4) to [out=150,in=150] (0,6);
\draw (0,5) -- (0,6);
\draw[fill] (0,7) circle(.1);
\draw (0,5) to [out=30,in=30] (0,7);
\draw[fill] (0,8) circle(.1);
\draw (0,7) -- (0,8);

\node[below] at (2,4) {$c_2'$};

\draw[fill] (2,4) circle(.1);
\draw[fill] (2,5) circle(.1);
\draw[fill] (2,6) circle(.1);
\draw (2,4) to [out=30,in=30] (2,6);
\draw (2,5) -- (2,6);
\draw[fill] (2,7) circle(.1);
\draw (2,5) to [out=150,in=150] (2,7);
\draw[fill] (2,8) circle(.1);
\draw (2,7) -- (2,8);
\draw[fill] (2,9) circle(.1);
\draw[fill] (2,10) circle(.1);
\draw (2,8) to [out=150,in=150] (2,10);
\draw (2,9) -- (2,10);
\draw[fill] (2,11) circle(.1);
\draw (2,9) to [out=30,in=30] (2,11);
\draw[fill] (2,12) circle(.1);
\draw (2,11) -- (2,12);

\node[below] at (4,4) {$c_2$};

\draw[fill] (4,4) circle(.1);
\draw[fill] (4,6) circle(.1);
\draw (4,4) to [out=150,in=150] (4,6);

\node[below] at (6,5) {$ac_2$};

\draw[fill] (6,5) circle(.1);
\draw[fill] (6,6) circle(.1);
\draw (6,5) -- (6,6);
\draw[fill] (6,7) circle(.1);
\draw (6,5) to [out=150,in=150] (6,7);
\draw[fill] (7,7) circle(.1);
\draw[fill] (7,8) circle(.1);
\draw (7,7) -- (7,8);
\draw (6,6) to [out=30,in=150] (7,8);
\draw[fill] (6,8) circle(.1);
\draw (6,7) -- (6,8);
\draw[fill] (6,9) circle(.1);
\draw[fill] (6,10) circle(.1);
\draw (6,9) -- (6,10);
\draw (6,8) to [out=150,in=150] (6,10);
\draw[fill] (7,9) circle(.1);
\draw (7,7) to [out=30,in=30] (7,9);
\draw[fill] (7,10) circle(.1);
\draw (7,9) -- (7,10);
\draw[fill] (6,11) circle(.1);
\draw (6,9) to [out=30,in=30] (6,11);
\draw[fill] (7,11) circle(.1);
\draw[fill] (7,12) circle(.1);
\draw (7,11) -- (7,12);
\draw (7,10) to [out=150,in=150] (7,12);
\draw[fill] (6,12) circle(.1);
\draw (6,11) -- (6,12);
\draw[fill] (7,13) circle(.1);
\draw[fill] (7,14) circle(.1);
\draw (7,13) -- (7,14);
\draw (7,11) to [out=30,in=30] (7,13);

\node[below] at (9,2) {$c_1$};

\draw[fill] (9,2) circle(.1);
\draw[fill] (9,4) circle(.1);
\draw (9,2) to [out=150,in=150] (9,4);

\node[below] at (11,3) {$ac_1$};

\draw[fill] (11,3) circle(.1);
\draw[fill] (11,4) circle(.1);
\draw (11,3) -- (11,4);
\draw[fill] (11,5) circle(.1);
\draw (11,3) to [out=150,in=150] (11,5);
\draw[fill] (12,5) circle(.1);
\draw[fill] (12,6) circle(.1);
\draw (12,5) -- (12,6);
\draw (11,4) to [out=30,in=150] (12,6);
\draw[fill] (11,6) circle(.1);
\draw (11,5) -- (11,6);
\draw[fill] (11,7) circle(.1);
\draw[fill] (11,8) circle(.1);
\draw (11,7) -- (11,8);
\draw (11,6) to [out=150,in=150] (11,8);
\draw[fill] (12,7) circle(.1);
\draw (12,5) to [out=30,in=30] (12,7);
\draw[fill] (12,8) circle(.1);
\draw (12,7) -- (12,8);
\draw[fill] (11,9) circle(.1);
\draw (11,7) to [out=30,in=30] (11,9);
\draw[fill] (12,9) circle(.1);
\draw[fill] (12,10) circle(.1);
\draw (12,9) -- (12,10);
\draw (12,8) to [out=150,in=150] (12,10);
\draw[fill] (11,10) circle(.1);
\draw (11,9) -- (11,10);
\draw[fill] (12,11) circle(.1);
\draw[fill] (12,12) circle(.1);
\draw (12,11) -- (12,12);
\draw (12,9) to [out=30,in=30] (12,11);

\node[below] at (14,6) {$c_1^3$};

\draw[fill] (14,6) circle(.1);
\draw[fill] (14,8) circle(.1);
\draw (14,6) to [out=150,in=150] (14,8);

\node[below] at (16,6) {$c_3$};

\draw[fill] (16,6) circle(.1);
\draw[fill] (16,8) circle(.1);
\draw (16,6) to [out=150,in=150] (16,8);

\node[below] at (18,6) {$c_1c_2'$};

\draw[fill] (18,6) circle(.1);
\draw[fill] (18,8) circle(.1);
\draw (18,6) to [out=150,in=150] (18,8);

\end{tikzpicture}
\end{center}
\caption{The $\A_2(1)$-module structure of $\H^{*+2}(\RP_2^{\infty},\Z_2)\otimes\H^*(\B(\U(3)\times\SU(2)),\Z_2)$ below degree 6. {Here, $c_i$ is the Chern class of U(3) bundle, and $c_i'$ is the Chern class of SU(2) bundle.}}
\label{fig:A_2(1)RP_2U3SU2}
\end{figure}

The $E_2$ page is shown in Figure \ref{fig:E_2Z4U3SU2}.
Here we have used the correspondence between $\A_2(1)$-module structure and the $E_2$ page shown in Figure \ref{fig:Ceta}, \ref{fig:RP_2} and \ref{fig:L_1}.

\begin{figure}[H]
\begin{center}
\begin{tikzpicture}
\node at (0,-1) {0};
\node at (1,-1) {1};
\node at (2,-1) {2};
\node at (3,-1) {3};
\node at (4,-1) {4};
\node at (5,-1) {5};
\node at (6,-1) {6};
\node at (7,-1) {$t-s$};
\node at (-1,0) {0};
\node at (-1,1) {1};
\node at (-1,2) {2};
\node at (-1,3) {3};
\node at (-1,4) {4};
\node at (-1,5) {5};
\node at (-1,6) {$s$};

\draw[->] (-0.5,-0.5) -- (-0.5,6);
\draw[->] (-0.5,-0.5) -- (7,-0.5);

\draw (0,0) -- (0,5);

\draw (1,0) -- (1,1);

\draw (2,0) -- (2,5);

\draw[fill] (3,0) circle(0.05);

\draw (4,1) -- (4,5);

\draw (4.1,2) -- (4.1,5);
\draw (4.2,0) -- (4.2,5);

\draw (4.3,0) -- (4.3,5);

\draw (5,0) -- (5,1);
\draw (5.1,0) -- (5.1,3);

\draw (5.2,0) -- (5.2,1);

\draw[fill] (5.3,0) circle(0.05);

\draw (6,1) -- (6,5);

\draw (6.1,2) -- (6.1,5);

\draw (6.2,0) -- (6.2,5);

\draw (6.3,0) -- (6.3,5);

\draw (6.4,0) -- (6.4,5);

\end{tikzpicture}
\end{center}
\caption{$\Omega_*^{\Spin\times_{\Z_2}\Z_4\times \U(3)\times\SU(2)}$.}
\label{fig:E_2Z4U3SU2}
\end{figure}

Thus we obtain the bordism group $\Omega^{\Spin \times_{\Z_2} \Z_4 \times {\frac{\SU(3)\times \SU(2)\times \U(1)}{\Z_3}}}_d$ shown in Table \ref{table:Z4SU3SU2U1Z3Bordism}.

\begin{table}[H]
\centering
\hspace*{-25mm}
\begin{tabular}{ c c c}
\hline
\multicolumn{3}{c}{Bordism group}\\
\hline
$d$ & 
$\Omega^{\Spin \times_{\Z_2} \Z_4 \times {\frac{\SU(3)\times \SU(2)\times \U(1)}{\Z_3}}}_d$
& bordism invariants \\
\hline
0&  $\Z$ \\
\hline
1&  $\Z_4$& $\eta'$\\
\hline
2&  $\Z$ & $c_1(\U(3))$ \\
\hline
3 & $\Z_2$ & $ac_1(\U(3))$ \\
\hline
4 & $\Z^4$ & ${\frac{\sigma-\rF\cdot \rF }{8},
\frac{a^2c_1(\U(3))+c_1(\U(3))^2}{2}}, c_2(\SU(2)),c_2(\U(3))$ \\
\hline
5 &  $\Z_2\times\Z_4^2\times\Z_{16}$ & $ac_2(\U(3)),c_2(\SU(2))\eta',c_1(\U(3))^2\eta',\eta(\text{PD}(a))$ \\
\hline
6 & $\Z^5$  & $\begin{matrix}c_1(\U(3))\frac{\sigma-\rF\cdot \rF }{8},c_1(\U(3))^3,c_1(\U(3))c_2(\SU(2)), 
\\\frac{a^2c_2(\U(3))+c_1(\U(3))c_2(\U(3))+c_3(\U(3))}{2}, c_3(\U(3))\end{matrix}$ \\
\hline
\end{tabular}
\caption{Bordism group. 
$\eta'$ is a $\Z_4$ valued 1d eta invariant explained in Table \ref{table:Z4SU3SU2U1Bordism}.
$\eta$ is the 4d eta invariant.
$\sigma$ is the signature of manifold.
$c_i(G)$ is the Chern class of the associated vector bundle of the principal $G$-bundle.
$a$ is the generator of $\H^1(\B\Z_2,\Z_2)$.
$\rF $ is the characteristic 2-surface \cite{Saveliev} in a 4-manifold $M^4$, it satisfies the condition $\rF\cdot x=x\cdot x\mod2$ for all $x\in\H_2(M^4,\Z)$.
Here $\cdot$ is the intersection form of $M^4$.
By the Freedman-Kirby theorem, $(\frac{\sigma-\rF\cdot\rF}{8} )(M^4)=\text{Arf}(M^4,\rF)\mod2$. 
The PD is the Poincar\'e dual.
Note that
$c_1(\U(3))^2=\Sq^2c_1(\U(3))=(w_2+w_1^2)c_1(\U(3))=a^2c_1(\U(3))\mod2$.
$c_1(\U(3))c_2(\U(3))+c_3(\U(3))=\Sq^2c_2(\U(3))=(w_2+w_1^2)c_2(\U(3))=a^2c_2(\U(3))\mod2$.
The $\eta(\text{PD}(a))$ is explained in Table \ref{table:Z4SU3SU2U1Bordism}.
}
\label{table:Z4SU3SU2U1Z3Bordism}
\end{table}

By \eqref{eq:TPexact}, we obtain the cobordism group $\TP_d
({\Spin \times_{\Z_2} \Z_4 \times {\frac{\SU(3)\times \SU(2)\times \U(1)}{\Z_3}}})
$ shown in Table \ref{table:Z4SU3SU2U1Z3TP}.

\begin{table}[H]
\centering
\hspace*{-14mm}
\begin{tabular}{ c c c}
\hline
\multicolumn{3}{c}{Cobordism group}\\
\hline
$d$ & 

$\begin{matrix}\TP_d\\
({\Spin \times_{\Z_2} \Z_4 \times {\frac{\SU(3)\times \SU(2)\times \U(1)}{\Z_3}}})
\end{matrix}$
& topological terms \\
\hline
0& 0 \\
\hline
1& $\Z\times\Z_4$ &  $\text{CS}_1^{\U(3)},\eta'$\\
\hline
2& 0 &  \\
\hline
3 & $\Z^4\times\Z_2$ & ${\mu, 
\frac{a^2\text{CS}_1^{\U(3)}+\text{CS}_1^{\U(3)}c_1(\U(3))}{2}},
\text{CS}_3^{\SU(2)},\text{CS}_3^{\U(3)},ac_1(\U(3))$ \\
\hline
4 & 0 & \\
\hline
5 & $\Z^5\times\Z_2\times\Z_4^2\times\Z_{16}$ & 
$\begin{matrix}\mu(\text{PD}(c_1(\U(3)))),
{c_1(\U(3))^2\text{CS}_1^{\U(3)}, 
\text{CS}_1^{\U(3)}c_2(\SU(2)) \sim c_1(\U(3))\text{CS}_3^{\SU(2)}},
\\
{\frac{a^2\text{CS}_3^{\U(3)}+\text{CS}_1^{\U(3)} c_2(\U(3))+\text{CS}_5^{\U(3)}}{2}
\sim}
\frac{a^2\text{CS}_3^{\U(3)}+c_1(\U(3))\text{CS}_3^{\U(3)}+\text{CS}_5^{\U(3)}}{2},\text{CS}_5^{\U(3)},
\\
ac_2(\U(3)), c_2(\SU(2))\eta',c_1(\U(3))^2\eta',\eta(\text{PD}(a))
\end{matrix}$\\
\hline
\end{tabular}
\caption{Topological phase classification ($\equiv$ TP) as a cobordism group, following Table \ref{table:Z4SU3SU2U1Z3Bordism}. 
$\eta'$ is a $\Z_4$ valued 1d eta invariant explained in Table \ref{table:Z4SU3SU2U1Bordism}.
$\eta$ is the 4d eta invariant.
$c_i(G)$ is the Chern class of the associated vector bundle of the principal $G$-bundle.
$\text{CS}_{2n-1}^{G}$ is the Chern-Simons form of the associated vector bundle of the principal $G$-bundle.
$a$ is the generator of $\H^1(\B\Z_2,\Z_2)$.
The PD is the Poincar\'e dual.
The $\mu$ is the 3d Rokhlin invariant.
If $\partial M^4=M^3$, then $\mu(M^3)=(\frac{\sigma-\rF\cdot\rF}{8} )(M^4)$, thus
$\mu(\text{PD}(c_1(\U(3))))$ is related to $\frac{c_1(\U(3))(\sigma-\rF \cdot \rF )}{8}$
in Table \ref{table:Z4SU3SU2U1Z3Bordism}.
The $\eta(\text{PD}(a))$ is explained in Table \ref{table:Z4SU3SU2U1Bordism}.
}
\label{table:Z4SU3SU2U1Z3TP}
\end{table}
{In Table \ref{table:Z4SU3SU2U1Z3TP}, note that $\text{CS}_1^{\U(3)} c_2(\U(3))= c_1(\U(3)) \text{CS}_3^{\U(3)}$ 
and $\text{CS}_1^{\U(3)}c_2(\SU(2))= c_1(\U(3)) \text{CS}_3^{\SU(2)}$ up to a total derivative term (vanishing on a closed 5-manifold). See footnote \ref{ft:Chern-Simons}.}

\subsection{
${\Spin \times_{\Z_2} \Z_4\times {\frac{\SU(3)\times \SU(2)\times \U(1)}{\Z_6}}}$ model
}
We consider $G={\Spin \times_{\Z_2} \Z_4 \times {\frac{\SU(3)\times \SU(2)\times \U(1)}{\Z_6}}}=\Spin \times_{\Z_2} \Z_4\times \rS(\U(3)\times\U(2))$, the Madsen-Tillmann spectrum $MTG$ of the group $G$ is 
\bea
MTG
&=&M\Spin\wedge(\B\Z_2)^{2\xi}\wedge (\B (\rS(\U(3)\times\U(2))))_+\nn\\
&=&M\Spin\wedge\Sigma^{-2}\RP_2^{\infty}\wedge (\B (\rS(\U(3)\times\U(2))))_+.
\eea
 The $(\B(\rS(\U(3)\times \U(2))))_+$ is the disjoint union of the classifying space $\B(\rS(\U(3)\times \U(2)))$ and a point, see footnote \ref{ft:X_+}.

Since 
\bea
\H^*(M\Spin\wedge(\B\Z_2)^{2\xi},\Z_3)=\H^*(M\Spin,\Z_3),
\eea
and
\bea
\H^*(\B(\frac{\SU(3)\times \SU(2)\times \U(1)}{\Z_6}),\Z_3)=\H^*(\B(\frac{\SU(3)\times \SU(2)\times \U(1)}{\Z_3}),\Z_3),
\eea
similarly as the discussion in \Sec{sec:SU3SU2U1Z3}, there is no 3-torsion in $\Omega_d^{\Spin \times_{\Z_2} \Z_4\times \rS(\U(3)\times\U(2))}$.

For the dimension $d=t-s<8$, since there is no odd torsion (see footnote \ref{ft:no-odd-torsion}), by \eqref{eq:ExtA_2(1)}, we have the Adams spectral sequence
\bea
\Ext_{\A_2(1)}^{s,t}(\H^{*+2}(\RP_2^{\infty},\Z_2)\otimes\H^*(\B (\rS(\U(3)\times \U(2))),\Z_2)\Rightarrow\Omega_{t-s}^{\Spin \times_{\Z_2} \Z_4\times \rS(\U(3)\times\U(2))}.
\eea

Based on Figure \ref{fig:A_2(1)S(U3U2)} and \ref{fig:A_2(1)RP_2Ceta}, we obtain the $\A_2(1)$-module structure of $\H^{*+2}(\RP_2^{\infty},\Z_2)\otimes\H^*(\B (\rS(\U(3)\times \U(2))),\Z_2)$ below degree 6, as shown in Figure \ref{fig:A_2(1)RP_2S(U3U2)}.

\begin{figure}[H]
\begin{center}
\begin{tikzpicture}[scale=0.5]

\draw[fill] (0,0) circle(.1);
\draw[fill] (0,1) circle(.1);
\draw[fill] (0,2) circle(.1);
\draw (0,0) to [out=30,in=30] (0,2);
\draw (0,1) -- (0,2);
\draw[fill] (0,3) circle(.1);
\draw (0,1) to [out=150,in=150] (0,3);
\draw[fill] (0,4) circle(.1);
\draw (0,3) -- (0,4);
\draw[fill] (0,5) circle(.1);
\draw[fill] (0,6) circle(.1);
\draw (0,4) to [out=150,in=150] (0,6);
\draw (0,5) -- (0,6);
\draw[fill] (0,7) circle(.1);
\draw (0,5) to [out=30,in=30] (0,7);
\draw[fill] (0,8) circle(.1);
\draw (0,7) -- (0,8);

\node[below] at (4,4) {$c_2$};

\draw[fill] (4,4) circle(.1);
\draw[fill] (4,6) circle(.1);
\draw (4,4) to [out=150,in=150] (4,6);

\node[below] at (6,5) {$ac_2$};

\draw[fill] (6,5) circle(.1);
\draw[fill] (6,6) circle(.1);
\draw (6,5) -- (6,6);
\draw[fill] (6,7) circle(.1);
\draw (6,5) to [out=150,in=150] (6,7);
\draw[fill] (7,7) circle(.1);
\draw[fill] (7,8) circle(.1);
\draw (7,7) -- (7,8);
\draw (6,6) to [out=30,in=150] (7,8);
\draw[fill] (6,8) circle(.1);
\draw (6,7) -- (6,8);
\draw[fill] (6,9) circle(.1);
\draw[fill] (6,10) circle(.1);
\draw (6,9) -- (6,10);
\draw (6,8) to [out=150,in=150] (6,10);
\draw[fill] (7,9) circle(.1);
\draw (7,7) to [out=30,in=30] (7,9);
\draw[fill] (7,10) circle(.1);
\draw (7,9) -- (7,10);
\draw[fill] (6,11) circle(.1);
\draw (6,9) to [out=30,in=30] (6,11);
\draw[fill] (7,11) circle(.1);
\draw[fill] (7,12) circle(.1);
\draw (7,11) -- (7,12);
\draw (7,10) to [out=150,in=150] (7,12);
\draw[fill] (6,12) circle(.1);
\draw (6,11) -- (6,12);
\draw[fill] (7,13) circle(.1);
\draw[fill] (7,14) circle(.1);
\draw (7,13) -- (7,14);
\draw (7,11) to [out=30,in=30] (7,13);

\node[below] at (9,2) {$c_1$};

\draw[fill] (9,2) circle(.1);
\draw[fill] (9,4) circle(.1);
\draw (9,2) to [out=150,in=150] (9,4);

\node[below] at (11,3) {$ac_1$};

\draw[fill] (11,3) circle(.1);
\draw[fill] (11,4) circle(.1);
\draw (11,3) -- (11,4);
\draw[fill] (11,5) circle(.1);
\draw (11,3) to [out=150,in=150] (11,5);
\draw[fill] (12,5) circle(.1);
\draw[fill] (12,6) circle(.1);
\draw (12,5) -- (12,6);
\draw (11,4) to [out=30,in=150] (12,6);
\draw[fill] (11,6) circle(.1);
\draw (11,5) -- (11,6);
\draw[fill] (11,7) circle(.1);
\draw[fill] (11,8) circle(.1);
\draw (11,7) -- (11,8);
\draw (11,6) to [out=150,in=150] (11,8);
\draw[fill] (12,7) circle(.1);
\draw (12,5) to [out=30,in=30] (12,7);
\draw[fill] (12,8) circle(.1);
\draw (12,7) -- (12,8);
\draw[fill] (11,9) circle(.1);
\draw (11,7) to [out=30,in=30] (11,9);
\draw[fill] (12,9) circle(.1);
\draw[fill] (12,10) circle(.1);
\draw (12,9) -- (12,10);
\draw (12,8) to [out=150,in=150] (12,10);
\draw[fill] (11,10) circle(.1);
\draw (11,9) -- (11,10);
\draw[fill] (12,11) circle(.1);
\draw[fill] (12,12) circle(.1);
\draw (12,11) -- (12,12);
\draw (12,9) to [out=30,in=30] (12,11);

\node[below] at (2,6) {$c_1^3$};

\draw[fill] (2,6) circle(.1);
\draw[fill] (2,8) circle(.1);
\draw (2,6) to [out=150,in=150] (2,8);

\node[below] at (19,6) {$c_3$};

\draw[fill] (19,6) circle(.1);
\draw[fill] (19,8) circle(.1);
\draw (19,6) to [out=150,in=150] (19,8);

\node[below] at (14,4) {$c_2'$};

\draw[fill] (14,4) circle(.1);
\draw[fill] (14,6) circle(.1);
\draw (14,4) to [out=150,in=150] (14,6);

\node[below] at (16,5) {$ac_2'$};

\draw[fill] (16,5) circle(.1);
\draw[fill] (16,6) circle(.1);
\draw (16,5) -- (16,6);
\draw[fill] (16,7) circle(.1);
\draw (16,5) to [out=150,in=150] (16,7);
\draw[fill] (17,7) circle(.1);
\draw[fill] (17,8) circle(.1);
\draw (17,7) -- (17,8);
\draw (16,6) to [out=30,in=150] (17,8);
\draw[fill] (16,8) circle(.1);
\draw (16,7) -- (16,8);
\draw[fill] (16,9) circle(.1);
\draw[fill] (16,10) circle(.1);
\draw (16,9) -- (16,10);
\draw (16,8) to [out=150,in=150] (16,10);
\draw[fill] (17,9) circle(.1);
\draw (17,7) to [out=30,in=30] (17,9);
\draw[fill] (17,10) circle(.1);
\draw (17,9) -- (17,10);
\draw[fill] (16,11) circle(.1);
\draw (16,9) to [out=30,in=30] (16,11);
\draw[fill] (17,11) circle(.1);
\draw[fill] (17,12) circle(.1);
\draw (17,11) -- (17,12);
\draw (17,10) to [out=150,in=150] (17,12);
\draw[fill] (16,12) circle(.1);
\draw (16,11) -- (16,12);
\draw[fill] (17,13) circle(.1);
\draw[fill] (17,14) circle(.1);
\draw (17,13) -- (17,14);
\draw (17,11) to [out=30,in=30] (17,13);

\end{tikzpicture}
\end{center}
\caption{The $\A_2(1)$-module structure of $\H^{*+2}(\RP_2^{\infty},\Z_2)\otimes\H^*(\B (\rS(\U(3)\times \U(2))),\Z_2)$ below degree 6. {Here, $c_i$ is the Chern class of U(3) bundle, and $c_i'$ is the Chern class of U(2) bundle, and $c_1=c_1'$.}}
\label{fig:A_2(1)RP_2S(U3U2)}
\end{figure}

The $E_2$ page is shown in Figure \ref{fig:E_2Z4S(U3U2)}.
Here we have used the correspondence between $\A_2(1)$-module structure and the $E_2$ page shown in Figure \ref{fig:Ceta}, \ref{fig:RP_2} and \ref{fig:L_1}.

\begin{figure}[H]
\begin{center}
\begin{tikzpicture}
\node at (0,-1) {0};
\node at (1,-1) {1};
\node at (2,-1) {2};
\node at (3,-1) {3};
\node at (4,-1) {4};
\node at (5,-1) {5};
\node at (6,-1) {6};
\node at (7,-1) {$t-s$};
\node at (-1,0) {0};
\node at (-1,1) {1};
\node at (-1,2) {2};
\node at (-1,3) {3};
\node at (-1,4) {4};
\node at (-1,5) {5};
\node at (-1,6) {$s$};

\draw[->] (-0.5,-0.5) -- (-0.5,6);
\draw[->] (-0.5,-0.5) -- (7,-0.5);

\draw (0,0) -- (0,5);

\draw (1,0) -- (1,1);

\draw (2,0) -- (2,5);

\draw[fill] (3,0) circle(0.05);

\draw (4,2) -- (4,5);

\draw (4.1,1) -- (4.1,5);
\draw (4.2,0) -- (4.2,5);

\draw (4.3,0) -- (4.3,5);

\draw (5,0) -- (5,3);
\draw (5.1,0) -- (5.1,1);

\draw[fill] (5.2,0) circle(0.05);

\draw[fill] (5.3,0) circle(0.05);

\draw (6,1) -- (6,5);

\draw (6.1,2) -- (6.1,5);

\draw (6.2,1) -- (6.2,5);

\draw (6.3,0) -- (6.3,5);

\draw (6.4,0) -- (6.4,5);

\end{tikzpicture}
\end{center}
\caption{$\Omega_*^{\Spin\times_{\Z_2}\Z_4\times \rS(\U(3)\times\U(2))}$.}
\label{fig:E_2Z4S(U3U2)}
\end{figure}

Thus we obtain the bordism group $\Omega^{\Spin \times_{\Z_2} \Z_4 \times {\frac{\SU(3)\times \SU(2)\times \U(1)}{\Z_6}}}_d$ shown in Table \ref{table:Z4SU3SU2U1Z6Bordism}.

\begin{table}[H]
\centering
\hspace*{-25mm}
\begin{tabular}{ c c c}
\hline
\multicolumn{3}{c}{Bordism group}\\
\hline
$d$ & 
$\Omega^{\Spin \times_{\Z_2} \Z_4 \times {\frac{\SU(3)\times \SU(2)\times \U(1)}{\Z_6}}}_d$
& bordism invariants \\
\hline
0&  $\Z$ \\
\hline
1& $\Z_4$  & $\eta'$ \\
\hline
2&  $\Z$  & $c_1(\U(3))$ \\
\hline
3 & $\Z_2$  & $ac_1(\U(3))$\\
\hline
4 & $\Z^4$ & ${\frac{\sigma-\rF\cdot \rF }{8},
\frac{a^2c_1(\U(3))+c_1(\U(3))^2}{2}},
c_2(\U(2)),c_2(\U(3))$ \\
\hline
5 & $\Z_2^2\times\Z_4\times\Z_{16}$  &$ac_2(\U(3)),ac_2(\U(2)),c_1(\U(3))^2\eta',\eta(\text{PD}(a))$ \\
\hline
6 & $\Z^5$  &$\begin{matrix}c_1(\U(3))\frac{\sigma-\rF\cdot \rF }{8},{c_1(\U(3))^3,
\frac{a^2c_2(\U(2))+c_1(\U(2))c_2(\U(2))}{2}},
\\
\frac{a^2c_2(\U(3))+c_1(\U(3))c_2(\U(3))+c_3(\U(3))}{2},
c_3(\U(3))\end{matrix}$ \\
\hline
\end{tabular}
\caption{Bordism group. 
$\eta'$ is a $\Z_4$ valued 1d eta invariant explained in Table \ref{table:Z4SU3SU2U1Bordism}.
$\eta$ is the 4d eta invariant.
$\sigma$ is the signature of manifold.
$c_i(G)$ is the Chern class of the associated vector bundle of the principal $G$-bundle.
$a$ is the generator of $\H^1(\B\Z_2,\Z_2)$.
$\rF $ is the characteristic 2-surface \cite{Saveliev} in a 4-manifold $M^4$, it satisfies the condition $\rF\cdot x=x\cdot x\mod2$ for all $x\in\H_2(M^4,\Z)$.
Here $\cdot$ is the intersection form of $M^4$.
By the Freedman-Kirby theorem, $(\frac{\sigma-\rF\cdot\rF}{8} )(M^4)=\text{Arf}(M^4,\rF)\mod2$. 
The PD is the Poincar\'e dual.
Here $c_1(\U(3))$ is identified with $c_1(\U(2))$.
Note that 
$c_1(\U(3))^2=\Sq^2c_1(\U(3))=(w_2+w_1^2)c_1(\U(3))=a^2c_1(\U(3))\mod2$.
$c_1(\U(3))c_2(\U(3))+c_3(\U(3))=\Sq^2c_2(\U(3))=(w_2+w_1^2)c_2(\U(3))=a^2c_2(\U(3))\mod2$.
$c_1(\U(2))c_2(\U(2))=\Sq^2c_2(\U(2))=(w_2+w_1^2)c_2(\U(2))=a^2c_2(\U(2))\mod2$. The $\eta(\text{PD}(a))$ is explained in Table \ref{table:Z4SU3SU2U1Bordism}.
}
\label{table:Z4SU3SU2U1Z6Bordism}
\end{table}

By \eqref{eq:TPexact}, we obtain the cobordism group $\TP_d
({\Spin \times_{\Z_2} \Z_4 \times {\frac{\SU(3)\times \SU(2)\times \U(1)}{\Z_6}}})
$ shown in Table \ref{table:Z4SU3SU2U1Z6TP}.

\begin{table}[H]
\centering
\hspace*{-19mm}
\begin{tabular}{ c c c}
\hline
\multicolumn{3}{c}{Cobordism group}\\
\hline
$d$ & 
$\begin{matrix}\TP_d\\
({\Spin \times_{\Z_2} \Z_4 \times {\frac{\SU(3)\times \SU(2)\times \U(1)}{\Z_6}}})
\end{matrix}$
& topological terms \\
\hline
0& 0 \\
\hline
1& $\Z\times\Z_4$  & $\text{CS}_1^{\U(3)},\eta'$  \\
\hline
2& 0 &  \\
\hline
3 & $\Z^4\times\Z_2$ &  ${\mu,\frac{a^2\text{CS}_1^{\U(3)}+\text{CS}_1^{\U(3)}c_1(\U(3))}{2}},
\text{CS}_3^{\U(2)},\text{CS}_3^{\U(3)},ac_1(\U(3))$ \\
\hline
4 & 0 & \\
\hline
5 & $\Z^5\times\Z_2^2\times\Z_4\times\Z_{16}$  &
$\begin{matrix}\mu(\text{PD}(c_1(\U(3)))),
{
c_1(\U(3))^2\text{CS}_1^{\U(3)},
\frac{a^2\text{CS}_3^{\U(2)}+\text{CS}_1^{\U(3)}c_2(\U(2))}{2}
\sim 
\frac{a^2\text{CS}_3^{\U(2)}+c_1(\U(3))\text{CS}_3^{\U(2)}}{2}},
\\
\frac{a^2\text{CS}_3^{\U(3)}+\text{CS}_1^{\U(3)} c_2(\U(3))+\text{CS}_5^{\U(3)}}{2}
{\sim 
\frac{a^2\text{CS}_3^{\U(3)}+c_1(\U(3))\text{CS}_3^{\U(3)}+\text{CS}_5^{\U(3)}}{2}},
\text{CS}_5^{\U(3)},\\
ac_2(\U(3)),ac_2(\U(2)),c_1(\U(3))^2\eta',\eta(\text{PD}(a))
\end{matrix}$ \\
\hline
\end{tabular}
\caption{Topological phase classification ($\equiv$ TP) as a cobordism group, following Table \ref{table:Z4SU3SU2U1Z6Bordism}. 
$\eta'$ is a $\Z_4$ valued 1d eta invariant explained in Table \ref{table:Z4SU3SU2U1Bordism}.
$\eta$ is the 4d eta invariant.
$c_i(G)$ is the Chern class of the associated vector bundle of the principal $G$-bundle.
$\text{CS}_{2n-1}^{G}$ is the Chern-Simons form of the associated vector bundle of the principal $G$-bundle.
$a$ is the generator of $\H^1(\B\Z_2,\Z_2)$.
The PD is the Poincar\'e dual.
The $\mu$ is the 3d Rokhlin invariant.
If $\partial M^4=M^3$, then $\mu(M^3)=(\frac{\sigma-\rF\cdot\rF}{8} )(M^4)$, thus
$\mu(\text{PD}(c_1(\U(3))))$ is related to $\frac{c_1(\U(3))(\sigma-\rF \cdot \rF )}{8}$
in Table \ref{table:Z4SU3SU2U1Z6Bordism}. The $\eta(\text{PD}(a))$ is explained in Table \ref{table:Z4SU3SU2U1Bordism}.
}
\label{table:Z4SU3SU2U1Z6TP}
\end{table}

{In Table \ref{table:Z4SU3SU2U1Z6TP}, note that 
$\text{CS}_1^{\U(3)} c_2(\U(3))= c_1(\U(3)) \text{CS}_3^{\U(3)}$ 
and $\text{CS}_1^{\U(2)}c_2(\U(2))= c_1(\U(2)) \text{CS}_3^{\U(2)}$ up to a total derivative term (vanishing on a closed 5-manifold). Note
that $c_1(\U(2))=c_1(\U(3))$ and
$\text{CS}_1^{\U(2)}=\text{CS}_1^{\U(3)}$.
See footnote \ref{ft:Chern-Simons}.}

\section{Pati-Salam models}\label{sec:PS}

Now we consider the co/bordism classes relevant for Pati-Salam GUT models \cite{Pati1974yyPatiSalamLeptonNumberastheFourthColor} .
There are actually two different cases for modding out different discrete normal subgroups.
\subsection{
$\frac{\Spin\times \frac{\SU(4)\times(\SU(2)\times \SU(2))}{\Z_2}}{\Z_2^F}$ Pati-Salam model
}
We consider $G=\frac{\Spin\times \frac{\SU(4)\times(\SU(2)\times \SU(2))}{\Z_2}}{\Z_2^F}$.

Note that $\frac{\SU(4)}{\Z_2}=\frac{\Spin(6)}{\Z_2}=\SO(6)$, 
and $\frac{\SU(2)\times \SU(2)}{\Z_2}=\frac{\Spin(4)}{\Z_2}=\SO(4)$.

We have a homotopy pullback square
\bea
\xymatrix{
\B G\ar[r]\ar[d]&\B(\SO(6)\times \SO(4))\ar[d]^{w_2'+w_2''}\\
\B\SO\ar[r]^{w_2}&\B^2\Z_2.}
\eea

Here $w_2'=w_2(\SO(6))$, $w_2''=w_2(\SO(4))$.


$\B G$ sits in a fibration sequence
\bea
\B G\to\B\SO\times\B\SO(6)\times\B\SO(4)\xrightarrow{w_2(TM)+w_2(V_1)+w_2(V_2)}\B^2\Z_2.
\eea
Map the three-fold product to itself by the matrix
\bea
\left(\begin{array}{ccc}1&-V_1+6&-V_2+4\\0&1&0\\0&0&1\end{array}\right),
\eea
we find that $\B G$ is homotopy equivalent to $\B\Spin\times\B\SO(6)\times\B\SO(4)$.

So we can identify 
$\B G\to \B \O$ with $\B \Spin\times \B \SO(6)\times \B \SO(4)\to \B \O$: $(W,V_1,V_2)\mapsto -W-(V_1+V_2-10)$.

Hence the Madsen-Tillmann spectrum $MTG$ of the group $G$ is 
\bea
MTG=M\Spin\wedge \Sigma^{-6}M\SO(6)\wedge\Sigma^{-4}M\SO(4).
\eea

For the dimension $d=t-s<8$, since there is no odd torsion (see footnote \ref{ft:no-odd-torsion}), by \eqref{eq:ExtA_2(1)}, we have the Adams spectral sequence
\bea
\Ext_{\A_2(1)}^{s,t}(\H^{*+6}(M\SO(6),\Z_2)\otimes\H^{*+4}(M\SO(4),\Z_2),\Z_2)\Rightarrow\Omega_{t-s}^{\frac{\Spin\times \frac{\SU(4)\times(\SU(2)\times \SU(2))}{\Z_2}}{\Z_2^F}}.
\eea

$\H^{*+6}(M\SO(6),\Z_2)=\Z_2[w_2',w_3',w_4',w_5',w_6']U$, $\H^{*+4}(M\SO(4),\Z_2)=\Z_2[w_2'',w_3'',w_4'']V$.
{Here in this subsection, $w_i'$ is the Stiefel-Whitney class of SO(6) bundle, and $w_i''$ is the Stiefel-Whitney class of SO(4) bundle.}
Here $U$ and $V$ are Thom classes with $\Sq^1U=0$, $\Sq^2U=w_2'U$, $\Sq^1V=0$ and $\Sq^2V=w_2''V$. 

\begin{figure}[H]
\begin{center}
\begin{tikzpicture}[scale=0.5]

\node[below] at (0,0) {$U$};

\draw[fill] (0,0) circle(.1);
\draw[fill] (0,2) circle(.1);
\draw (0,0) to [out=150,in=150] (0,2);
\draw[fill] (0,3) circle(.1);
\draw (0,2) -- (0,3);
\draw[fill] (0,4) circle(.1);
\draw[fill] (0,5) circle(.1);
\draw[fill] (0,6) circle(.1);
\draw (0,3) to [out=150,in=150] (0,5);
\draw (0,4) -- (0,5);
\draw (0,4) to [out=30,in=30] (0,6);

\node[below] at (2,4) {$w_2'^2U$};

\draw[fill] (2,4) circle(.1);
\draw[fill] (2,6) circle(.1);
\draw (2,4) to [out=150,in=150] (2,6);
\draw[fill] (2,7) circle(.1);
\draw (2,6) -- (2,7);
\draw[fill] (2,8) circle(.1);
\draw[fill] (2,9) circle(.1);
\draw[fill] (2,10) circle(.1);
\draw (2,7) to [out=150,in=150] (2,9);
\draw (2,8) -- (2,9);
\draw (2,8) to [out=30,in=30] (2,10);
\draw[fill] (2,11) circle(.1);
\draw (2,10) -- (2,11);

\node[below] at (4,5) {$w_2'w_3'U$};

\draw[fill] (4,5) circle(.1);
\draw[fill] (4,6) circle(.1);
\draw (4,5) -- (4,6);
\draw[fill] (4,7) circle(.1);
\draw (4,5) to [out=150,in=150] (4,7);
\draw[fill] (4,8) circle(.1);
\draw[fill] (5,8) circle(.1);
\draw (4,7) -- (4,8);
\draw (4,6) to [out=30,in=150] (5,8);
\draw[fill] (5,9) circle(.1);
\draw (5,8) -- (5,9);
\draw (4,7) to [out=30,in=150] (5,9);
\draw[fill] (5,10) circle(.1);
\draw (4,8) to [out=30,in=150] (5,10);
\draw[fill] (5,11) circle(.1);
\draw (5,10) -- (5,11);
\draw (5,9) to [out=30,in=30] (5,11);

\node[below] at (6,6) {$w_2'w_4'U$};

\draw[fill] (6,6) circle(.1);
\draw[fill] (6,7) circle(.1);
\draw (6,6) -- (6,7);
\draw[fill] (6,8) circle(.1);
\draw (6,6) to [out=150,in=150] (6,8);
\draw[fill] (6,9) circle(.1);
\draw[fill] (7,9) circle(.1);
\draw (6,8) -- (6,9);
\draw (6,7) to [out=30,in=150] (7,9);
\draw[fill] (7,10) circle(.1);
\draw (7,9) -- (7,10);
\draw (6,8) to [out=30,in=150] (7,10);
\draw[fill] (7,11) circle(.1);
\draw (6,9) to [out=30,in=150] (7,11);
\draw[fill] (7,12) circle(.1);
\draw (7,11) -- (7,12);
\draw (7,10) to [out=30,in=30] (7,12);

\node at (6,3) {$\bigotimes$};

\node[below] at (8,0) {$V$};

\draw[fill] (8,0) circle(.1);
\draw[fill] (8,2) circle(.1);
\draw (8,0) to [out=150,in=150] (8,2);
\draw[fill] (8,3) circle(.1);
\draw (8,2) -- (8,3);

\node[left] at (8,4) {$w_4''V$};

\draw[fill] (8,4) circle(.1);

\node[below] at (10,4) {$w_2''^2V$};

\draw[fill] (10,4) circle(.1);
\draw[fill] (10,5) circle(.1);
\draw[fill] (10,6) circle(.1);
\draw[fill] (11,6) circle(.1);
\draw (10,4) to [out=30,in=150] (11,6);
\draw (10,5) -- (10,6);
\draw[fill] (11,7) circle(.1);
\draw (11,6) -- (11,7);
\draw (10,5) to [out=30,in=150] (11,7);
\draw[fill] (11,8) circle(.1);
\draw[fill] (11,9) circle(.1);
\draw (10,6) to [out=30,in=150] (11,8);
\draw (11,7) to [out=30,in=30] (11,9);
\draw (11,8) -- (11,9);

\node[below] at (12,6) {$w_2''w_4''V$};

\draw[fill] (12,6) circle(.1);
\draw[fill] (12,7) circle(.1);
\draw (12,6) -- (12,7);
\draw[fill] (12,8) circle(.1);
\draw (12,6) to [out=150,in=150] (12,8);
\draw[fill] (13,9) circle(.1);
\draw (12,7) to [out=30,in=150] (13,9);
\draw[fill] (13,10) circle(.1);
\draw (13,9) -- (13,10);
\draw (12,8) to [out=30,in=150] (13,10);

\node at (-2,-10) {$=$};

\node[below] at (0,-15) {$UV$};

\draw[fill] (0,-15) circle(.1);
\draw[fill] (0,-13) circle(.1);
\draw (0,-15) to [out=150,in=150] (0,-13);
\draw[fill] (0,-12) circle(.1);
\draw (0,-13) -- (0,-12);
\draw[fill] (0,-11) circle(.1);
\draw[fill] (0,-10) circle(.1);
\draw[fill] (0,-9) circle(.1);
\draw (0,-12) to [out=150,in=150] (0,-10);
\draw (0,-11) -- (0,-10);
\draw (0,-11) to [out=30,in=30] (0,-9);
\draw[fill] (0,-8) circle(.1);
\draw (0,-9) -- (0,-8);
\draw[fill] (0,-7) circle(.1);
\draw[fill] (0,-6) circle(.1);
\draw[fill] (1,-9) circle(.1);
\draw (0,-8) to [out=150,in=150] (0,-6);
\draw (0,-7) -- (0,-6);
\draw (1,-9) to [out=150,in=30] (0,-7);

\node[below] at (2,-13) {$w_2'UV$};

\draw[fill] (2,-13) circle(.1);
\draw[fill] (2,-12) circle(.1);
\draw (2,-13) -- (2,-12);
\draw[fill] (2,-11) circle(.1);
\draw (2,-13) to [out=150,in=150] (2,-11);
\draw[fill] (2,-10) circle(.1);
\draw[fill] (3,-10) circle(.1);
\draw (2,-11) -- (2,-10);
\draw (2,-12) to [out=30,in=150] (3,-10);
\draw[fill] (3,-9) circle(.1);
\draw (3,-10) -- (3,-9);
\draw (2,-11) to [out=30,in=150] (3,-9);
\draw[fill] (3,-8) circle(.1);
\draw (2,-10) to [out=30,in=150] (3,-8);
\draw[fill] (3,-7) circle(.1);
\draw (3,-8) -- (3,-7);
\draw (3,-9) to [out=30,in=30] (3,-7);

\node[below] at (4,-11) {$w_4''UV$};

\draw[fill] (4,-11) circle(.1);
\draw[fill] (4,-9) circle(.1);
\draw (4,-11) to [out=150,in=150] (4,-9);
\draw[fill] (4,-8) circle(.1);
\draw (4,-9) -- (4,-8);
\draw[fill] (4,-7) circle(.1);
\draw[fill] (4,-6) circle(.1);
\draw[fill] (4,-5) circle(.1);
\draw (4,-8) to [out=150,in=150] (4,-6);
\draw (4,-7) -- (4,-6);
\draw (4,-7) to [out=30,in=30] (4,-5);

\node[below] at (6,-11) {$w_2''^2UV$};

\draw[fill] (6,-11) circle(.1);
\draw[fill] (6,-9) circle(.1);
\draw (6,-11) to [out=150,in=150] (6,-9);
\draw[fill] (6,-8) circle(.1);
\draw (6,-9) -- (6,-8);
\draw[fill] (6,-7) circle(.1);
\draw[fill] (6,-6) circle(.1);
\draw[fill] (6,-5) circle(.1);
\draw (6,-8) to [out=150,in=150] (6,-6);
\draw (6,-7) -- (6,-6);
\draw (6,-7) to [out=30,in=30] (6,-5);
\draw[fill] (6,-4) circle(.1);
\draw (6,-5) -- (6,-4);
\draw[fill] (6,-3) circle(.1);
\draw[fill] (6,-2) circle(.1);
\draw[fill] (6,-1) circle(.1);
\draw (6,-4) to [out=150,in=150] (6,-2);
\draw (6,-3) -- (6,-2);
\draw (6,-3) to [out=30,in=30] (6,-1);
\draw[fill] (6,0) circle(.1);
\draw (6,-1) -- (6,0);

\node[below] at (8,-10) {$w_2''w_3''UV$};

\draw[fill] (8,-10) circle(.1);
\draw[fill] (8,-9) circle(.1);
\draw (8,-10) -- (8,-9);
\draw[fill] (8,-8) circle(.1);
\draw (8,-10) to [out=150,in=150] (8,-8);
\draw[fill] (8,-7) circle(.1);
\draw[fill] (9,-7) circle(.1);
\draw (8,-8) -- (8,-7);
\draw (8,-9) to [out=30,in=150] (9,-7);
\draw[fill] (9,-6) circle(.1);
\draw (9,-7) -- (9,-6);
\draw (8,-8) to [out=30,in=150] (9,-6);
\draw[fill] (9,-5) circle(.1);
\draw (8,-7) to [out=30,in=150] (9,-5);
\draw[fill] (9,-4) circle(.1);
\draw (9,-5) -- (9,-4);
\draw (9,-6) to [out=30,in=30] (9,-4);

\node[below] at (10,-11) {$w_2'^2UV$};

\draw[fill] (10,-11) circle(.1);
\draw[fill] (10,-9) circle(.1);
\draw (10,-11) to [out=150,in=150] (10,-9);
\draw[fill] (10,-8) circle(.1);
\draw (10,-9) -- (10,-8);
\draw[fill] (10,-7) circle(.1);
\draw[fill] (10,-6) circle(.1);
\draw[fill] (10,-5) circle(.1);
\draw (10,-8) to [out=150,in=150] (10,-6);
\draw (10,-7) -- (10,-6);
\draw (10,-7) to [out=30,in=30] (10,-5);
\draw[fill] (10,-4) circle(.1);
\draw (10,-5) -- (10,-4);
\draw[fill] (10,-3) circle(.1);
\draw[fill] (10,-2) circle(.1);
\draw[fill] (10,-1) circle(.1);
\draw (10,-4) to [out=150,in=150] (10,-2);
\draw (10,-3) -- (10,-2);
\draw (10,-3) to [out=30,in=30] (10,-1);

\node[below] at (12,-10) {$w_2'w_3'UV$};

\draw[fill] (12,-10) circle(.1);
\draw[fill] (12,-9) circle(.1);
\draw (12,-10) -- (12,-9);
\draw[fill] (12,-8) circle(.1);
\draw (12,-10) to [out=150,in=150] (12,-8);
\draw[fill] (12,-7) circle(.1);
\draw[fill] (13,-7) circle(.1);
\draw (12,-8) -- (12,-7);
\draw (12,-9) to [out=30,in=150] (13,-7);
\draw[fill] (13,-6) circle(.1);
\draw (13,-7) -- (13,-6);
\draw (12,-8) to [out=30,in=150] (13,-6);
\draw[fill] (13,-5) circle(.1);
\draw (12,-7) to [out=30,in=150] (13,-5);
\draw[fill] (13,-4) circle(.1);
\draw (13,-5) -- (13,-4);
\draw (13,-6) to [out=30,in=30] (13,-4);

\node[below] at (14,-9) {$w_2'w_2''^2UV$};

\draw[fill] (14,-9) circle(.1);
\draw[fill] (14,-8) circle(.1);
\draw (14,-9) -- (14,-8);
\draw[fill] (14,-7) circle(.1);
\draw (14,-9) to [out=150,in=150] (14,-7);
\draw[fill] (14,-6) circle(.1);
\draw[fill] (15,-6) circle(.1);
\draw (14,-7) -- (14,-6);
\draw (14,-8) to [out=30,in=150] (15,-6);
\draw[fill] (15,-5) circle(.1);
\draw (15,-6) -- (15,-5);
\draw (14,-7) to [out=30,in=150] (15,-5);
\draw[fill] (15,-4) circle(.1);
\draw (14,-6) to [out=30,in=150] (15,-4);
\draw[fill] (15,-3) circle(.1);
\draw (15,-4) -- (15,-3);
\draw (15,-5) to [out=30,in=30] (15,-3);

\node[below] at (16,-11) {$w_2''w_4''UV$};

\draw[color=blue,dotted, ->] (16,-11) -- (16,-9);

\draw[fill] (16,-9) circle(.1);
\draw[fill] (16,-8) circle(.1);
\draw (16,-9) -- (16,-8);
\draw[fill] (16,-7) circle(.1);
\draw (16,-9) to [out=150,in=150] (16,-7);
\draw[fill] (16,-6) circle(.1);
\draw[fill] (17,-6) circle(.1);
\draw (16,-7) -- (16,-6);
\draw (16,-8) to [out=30,in=150] (17,-6);
\draw[fill] (17,-5) circle(.1);
\draw (17,-6) -- (17,-5);
\draw (16,-7) to [out=30,in=150] (17,-5);
\draw[fill] (17,-4) circle(.1);
\draw (16,-6) to [out=30,in=150] (17,-4);
\draw[fill] (17,-3) circle(.1);
\draw (17,-4) -- (17,-3);
\draw (17,-5) to [out=30,in=30] (17,-3);

\node[below] at (18,-10) {$w_2'^2w_2''UV$};

\draw[color=blue,dotted,->] (18,-10) -- (18,-9);

\draw[fill] (18,-9) circle(.1);
\draw[fill] (18,-8) circle(.1);
\draw (18,-9) -- (18,-8);
\draw[fill] (18,-7) circle(.1);
\draw (18,-9) to [out=150,in=150] (18,-7);
\draw[fill] (18,-6) circle(.1);
\draw[fill] (19,-6) circle(.1);
\draw (18,-7) -- (18,-6);
\draw (18,-8) to [out=30,in=150] (19,-6);
\draw[fill] (19,-5) circle(.1);
\draw (19,-6) -- (19,-5);
\draw (18,-7) to [out=30,in=150] (19,-5);
\draw[fill] (19,-4) circle(.1);
\draw (18,-6) to [out=30,in=150] (19,-4);
\draw[fill] (19,-3) circle(.1);
\draw (19,-4) -- (19,-3);
\draw (19,-5) to [out=30,in=30] (19,-3);

\node[below] at (20,-9) {$w_2'w_4'UV$};

\draw[fill] (20,-9) circle(.1);
\draw[fill] (20,-8) circle(.1);
\draw (20,-9) -- (20,-8);
\draw[fill] (20,-7) circle(.1);
\draw (20,-9) to [out=150,in=150] (20,-7);
\draw[fill] (20,-6) circle(.1);
\draw[fill] (21,-6) circle(.1);
\draw (20,-7) -- (20,-6);
\draw (20,-8) to [out=30,in=150] (21,-6);
\draw[fill] (21,-5) circle(.1);
\draw (21,-6) -- (21,-5);
\draw (20,-7) to [out=30,in=150] (21,-5);
\draw[fill] (21,-4) circle(.1);
\draw (20,-6) to [out=30,in=150] (21,-4);
\draw[fill] (21,-3) circle(.1);
\draw (21,-4) -- (21,-3);
\draw (21,-5) to [out=30,in=30] (21,-3);

\end{tikzpicture}
\end{center}
\caption{The $\A_2(1)$-module structure of $\H^{*+6}(M\SO(6),\Z_2)\otimes\H^{*+4}(M\SO(4),\Z_2)$ below degree 6.}
\label{fig:A_2(1)MSO(6)MSO(4)}
\end{figure}
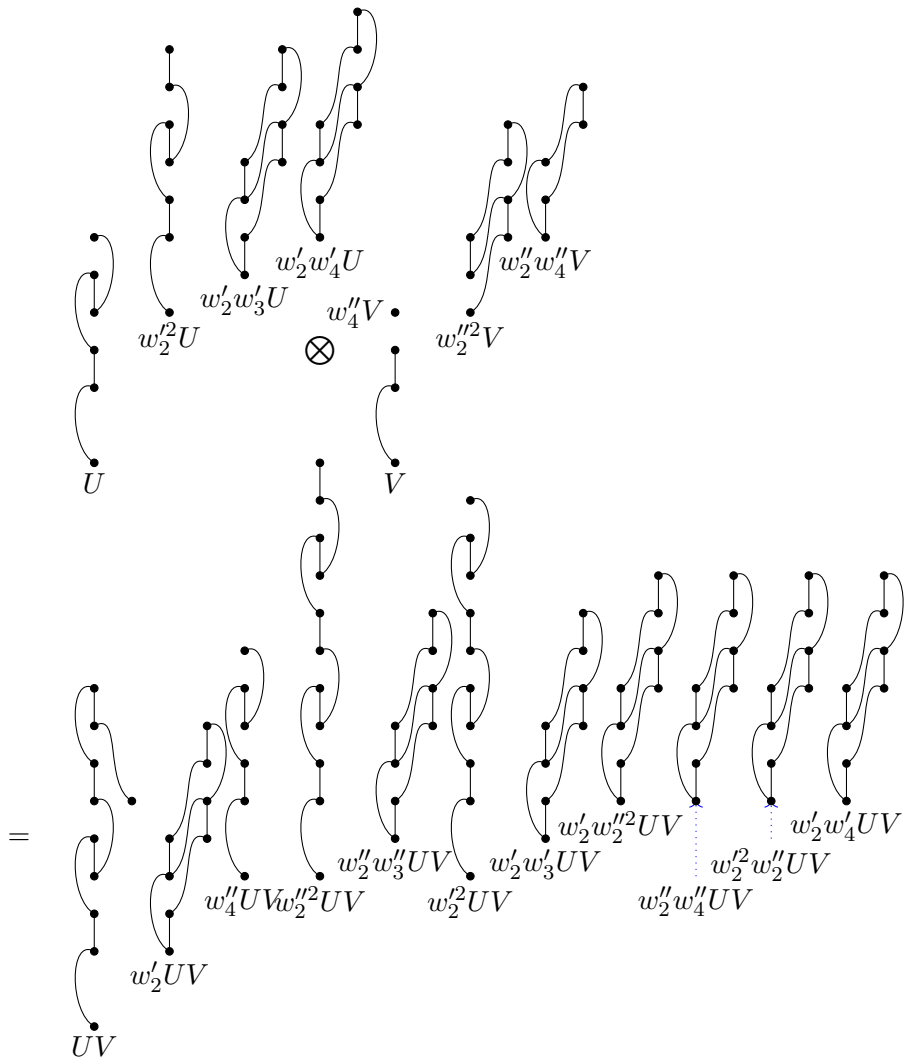

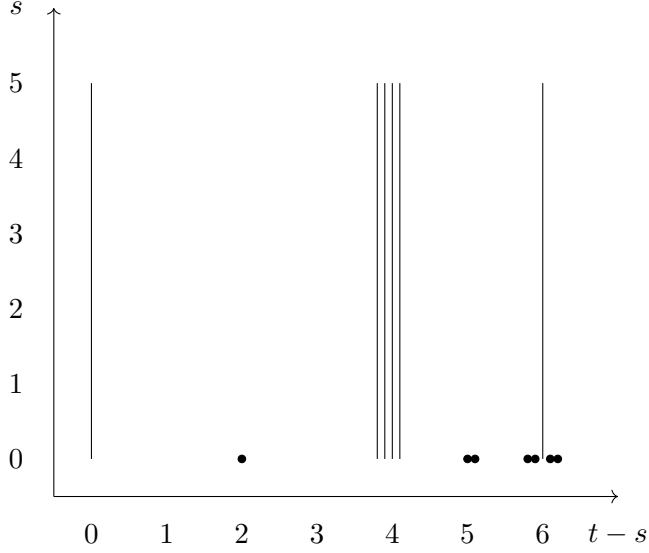
\begin{figure}[H]
\begin{center}
\begin{tikzpicture}
\node at (0,-1) {0};
\node at (1,-1) {1};
\node at (2,-1) {2};
\node at (3,-1) {3};
\node at (4,-1) {4};
\node at (5,-1) {5};
\node at (6,-1) {6};
\node at (7,-1) {$t-s$};
\node at (-1,0) {0};
\node at (-1,1) {1};
\node at (-1,2) {2};
\node at (-1,3) {3};
\node at (-1,4) {4};
\node at (-1,5) {5};
\node at (-1,6) {$s$};

\draw[->] (-0.5,-0.5) -- (-0.5,6);
\draw[->] (-0.5,-0.5) -- (7,-0.5);

\draw (0,0) -- (0,5);
\draw[fill] (2,0) circle(0.05);
\draw (3.8,0) -- (3.8,5);
\draw (3.9,0) -- (3.9,5);
\draw (4,0) -- (4,5);
\draw (4.1,0) -- (4.1,5);
\draw[fill] (5,0) circle(0.05);
\draw[fill] (5.1,0) circle(0.05);
\draw (6,0) -- (6,5);
\draw[fill] (5.8,0) circle(0.05);
\draw[fill] (5.9,0) circle(0.05);
\draw[fill] (6.1,0) circle(0.05);
\draw[fill] (6.2,0) circle(0.05);
\end{tikzpicture}
\end{center}
\caption{$\Omega_*^{\frac{\Spin\times \frac{\SU(4)\times(\SU(2)\times \SU(2))}{\Z_2}}{\Z_2^F}}$.}
\label{fig:E_2SU4SU2SU2Z2}
\end{figure}

The $\A_2(1)$-module structure of $\H^{*+6}(M\SO(6),\Z_2)\otimes\H^{*+4}(M\SO(4),\Z_2)$ and the $E_2$ page are shown in Figure \ref{fig:A_2(1)MSO(6)MSO(4)}, \ref{fig:E_2SU4SU2SU2Z2}.
Here we have used the correspondence between $\A_2(1)$-module structure and the $E_2$ page shown in Figure \ref{fig:A_2(1)}, \ref{fig:L_2}, \ref{fig:L_3}, \ref{fig:L_5} and \ref{fig:L_8}.

Thus we obtain the bordism group $\Omega^{\frac{\Spin\times \frac{\SU(4)\times(\SU(2)\times \SU(2))}{\Z_2}}{\Z_2^F}}_d$ shown in Table \ref{table:SU4SU2SU2Z2Bordism}.

\begin{table}[H]
\centering
\begin{tabular}{ c c c}
\hline
\multicolumn{3}{c}{Bordism group}\\
\hline
$d$ & 
$\Omega^{\frac{\Spin\times \frac{\SU(4)\times(\SU(2)\times \SU(2))}{\Z_2}}{\Z_2^F}}_d$
& bordism invariants \\
\hline
0& $\Z$\\
1& $0$\\
2& $\Z_2$  & $w_2' \sim w_2''$\\
3 & $0$\\
4 & $\Z^4$ & $p_1'$ from $w_2'^2$, $e_4'$ from $w_4'+w_2'w_2''$, $p_1''$ from $w_2''^2$, $e_4''$ from $w_4''$ \\
5 & $\Z_2^2$  & $w_2'w_3',w_2''w_3''$ \\
6 & $\Z\times\Z_2^4$ & $e_6'$ from $w_6'$, $w_2'w_2''^2,w_2''w_4'',w_2'^2w_2'' \sim w_2'^3+w_3'^2,w_2'w_4'$ \\
\hline
\end{tabular}
\caption{Bordism group. Here $w_i'$ is the Stiefel-Whitney class of the $\SO(6)=\frac{\SU(4)}{\Z_2}$ bundle, $w_i''$ is the Stiefel-Whitney class of the $\SO(4)=\frac{\SU(2)\times\SU(2)}{\Z_2}$ bundle.
$p_i'$ is the Pontryagin class of the $\SO(6)=\frac{\SU(4)}{\Z_2}$ bundle, $p_i''$ is the Pontryagin class of the $\SO(4)=\frac{\SU(2)\times\SU(2)}{\Z_2}$ bundle.
$e_i'$ is the Euler class of the $\SO(6)=\frac{\SU(4)}{\Z_2}$ bundle, $e_i''$ is the Euler class of the $\SO(4)=\frac{\SU(2)\times\SU(2)}{\Z_2}$ bundle.
Here $w_2=w_2'+w_2''$, $w_3=w_3'+w_3''$.
Since $w_2+w_1^2=0$ on 2-manifold, we have $w_2=w_2'+w_2''=w_1^2=0$ on oriented 2-manifold.
Since $\Sq^2(w_2'^2)=w_3'^2=(w_2+w_1^2)w_2'^2=(w_2'+w_2'')w_2'^2$ on oriented 6-manifold by Wu formula, we have $w_2''w_2'^2=w_2'^3+w_3'^2$. Since $w_2'^2=\Sq^2(w_2')=(w_2+w_1^2)w_2'=(w_2'+w_2'')w_2'$ on oriented 4-manifold by Wu formula, we have $w_2'w_2''=0$ on oriented 4-manifold.
On a 4-manifold, the oriented bundle of rank 6 splits as the direct sum of an oriented bundle of rank 4 and a trivial plane bundle, the Euler class $e_4'$ is the Euler class of the subbundle of rank 4.
}
\label{table:SU4SU2SU2Z2Bordism}
\end{table}

By \eqref{eq:TPexact}, we obtain the cobordism group $\TP_d(\frac{\Spin\times \frac{\SU(4)\times(\SU(2)\times \SU(2))}{\Z_2}}{\Z_2^F})$ shown in Table \ref{table:SU4SU2SU2Z2TP}.

\begin{table}[H]
\centering
\begin{tabular}{ c c c}
\hline
\multicolumn{3}{c}{Cobordism group}\\
\hline
$d$ & 
$\TP_d(\frac{\Spin\times \frac{\SU(4)\times(\SU(2)\times \SU(2))}{\Z_2}}{\Z_2^F})$
& topological terms \\
\hline
0& $0$\\
1& $0$\\
2& $\Z_2$ &$w_2'\sim w_2''$ \\
3 & $\Z^4$ & $\text{CS}_3^{\SO(6)},\text{CS}_3^{\SO(4)},\text{CS}_{3,e}^{\SO(6)},\text{CS}_{3,e}^{\SO(4)}$\\
4 & $0$ \\
5 & $\Z\times\Z_2^2$  & $\text{CS}_{5,e}^{\SO(6)}, w_2'w_3',w_2''w_3''$\\
\hline
\end{tabular}
\caption{Topological phase classification ($\equiv$ TP) as a cobordism group, following Table \ref{table:SU4SU2SU2Z2Bordism}. 
$\text{CS}_{2n-1}^G$ is the Chern-Simons form of the associated vector bundle of the principal $G$-bundle (associated to the Chern/Pontryagin class).
$\text{CS}_{2n-1,e}^G$ is the Chern-Simons form of the associated vector bundle of the principal $G$-bundle (associated to the Euler class).
$w_i'$ is the Stiefel-Whitney class of the $\SO(6)=\frac{\SU(4)}{\Z_2}$ bundle, $w_i''$ is the Stiefel-Whitney class of the $\SO(4)=\frac{\SU(2)\times\SU(2)}{\Z_2}$ bundle.
}
\label{table:SU4SU2SU2Z2TP}
\end{table}

\subsection{
$\frac{\Spin\times {\SU(4)\times(\SU(2)\times \SU(2))}}{\Z_2^F}$ Pati-Salam model
}

We consider $G=\frac{\Spin\times {\SU(4)\times(\SU(2)\times \SU(2))}}{\Z_2^F}$, let $G'=\frac{{\SU(4)\times(\SU(2)\times \SU(2))}}{\Z_2}$, then we have a fibration
\bea
\xymatrix{\B G'\ar[d]&\\
\B\SO(6)\times\B\SO(4)\ar[r]^-{w_2'+w_2''}&\B^2\Z_2.}
\eea
Here $w_2'=w_2(\SO(6))$, $w_2''=w_2(\SO(4))$.

We have a homotopy pullback square
\bea
\xymatrix{
\B G\ar[r]\ar[d]&\B G'\ar[d]^{w_2'=w_2''}\\
\B\SO\ar[r]^{w_2}&\B^2\Z_2.
}
\eea

There is a homotopy equivalence
$f:\B \SO\times \B G'\xrightarrow{\sim}\B \SO\times \B G'$ by $(V,W)\mapsto(V-W+10,W)$, and there is obviously also an inverse map.
Note that the pullback $f^*(w_2)=w_2(V-W)=w_2(V)+w_1(V)w_1(W)+w_2(W)=w_2+w_2'$.
Then we have the following homotopy pullback
\bea
\xymatrix{
\B G\ar[r]^-{\sim}\ar[d]&\B \Spin\times \B G'\ar[d]&\\
\B \SO\times \B G'\ar[r]^{f}\ar[d]_{(V,W)\mapsto V}\ar@/_1pc/[rr]_{w_2+w_2'}&\B \SO\times \B G'\ar[r]^-{w_2+0}\ar[ld]^{(V,W)\mapsto V+W-10}&\B ^2\Z_2\\
\B \SO&&}.
\eea
This implies that the two classifying spaces $\B G\sim \B \Spin\times \B G'$ are homotopy equivalent.

By definition, the Madsen-Tillmann spectrum $MTG$ of the group $G$ is 
$MTG=\text{Thom}(\B G;-V)$, where $V$ is the induced virtual bundle (of dimension $0$) by the map $\B G\to \B \O$.

We can identify $\B G\to \B \O$ with
$\B \Spin\times \B G'\xrightarrow{V-W+10}\B \SO\hookrightarrow \B \O$.

So the spectrum $MTG$ is homotopy equivalent to $\text{Thom}(\B \Spin\times \B G';-(V-W+10))$, which is 
\bea
MTG=M\Spin\wedge\Sigma^{-10}MG'.
\eea


For the dimension $d=t-s<8$, since there is no odd torsion (see footnote \ref{ft:no-odd-torsion}), by \eqref{eq:ExtA_2(1)}, we have the Adams spectral sequence
\bea
\Ext_{\A_2(1)}^{s,t}(\H^{*+10}(MG',\Z_2),\Z_2)\Rightarrow\Omega_{t-s}^{\frac{\Spin\times \SU(4)\times(\SU(2)\times \SU(2))}{\Z_2^F}}.
\eea

There is a fibration
\bea
\xymatrix{\B\Z_2\ar[r]&\B G'\ar[d]\\&\B\SO(6)\times\B\SO(4).}
\eea
So we have the Serre spectral sequence
\bea
\H^p(\B\SO(6)\times\B\SO(4),\H^q(\B\Z_2,\Z_2))\Rightarrow \H^{p+q}(\B G',\Z_2).
\eea
Let $a$ be the generator of $\H^1(\B\Z_2,\Z_2)$, then we have the differentials 
$d_2(a)=w_2'+w_2''$, $d_3(a^2)=\Sq^1d_2(a)=w_3'+w_3''$, $d_5(a^4)=\Sq^2d_3(a^2)=w_2'w_3'+w_5'+w_2''w_3''$ and so on.
So in $\H^*(\B G',\Z_2)$, we have $w_2'=w_2''$, $w_3'=w_3''$ and $w_5'=0$.

So below degree 6, we have
\bea
\H^{*+10}(MG',\Z_2)=((\Z_2[w_2',w_3',w_4',w_5',w_6']\otimes\Z_2[w_2'',w_3'',w_4''])/(w_2'=w_2'',w_3'=w_3'',w_5'=0))U
\eea
where $U$ is the Thom class with $\Sq^1U=0$, $\Sq^2U=w_2'U=w_2''U$. {Here in this subsection, $w_i'$ is the Stiefel-Whitney class of SO(6) bundle, and $w_i''$ is the Stiefel-Whitney class of SO(4) bundle.}

\begin{figure}[H]
\begin{center}
\begin{tikzpicture}
\node at (0,-1) {0};
\node at (1,-1) {1};
\node at (2,-1) {2};
\node at (3,-1) {3};
\node at (4,-1) {4};
\node at (5,-1) {5};
\node at (6,-1) {6};
\node at (7,-1) {$t-s$};
\node at (-1,0) {0};
\node at (-1,1) {1};
\node at (-1,2) {2};
\node at (-1,3) {3};
\node at (-1,4) {4};
\node at (-1,5) {5};
\node at (-1,6) {$s$};

\draw[->] (-0.5,-0.5) -- (-0.5,6);
\draw[->] (-0.5,-0.5) -- (7,-0.5);

\draw (0,0) -- (0,5);

\draw (3.8,0) -- (3.8,5);
\draw (3.9,0) -- (3.9,5);
\draw (4,0) -- (4,5);
\draw (4,0) -- (6,2);
\draw[color=red] (4.1,1) -- (4.1,5);
\draw[color=red] (4.1,1) -- (6.1,3);

\draw (5,0) -- (6,1);

\draw (6.2,1) -- (6.2,5);

\draw[fill] (6,0) circle(0.05);
\draw[fill] (6.1,0) circle(0.05);
\end{tikzpicture}
\end{center}
\caption{$\Omega_*^{\frac{\Spin\times {\SU(4)\times(\SU(2)\times \SU(2))}}{\Z_2^F}}$. The red part will be explained in Appendix \ref{sec:explanation}.}
\label{fig:E_2SU4SU2SU2}
\end{figure}
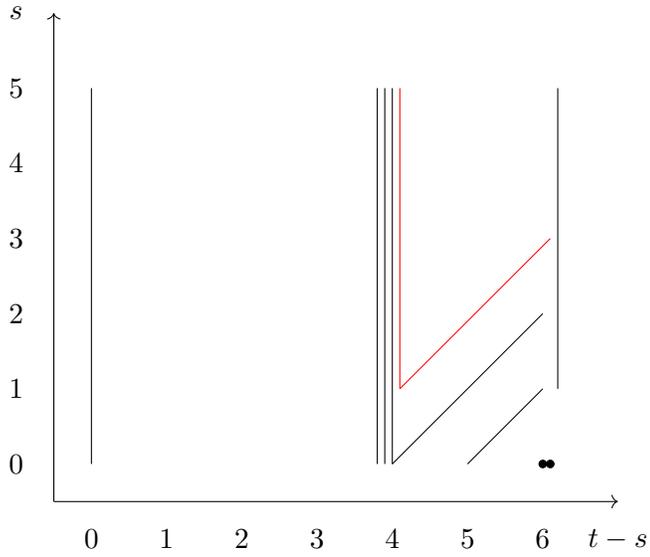

The $\A_2(1)$-module structure of $\H^{*+10}(MG',\Z_2)$ below degree 6 and the $E_2$ page are shown in Figure \ref{fig:A_2(1)MG}, \ref{fig:E_2SU4SU2SU2}.
Here we have used the correspondence between $\A_2(1)$-module structure and the $E_2$ page shown in Figure \ref{fig:Z_2}, \ref{fig:A_2(1)}, \ref{fig:Ceta}, \ref{fig:L_4},
 \ref{fig:L_6} and \ref{fig:L_7}.

\begin{figure}[H]
\begin{center}
\begin{tikzpicture}[scale=0.5]

\node[below] at (0,0) {$U$};

\draw[fill] (0,0) circle(.1);
\draw[fill] (0,2) circle(.1);
\draw (0,0) to [out=150,in=150] (0,2);
\draw[fill] (0,3) circle(.1);
\draw (0,2) -- (0,3);
%
%
%
%

\node[below] at (2,4) {$w_4'U$};

\draw[fill] (2,4) circle(.1);
\draw[fill] (2,6) circle(.1);
\draw (2,4) to [out=150,in=150] (2,6);

\node[below] at (4,4) {$w_2'^2U$};

\draw[fill] (4,4) circle(.1);
\draw[fill] (4,5) circle(.1);
\draw[fill] (4,6) circle(.1);
\draw[fill] (5,6) circle(.1);
\draw (4,4) to [out=30,in=150] (5,6);
\draw (4,5) -- (4,6);
\draw[fill] (5,7) circle(.1);
\draw (5,6) -- (5,7);
\draw (4,5) to [out=30,in=150] (5,7);
\draw[fill] (5,8) circle(.1);
\draw[fill] (5,9) circle(.1);
\draw (4,6) to [out=30,in=150] (5,8);
\draw (5,7) to [out=30,in=30] (5,9);
\draw (5,8) -- (5,9);

\node[below] at (6,6) {$w_2'w_4'U$};

\draw[fill] (6,6) circle(.1);
\draw[fill] (6,7) circle(.1);
\draw (6,6) -- (6,7);
\draw[fill] (6,8) circle(.1);
\draw (6,6) to [out=150,in=150] (6,8);
\draw[fill] (6,9) circle(.1);
\draw[fill] (7,9) circle(.1);
\draw (6,8) -- (6,9);
\draw (6,7) to [out=30,in=150] (7,9);
\draw[fill] (7,10) circle(.1);
\draw (7,9) -- (7,10);
\draw (6,8) to [out=30,in=150] (7,10);
\draw[fill] (7,11) circle(.1);
\draw (6,9) to [out=30,in=150] (7,11);
\draw[fill] (7,12) circle(.1);
\draw (7,11) -- (7,12);
\draw (7,10) to [out=30,in=30] (7,12);

\node[below] at (8,4) {$w_4''U$};

\draw[fill] (8,4) circle(.1);

\node[below] at (10,6) {$w_2'w_4''U$};

\draw[fill] (10,6) circle(.1);
\draw[fill] (10,7) circle(.1);
\draw (10,6) -- (10,7);
\draw[fill] (10,8) circle(.1);
\draw (10,6) to [out=150,in=150] (10,8);
\draw[fill] (11,9) circle(.1);
\draw (10,7) to [out=30,in=150] (11,9);
\draw[fill] (11,10) circle(.1);
\draw (11,9) -- (11,10);
\draw (10,8) to [out=30,in=150] (11,10);

\end{tikzpicture}
\end{center}
\caption{The $\A_2(1)$-module structure of $\H^{*+10}(MG',\Z_2)$ below degree 6.}
\label{fig:A_2(1)MG}
\end{figure}
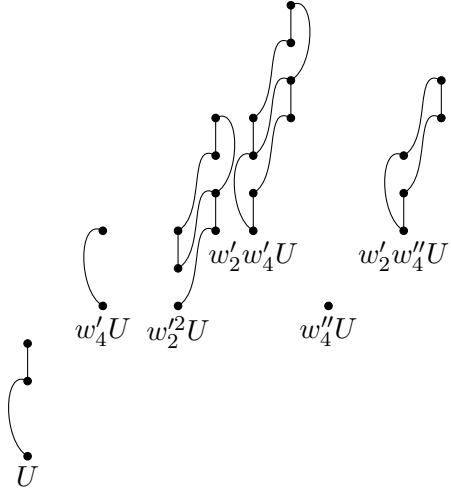

Thus we obtain the bordism group $\Omega^{\frac{\Spin\times {\SU(4)\times(\SU(2)\times \SU(2))}}{\Z_2^F}}_d$ shown in Table \ref{table:SU4SU2SU2Bordism}.

By \eqref{eq:TPexact}, we obtain the cobordism group $\TP_d(\frac{\Spin\times {\SU(4)\times(\SU(2)\times \SU(2))}}{\Z_2^F})$ shown in Table \ref{table:SU4SU2SU2TP}.

\begin{table}[H]
\centering
\begin{tabular}{ c c c}
\hline
\multicolumn{3}{c}{Bordism group}\\
\hline
$d$ & 
$\Omega^{\frac{\Spin\times {\SU(4)\times(\SU(2)\times \SU(2))}}{\Z_2^F}}_d$
& bordism invariants \\
\hline
0& $\Z$\\
1& $0$\\
2& $0$  \\
3 & $0$\\
4 & $\Z^4$ & ($p_1'$ from $w_2'^2$, $e_4'$ from $w_4'$, $e_4''$ from $w_4''$, $(f^*)^{-1}(c_2(\SU(2)))$) \\
5 & $\Z_2^3$  & $(w_2'w_3',w_4''\tilde\eta,(f^*)^{-1}(c_2(\SU(2))\tilde\eta))$ \\
6 & $\Z\times\Z_2^5$ & ($\frac{e_6'}{2}$ from $w_6'$, $w_2'w_4',w_2'w_4'',w_4''\text{Arf},w_2'w_3'\tilde\eta,(f^*)^{-1}(c_2(\SU(2))\text{Arf})$) \\
\hline
\end{tabular}
\caption{Bordism group. Here $w_i'$ is the Stiefel-Whitney class of the $\SO(6)=\frac{\SU(4)}{\Z_2}$ bundle, $w_i''$ is the Stiefel-Whitney class of the $\SO(4)=\frac{\SU(2)\times\SU(2)}{\Z_2}$ bundle.
$p_i'$ is the Pontryagin class of the $\SO(6)=\frac{\SU(4)}{\Z_2}$ bundle, $p_i''$ is the Pontryagin class of the $\SO(4)=\frac{\SU(2)\times\SU(2)}{\Z_2}$ bundle.
$e_i'$ is the Euler class of the $\SO(6)=\frac{\SU(4)}{\Z_2}$ bundle, $e_i''$ is the Euler class of the $\SO(4)=\frac{\SU(2)\times\SU(2)}{\Z_2}$ bundle.
Here $w_2=w_2'=w_2''$, $w_3=w_3'=w_3''$.
$\tilde\eta$ is a mod 2 index of 1d Dirac operator.
Arf is a 2d Arf invariant.
Since $\Sq^2w_4'=w_2'w_4'+w_6'=(w_2+w_1^2)w_4'=w_2'w_4'$ on oriented 6-manifold by Wu formula, we have $e_6'=w_6'=0\mod2$.
On a 4-manifold, the oriented bundle of rank 6 splits as the direct sum of an oriented bundle of rank 4 and a trivial plane bundle, the Euler class $e_4'$ is the Euler class of the subbundle of rank 4.
Here $f:\Omega_d^{\Spin\times\SU(2)\times\SU(2)\times\SU(4)}\to\Omega_d^{\Spin\times_{\Z_2}(\SU(2)\times\SU(2)\times\SU(4))}$ is the natural group homomorphism, $f^*$ is the induced map between bordism invariants. $c_2(\SU(2))$ is one of the bordism invariants of $\Omega_4^{\Spin\times \SU(2)\times\SU(2)\times\SU(4)}$, $c_2(\SU(2))\tilde\eta$ is one of the bordism invariants 
{(related to Witten anomaly)}
of $\Omega_5^{\Spin\times\SU(2)\times\SU(2)\times\SU(4)}$, 
while $c_2(\SU(2))\text{Arf}$ is one of the bordism invariants of $\Omega_6^{\Spin\times\SU(2)\times\SU(2)\times\SU(4)}$. Here $\tilde\eta$ is the mod 2 index of 1d Dirac operator, Arf is the Arf invariant. The three bordism invariants which involve $f$ correspond to the red part of Figure \ref{fig:E_2SU4SU2SU2} and will be explained in Appendix \ref{sec:explanation}.
}
\label{table:SU4SU2SU2Bordism}
\end{table}

\begin{table}[H]
\centering
\begin{tabular}{ c c c}
\hline
\multicolumn{3}{c}{Cobordism group}\\
\hline
$d$ & 
$\TP_d(\frac{\Spin\times {\SU(4)\times(\SU(2)\times \SU(2))}}{\Z_2^F})$
& topological terms \\
\hline
0& $0$\\
1& $0$\\
2& $0$ \\
3 & $\Z^4$ & $\text{CS}_3^{\SO(6)},\text{CS}_{3,e}^{\SO(6)},\text{CS}_{3,e}^{\SO(4)},g^{-1}(\text{CS}_3^{\SU(2)})$\\
4 & $0$ \\
5 & $\Z\times\Z_2^3$ & $\frac{1}{2}\text{CS}_{5,e}^{\SO(6)},w_2'w_3',w_4''\tilde\eta,(f^*)^{-1}(c_2(\SU(2))\tilde\eta)$ \\
\hline
\end{tabular}
\caption{Topological phase classification ($\equiv$ TP) as a cobordism group, following Table \ref{table:SU4SU2SU2Bordism}. 
$\text{CS}_{2n-1}^G$ is the Chern-Simons form of the associated vector bundle of the principal $G$-bundle (associated to the Chern/Pontryagin class).
$\text{CS}_{2n-1,e}^G$ is the Chern-Simons form of the associated vector bundle of the principal $G$-bundle (associated to the Euler class).
$w_i'$ is the Stiefel-Whitney class of the $\SO(6)=\frac{\SU(4)}{\Z_2}$ bundle, $w_i''$ is the Stiefel-Whitney class of the $\SO(4)=\frac{\SU(2)\times\SU(2)}{\Z_2}$ bundle, $\tilde\eta$ is the mod 2 index of 1d Dirac operator.
Here $g:\TP_3(\Spin\times_{\Z_2}(\SU(2)\times\SU(2)\times\SU(4)))\to\TP_3(\Spin\times\SU(2)\times\SU(2)\times\SU(4))$ is the induced group homomorphism from the $f$ explained in Table \ref{table:SU4SU2SU2Bordism}. Both $f$ and $g$ are examined in Appendix \ref{sec:explanation}. 
$\text{CS}_3^{\SU(2)}$ is one of the topological terms of $\TP_3(\Spin\times\SU(2)\times\SU(2)\times\SU(4))$.
}
\label{table:SU4SU2SU2TP}
\end{table}

\section{SO(10),  SO(18) and SO($n$) Grand Unifications}\label{sec:GUSO}

Now we consider the co/bordism classes relevant for 
Fritzsch-Minkowski SO(10) GUT \cite{Fritzsch1974nnMinkowskiUnifiedInteractionsofLeptonsandHadrons}.
There are actually two cases, depending on whether the gauged SO(10) GUT allows gauge-invariant fermions, or whether 
the gauged SO(10) GUT only allows gauge-invariant bosons.
The first case requires the bordism group computation of ${\Spin\times \Spin(n)}$ structure shown in \Sec{sec:SpinSpin10}, while the
second case  requires the bordism group computation of ${\frac{\Spin \times \Spin(n)}{\Z_2^F}}$ structure shown in \Sec{sec:SpinSpin10modZ2}.
Because the matter fields are in the 16-dimensional complex spinor representation of Spin(10) instead of the 10-dimensional vector representation of SO(10), 
we should \emph{not} consider the ${\Spin\times \SO(n)}$ structure for SO(10) GUT.
However, we list down bordism group computation of ${\Spin\times \SO(n)}$ in \Sec{sec:SpinSO10} merely for the convenience of comparison.

\subsection{${\Spin\times \Spin(n)}$ for $n\ge7$:
Spin $\times$ Spin(10) and
Spin $\times$ Spin(18)}
\label{sec:SpinSpin10}

Here we consider $G={\Spin\times \Spin(n)}$ for $n\ge7$, especially for
Spin(10) and
Spin(18).
The Madsen-Tillmann spectrum $MTG$ of the group $G$ is 
\bea
MTG=M\Spin\wedge(\B\Spin(n))_+.
\eea
The $(\B\Spin(n))_+$ is the disjoint union of the classifying space $\B\Spin(n)$ and a point, see footnote \ref{ft:X_+}.

For the dimension $d=t-s<8$, since there is no odd torsion (see footnote \ref{ft:no-odd-torsion}), by \eqref{eq:ExtA_2(1)}, we have the Adams spectral sequence
\bea
\Ext_{\A_2(1)}^{s,t}(\H^*(\B\Spin(n),\Z_2),\Z_2)\Rightarrow\Omega_{t-s}^{\Spin\times\Spin(n)}.
\eea

By \cite{Quillen1971}'s Theorem 6.5, we have
\bea
\H^*(\B\Spin(n),\Z_2)=\H^*(\B\SO(n),\Z_2)/(r_1,r_2,\dots,r_{h(n)})\otimes\Z_2[e]
\eea
where 
\bea
\H^*(\B\SO(n),\Z_2)=\Z_2[w_2',w_3',\dots,w_n'],
\eea
and
\bea
r_1=w_2',\ r_2=\Sq^1w_2',\ r_3=\Sq^2\Sq^1w_2',\ \dots,\ r_h=\Sq^{2^{h-2}}\cdots\Sq^2\Sq^1w_2',\ \dots.
\eea
The value of the function $h$ is
\bea
&&h(8k+1)=4k,\ h(8k+2)=4k+1,\ h(8k+3)=h(8k+4)=4k+2,\nn\\
&&h(8k+5)=h(8k+6)=h(8k+7)=h(8k+8)=4k+3.
\eea
Here $2^{h(n)}$ is the dimension of the real spinor representation of $\Spin(n)$, it is also called the Hurwitz-Radon number.
$e\in\H^{2^{h(n)}}(\B\Spin(n),\Z_2)$ is the $2^{h(n)}$-th Stiefel-Whitney class of the real spinor representation $\Delta:\B\Spin(n)\to\B\O(2^{h(n)})$.
{Here in this subsection, $w_i'$ is the Stiefel-Whitney class of $\SO(n)$ bundle.}

Since $\Sq^1w_2'=w_3'$, $\Sq^2\Sq^1w_2'=\Sq^2w_3'=w_2'w_3'+w_5'$,
for $n\ge7$, we have
\bea
\H^*(\B\Spin(n),\Z_2)=\Z_2[w_4',w_6',w_7',\dots]
\eea
where the $\dots$ are generators in higher degrees.

For $n\ge7$, the $\A_2(1)$-module structure of $\H^*(\B\Spin(n),\Z_2)$ below degree 6 is shown in Figure \ref{fig:A_2(1)Spinn}.

\begin{figure}[H]
\begin{center}
\begin{tikzpicture}[scale=0.5]

\node[below] at (0,0) {$1$};

\draw[fill] (0,0) circle(.1);

\node[right] at (0,4) {$w_4'$};

\draw[fill] (0,4) circle(.1);

\node[right] at (0,6) {$w_6'$};

\draw[fill] (0,6) circle(.1);

\draw (0,4) to [out=150,in=150] (0,6);

\node[right] at (0,7) {$w_7'$};

\draw[fill] (0,7) circle(.1);

\draw (0,6) -- (0,7);

\end{tikzpicture}
\end{center}
\caption{The $\A_2(1)$-module structure of $\H^*(\B\Spin(n),\Z_2)$ below degree 6 for $n\ge7$.}
\label{fig:A_2(1)Spinn}
\end{figure}

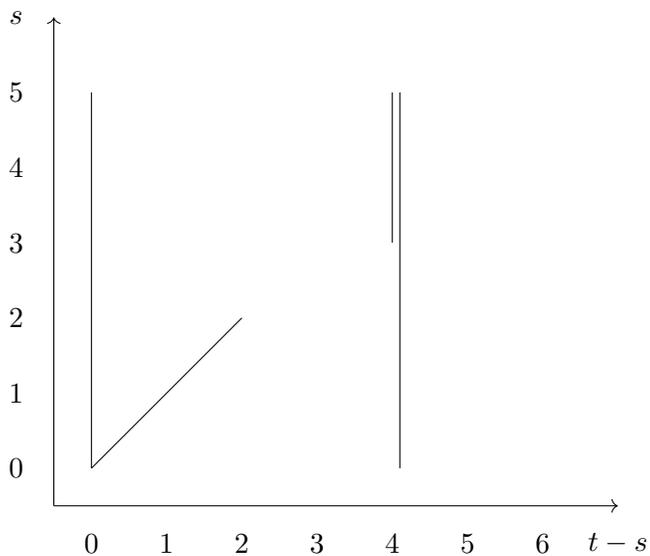
\begin{figure}[H]
\begin{center}
\begin{tikzpicture}
\node at (0,-1) {0};
\node at (1,-1) {1};
\node at (2,-1) {2};
\node at (3,-1) {3};
\node at (4,-1) {4};
\node at (5,-1) {5};
\node at (6,-1) {6};
\node at (7,-1) {$t-s$};
\node at (-1,0) {0};
\node at (-1,1) {1};
\node at (-1,2) {2};
\node at (-1,3) {3};
\node at (-1,4) {4};
\node at (-1,5) {5};
\node at (-1,6) {$s$};

\draw[->] (-0.5,-0.5) -- (-0.5,6);
\draw[->] (-0.5,-0.5) -- (7,-0.5);

\draw (0,0) -- (0,5);

\draw (0,0) -- (2,2);

\draw (4,3) -- (4,5);

\draw (4.1,0) -- (4.1,5);

\end{tikzpicture}
\end{center}
\caption{$\Omega_*^{\Spin\times\Spin(n)}$ for $n\ge7$.}
\label{fig:E_2Spinn}
\end{figure}

The $E_2$ page is shown in Figure \ref{fig:E_2Spinn}.
Here we have used the correspondence between $\A_2(1)$-module structure and the $E_2$ page shown in Figure \ref{fig:Z_2} and \ref{fig:L_6}.

Thus we obtain the bordism group $\Omega^{\Spin\times\Spin(n)}_d$ for $n\ge7$ shown in Table \ref{table:SpinnBordism}.

\begin{table}[H]
\centering
\hspace*{-25mm}
\begin{tabular}{ c c c}
\hline
\multicolumn{3}{c}{Bordism group}\\
\hline
$d$ & 
$\Omega^{\Spin\times\Spin(n)}_d$ for $n\ge7$
& bordism invariants \\
\hline
0&  $\Z$ \\
\hline
1& $\Z_2$  & $\tilde\eta$ \\
\hline
2&  $\Z_2$  & Arf \\
\hline
3 & $0$ \\
\hline
4 & $\Z^2$ & $\frac{\sigma}{16}$, $e_4'$  \\
\hline
5 & $0$  & \\
\hline
6 & $0$  & \\
\hline
\end{tabular}
\caption{Bordism group. 
Here $e_4'$ is the Euler class of the $\Spin(n)$ bundle.
$\tilde\eta$ is a mod 2 index of 1d Dirac operator.
Arf is a 2d Arf invariant.
$\sigma$ is the signature of manifold.
On a 4-manifold, the oriented bundle of rank $n$ splits as the direct sum of an oriented bundle of rank 4 and a trivial bundle of rank $n-4$, the Euler class $e_4'$ is the Euler class of the subbundle of rank 4.
}
\label{table:SpinnBordism}
\end{table}

By \eqref{eq:TPexact}, we obtain the cobordism group $\TP_d(\Spin \times\Spin(n))$ for $n\ge7$ shown in Table \ref{table:SpinnTP}.

\begin{table}[H]
\centering
\hspace*{-14mm}
\begin{tabular}{ c c c}
\hline
\multicolumn{3}{c}{Cobordism group}\\
\hline
$d$ & 

$\TP_d(\Spin \times\Spin(n))$ for $n\ge7$
& topological terms \\
\hline
0& 0 \\
\hline
1& $\Z_2$  & $\tilde\eta$ \\
\hline
2& $\Z_2$ & Arf \\
\hline
3 & $\Z^2$ & $\frac{1}{48}\text{CS}_3^{TM}$, $\text{CS}_{3,e}^{\Spin(n)}$ \\
\hline
4 & $0$ &  \\
\hline
5 & $0$  & \\
\hline
\end{tabular}
\caption{Topological phase classification ($\equiv$ TP) as a cobordism group, following Table \ref{table:SpinnBordism}. 
$\tilde\eta$ is a mod 2 index of 1d Dirac operator.
Arf is a 2d Arf invariant.
The $TM$ is the spacetime tangent bundle.
$\text{CS}_{2n-1}^V$ is the Chern-Simons form of the vector bundle $V$ (associated to the Chern/Pontryagin class).
$\text{CS}_{2n-1,e}^G$ is the Chern-Simons form of the associated vector bundle of the principal $G$-bundle (associated to the Euler class).
}
\label{table:SpinnTP}
\end{table}

\subsection{${\Spin\times \SO(n)}$ for $n\ge7$:
Spin $\times$ SO(10) and
Spin $\times$ SO(18)}

\label{sec:SpinSO10}

Here we consider $G={\Spin\times \SO(n)}$ for $n\ge7$, especially for
SO(10) and
SO(18).
The Madsen-Tillmann spectrum $MTG$ of the group $G$ is 
\bea
MTG=M\Spin\wedge(\B\SO(n))_+.
\eea
The $(\B\SO(n))_+$ is the disjoint union of the classifying space $\B\SO(n)$ and a point, see footnote \ref{ft:X_+}.

For the dimension $d=t-s<8$, since there is no odd torsion (see footnote \ref{ft:no-odd-torsion}), by \eqref{eq:ExtA_2(1)}, we have the Adams spectral sequence
\bea
\Ext_{\A_2(1)}^{s,t}(\H^*(\B\SO(n),\Z_2),\Z_2)\Rightarrow\Omega_{t-s}^{\Spin\times\SO(n)}.
\eea

We have
\bea
\H^*(\B\SO(n),\Z_2)=\Z_2[w_2',w_3',\dots,w_n'].
\eea
{Here in this subsection, $w_i'$ is the Stiefel-Whitney class of $\SO(n)$ bundle.}

We also have the Wu formula
\bea
\Sq^jw_i'=\sum_{k=0}^j\binom{i-k-1}{j-k}w_{i+j-k}'w_k'\text{ for }0\le j\le i.
\eea

For $n\ge7$, the $\A_2(1)$-module structure of $\H^*(\B\SO(n),\Z_2)$ below degree 6 is shown in Figure \ref{fig:A_2(1)SOn}.

\begin{figure}[H]
\begin{center}
\begin{tikzpicture}[scale=0.5]

\node[below] at (0,0) {$1$};

\draw[fill] (0,0) circle(.1);

\node[below] at (0,2) {$w_2'$};

\draw[fill] (0,2) circle(.1);

\draw[fill] (0,3) circle(.1);

\draw (0,2) -- (0,3);

\draw[fill] (0,4) circle(.1);

\draw[fill] (1,5) circle(.1);

\draw[fill] (1,6) circle(.1);
\draw (1,5) -- (1,6);

\draw (0,2) to [out=150,in=150] (0,4);

\draw (0,3) to [out=30,in=150] (1,5);

\draw (0,4) to [out=30,in=150] (1,6);

\node[below] at (2,4) {$w_4'$};

\draw[fill] (2,4) circle(.1);

\draw[fill] (2,5) circle(.1);

\draw (2,4) -- (2,5);

\draw[fill] (2,6) circle(.1);

\draw[fill] (3,7) circle(.1);

\draw[fill] (3,8) circle(.1);
\draw (3,7) -- (3,8);

\draw (2,4) to [out=150,in=150] (2,6);

\draw (2,5) to [out=30,in=150] (3,7);

\draw (2,6) to [out=30,in=150] (3,8);

\draw[fill] (2,7) circle(.1);

\draw[fill] (3,9) circle(.1);

\draw[fill] (3,10) circle(.1);
\draw (3,9) -- (3,10);
\draw (2,6) -- (2,7);

\draw (2,7) to [out=30,in=150] (3,9);
\draw (3,8) to [out=30,in=30] (3,10);

\node[below] at (4,6) {$w_2'^3$};

\draw[fill] (4,6) circle(.1);

\draw[fill] (4,7) circle(.1);

\draw (4,6) -- (4,7);

\draw[fill] (4,8) circle(.1);

\draw[fill] (5,9) circle(.1);

\draw[fill] (5,10) circle(.1);
\draw (5,9) -- (5,10);

\draw (4,6) to [out=150,in=150] (4,8);

\draw (4,7) to [out=30,in=150] (5,9);

\draw (4,8) to [out=30,in=150] (5,10);

\draw[fill] (4,9) circle(.1);

\draw[fill] (5,11) circle(.1);

\draw[fill] (5,12) circle(.1);
\draw (5,11) -- (5,12);
\draw (4,8) -- (4,9);

\draw (4,9) to [out=30,in=150] (5,11);
\draw (5,10) to [out=30,in=30] (5,12);

\node[below] at (6,6) {$w_6'$};

\draw[fill] (6,6) circle(.1);

\draw[fill] (6,7) circle(.1);

\draw (6,6) -- (6,7);

\draw[fill] (6,8) circle(.1);

\draw[fill] (7,9) circle(.1);

\draw[fill] (7,10) circle(.1);
\draw (7,9) -- (7,10);

\draw (6,6) to [out=150,in=150] (6,8);

\draw (6,7) to [out=30,in=150] (7,9);

\draw (6,8) to [out=30,in=150] (7,10);

\draw[fill] (6,9) circle(.1);

\draw[fill] (7,11) circle(.1);

\draw[fill] (7,12) circle(.1);
\draw (7,11) -- (7,12);
\draw (6,8) -- (6,9);

\draw (6,9) to [out=30,in=150] (7,11);
\draw (7,10) to [out=30,in=30] (7,12);

\end{tikzpicture}
\end{center}
\caption{The $\A_2(1)$-module structure of $\H^*(\B\SO(n),\Z_2)$ below degree 6 for $n\ge7$.}
\label{fig:A_2(1)SOn}
\end{figure}

The $E_2$ page is shown in Figure \ref{fig:E_2SOn}.
Here we have used the correspondence between $\A_2(1)$-module structure and the $E_2$ page shown in Figure \ref{fig:Z_2}, \ref{fig:A_2(1)} and \ref{fig:L_4}.

\begin{figure}[H]
\begin{center}
\begin{tikzpicture}
\node at (0,-1) {0};
\node at (1,-1) {1};
\node at (2,-1) {2};
\node at (3,-1) {3};
\node at (4,-1) {4};
\node at (5,-1) {5};
\node at (6,-1) {6};
\node at (7,-1) {$t-s$};
\node at (-1,0) {0};
\node at (-1,1) {1};
\node at (-1,2) {2};
\node at (-1,3) {3};
\node at (-1,4) {4};
\node at (-1,5) {5};
\node at (-1,6) {$s$};

\draw[->] (-0.5,-0.5) -- (-0.5,6);
\draw[->] (-0.5,-0.5) -- (7,-0.5);

\draw (0,0) -- (0,5);

\draw (0,0) -- (2,2);

\draw[fill] (2,0) circle(0.05);

\draw (4,3) -- (4,5);

\draw (4.1,1) -- (4.1,5);

\draw[fill] (4,0) circle(0.05);

\draw[fill] (6,0) circle(0.05);

\draw[fill] (6.1,0) circle(0.05);

\end{tikzpicture}
\end{center}
\caption{$\Omega_*^{\Spin\times\SO(n)}$ for $n\ge7$.}
\label{fig:E_2SOn}
\end{figure}

Thus we obtain the bordism group $\Omega^{\Spin\times\SO(n)}_d$ for $n\ge7$
shown in Table \ref{table:SOnBordism}.

\begin{table}[H]
\centering
\hspace*{-25mm}
\begin{tabular}{ c c c}
\hline
\multicolumn{3}{c}{Bordism group}\\
\hline
$d$ & 
$\Omega^{\Spin\times\SO(n)}_d$ for $n\ge7$
& bordism invariants \\
\hline
0&  $\Z$ \\
\hline
1& $\Z_2$  & $\tilde\eta$ \\
\hline
2&  $\Z_2^2$  & Arf, $w_2'$ \\
\hline
3 & $0$ \\
\hline
4 & $\Z^2\times\Z_2$ & $\frac{\sigma}{16}$, $\frac{p_1'}{2}$, $w_4'$  \\
\hline
5 & $0$  & \\
\hline
6 & $\Z_2^2$  & $w_2'^3$, $w_6'$\\
\hline
\end{tabular}
\caption{Bordism group. 
Here $w_i'$ is the Stiefel-Whitney class of the $\SO(10)$ bundle, $p_1'$ is the Pontryagin class of the $\SO(10)$ bundle.
$\tilde\eta$ is a mod 2 index of 1d Dirac operator.
Arf is a 2d Arf invariant.
$\sigma$ is the signature of manifold.
}
\label{table:SOnBordism}
\end{table}

By \eqref{eq:TPexact}, we obtain the cobordism group $\TP_d(\Spin \times\SO(n))$ for $n\ge7$ shown in Table \ref{table:SOnTP}.

\begin{table}[H]
\centering
\hspace*{-14mm}
\begin{tabular}{ c c c}
\hline
\multicolumn{3}{c}{Cobordism group}\\
\hline
$d$ & 

$\TP_d(\Spin \times\SO(n))$ for $n\ge7$
& topological terms \\
\hline
0& 0 \\
\hline
1& $\Z_2$  & $\tilde\eta$ \\
\hline
2& $\Z_2^2$ & Arf, $w_2'$ \\
\hline
3 & $\Z^2$ & $\frac{1}{48}\text{CS}_3^{TM}$, $\frac{1}{2}\text{CS}_3^{\SO(n)}$ \\
\hline
4 & $\Z_2$ & $w_4'$ \\
\hline
5 & $0$  & \\
\hline
\end{tabular}
\caption{Topological phase classification ($\equiv$ TP) as a cobordism group, following Table \ref{table:SOnBordism}. 
Here $w_i'$ is the Stiefel-Whitney class of the $\SO(10)$ bundle,
$\tilde\eta$ is a mod 2 index of 1d Dirac operator.
Arf is a 2d Arf invariant.
The $TM$ is the spacetime tangent bundle.
$\text{CS}_{2n-1}^V$ or $\text{CS}_{2n-1}^G$ is the Chern-Simons form of the vector bundle $V$ or the associated vector bundle of the principal $G$-bundle (associated to the Chern/Pontryagin class).
}
\label{table:SOnTP}
\end{table}

\subsection{
${\frac{\Spin \times \Spin(n)}{\Z_2^F}}$:
${\frac{\Spin \times \Spin(10)}{\Z_2^F}}$ and
${\frac{\Spin \times \Spin(18)}{\Z_2^F}}$
}

\label{sec:SpinSpin10modZ2}

Let $G=\frac{\Spin\times \Spin(n)}{\Z_2^F}$, we have a homotopy pullback square
\bea
\xymatrix{
\B G\ar[r]\ar[d]&\B \SO(n)\ar[d]^{w_2'}\\
\B \SO\ar[r]^{w_2}&\B ^2\Z_2.}
\eea
Here $w_2'=w_2(\SO(n))$.

There is a homotopy equivalence
$f:\B \SO\times \B \SO(n)\xrightarrow{\sim}\B \SO\times \B \SO(n)$ by $(V,W)\mapsto(V-W+n,W)$, and there is obviously also an inverse map.
Note that the pullback $f^*(w_2)=w_2(V-W)=w_2(V)+w_1(V)w_1(W)+w_2(W)=w_2+w_2'$.
Then we have the following homotopy pullback
\bea
\xymatrix{
\B G\ar[r]^-{\sim}\ar[d]&\B \Spin\times \B \SO(n)\ar[d]&\\
\B \SO\times \B \SO(n)\ar[r]^{f}\ar[d]_{(V,W)\mapsto V}\ar@/_1pc/[rr]_{w_2+w_2'}&\B \SO\times \B \SO(n)\ar[r]^-{w_2+0}\ar[ld]^{(V,W)\mapsto V+W-n}&\B ^2\Z_2\\
\B \SO&&}.
\eea
This implies that the two classifying spaces $\B G\sim \B \Spin\times \B \SO(n)$ are homotopy equivalent.

By definition, the Madsen-Tillmann spectrum $MTG$ of the group $G$ is 
 $MTG=\text{Thom}(\B G;-V)$, where $V$ is the induced virtual bundle (of dimension $0$) by the map $\B G\to \B \O$.

We can identify $\B G\to \B \O$ with
$\B \Spin\times \B \SO(n)\xrightarrow{V-V_n+n}\B \SO\hookrightarrow \B \O$.

So the spectrum $MTG$ is homotopy equivalent to $\text{Thom}(\B \Spin\times \B \SO(n);-(V-V_n+n))$, which is $M\Spin\wedge\Sigma^{-n}M\SO(n)$.

We consider $G=\frac{\Spin \times \Spin(10)}{\Z_2^F}$, the Madsen-Tillmann spectrum $MTG$ of the group $G$ is 
\bea
MTG=M\Spin\wedge\Sigma^{-10}M\SO(10).
\eea

We have $w_2=w_2'$, namely $w_2(TM)=w_2(\SO(10))$.

For the dimension $d=t-s<8$, since there is no odd torsion (see footnote \ref{ft:no-odd-torsion}), by \eqref{eq:ExtA_2(1)}, we have the Adams spectral sequence
\bea
\Ext_{\A_2(1)}^{s,t}(\H^{*+10}(M\SO(10),\Z_2),\Z_2)\Rightarrow\Omega_{t-s}^{\frac{\Spin \times \Spin(10)}{\Z_2^F}}.
\eea

We have
\bea
\H^{*+10}(M\SO(10),\Z_2)=\Z_2[w_2',w_3',\dots,w_{10}']U
\eea
where $U$ is the Thom class with $\Sq^1U=0$, $\Sq^2U=w_2'U$. {Here in this subsection, $w_i'$ is the Stiefel-Whitney class of $\SO(n)$ bundle.}

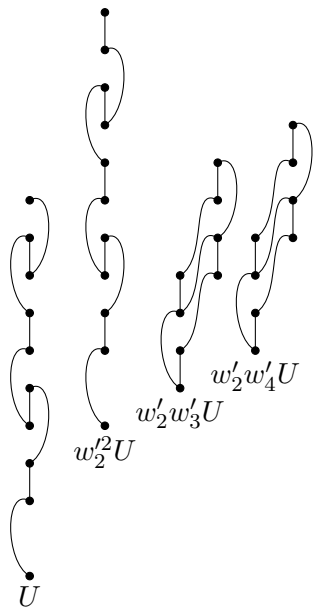
\begin{figure}[H]
\begin{center}
\begin{tikzpicture}[scale=0.5]

\node[below] at (0,0) {$U$};

\draw[fill] (0,0) circle(.1);
\draw[fill] (0,2) circle(.1);
\draw (0,0) to [out=150,in=150] (0,2);
\draw[fill] (0,3) circle(.1);
\draw (0,2) -- (0,3);
\draw[fill] (0,4) circle(.1);
\draw[fill] (0,5) circle(.1);
\draw (0,3) to [out=30,in=30] (0,5);
\draw (0,4) -- (0,5);
\draw[fill] (0,6) circle(.1);
\draw (0,4) to [out=150,in=150] (0,6);
\draw[fill] (0,7) circle(.1);
\draw (0,6) -- (0,7);
\draw[fill] (0,8) circle(.1);
\draw[fill] (0,9) circle(.1);
\draw (0,7) to [out=150,in=150] (0,9);
\draw (0,8) -- (0,9);
\draw[fill] (0,10) circle(.1);
\draw (0,8) to [out=30,in=30] (0,10);

\node[below] at (2,4) {$w_2'^2U$};

\draw[fill] (2,4) circle(.1);
\draw[fill] (2,6) circle(.1);
\draw (2,4) to [out=150,in=150] (2,6);
\draw[fill] (2,7) circle(.1);
\draw (2,6) -- (2,7);
\draw[fill] (2,8) circle(.1);
\draw[fill] (2,9) circle(.1);
\draw (2,7) to [out=30,in=30] (2,9);
\draw (2,8) -- (2,9);
\draw[fill] (2,10) circle(.1);
\draw (2,8) to [out=150,in=150] (2,10);
\draw[fill] (2,11) circle(.1);
\draw (2,10) -- (2,11);
\draw[fill] (2,12) circle(.1);
\draw[fill] (2,13) circle(.1);
\draw (2,11) to [out=150,in=150] (2,13);
\draw (2,12) -- (2,13);
\draw[fill] (2,14) circle(.1);
\draw (2,12) to [out=30,in=30] (2,14);
\draw[fill] (2,15) circle(.1);
\draw (2,14) -- (2,15);

\node[below] at (4,5) {$w_2'w_3'U$};

\draw[fill] (4,5) circle(.1);
\draw[fill] (4,6) circle(.1);
\draw (4,5) -- (4,6);
\draw[fill] (4,7) circle(.1);
\draw (4,5) to [out=150,in=150] (4,7);
\draw[fill] (4,8) circle(.1);
\draw[fill] (5,8) circle(.1);
\draw (4,7) -- (4,8);
\draw (4,6) to [out=30,in=150] (5,8);
\draw[fill] (5,9) circle(.1);
\draw (5,8) -- (5,9);
\draw (4,7) to [out=30,in=150] (5,9);
\draw[fill] (5,10) circle(.1);
\draw (4,8) to [out=30,in=150] (5,10);
\draw[fill] (5,11) circle(.1);
\draw (5,10) -- (5,11);
\draw (5,9) to [out=30,in=30] (5,11);

\node[below] at (6,6) {$w_2'w_4'U$};

\draw[fill] (6,6) circle(.1);
\draw[fill] (6,7) circle(.1);
\draw (6,6) -- (6,7);
\draw[fill] (6,8) circle(.1);
\draw (6,6) to [out=150,in=150] (6,8);
\draw[fill] (6,9) circle(.1);
\draw[fill] (7,9) circle(.1);
\draw (6,8) -- (6,9);
\draw (6,7) to [out=30,in=150] (7,9);
\draw[fill] (7,10) circle(.1);
\draw (7,9) -- (7,10);
\draw (6,8) to [out=30,in=150] (7,10);
\draw[fill] (7,11) circle(.1);
\draw (6,9) to [out=30,in=150] (7,11);
\draw[fill] (7,12) circle(.1);
\draw (7,11) -- (7,12);
\draw (7,10) to [out=30,in=30] (7,12);

\end{tikzpicture}
\end{center}
\caption{The $\A_2(1)$-module structure of $\H^{*+10}(M\SO(10),\Z_2)$ below degree 6.
Here $U$ is a Thom class.}
\label{fig:A_2(1)MSO10}
\end{figure}

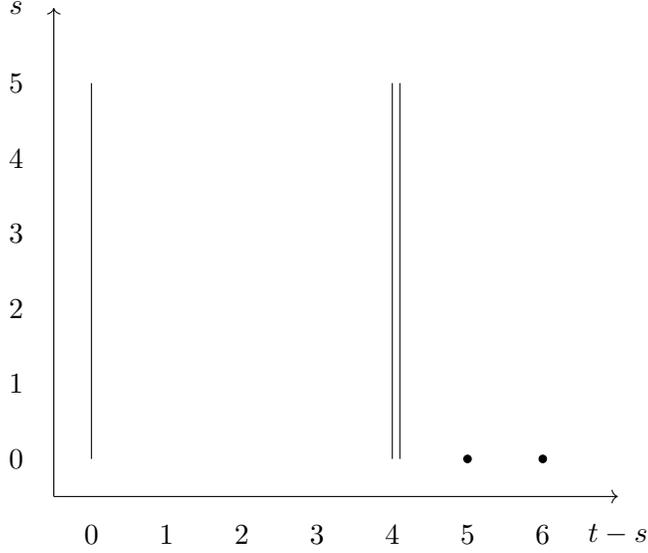
\begin{figure}[H]
\begin{center}
\begin{tikzpicture}
\node at (0,-1) {0};
\node at (1,-1) {1};
\node at (2,-1) {2};
\node at (3,-1) {3};
\node at (4,-1) {4};
\node at (5,-1) {5};
\node at (6,-1) {6};
\node at (7,-1) {$t-s$};
\node at (-1,0) {0};
\node at (-1,1) {1};
\node at (-1,2) {2};
\node at (-1,3) {3};
\node at (-1,4) {4};
\node at (-1,5) {5};
\node at (-1,6) {$s$};

\draw[->] (-0.5,-0.5) -- (-0.5,6);
\draw[->] (-0.5,-0.5) -- (7,-0.5);

\draw (0,0) -- (0,5);
\draw (4,0) -- (4,5);

\draw (4.1,0) -- (4.1,5);
\draw[fill] (5,0) circle(0.05);
\draw[fill] (6,0) circle(0.05);

\end{tikzpicture}
\end{center}
\caption{$\Omega_*^{\frac{\Spin \times \Spin(10)}{\Z_2^F}}$.}
\label{fig:E_2MSO10}
\end{figure}

The $\A_2(1)$-module structure of $\H^{*+10}(M\SO(10),\Z_2)$ below degree 6 and the $E_2$ page are shown in Figure \ref{fig:A_2(1)MSO10}, \ref{fig:E_2MSO10}.
Here we have used the correspondence between $\A_2(1)$-module structure and the $E_2$ page shown in Figure \ref{fig:A_2(1)}, \ref{fig:L_5}, and \ref{fig:L_8}.

Thus we obtain the bordism group $\Omega^{\frac{\Spin \times \Spin(10)}{\Z_2^F}}_d$ shown in Table \ref{table:Spin10Z2Bordism}.

\begin{table}[H]
\centering
\begin{tabular}{ c c c}
\hline
\multicolumn{3}{c}{Bordism group}\\
\hline
$d$ & 
$\Omega^{\frac{\Spin \times \Spin(10)}{\Z_2^F}}_d$
& bordism invariants \\
\hline
0& $\Z$\\
1& $0$\\
2& $0$\\
3 & $0$\\
4 & $\Z^2$ & $p_1'$ (from $w_2'^2$), $e_4'$ (from $w_4'$)\\
5 & $\Z_2$ & $w_2 w_3=w_2' w_3'$\\
6 & $\Z_2$ & $w_2 w_4'=w_2'w_4'$ \\
\hline
\end{tabular}
\caption{Bordism group. Here $w_i$ is the Stiefel-Whitney class of the tangent bundle, $w_i'$ is the Stiefel-Whitney class of the $\SO(10)$ bundle,
 $p_1'$ is the Pontryagin class of the $\SO(10)$ bundle. $e_4'$ is the Euler class of the $\SO(10)$ bundle, on a 4-manifold, the oriented bundle of rank 10 splits as the direct sum of an oriented bundle of rank 4 and a trivial bundle of rank 6, the Euler class $e_4'$ is the Euler class of the subbundle of rank 4.
Same result for $\Omega^{\frac{\Spin \times \Spin(18)}{\Z_2^F}}_d$ and
$\Omega^{\frac{\Spin \times \Spin(n)}{\Z_2^F}}_d$ with $n\ge7$ and $0\le d\le 6$.
Here, the $p_1'$ has another form, see Appendix \ref{sec:another}.
}
\label{table:Spin10Z2Bordism}
\end{table}

Actually $\Omega^{\frac{\Spin \times \Spin(n)}{\Z_2^F}}_d=\Omega^{\frac{\Spin \times \Spin(n+1)}{\Z_2^F}}_d$ for $n\ge7$ and $0\le d\le 6$.

By \eqref{eq:TPexact}, we obtain the cobordism group $\TP_d({\frac{\Spin \times \Spin(10)}{\Z_2^F}})$ shown in Table \ref{table:Spin10Z2TP}.

\begin{table}[H]
\centering
\begin{tabular}{ c c c}
\hline
\multicolumn{3}{c}{Cobordism group}\\
\hline
$d$ & 
$\TP_d({\frac{\Spin \times \Spin(10)}{\Z_2^F}})$
& topological terms \\
\hline
0& $0$\\
1& $0$\\
2& $0$\\
3 & $\Z^2$ & $\text{CS}_3^{\SO(10)},\text{CS}_{3,e}^{\SO(10)}$\\
4 & $0$\\ 
5 & $\Z_2$ & {$w_2 w_3=w_2' w_3'$}\\
%
\hline
\end{tabular}
\caption{Topological phase classification ($\equiv$ TP) as a cobordism group, following Table \ref{table:Spin10Z2Bordism}. 
Here $w_i$ is the Stiefel-Whitney class of the tangent bundle, $w_i'$ is the Stiefel-Whitney class of the $\SO(10)$ bundle, 
$\text{CS}_{2n-1}^G$ is the Chern-Simons form of the associated vector bundle of the principal $G$-bundle (associated to the Chern/Pontryagin class).
$\text{CS}_{2n-1,e}^G$ is the Chern-Simons form of the associated vector bundle of the principal $G$-bundle (associated to the Euler class).
Same result for $\TP_d({\frac{\Spin \times \Spin(18)}{\Z_2^F}})$ and
$\TP_d({\frac{\Spin \times \Spin(n)}{\Z_2^F}})$ with $n\ge7$  and $0\le d\le 5$.
}
\label{table:Spin10Z2TP}
\end{table}

\section{SU(5) and SU($n$) Grand Unifications: $\Spin \times \SU(n)$:
$\Spin \times \SU(5)$
}\label{sec:GUSU}

Now we consider the co/bordism classes relevant for Georgi-Glashow SU(5) GUT \cite{Georgi1974syUnityofAllElementaryParticleForces}.

We consider $G=\Spin \times \SU(5)$, the Madsen-Tillmann spectrum $MTG$ of the group $G$ is 
\bea
MTG=M\Spin\wedge (\B \SU(5))_+.
\eea
The $(\B\SU(5))_+$ is the disjoint union of the classifying space $\B\SU(5)$ and a point, see footnote \ref{ft:X_+}.

For the dimension $d=t-s<8$, since there is no odd torsion (see footnote \ref{ft:no-odd-torsion}), by \eqref{eq:ExtA_2(1)}, we have the Adams spectral sequence
\bea
\Ext_{\A_2(1)}^{s,t}(\H^*(\B \SU(5),\Z_2),\Z_2)\Rightarrow\Omega_{t-s}^{\Spin \times \SU(5)}.
\eea

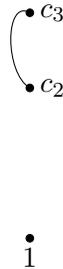
\begin{figure}[H]
\begin{center}
\begin{tikzpicture}[scale=0.5]

\node[below] at (0,0) {$1$};
\draw[fill] (0,0) circle(.1);

\node[right] at (0,4) {$c_2$};
\node[right] at (0,6) {$c_3$};
\draw[fill] (0,4) circle(.1);
\draw[fill] (0,6) circle(.1);

\draw (0,4) to [out=150,in=150] (0,6);

\end{tikzpicture}
\end{center}
\caption{The $\A_2(1)$-module structure of $\H^*(\B \SU(5),\Z_2)$ below degree 6. {Here, $c_i$ is the Chern class of SU(5) bundle.}}
\label{fig:A_2(1)SU5}
\end{figure}

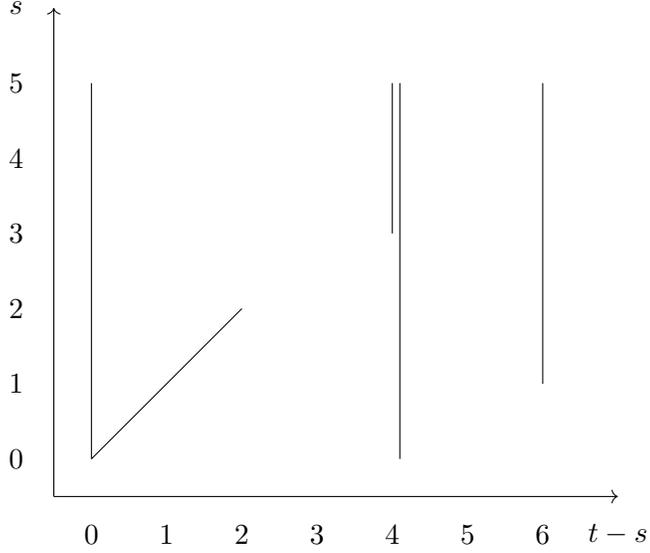
\begin{figure}[H]
\begin{center}
\begin{tikzpicture}
\node at (0,-1) {0};
\node at (1,-1) {1};
\node at (2,-1) {2};
\node at (3,-1) {3};
\node at (4,-1) {4};
\node at (5,-1) {5};
\node at (6,-1) {6};
\node at (7,-1) {$t-s$};
\node at (-1,0) {0};
\node at (-1,1) {1};
\node at (-1,2) {2};
\node at (-1,3) {3};
\node at (-1,4) {4};
\node at (-1,5) {5};
\node at (-1,6) {$s$};

\draw[->] (-0.5,-0.5) -- (-0.5,6);
\draw[->] (-0.5,-0.5) -- (7,-0.5);

\draw (0,0) -- (0,5);
\draw (0,0) -- (2,2);
\draw (4,3) -- (4,5);
\draw (4.1,0) -- (4.1,5);
\draw (6,1) -- (6,5);

\end{tikzpicture}
\end{center}
\caption{$\Omega_*^{\Spin \times \SU(5)}$.}
\label{fig:E_2SU5}
\end{figure}

The $\A_2(1)$-module structure of $\H^*(\B \SU(5),\Z_2)$ below degree 6 and the $E_2$ page are shown in Figure \ref{fig:A_2(1)SU5}, \ref{fig:E_2SU5}.
Here we have used the correspondence between $\A_2(1)$-module structure and the $E_2$ page shown in Figure \ref{fig:Z_2} and \ref{fig:Ceta}.

Thus we obtain the bordism group $\Omega^{\Spin \times \SU(5)}_d$ shown in Table \ref{table:SU5Bordism}.

\begin{table}[H]
\centering
\begin{tabular}{ c c c}
\hline
\multicolumn{3}{c}{Bordism group}\\
\hline
$d$ & 
$\Omega^{\Spin \times \SU(5)}_d$
& bordism invariants \\
\hline
0& $\Z$\\
1& $\Z_2$ & $\tilde\eta$\\
2& $\Z_2$ & Arf\\
3 & $0$\\
4 & $\Z^2$ & $\frac{\sigma}{16},c_2$\\
5 & $0$ \\
6 & $\Z$ & $\frac{c_3}{2}$ \\
\hline
\end{tabular}
\caption{Bordism group. 
$\tilde\eta$ is a mod 2 index of 1d Dirac operator.
Arf is a 2d Arf invariant.
$\sigma$ is the signature of manifold.
Here $c_i$ is the Chern class of the $\SU(5)$ bundle.
Note that $c_3=\Sq^2c_2=(w_2+w_1^2)c_2=0\mod2$ on Spin 6-manifolds.
}
\label{table:SU5Bordism}
\end{table}

Actually $\Omega^{\Spin \times \SU(n)}_d=\Omega^{\Spin \times \SU(n+1)}_d$ for $n\ge3$ and $0\le d\le 6$.

By \eqref{eq:TPexact}, we obtain the cobordism group $\TP_d(\Spin \times \SU(5))$ shown in Table \ref{table:SU5TP}.

\begin{table}[H]
\centering
\begin{tabular}{ c c c}
\hline
\multicolumn{3}{c}{Cobordism group}\\
\hline
$d$ & 
$\TP_d(\Spin \times \SU(5))$
& topological terms \\
\hline
0& $0$\\
1& $\Z_2$ & $\tilde\eta$\\
2& $\Z_2$ & Arf \\
3 & $\Z^2$ & $\frac{1}{48}$CS$_3^{TM}$, CS$_3^{\SU(5)}$\\
4 & $0$ \\
5 & $\Z$ & $\frac{1}{2}$CS$_5^{\SU(5)}$  \\
\hline
\end{tabular}
\caption{Topological phase classification ($\equiv$ TP) as a cobordism group, following Table \ref{table:SU5Bordism}. 
$\tilde\eta$ is a mod 2 index of 1d Dirac operator.
Arf is a 2d Arf invariant.
The $TM$ is the spacetime tangent bundle.
$\text{CS}_{2n-1}^V$ or $\text{CS}_{2n-1}^G$ is the Chern-Simons form of the vector bundle $V$ or the associated vector bundle of the principal $G$-bundle (associated to the Chern/Pontryagin class).
}
\label{table:SU5TP}
\end{table}

\section{Physics interpretations of topological terms, anomalies and invertible topological orders}
\label{sec:interpretations}

Let us present some explicit examples of topological terms, anomalies and invertible topological orders, and their interpretations (see also \cite{WanWang2018bns1812.11967}). 
We shall interpret the new results obtained in our Tables in Section \ref{sec:SM}-\ref{sec:GUSU} in a
similar manner.

\subsection{Interpretations of
 the $\Z$ classes of co/bordism invariants}

\begin{enumerate}[leftmargin=2mm, label=\textcolor{blue}{(\arabic*)}., ref={(\arabic*)}]
\item The bordism generator of bordism group $\Omega_4^{\SO}=\Z$ is the complex projective space $\CP^2$, 
$\overline{\CP^2}$ and their connected-sum manifolds,
whose bordism invariant is the signature $\sigma=\frac{1}{3}\int_{M^3}p_1(TM)$ related to Pontryagin class of tangent spacetime $TM$ of manifold $M$.
Since $\sigma(\CP^2)=1$ and all other SO-manifolds have quantized signatures,
we can define a so-called $\theta$-term whose partition function 
$${\bf Z}=\exp(\ii \frac{\theta}{3}\int_{M^4}p_1(TM))$$  
on non-spin manifolds
with a compact periodic $\theta \in [0, 2 \pi)$, with a different $\theta$ specifying a different theory on non-spin manifolds.

The cobordism generator of cobordism group $\TP_3(\SO)=\Z$ is the 3d gravitational Chern-Simons (CS) theory given by 
a partition function 
$${\bf Z}= \exp(\ii \frac{k}{3}\int_{M^3} {\rm{CS}}_{\rm{grav}})=\exp(\ii \frac{k}{3} \frac{1}{4 \pi} \int_{M^3} \Tr( \omega \dd \omega + \frac{2}{3} \omega^3)),$$
here $k \in \Z$ and $\omega$ is a connection of the tangent bundle of an oriented non-spin manifold $M$. 
The $\exp(\ii \frac{k}{3} \rm{CS}_{\rm{grav}})$ on the non-spin $M^3$ can be defined by extending the 3-manifold $M^3$ as a boundary of 4-manifold $M^4$ with a {stable tangential SO structure. 
{The extension from 3d to 4d is doable, thanks to $\Omega_3^{\SO}=\Omega_3^{\SO}(pt)=0$.} 
So $\left.\exp(\ii \frac{k}{3}\int_{M^3} \rm{CS}_{\rm{grav}})\right|_{{M^3=\partial(M^4) }}
=\exp({\ii}\frac{\nu}{3} \frac{1}{4 \pi} \int_{M^4} \Tr( R(\omega) \wedge R(\omega) )
=
\left.\exp(\ii \frac{k}{3} (2 \pi)\int_{M^4}p_1(TM))\right.$,
with $R(\omega)=\dd \omega + \omega \wedge \omega$ as the curvature 2-form of $\omega$.
We can show this definition is independent for any suitable $M^4$ extension. 
If ${M^3=\partial(M^4)=\partial(M'^4) }$, then the two definitions are differed by an identity 
$\exp(\ii \frac{k}{3} (2 \pi)\int_{(M^4) \cup (M'^4)}p_1(TM))
=\exp(\ii {k} (2 \pi  )\sigma )=1$ on a closed SO manifold ${(M^4) \cup (M'^4)}$.}\\

When $k=1$, the 3d gravitational CS theory is the gravitational background field theory,
which can probe
the dynamical internal 3d CS gauge theory of gauge group $(\rm{E}_8)_1$.
Namely, by integrating out the internal gauge fields of $(\rm{E}_8)_1$, 
we should obtain back the 3d gravitational CS background field theory.
Also this bulk internal field theory can be equivalently written as at least three CS descriptions: 
\begin{itemize} 
\item  First, the 3d $(\rm{E}_8)_1$ CS theory with a gauge group $\rm{E}_8$ at the level 1.
\item Second, a  $\SO(16)_1$ CS $\otimes$ (a spin TQFT  $\{1,f\}$ with only a trivial line operator 1 and a single fermionic line $f$).
\item Third, a rank-8 $K_{\rm{E}_8}$ matrix abelian CS theory ${\bf Z}=\int[Da]\exp(\frac{(K_{\rm{E}_8})_{IJ}}{4 \pi}\int_{M^3} a_I \dd a_J)$ 
with 
an $\rm{E}_8$
Cartan matrix $K_{\rm{E}_8}=\left(
\begin{smallmatrix}
 2 & -1 &  0 &  0 &  0 &  0 &  0 & 0 \\
-1 &  2 & -1&  0 &  0 &  0 &  0 & 0 \\
 0 & -1 &  2 & -1 &  0 &  0 &  0 & 0 \\
 0 &  0 & -1 &  2 & -1 &  0 &  0 & 0 \\
 0 &  0 &  0 & -1 &  2 & -1 &  0 & -1 \\
 0 &  0 &  0 &  0 & -1 &  2 & -1 & 0 \\
 0 &  0 &  0 &  0 &  0 & -1 &  2 & 0 \\
 0 &  0 & 0 &  0 &  -1 &  0 &  0 & 2
\end{smallmatrix}\right)$
as the symmetric bilinear form pairing the 1-form abelian \emph{dynamical} gauge fields $a_I$ and $a_J$, for $I,J \in \{1,2,\dots, 8\}$. 
\end{itemize} 
Note that all these internal 3d CS  theories
can have a 2d boundary CFT with a chiral central charge $c_-=8$, namely associated with 
a ${\rm{E}_8}$ chiral boson theory with 8 chiral modes each with $c_-=1$.
This is a rank-8 $K$-matrix complex chiral boson theory whose $K=K_{\rm{E}_8}$.
Alternatively, this CFT can also be formed by gapless 16 multiple of chiral Majorana-Weyl modes, each mode with a $c_-=1/2$.
\\
\item The bordism generator of bordism group $\Omega_4^{\Spin}=\Z$ is the complex $K_3$ manifold (belonging to Calabi-Yau manifold and hyperk\"ahler manifold),
whose bordism invariant is the signature $\frac{1}{16}\sigma=\frac{1}{16}\frac{1}{3}\int_{M^4}p_1(TM)$ related to Pontryagin class of tangent spacetime $TM$ of manifold $M$.
Since $\frac{1}{16}\sigma(K_3)=-1$ and all other spin manifolds have quantized signatures,
we can define a so-called $\theta$-term whose partition function is
$${\bf Z}=\exp(\ii \frac{\theta}{48}\int_{M^4}p_1(TM))$$ on spin manifolds
with a compact periodic $\theta \in [0, 2 \pi)$, with any different $\theta$ value specifying a different theory on spin manifolds.
The cobordism generator of cobordism group $\TP_3(\Spin)=\Z$ is the 3d gravitational Chern-Simons theory given by 
$${\bf Z}=\exp(\frac{\ii}{16}\frac{\nu}{3}\int_{M^3} {\rm{CS}}_{\rm{grav}})=\exp(\frac{\ii}{16}\frac{\nu}{3} \frac{1}{4 \pi} \int_{M^3} \Tr( \omega \dd \omega + \frac{2}{3} \omega^3)),$$
here $\nu \in \Z$.
Similarly, the $\exp(\ii \frac{\nu}{16 \cdot 3}\int_{M^3} \rm{CS}_{\rm{grav}})$ on the spin $M^3$ can be defined by extending the 3-manifold $M^3$ as a boundary of 4-manifold $M^4$ with a {stable tangential spin structure. 
{The extension from 3d to 4d is doable, thanks to $\Omega_3^{\Spin}=\Omega_3^{\Spin}(pt)=0$.}
So
$\left.\exp(\ii \frac{\nu}{16 \cdot 3} \int_{M^3} \rm{CS}_{\rm{grav}})\right|_{{M^3=\partial(M^4) }}
=\exp(\frac{\ii}{16}\frac{\nu}{3} \frac{1}{4 \pi} \int_{M^4} \Tr( R(\omega) \wedge R(\omega) )
=
\left.\exp(\ii \frac{\nu}{16 \cdot 3}  (2 \pi)\int_{M^4}p_1(TM))\right.$.
We can show this definition is independent for any suitable $M^4$ extension. 
If ${M^3=\partial(M^4)=\partial(M'^4) }$, then the two definitions are differed by an identity 
$\exp(\ii \frac{\nu}{16 \cdot 3}  (2 \pi)\int_{(M^4) \cup (M'^4)}p_1(TM))
=\exp(\ii {\nu} (2 \pi  )\frac{\sigma}{16} )=1$ on a closed spin manifold ${(M^4) \cup (M'^4)}$.}
Note that the 3d CS theory
can have a 2d boundary CFT with a chiral central charge  $c_-=1/2$, namely associated with a gapless real-valued chiral Majorana-Weyl mode $\chi$ with an action
$\int \dd t  \dd  x    \chi  \ii( \partial_t - \partial_x) \chi$.
\\
\item Next let us consider the bordism group $\Omega_4^{\SO}(\B\SU(2))=\Z^2$ and $\Omega_4^{\Spin}(\B\SU(2))=\Z^2$ for the internal symmetry SU(2).
The two bordism invariants of $\Omega_4^{\SO}(\B\SU(2))=\Z^2$ are the signature $\sigma$ and the second Chern class $c_2(V_{\SU(2)})$ of the $\SU(2)$ bundle.
The two bordism invariants of  $\Omega_4^{\Spin}(\B\SU(2))=\Z^2$ are $\frac{\sigma}{16}$ where $\sigma$ is the signature and the second Chern class $c_2(V_{\SU(2)})$ of the $\SU(2)$ bundle. We had already discussed the invariants related to signature $\sigma$.
Let us focus on the other, but common, bordism invariant $c_2(V_{\SU(2)})$ of $\Omega_4^{\SO}(\B\SU(2))$ and $\Omega_4^{\Spin}(\B\SU(2))$.
This, in fact, specifies the so-called $\theta$-term of particle physics 
whose partition function contains
$${\bf Z}=\exp(\ii {\theta} \int_{M^4} c_2(V_{\SU(2)}) )=\exp(  \frac{{\ii\theta}}{8 \pi^2}  \int_{M^4} \Tr[F_{V_{\SU(2)}} \wedge F_{V_{\SU(2)}}]) $$ 
on both non-spin and spin 
manifolds
with a compact periodic $\theta \in [0, 2 \pi)$,
because the instanton number 
$\frac{{\ii\theta}}{8 \pi^2} \int_{M^4} \Tr(F_{}\wedge F_{})$
for the SU(2) gauge bundle on 4-manifolds is in $\Z$ value. 
Here the curvature 2-form is $$F= \dd A +  A \wedge A$$ where
$A=A_\mu^\alpha T^\alpha$  is Lie algebra valued 1-form and we choose the Lie algebra
$[T^\alpha, T^\beta]= f^{\alpha \beta \gamma} T^\gamma$ with an \emph{anti-Hermitian} $T^\alpha$ and a structure constant $f^{\alpha \beta \gamma}$.\footnote{We
warn the readers that mathematicians and physicists may prefer different conventions.
For a vector bundle $V$:
\begin{enumerate}
\item Mathematicians may choose the curvature 2-form as $$F= \dd A +  A \wedge A$$ where
$A=A_\mu^\alpha T^\alpha$ is Lie algebra $\mathfrak{g}$ valued 1-form. The Lie algebra can be $[T^\alpha, T^\beta]= f^{\alpha \beta \gamma} T^\gamma$ with an \emph{anti-Hermitian} $T^\alpha$ and a structure constant $f^{\alpha \beta \gamma}$. For some variable ${\rm{X}}$, 
the following polynomial expansion computes the Chern classes of the vector bundle $V$: 
\bea\hspace*{-14mm}
 \sum_k c_k(V){\rm{X}}^k &=& \det(\mathbb{I}+{\rm{X}} (\ii \frac{F}{2\pi}))\quad \quad \\
 &=& \left[ 1 
       +  \frac{\mathrm{Tr}(F)}{2\pi} {(\ii\rm{X})}
       +(-1)\frac{\mathrm{Tr}(F^2)-(\mathrm{Tr}(F))^2}{8\pi^2} {(\ii \rm{X})}^2
       +  \frac{2\mathrm{Tr}(F^3)-3\mathrm{Tr}(F^2)\mathrm{Tr}(F)+\mathrm{Tr}(F)^3}{48\pi^3} {(\ii\rm{X})}^3
       + \cdots
       \right].\quad \quad\quad 
\eea
The $\mathbb{I}$ is an identity matrix whose rank is equal to the rank
of the matrix representation of Lie algebra $\mathfrak{g}$.
We use $\det(\e^{\rm M})=\e^{\Tr(\rm M)}$
and take ${\rm M}=\log (\mathbb{I}+{\rm{X}} (\ii \frac{F}{2\pi}))$.
In that case,
\bea
c_1 &=& \ii\frac{\mathrm{Tr}(F)}{2\pi}, \\
c_2 &=& \frac{\mathrm{Tr}(F^2)-(\mathrm{Tr}(F))^2}{8\pi^2}, \\
c_3 &=& - \ii\frac{2\mathrm{Tr}(F^3)-3\mathrm{Tr}(F^2)\mathrm{Tr}(F)+\mathrm{Tr}(F)^3}{48\pi^3}, etc., 
\eea
also
${\bf Z}=\exp(\ii k (2 \pi) \int_{M^3}\text{CS}_3^{\SU(2)})$ is related to $\exp(\frac{\ii k }{4\pi} \int_{M^3} \text{Tr}(A\dd A + \frac{2}{3}A^3))$.
\item
Physicists may choose the curvature 2-form as $$F= \dd A - \ii A \wedge A$$ where
$A=A_\mu^\alpha T^\alpha$ is Lie algebra valued 1-form. The Lie algebra can be $[T^\alpha, T^\beta]=\ii f^{\alpha \beta \gamma} T^\gamma$ with a \emph{Hermitian} $T^\alpha$ and a structure constant $f^{\alpha \beta \gamma}$. 
For some variable ${\rm{X}}$, 
the following polynomial expansion computes the Chern classes of the vector bundle $V$: 
\bea
\hspace*{-14mm}
 \sum_k c_k(V){\rm{X}}^k &=& \det( \mathbb{I}+{\rm{X}} ( \frac{F}{2\pi}))\quad \quad \\
 &=& \left[ 1 
       +  \frac{\mathrm{Tr}(F)}{2\pi} {(\rm{X})}
       +(-1)\frac{\mathrm{Tr}(F^2)-(\mathrm{Tr}(F))^2}{8\pi^2} {( \rm{X})}^2
       +  \frac{{2\mathrm{Tr}(F^3)-3\mathrm{Tr}(F^2)\mathrm{Tr}(F)+\mathrm{Tr}(F)^3}}{48\pi^3} {(\rm{X})}^3
       + \cdots
       \right].\quad \quad 
\eea
\pagebreak
In that case,
\bea
c_1 &=& \frac{\mathrm{Tr}(F)}{2\pi}, \\
c_2 &=& - \frac{\mathrm{Tr}(F^2)-(\mathrm{Tr}(F))^2}{8\pi^2}, \\
c_3 &=&  \frac{2\mathrm{Tr}(F^3)-3\mathrm{Tr}(F^2)\mathrm{Tr}(F)+\mathrm{Tr}(F)^3}{48\pi^3}, etc.,
\eea
also
${\bf Z}=\exp(\ii k (2 \pi)\int_{M^3} \text{CS}_3^{\SU(2)})$ is related to instead $\exp(\frac{-\ii k }{4\pi} \int_{M^3} \text{Tr}(A\dd A - \ii \frac{2}{3}A^3))$.
\end{enumerate}
 }

Also $\TP_3(\SO\times\SU(2))=\Z^2$ and $\TP_3(\Spin\times\SU(2))=\Z^2$.
We had discussed one cobordism generator associated to a multiple of 3d ${\rm{CS}}_{\rm{grav}}$.
There is another cobordism generator $\text{CS}_3^{\SU(2)}$ common to both $\TP_3(\SO\times\SU(2))$ and $\TP_3(\Spin\times\SU(2))$. Here $\text{CS}_3^{\SU(2)}$ is the Chern-Simons 3-form of the $\SU(2)$ gauge bundle given by 
$${\bf Z}=\exp(\ii k (2 \pi) \int_{M^3}\text{CS}_3^{\SU(2)})=\exp(\frac{\ii k }{4\pi} \int_{M^3} \text{Tr}(A\dd A + \frac{2}{3}A^3)),$$ 
Here $A$ is the connection of the $\SU(2)$ gauge bundle.
Similarly, the $\exp(\ii k (2 \pi) \int_{M^3}\text{CS}_3^{\SU(2)})$ 
on the spin $M^3$ can be defined by extending the 3-manifold $M^3$ as a boundary of 4-manifold $M^4$ with a {principal SU(2) structure,
thanks to. So
$\left.\exp(\ii k  (2 \pi) \int_{M^3}\text{CS}_3^{\SU(2)})\right|_{{M^3=\partial(M^4) }}=
\left.\exp(\ii 
\frac{{k (2 \pi)}}{8 \pi^2}  \int_{M^4} \Tr[F_{V_{\SU(2)}} \wedge F_{V_{\SU(2)}}])
  )\right.
=
\left.\exp(\ii k (2 \pi) \int_{M^4} c_2(V_{\SU(2)})  
\right.$,
we can show this definition is independent for any suitable $M^4$ extension, thanks to
$\Omega_3^{\Spin}(\B\SU(2))=\Omega_3^{\SO}(\B\SU(2))=0$. 
If ${M^3=\partial(M^4)=\partial(M'^4) }$, then the two definitions are differed by an identity 
$\exp(\ii k (2 \pi) \int_{(M^4) \cup (M'^4)}c_2(V_{\SU(2)}))=1$
 on ${(M^4) \cup (M'^4)}$,
since the second Chern class is precisely the instanton number such that
$c_2(V_{\SU(2)})\in \Z$ on both closed spin and non-spin manifolds.}
\end{enumerate} 
We can interpret the $\Z$ class co/bordism invariants  obtained in our Tables in Section \ref{sec:SM}-\ref{sec:GUSU} in a
similar manner. In this next subsection, we
enumerate the interpretations of the
4d anomalies (and 5d cobordism invariants)
for various SM, BSM, and GUT
obtained in Section \ref{sec:SM}-\ref{sec:GUSU}.

\subsection{Interpretations of
 the 4d anomalies of
 SM and GUT
from co/bordism invariants:\\ 
Beyond the Witten anomaly}

We can interpret the classifications of
all possible 4d perturbative local and
non-perturbative global invertible
anomalies for SM, BSM, and GUTs obtained in Section \ref{sec:SM}-\ref{sec:GUSU} as follows: 

\begin{enumerate}

\item
In Sec 2, we find that their 5d cobordism invariants contain the following:
\begin{itemize} 
\item A $\Z_2$ class from the well-known Witten anomaly $c_2(\SU(2))\tilde\eta$ for $q=1,3$.
This familiar Witten SU(2) anomaly disappears from the global anomaly and can be cancelled by a local anomaly for $q=2,6$ with an internal U(2) symmetry group, see \cite{Davighi2020bvi2001.07731} and more discussions below.

\item Other $\Z^4$ classes from Chern-Simons 5 form like terms from the gauge bundles $E$ of 
gauge group $\frac{\SU(3)\times\SU(2)\times\U(1)}{\Z_q}$.
\item  A $\Z$ class from
the more exotic $\mu(\text{PD}(c_1(\U(p))))$
constructed from the 3d Rokhlin invariant $\mu$ of the Poincar\'e dual of $c_1(\U(p))$,
where $p=1$ for $q=1$,
while $p=2$ for $q=2$ or $6$,
and $p=3$ for $q=3$ or $6$.
Note that for $q=6$, 
we identify $c_1(\U(2))=c_1(\U(3))$ by \Eq{eq:U3U2detmap}.

\end{itemize}

\item
In Sec 3,  we find that their 5d cobordism invariants contain the following:
\begin{itemize}

\item The $\Z^5$ classes are similar to  Sec 2.
Again one of a $\Z$ class is from
the more exotic $\mu(\text{PD}(c_1(\U(3))))$. The other 
four $\Z^4$ class are from Chern-Simons 5 form like terms from the gauge bundles $E$.

\item
For $q=2,6$,
there is a new $\Z_2$ class anomaly  as $a c_2(\U(2))$.\\
For $q=1,3$,
there is a new $\Z_4$ class $c_2(\SU(2))\eta'$
via a short exact sequence $0 \to \Z_2 \to \Z_4 \to \Z_2 \to 0$.
The original Witten anomaly $c_2(\SU(2))\tilde\eta$ 
sits at the $\Z_2$ normal subgroup of $\Z_4$.
The $ c_2(\SU(2)) a= a c_2(\SU(2))$ with $a \in \H^1(\B\Z_2,\Z_2)$ 
sits at the $\Z_2$ quotient.
Thus the $\Z_4$ class generator $c_2(\SU(2))\eta'$ sitting at 
the extended total group is beyond the Witten anomaly.
{The same structure occurs for 
the classes of the levels,
$$0 \to k \in \Z_2 \to  k_{\text{total}} \in \Z_4 \to 
k_{Q} \in\Z_2 \to 0,$$
where $k,k_{\text{total}}$, and $k_{Q}$
label 
the levels of those cobordism invariants
respectively.\footnote{Say in terms of the partition functions, they are
$\exp(\ii k_{} \frac{2 \pi}{2} c_2(\SU(2))\tilde\eta)$,
$\exp(\ii k_{\text{total}} \frac{2 \pi}{4} c_2(\SU(2))\eta')$.
and 
$\exp(\ii k_{Q} \frac{2 \pi}{2} c_2(\SU(2)) a)$.}}

\item A $\Z_{16}$ is $\eta(\text{PD}(a))$ from
$\Omega_5^{\Spin\times_{\Z_2}\Z_4}=\Z_{16}$.
On 5-manifolds with $\Spin\times_{\Z_2}\Z_4$ structure, 
there is a $\Pin^+$ structure on $\text{PD}(a)$ and there is an isomorphism $\Omega_5^{\Spin\times_{\Z_2}\Z_4}\xrightarrow{\cap a}\Omega_4^{\Pin^+}$ \cite{2018arXiv180502772T}.

\item A $\Z_{4}$ class is from $c_1(\U(1))^2\eta'$ (for $q=1$), or
$c_1(\U(2))^2\eta'$ (for $q=2,6$), or $c_1(\U(3))^2\eta'$ (for $q=3,6$).
The $\eta'$ is a $\Z_4$ valued 1d eta invariant. 
Note that for $q=6$, we have $c_1(\U(2))^2\eta'=c_1(\U(3))^2\eta'$
because we identify $c_1(\U(2))=c_1(\U(3))$ by \Eq{eq:U3U2detmap}.

\item A $\Z_{2}$ class is from $ac_2(\SU(3))$ (for $q=1,2$) or
$ac_2(\U(3))$ (for $q=3,6$). 

The last three cobordism invariants in 5d
of $\Z_{16}$, $\Z_4$, and $\Z_2$, are also beyond the Witten anomaly.

\end{itemize}

\item
In Sec 4,
for $\Spin\times_{\Z_2}(\SU(4)\times_{\Z_q}(\SU(2)\times\SU(2)))$, 
\begin{itemize} 
\item With $q=2$, 
we have a $\Z$ class of a Chern-Simons 5-form in 5d from Euler class $e_6(\SO(6))$ in 6d. We also have other two $\Z_2$ classes: $w_2(\SO(6))w_3(\SO(6))$ and $w_2(\SO(4))w_3(\SO(4))$ in 5d. Since $w_2=w_2(\SO(6))+w_2(\SO(4))$, $w_3=w_3(\SO(6))+w_3(\SO(4))$ and $w_2(\SO(6))w_3(\SO(4))+w_2(\SO(4))w_3(\SO(6))=\Sq^1(w_2(\SO(6))w_2(\SO(4)))=w_1w_2(\SO(6))w_2(\SO(4))=0$ by Wu formula, we have
$w_2w_3=w_2(\SO(6))w_3(\SO(6))+w_2(\SO(4))w_3(\SO(4))$. 
\item With $q=1$, we also have a $\Z$ class of a Chern-Simons 5-form in 5d from Euler class $e_6(\SO(6))$ in 6d, 
and other three $\Z_2$ classes: $w_2w_3=w_2(\SO(6))w_3(\SO(6))=w_2(\SO(4))w_3(\SO(4))$, $w_4(\SO(4))\tilde\eta$, and $(f^*)^{-1}(c_2(\SU(2))\tilde\eta)$ in 5d. Here $f:\Omega_d^{\Spin\times\SU(2)\times\SU(2)\times\SU(4)}\to\Omega_d^{\Spin\times_{\Z_2}(\SU(2)\times\SU(2)\times\SU(4))}$ is the natural group homomorphism, $f^*$ is the induced map between bordism invariants. 
\end{itemize}

\item
In Sec 5, for $n\ge7$,
for $\Spin\times\Spin(n)$ and $\Spin\times\SO(n)$, 
we have no cobordism invariants in 5d.
For $\Spin\times_{\Z_2}\Spin(n)$, we have 
$w_2w_3=w_2(TM) w_3(TM) =w_2(V_{\SO(n)}) w_3(V_{\SO(n)})$ in 5d.
\item
In Sec 6,
for $\Spin\times\SU(5)$, we have Chern-Simons 5-form in 5d from the Chern class $c_3(\SU(5))$ in 6d.
\end{enumerate}

The difference between $q=1$ and $q=2$ cases 
in \Sec{sec:SM} is that 
the Witten anomaly $c_2(\SU(2))\tilde\eta$ of $\Z_2$ for $q=1$ vanishes in $q=2$, while in contrast the {$\text{CS}_1^{\U(1)}c_2(\SU(2))
\sim
c_1(\U(1))\text{CS}_3^{\SU(2)}
$} of $\Z$ for $q=1$ becomes {
$\frac{1}{2}\text{CS}_1^{\U(2)}c_2(\U(2))
\sim
\frac{1}{2}c_1(\U(2))\text{CS}_3^{\U(2)}
$} 
in $q=2$. The $\text{CS}_1^{\U(1)}c_2(\SU(2))$ in 5d comes from $c_1(\U(1))c_2(\U(2))$ in 6d.
Due to the nontrivial differential $d_1$ in Figure \ref{fig:detail} for $q=2$, the $c_1(\U(2))c_2(\U(2))\mod2$ in 6d and the Witten anomaly from the 5d term $c_2(\U(2))\tilde\eta$ cancel with each other, while the $\frac{1}{2}c_1(\U(2))c_2(\U(2))$ remains for $q=2$. Here the
$c_1(\U(2))c_2(\U(2))\mod2$ is the quotient $\Z_2$, the $c_1(\U(2))c_2(\U(2))$ is the total $\Z$, and the $\frac{1}{2}c_1(\U(2))c_2(\U(2))$ is the normal $\Z$ in the short exact sequence $0 \to \Z \overset{2}{\to} \Z \to \Z_2 \to 0$. 
Let us 
express the levels of those cobordism invariants as
$k_{q=2}$, $k_{q=1}$, and ${k}_{q=1}'$
respectively.\footnote{In terms of their partition functions, they are
$\exp(\ii (k_{q=2}) (2 \pi) \Big(\frac{1}{2}\text{CS}_1^{\U(2)}c_2(\U(2))\Big))$ of 
$k_{q=2} \in \Z$ 
(meanwhile 
$\text{CS}_1^{\U(2)}c_2(\U(2)) \in 2\Z$,
thus
$\frac{1}{2}\text{CS}_1^{\U(2)}c_2(\U(2)) \in \Z$)
for $q=2$,
$\exp(\ii (k_{q=1}) (2 \pi) \Big(\text{CS}_1^{\U(1)}c_2(\SU(2))\Big)$ of 
$k_{q=1} \in \Z$
and 
$\exp(\ii ({k}_{q=1}') \frac{2 \pi}{2}
\Big(c_2(\SU(2))\tilde\eta\Big)$ of ${k}_{q=1}' 
\in \Z_2$ for $q=1$.
} In fact, we can also understand
the short exact sequence as the classes of the levels:
$0 \to 
k_{q=1} \in \Z \overset{2}{\to} 
k_{q=2} \in  \Z 
\to k_{q=1}'  \in  \Z_2 \to 0$.
There is yet another way to explain why $c_1(\U(2))c_2(\U(2))\mod2$ vanishes for $q=2$:
Namely by Wu formula, the $c_1(\U(2))c_2(\U(2))=\Sq^2c_2(\U(2))=(w_2+w_1^2)c_2(\U(2))=0\mod2$ on the Spin 6-manifolds.

The difference between $q=3$ and $q=6$ 
cases 
in \Sec{sec:SM}
is that the Witten anomaly
$c_2(\SU(2))\tilde\eta$ of $\Z_2$ for $q=3$ vanishes in $q=6$, while in contrast the
{$\text{CS}_1^{\U(3)}c_2(\SU(2))
\sim
c_1(\U(3))\text{CS}_3^{\SU(2)}
$}
of $\Z$ for $q=3$ becomes
the
{$\frac{1}{2}\text{CS}_1^{\U(3)}c_2(\U(2))
\sim
\frac{1}{2}
c_1(\U(3))\text{CS}_3^{\U(2)}
$} 
in $q=6$. 
There is yet another way to explain why $c_1(\U(3))c_2(\U(2))\mod2$ vanishes for $q=6$:
Namely by
Wu formula, $c_1(\U(3))c_2(\U(2))=c_1(\U(2))c_2(\U(2))=\Sq^2c_2(\U(2))=(w_2+w_1^2)c_2(\U(2))=0\mod2$ on Spin 6-manifolds.
In short, the situations of $q=3$ and $q=6$
are exactly 
parallel to the situations of $q=1$ and $q=2$
discussed above.

The difference between $q=1$ and $q=2$ cases 
in \Sec{sec:SM4} is that
the global Witten SU(2) anomaly of $\Z_2$ class in $q=1$
becomes a local anomaly
${\frac{a^2\text{CS}_3^{\U(2)}+\text{CS}_1^{\U(2)}c_2(\U(2)) }{2}\sim
\frac{a^2\text{CS}_3^{\U(2)}+c_1(\U(2))\text{CS}_3^{\U(2)}}{2}}$ of $\Z$ class
in $q=2$. See Appendix \ref{sec:comment-2} for more details. As shown in Figure \ref{fig:difference-2},
the left part which starts at $c_2(\SU(2))$ for $q=1$ becomes the right two part which start from $c_2(\U(2))$ and $c_2(\U(2))a$ respectively for $q=2$. 
For $q=1$, the left part which starts from $c_2(\SU(2))$ contributes a $\Z_4$ class $c_2(\SU(2))\eta'$ in 5d.
For $q=2$, the right part which starts from $c_2(\U(2))a$ contributes a $\Z_2$ class $c_2(\U(2))a$ in 5d, while the right part which starts from $c_2(\U(2))$ contributes a $\Z$ class $\frac{a^2c_2(\U(2))+c_1(\U(2))c_2(\U(2))}{2}$ in 6d. There is yet another way to explain why $a^2c_2(\U(2))+c_1(\U(2))c_2(\U(2))\mod2$ vanishes for $q=2$. Namely by Wu formula, the $c_1(\U(2))c_2(\U(2))=\Sq^2(c_2(\U(2)))=(w_2+w_1^2)c_2(\U(2))=a^2c_2(\U(2))\mod2$ on Spin 6-manifolds.

The difference between $q=3$ and $q=6$ 
cases 
in \Sec{sec:SM4} is that
the global Witten SU(2) anomaly of $\Z_2$ class
in $q=3$
becomes a local anomaly
${
\frac{a^2\text{CS}_3^{\U(2)}+\text{CS}_1^{\U(3)}c_2(\U(2))}{2} 
\sim 
\frac{a^2\text{CS}_3^{\U(2)}+c_1(\U(3))\text{CS}_3^{\U(2)}}{2}}$ of $\Z$ class
in $q=6$. In short, the situations of $q=3$ and $q=6$
are exactly parallel to the situations of $q=1$ and $q=2$
discussed above.

What we have found here is consistent with the fact that the Witten anomaly as a \emph{global non-perturbative anomaly} becomes a \emph{local
perturbative anomaly}, whenever we change from $q=1$ to $q=2$, or change from $q=3$ to $q=6$, which trades the SU(2) internal symmetry for the U(2) internal symmetry \cite{Davighi2020bvi2001.07731}.


\section{Conclusions, and Explorations on Non-Perturbative and Topological Sectors of BSM}
\label{sec:ConclusionsExplorationsonNon-PerturbativeBSM}

We have explored the cobordism theory relevant for SM and GUTs in Section \ref{sec:SM}-\ref{sec:GUSU}. We also have interpreted the classifications of
various 4d perturbative local and
non-perturbative global invertible
anomalies for SM and GUTs in \Sec{sec:interpretations}.
Below we summarize the results of co/bordism groups
and conclude with further implications. 

\subsection{Summary}

\begin{itemize}

\item
In \Sec{sec:SM}, for $q =1,2,3,6$, we compute the bordism groups $\Omega_d^{\Spin\times\frac{\SU(3)\times\SU(2)\times\U(1)}{\Z_q}}$
and the bordism invariants for $0\le d\le 6$. We also determine the group 
 $\TP_d(\Spin\times\frac{\SU(3)\times\SU(2)\times\U(1)}{\Z_q})$ and the topological terms for $0\le d\le 5$. 
 
 We find that there is only 2-torsion in these bordism groups, and the bordism groups
 $\Omega_d^{\Spin\times\SU(3)\times\SU(2)\times\U(1)}$ and $\Omega_d^{\Spin\times\frac{\SU(3)\times\SU(2)\times\U(1)}{\Z_3}}$ are isomorphic,
 while
 $\Omega_d^{\Spin\times\frac{\SU(3)\times\SU(2)\times\U(1)}{\Z_2}}$ and 
$\Omega_d^{\Spin\times\frac{\SU(3)\times\SU(2)\times\U(1)}{\Z_6}}$ are isomorphic.

We use the 3d Rokhlin invariant and Chern-Simons forms to express the topological terms.

\item
In \Sec{sec:SM4}, for  $q =1,2,3,6$, we compute the bordism groups $\Omega_d^{\Spin\times_{\Z_2}\Z_4\times\frac{\SU(3)\times\SU(2)\times\U(1)}{\Z_q}}$
and the bordism invariants for $0\le d\le 6$. We also determine the group 
 $\TP_d(\Spin\times_{\Z_2}\Z_4\times\frac{\SU(3)\times\SU(2)\times\U(1)}{\Z_q})$ and the topological terms for $0\le d\le 5$.

 We find that there is only 2-torsion in these bordism groups, and the bordism groups
 $\Omega_d^{\Spin\times_{\Z_2}\Z_4\times\SU(3)\times\SU(2)\times\U(1)}$ and $\Omega_d^{\Spin\times_{\Z_2}\Z_4\times\frac{\SU(3)\times\SU(2)\times\U(1)}{\Z_3}}$ are isomorphic,
 while
 $\Omega_d^{\Spin\times_{\Z_2}\Z_4\times\frac{\SU(3)\times\SU(2)\times\U(1)}{\Z_2}}$ and 
$\Omega_d^{\Spin\times_{\Z_2}\Z_4\times\frac{\SU(3)\times\SU(2)\times\U(1)}{\Z_6}}$ are isomorphic.

We also use the 3d Rokhlin invariant and Chern-Simons forms to express the topological terms. Compared to \Sec{sec:SM}, there are new bordism invariants. For example, in 1d, $\tilde\eta$ becomes $\eta'$ which is $\Z_4$ valued. 
{In 4d, the $\frac{\sigma}{16}$ becomes $\frac{\sigma-\rF \cdot \rF }{8}$ where $\rF $ is the characteristic 2-surface \cite{Saveliev} in a 4-manifold $M^4$, it satisfies the condition $\rF\cdot x=x\cdot x\mod2$ for all $x\in\H_2(M^4,\Z)$.
Here $\cdot$ is the intersection form of $M^4$.
By the Freedman-Kirby theorem, $(\frac{\sigma-\rF\cdot\rF}{8} )(M^4)=\text{Arf}(M^4,\rF)\mod2$. When $M^4$ is Spin, $\rF$ can be chosen to be 0 since the intersection form of $M^4$ is even ($x\cdot x=0\mod2$). So in this case, $\frac{\sigma}{8}(M^4)=0\mod2$, thus the bordism invariant is $\frac{\sigma}{16}$.} 
In 5d, there is a new $\eta(\text{PD}(a))$ since there is an induced Pin$^+$ structure on $\text{PD}(a)$ where $\eta$ is the 4d eta invariant, $\text{PD}$ is the Poincar\'e dual. 

\item
In \Sec{sec:PS}, we compute the bordism groups $\Omega_d^{\frac{\Spin\times \frac{\SU(4)\times(\SU(2)\times \SU(2))}{\Z_2}}{\Z_2^F}}$ and $\Omega_d^{\frac{\Spin\times \SU(4)\times(\SU(2)\times \SU(2))}{\Z_2^F}}$ for $0\le d\le 6$. We also determine the group $\TP_d(\frac{\Spin\times \frac{\SU(4)\times(\SU(2)\times \SU(2))}{\Z_2}}{\Z_2^F})$ and $\TP_d(\frac{\Spin\times \SU(4)\times(\SU(2)\times \SU(2))}{\Z_2^F})$ for $0\le d\le 5$.

{We find that there are three bordism invariants of $\Omega_d^{\frac{\Spin\times \SU(4)\times(\SU(2)\times \SU(2))}{\Z_2^F}}$ in dimensions 4, 5 and 6 which are hard to describe directly, while we express them in terms of the bordism invariants of $\Omega_d^{\Spin\times\SU(2)\times\SU(2)\times\SU(4)}$ and the group homomorphism $f:\Omega_d^{\Spin\times\SU(2)\times\SU(2)\times\SU(4)}\to \Omega_d^{\Spin\times_{\Z_2}(\SU(2)\times\SU(2)\times\SU(4))}$. See Appendix \ref{sec:explanation} for more details.}

We also find that Euler class appears in the bordism invariants and the Chern-Simons form for Euler class appears in the topological terms.

\item
In \Sec{sec:GUSO}, we compute the bordism groups $\Omega_d^{\Spin\times\Spin(n)}$, $\Omega_d^{\Spin\times\SO(n)}$ and $\Omega_d^{\frac{\Spin\times\Spin(n)}{\Z_2^F}}$ for $n\ge7$ and $0\le d\le 6$. 
We also determine the group $\TP_d(\Spin\times\Spin(n))$, $\TP_d(\Spin\times\SO(n))$ and $\TP_d(\frac{\Spin\times\Spin(n)}{\Z_2^F})$ for $n\ge7$ and $0\le d\le 5$.

We find that for $n\ge7$, $\Omega_d^{\Spin\times\Spin(n)}=\Omega_d^{\Spin\times\Spin(n+1)}$, $\Omega_d^{\Spin\times\SO(n)}=\Omega_d^{\Spin\times\SO(n+1)}$ and $\Omega_d^{\frac{\Spin\times\Spin(n)}{\Z_2^F}}=\Omega_d^{\frac{\Spin\times\Spin(n+1)}{\Z_2^F}}$ for $0\le d\le 6$. 

We also find that Euler class appears in the bordism invariants of
$\Omega_d^{\Spin\times\Spin(n)}$ and
 $\Omega_d^{\frac{\Spin\times\Spin(n)}{\Z_2^F}}$ and the Chern-Simons form for Euler class appears in the topological terms of $\TP_d(\Spin\times\Spin(n))$ and $\TP_d(\frac{\Spin\times\Spin(n)}{\Z_2^F})$.

\item
In \Sec{sec:GUSU}, we compute the bordism groups $\Omega_d^{\Spin\times\SU(5)}$ for $0\le d\le 6$. We also determine the group $\TP_d(\Spin\times \SU(5))$ for $0\le d\le 5$.

We find that for $n\ge3$, $\Omega_d^{\Spin\times\SU(n)}=\Omega_d^{\Spin\times\SU(n+1)}$ for $0\le d\le 6$.

\end{itemize}

\subsection{Constraints on Quantum Dynamics}
\label{sec:ConstraintsQuantumDynamics}


We have discussed and summarized potential anomalies and topological terms in SM, GUT and BSM. There are actually two versions of anomalies we are speaking of:
One is the Anomaly \ref{a2}
for the ungauged SM, GUT and BSM with the $G$ is simply a global symmetry.
Another is the Anomaly \ref{a3}
for the gauged SM, GUT and BSM with the ${\mathbb{G}_{\text{internal}} }$ dynamically gauged inside $G$.
In more details,
\begin{enumerate}[leftmargin=2mm, label=\textcolor{blue}{\Roman*}., ref={\Roman*}]
\item  \label{ungaugedSMGUTandBSM}
For ungauged SM, GUT and BSM, we can use the 't Hooft anomaly (Anomaly \ref{a2}) to the gappability of these models' matter field sectors.
For ungauged SM, GUT and BSM, we can simply have matter field contents (e.g. fermions: quarks and leptons) without dynamical gauge fields.
In fact,  \Ref{Wen2013oza1303.1803, Wen2013ppa1305.1045, WangWen2018cai1809.11171} use the all anomaly free conditions
to support that a non-perturbative definition (lattice regularization) of SO(10) GUT is doable, by checking.
By anomaly free, there exists non-perturbative interactions for gapping the mirror world chiral fermions \cite{JWangmirror}. 
This mirror-fermion gapping can help to get rid of the mirror world chiral fermion doublers,
surpassing the Nielsen-Ninomiya fermion doubling theorem (which is only true for the free non-interacting systems).
The fact that all anomaly free gapless theories can be deformed to a fully gapped trivial vacuum is also consistent with 
the concept of Seiberg's deformation class \cite{NSeiberg-Strings-2019-talk}. More details can be found in \cite{JWangmirror}.

\item \label{gaugedSMGUTandBSM}
For gauged SM, GUT and BSM, we can use the dynamical gauge anomaly matching conditions (Anomaly \ref{a3}) to rule out inconsistent theories.
Importantly, depending on the matter contents and their representations in ${\mathbb{G}_{\text{internal}} }$,
we may gain or loss some global symmetries.  For example, for fermions in the adjoint representation, we can have
a 1-form center symmetry for the gauge theory. Thus, we should beware 
potentially additional new higher 't Hooft anomalies (see \cite{WanWang2018bns1812.11967, WanWangv2})
for gauged SM, GUT and BSM 
can help us to constrain quantum dynamics (also more discussions below).

\end{enumerate} 
We use the path integral and the action to understand the basic \emph{kinematics} and the \emph{global symmetry} of the QFTs.
We can apply the spacetime geometric topology properties to constrain QFTs, such as doing the spacetime surgery for QFTs \cite{Witten1988hfJonesQFT, JWangthesis1602.05569, 1602.05951WWY, Wang2019diz1901.11537}. We can also determine the \emph{anomalies} of QFTs at UV.
However, given the potentially complete  anomalies,
we can constrain the IR dynamics by UV-IR anomaly matching.
The consequence of anomaly matching implies that the IR theories with 't Hooft anomalies in $G$-symmetry must be matched by at least one of the following \emph{dynamics} 
scenarios: 

\noindent
\begin{enumerate} 
\item Symmetry-breaking:\\ 
$\bullet$ (say, discrete or continuous $G$-symmetry breaking).\\
\item Symmetry-preserving: \\
$\bullet$ Degenerate ground states (like the ``Lieb-Schultz-Mattis theorem \cite{Lieb1961frLiebSchultzAOP, Hastings2003zx}''),\\
$\bullet$ Gapless, e.g., conformal field theory (CFT),\\
$\bullet$ 
Symmetry-preserving TQFT: Intrinsic topological orders. 
\item Symmetry-extension \cite{Wang2017locWWW1705.06728}:
Symmetry-extension is a rather exotic possibility, which does not occur naturally without fine-tuning or artificial designed, explained in \cite{Wang2017locWWW1705.06728}.
However, the symmetry-extension approach is a useful intermediate stepstone, 
to construct another earlier scenario: \emph{symmetry-preserving TQFT}, via gauging the extended-symmetry \cite{Wang2017locWWW1705.06728}.
\end{enumerate}

In more details, suppose there are mixed anomalies between the ordinary 0-form global symmetries and higher-form symmetries (say 1-form symmetries)
in the certain gauged SM, GUT, and BSM as in Model \ref{gaugedSMGUTandBSM}, 
we have further refinement of possibilities: 
\begin{enumerate}[label= (\roman*)]
\item 
Ordinary 0-form symmetry broken (spontaneously or explicitly).
\item 
1-form center discrete $\Z_{N,[1]}^e$ symmetry broken (spontaneously or explicitly) as \emph{deconfinement}:
\begin{enumerate}[label=(2-\roman*)]
    \item 1-form $\Z_{N,[1]}^e$-symmetry breaking and deconfined TQFTs, i.e., \emph{topological order} in condensed matter terminology. 
    \item 1-form $\Z_{N,[1]}^e$-symmetry breaking and deconfined gapless theories (e.g. CFTs).
\end{enumerate}

\item 
1-form center discrete $\Z_{N,[1]}^e$ symmetry unbroken as \emph{confinement}:
\begin{enumerate}[label=(3-\roman*)]
    \item 1-form symmetry-extended invertible TQFT: The exotic scenario of symmetry-extended invertible TQFT is systematically studied in \Ref{Wang2017locWWW1705.06728}. 
    This idea is generalized to higher symmetries for higher-symmetry-extended invertible TQFT in \Ref{Wan2018djlW2.1812.11955}.
    \item 1-form symmetry-preserving TQFT: In fact, under certain criteria, there is a no go theorem to match some specific anomalies by this scenario.
    Hinted by the obstruction of constructing the symmetry-extended TQFTs  \cite{Wan2018djlW2.1812.11955, Wan2019oyr1904.00994},
    Cordova-Ohmori \cite{Cordova2019bsd1910.04962} proved a 
    no go theorem for TQFTs preserving both the 1-form symmetry and time-reversal symmetry while
    saturating the 4d SU(N) YM anomaly with ${\theta=\pi}$. 
\end{enumerate}
\item 
Full symmetry-preserving gapless theory (e.g. CFTs). 
\end{enumerate}

In summary,  based on the kinematics, global symmetries (ordinary or higher symmetries),
anomalies and spacetime topology constraints, we may suggest new quantum dynamical constraints.

\subsection{Anomaly Cancellation or 
Anomaly-Matched Hidden Sectors for Beyond Standard Models 
}

The previous section also has phenomenology implications of the gauged SM, GUT, and BSM as in Model \ref{gaugedSMGUTandBSM}. 
\begin{itemize}
\item Suppose we prove that the theory is fully 't Hooft anomaly free (for global symmetries, not the dynamical gauge anomalies), then
there is an application for gapping the mirror world chiral fermions (see \Ref{Wen2013oza1303.1803, Wen2013ppa1305.1045, WangWen2018cai1809.11171, JWangmirror}
and References therein).
\item Suppose we prove that the theory has a certain 't Hooft anomaly (for global symmetries, not the dynamical gauge anomalies), then
we discover either of the following (see more details in):
\end{itemize}
\noindent
\begin{enumerate}[leftmargin=2mm, label=\textcolor{blue}{\arabic*}., ref={\arabic*}] 
\item  Anomaly-Matched Topological Field Theories: Anomaly-Matched Hidden Topological Sectors. The gapped topological sector
has its low energy physics described by a certain TQFT.
But the gapped topological sector may have a finite energy gap $\Delta_E$ (instead of infinite large gap $\Delta_E \to \infty$) which has experimental measurable consequences.
\item Anomaly-Matched Hidden Gapless Sectors: This may be more surprising for phenomenology grounds with additional massless or gapless beyond SM around the energy scale
$\Delta_E$ with anyonic excitations of particles, strings or branes (e.g. see \cite{1602.05951WWY,Putrov2016qdo1612.09298PWY,Wang2019diz1901.11537} and References therein).
\item Anomaly-Matched 
Symmetry-Breaking Sectors: Global symmetries can be spontaneously broken.
\item All Anomaly-Free conditions are matched: This suggests the relation between the gapping criteria and defining fermion/gauge theories. In particular, 
the challenge of defining chiral fermion/gauge theories can be overcome by gapping the mirror world chiral fermions
\cite{Wen2013oza1303.1803, Wen2013ppa1305.1045, WangWen2018cai1809.11171, JWangmirror}.
\item Swapland Implications and Defects:
In the recent work of McNamara-Vafa \cite{McNamara2019rupVafa1909.10355}, by 
combing the ideas of Quantum Gravity/String Landscape and Swampland (e.g. \Ref{Vafa2005uiswampland0509212, ArkaniHamed2006dzNima0601001}, see a review \cite{Palti2019pcaSwampland1903.06239})
together 
with no internal global symmetry for quantum gravity \cite{Misner1957mtWheeler, Polchinski2003bq0304042, Banks2010zn1011.5120, Harlow2018tngOoguri1810.05338},
\Ref{McNamara2019rupVafa1909.10355} argues that
all the cobordism classes in the full Quantum Gravity (QG) must be vanished as in the trivial class.
Effectively, the effective cobordism classes of QG named $\Omega^{\widetilde{\text{QG}}}_k$ must be vanished.
This gives several powerful constraints for the cobordism class data in any dimension.
If we find any anomalies or topological terms in lower dimensions, say $k=1,2,3,4,5$ in our case,
this suggests that in the full QG, their cobordism classes must be cancelled by some other new objects/excitations/extended operators.
Two particular interesting possibilities are \cite{McNamara2019rupVafa1909.10355}:  
\begin{enumerate}[leftmargin=6mm, label=\textcolor{blue}{(\arabic*)}., ref={(\arabic*)}] 
\item Symmetry is broken: Mathematically, this suggests a map
\bea
\Omega^{\widetilde{\text{QG}}}_k \to \Omega^{\widetilde{\text{QG}}+\text{defects}}_k,
\eea
which means the cancellation of cobordism classes via new additional symmetry defects.
Namely, the kernel in $\Omega^{\widetilde{\text{QG}}}_k$ of this map is mapped to a trivial class in $\Omega^{\widetilde{\text{QG}}+\text{defects}}_k$.
\item Symmetry is gauged: Mathematically, this suggests a map
\bea
 \Omega^{\widetilde{\text{QG}}+\text{gauge sectors}}_k
\to
\Omega^{\widetilde{\text{QG}}}_k,
\eea
which means the cancellation of cobordism classes via  additional new gauge sectors.
Namely, the cokernel in $\Omega^{\widetilde{\text{QG}}}_k$ 
of this map is mapped reversely to a trivial class in $ \Omega^{\widetilde{\text{QG}}+\text{gauge sectors}}_k$. 
 
\end{enumerate} 
\end{enumerate} 
We leave the further detailed explorations of the above predictions in upcoming works \cite{WanWangv2, MMVWangYau, toappear}.

\section{Acknowledgements} 

The authors are listed in the alphabetical order by the standard convention.
During the completion of this manuscript, we
become aware that a recent work has obtained related results \cite{2019arXiv191011277D} on global anomalies of SMs and GUTs;
also 
\Ref{Ang2019txyAngRoumpedakis} studies various topological terms of 4d gauge theories analogous to \cite{AharonyASY2013hdaSeiberg1305.0318} but 
on generic non-spin manifolds.\footnote{The  topological terms of 4d gauge theories in \Ref{Ang2019txyAngRoumpedakis} are related to the bordism invariants that we enumerate in our present work, and also those bordism invariants enumerated in \Ref{Wan2019oyr1904.00994}.}  
JW is grateful to Miguel Montero \cite{GarciaEtxebarriaMontero2018ajm1808.00009} for informing his unpublished note \cite{Montero}, 
and thanks colleagues for warmly encouragements \cite{MMVWangYau}.\footnote{Although the methods of ours \cite{WanWang2018bns1812.11967, WanWangv2} and 
theirs \cite{GarciaEtxebarriaMontero2018ajm1808.00009, 2019arXiv191011277D} are rather different:
ours is via Adams spectral sequence, while
theirs is via Atiyah-Hirzebruch spectral sequence. In addition, we can conveniently read the iTQFT (namely, co/bordism invariants) from the Adams chart  \cite{WanWang2018bns1812.11967, WanWangv2} ,
while iTQFT cannot be extracted easily from \Ref{GarciaEtxebarriaMontero2018ajm1808.00009, 2019arXiv191011277D}.}   
Major tools of this work were built in Fall 2018 in a prior work \Ref{WanWang2018bns1812.11967},
and a major part of calculations of this work were done in Fall 2018 - Summer 2019, 
when ZW was at USTC and when JW was at  Institute for Advanced Study.
Part of this work is also presented by JW in the workshop Lattice for Beyond the Standard Model physics 2019, on May 2-3, 2019 at Syracuse University \cite{JW-BSM-2019-talk}.
JW appreciates the support and feedback from the workshop.
JW thanks the authors of \Ref{2019arXiv191011277D} for a helpful Email correspondence on our \Sec{sec:Comparison}.
%
ZW  acknowledges previous supports from NSFC grants 11431010 and 11571329. 
ZW is supported by the Shuimu Tsinghua Scholar Program.
JW was supported by
NSF Grant PHY-1606531. 
This work is also supported by 
NSF Grant DMS-1607871 ``Analysis, Geometry and Mathematical Physics'' 
and Center for Mathematical Sciences and Applications at Harvard University.

\appendix
\section{The correspondence between $\A_2(1)$-module structure and the $E_2$ page}

In this Appendix, we list the correspondence between $\A_2(1)$-module $L$ and its $E_2$ page used in our computation before.

     \begin{figure}[H]
\begin{multicols}{2}
 
\begin{center}
\begin{tikzpicture}[scale=0.5]

\draw[fill] (0,0) circle(.1);

\end{tikzpicture}
\end{center}

\begin{center}
\begin{tikzpicture}[scale=0.5]
\node at (0,-1) {0};
\node at (1,-1) {1};
\node at (2,-1) {2};
\node at (3,-1) {3};
\node at (4,-1) {4};
\node at (5,-1) {5};
\node at (6,-1) {6};
\node at (7.5,-1) {$t-s$};
\node at (-1,0) {0};
\node at (-1,1) {1};
\node at (-1,2) {2};
\node at (-1,3) {3};
\node at (-1,4) {4};
\node at (-1,5) {5};
\node at (-1,6) {$s$};

\draw[->] (-0.5,-0.5) -- (-0.5,6);
\draw[->] (-0.5,-0.5) -- (7,-0.5);

\draw (0,0) -- (0,5);
\draw (0,0) -- (2,2);

\draw (4,3) -- (4,5);

\end{tikzpicture}
\end{center}
\end{multicols}

\caption{The $\A_2(1)$-module $L=\Z_2$ and its $E_2$ page.}
\label{fig:Z_2}
 \end{figure}
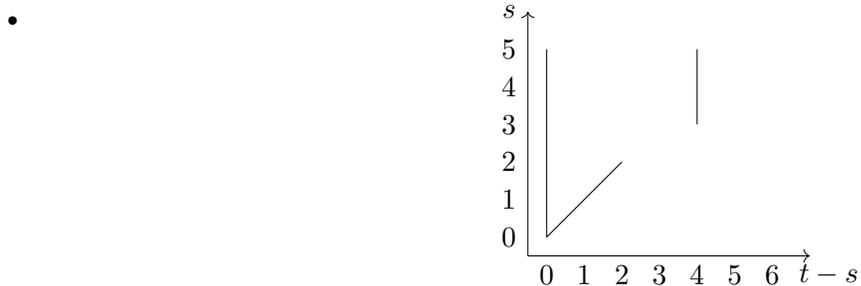

       \begin{figure}[H]
\begin{multicols}{2}
 
\begin{center}
\begin{tikzpicture}[scale=0.5]

\draw[fill] (2,4) circle(.1);

\draw[fill] (2,5) circle(.1);

\draw (2,4) -- (2,5);

\draw[fill] (2,6) circle(.1);

\draw[fill] (3,7) circle(.1);

\draw[fill] (3,8) circle(.1);
\draw (3,7) -- (3,8);

\draw (2,4) to [out=150,in=150] (2,6);

\draw (2,5) to [out=30,in=150] (3,7);

\draw (2,6) to [out=30,in=150] (3,8);

\draw[fill] (2,7) circle(.1);

\draw[fill] (3,9) circle(.1);

\draw[fill] (3,10) circle(.1);
\draw (3,9) -- (3,10);
\draw (2,6) -- (2,7);

\draw (2,7) to [out=30,in=150] (3,9);
\draw (3,8) to [out=30,in=30] (3,10);

\end{tikzpicture}
\end{center}

\begin{center}
\begin{tikzpicture}[scale=0.5]
\node at (0,-1) {0};
\node at (1,-1) {1};
\node at (2,-1) {2};
\node at (3,-1) {3};
\node at (4,-1) {4};
\node at (5,-1) {5};
\node at (6,-1) {6};
\node at (7.5,-1) {$t-s$};
\node at (-1,0) {0};
\node at (-1,1) {1};
\node at (-1,2) {2};
\node at (-1,3) {3};
\node at (-1,4) {4};
\node at (-1,5) {5};
\node at (-1,6) {$s$};

\draw[->] (-0.5,-0.5) -- (-0.5,6);
\draw[->] (-0.5,-0.5) -- (7,-0.5);

\draw[fill] (0,0) circle(.1);

\end{tikzpicture}
\end{center}
\end{multicols}

\caption{The $\A_2(1)$-module $L=\A_2(1)$ and its $E_2$ page.}
\label{fig:A_2(1)}
 \end{figure}
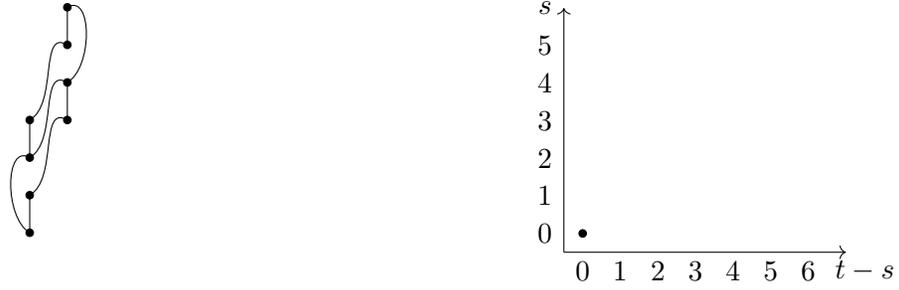

    \begin{figure}[H]
\begin{multicols}{2}
 
\begin{center}
\begin{tikzpicture}[scale=0.5]

 \draw[fill] (0,-12) circle(.1);
\draw[fill] (0,-10) circle(.1);
\draw (0,-12) to [out=150,in=150] (0,-10);

\end{tikzpicture}
\end{center}

\begin{center}
\begin{tikzpicture}[scale=0.5]
\node at (0,-1) {0};
\node at (1,-1) {1};
\node at (2,-1) {2};
\node at (3,-1) {3};
\node at (4,-1) {4};
\node at (5,-1) {5};
\node at (6,-1) {6};
\node at (7.5,-1) {$t-s$};
\node at (-1,0) {0};
\node at (-1,1) {1};
\node at (-1,2) {2};
\node at (-1,3) {3};
\node at (-1,4) {4};
\node at (-1,5) {5};
\node at (-1,6) {$s$};

\draw[->] (-0.5,-0.5) -- (-0.5,6);
\draw[->] (-0.5,-0.5) -- (7,-0.5);

\draw (0,0) -- (0,5);
\draw (2,1) -- (2,5);
\draw (4,2) -- (4,5);

\draw (6,3) -- (6,5);
\end{tikzpicture}
\end{center}
\end{multicols}

\caption{The $\A_2(1)$-module $L=\H^{*+2}(C\upeta,\Z_2)$ and its $E_2$ page.}
\label{fig:Ceta}
 \end{figure}
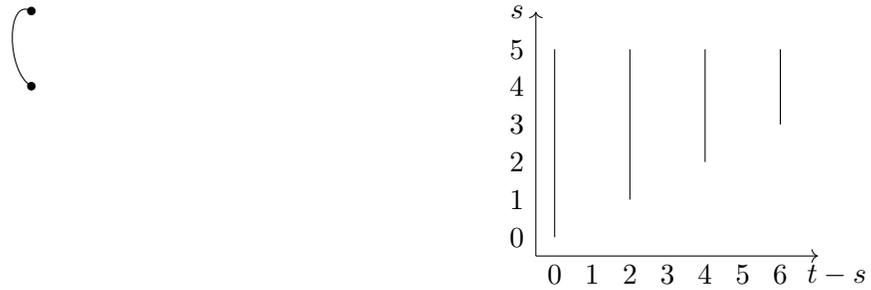

         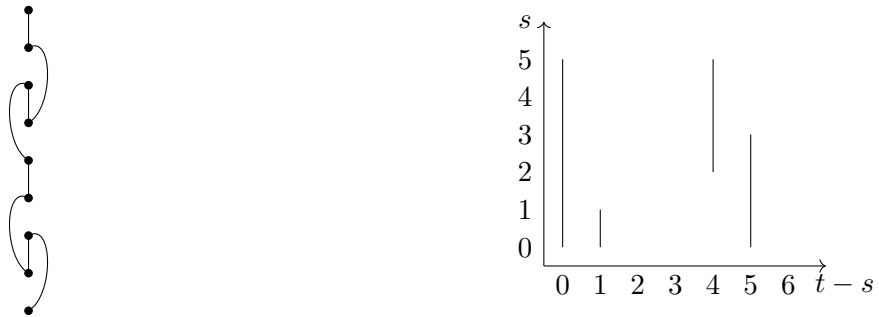
\begin{figure}[H]
\begin{multicols}{2}
 
\begin{center}
\begin{tikzpicture}[scale=0.5]

\draw[fill] (0,0) circle(.1);
\draw[fill] (0,1) circle(.1);
\draw[fill] (0,2) circle(.1);
\draw (0,0) to [out=30,in=30] (0,2);
\draw (0,1) -- (0,2);
\draw[fill] (0,3) circle(.1);
\draw (0,1) to [out=150,in=150] (0,3);
\draw[fill] (0,4) circle(.1);
\draw (0,3) -- (0,4);
\draw[fill] (0,5) circle(.1);
\draw[fill] (0,6) circle(.1);
\draw (0,4) to [out=150,in=150] (0,6);
\draw (0,5) -- (0,6);
\draw[fill] (0,7) circle(.1);
\draw (0,5) to [out=30,in=30] (0,7);
\draw[fill] (0,8) circle(.1);
\draw (0,7) -- (0,8);

\end{tikzpicture}
\end{center}

\begin{center}
\begin{tikzpicture}[scale=0.5]
\node at (0,-1) {0};
\node at (1,-1) {1};
\node at (2,-1) {2};
\node at (3,-1) {3};
\node at (4,-1) {4};
\node at (5,-1) {5};
\node at (6,-1) {6};
\node at (7.5,-1) {$t-s$};
\node at (-1,0) {0};
\node at (-1,1) {1};
\node at (-1,2) {2};
\node at (-1,3) {3};
\node at (-1,4) {4};
\node at (-1,5) {5};
\node at (-1,6) {$s$};

\draw[->] (-0.5,-0.5) -- (-0.5,6);
\draw[->] (-0.5,-0.5) -- (7,-0.5);

\draw (0,0) -- (0,5);

\draw (1,0) -- (1,1);

\draw (4,2) -- (4,5);

\draw (5,0) -- (5,3);

\end{tikzpicture}
\end{center}
\end{multicols}

\caption{The $\A_2(1)$-module $L=\H^{*+2}(\RP_2^{\infty},\Z_2)$ and its $E_2$ page.}
\label{fig:RP_2}
 \end{figure}

    \begin{figure}[H]
\begin{multicols}{2}
 
\begin{center}
\begin{tikzpicture}[scale=0.5]

\draw[fill] (2,-11) circle(.1);
\draw[fill] (2,-10) circle(.1);
\draw (2,-11) -- (2,-10);
\draw[fill] (2,-9) circle(.1);
\draw (2,-11) to [out=150,in=150] (2,-9);
\draw[fill] (3,-9) circle(.1);
\draw[fill] (3,-8) circle(.1);
\draw (3,-9) -- (3,-8);
\draw (2,-10) to [out=30,in=150] (3,-8);
\draw[fill] (2,-8) circle(.1);
\draw (2,-9) -- (2,-8);
\draw[fill] (2,-7) circle(.1);
\draw[fill] (2,-6) circle(.1);
\draw (2,-7) -- (2,-6);
\draw (2,-8) to [out=150,in=150] (2,-6);
\draw[fill] (3,-7) circle(.1);
\draw (3,-9) to [out=30,in=30] (3,-7);
\draw[fill] (3,-6) circle(.1);
\draw (3,-7) -- (3,-6);
\draw[fill] (2,-5) circle(.1);
\draw (2,-7) to [out=30,in=30] (2,-5);
\draw[fill] (3,-5) circle(.1);
\draw[fill] (3,-4) circle(.1);
\draw (3,-5) -- (3,-4);
\draw (3,-6) to [out=150,in=150] (3,-4);
\draw[fill] (2,-4) circle(.1);
\draw (2,-5) -- (2,-4);
\draw[fill] (3,-3) circle(.1);
\draw[fill] (3,-2) circle(.1);
\draw (3,-3) -- (3,-2);
\draw (3,-5) to [out=30,in=30] (3,-3);

\end{tikzpicture}
\end{center}

\begin{center}
\begin{tikzpicture}[scale=0.5]
\node at (0,-1) {0};
\node at (1,-1) {1};
\node at (2,-1) {2};
\node at (3,-1) {3};
\node at (4,-1) {4};
\node at (5,-1) {5};
\node at (6,-1) {6};
\node at (7.5,-1) {$t-s$};
\node at (-1,0) {0};
\node at (-1,1) {1};
\node at (-1,2) {2};
\node at (-1,3) {3};
\node at (-1,4) {4};
\node at (-1,5) {5};
\node at (-1,6) {$s$};

\draw[->] (-0.5,-0.5) -- (-0.5,6);
\draw[->] (-0.5,-0.5) -- (7,-0.5);

\draw[fill] (0,0) circle(.1);
\draw (2,0) -- (2,1);
\draw (4,0) -- (4,2);

\draw (6,0) -- (6,3);
\end{tikzpicture}
\end{center}
\end{multicols}

\caption{The $\A_2(1)$-module $L_1$ which appears in Figure \ref{fig:A_2(1)RP_2Ceta} and its $E_2$ page.}
\label{fig:L_1}
 \end{figure}

 \begin{figure}[H]
\begin{multicols}{2}
 
\begin{center}
\begin{tikzpicture}[scale=0.5]
\draw[fill] (0,-15) circle(.1);
\draw[fill] (0,-13) circle(.1);
\draw (0,-15) to [out=150,in=150] (0,-13);
\draw[fill] (0,-12) circle(.1);
\draw (0,-13) -- (0,-12);
\draw[fill] (0,-11) circle(.1);
\draw[fill] (0,-10) circle(.1);
\draw[fill] (0,-9) circle(.1);
\draw (0,-12) to [out=150,in=150] (0,-10);
\draw (0,-11) -- (0,-10);
\draw (0,-11) to [out=30,in=30] (0,-9);
\draw[fill] (0,-8) circle(.1);
\draw (0,-9) -- (0,-8);
\draw[fill] (0,-7) circle(.1);
\draw[fill] (0,-6) circle(.1);
\draw[fill] (1,-9) circle(.1);
\draw (0,-8) to [out=150,in=150] (0,-6);
\draw (0,-7) -- (0,-6);
\draw (1,-9) to [out=150,in=30] (0,-7);
\end{tikzpicture}
\end{center}

\begin{center}
\begin{tikzpicture}[scale=0.5]
\node at (0,-1) {0};
\node at (1,-1) {1};
\node at (2,-1) {2};
\node at (3,-1) {3};
\node at (4,-1) {4};
\node at (5,-1) {5};
\node at (6,-1) {6};
\node at (7.5,-1) {$t-s$};
\node at (-1,0) {0};
\node at (-1,1) {1};
\node at (-1,2) {2};
\node at (-1,3) {3};
\node at (-1,4) {4};
\node at (-1,5) {5};
\node at (-1,6) {$s$};

\draw[->] (-0.5,-0.5) -- (-0.5,6);
\draw[->] (-0.5,-0.5) -- (7,-0.5);

\draw (0,0) -- (0,5);
\draw (4,0) -- (4,5);
\draw (6,0) -- (6,5);
\end{tikzpicture}
\end{center}
\end{multicols}

\caption{The $\A_2(1)$-module $L_2$ which appears in Figure \ref{fig:A_2(1)MSO(6)MSO(4)} and its $E_2$ page.}
\label{fig:L_2}
 \end{figure}

  \begin{figure}[H]
\begin{multicols}{2}
 
\begin{center}
\begin{tikzpicture}[scale=0.5]
\draw[fill] (0,0) circle(.1);
\draw[fill] (0,2) circle(.1);
\draw (0,0) to [out=150,in=150] (0,2);
\draw[fill] (0,3) circle(.1);
\draw (0,2) -- (0,3);
\draw[fill] (0,4) circle(.1);
\draw[fill] (0,5) circle(.1);
\draw[fill] (0,6) circle(.1);
\draw (0,3) to [out=150,in=150] (0,5);
\draw (0,4) -- (0,5);
\draw (0,4) to [out=30,in=30] (0,6);
\end{tikzpicture}
\end{center}

\begin{center}
\begin{tikzpicture}[scale=0.5]
\node at (0,-1) {0};
\node at (1,-1) {1};
\node at (2,-1) {2};
\node at (3,-1) {3};
\node at (4,-1) {4};
\node at (5,-1) {5};
\node at (6,-1) {6};
\node at (7.5,-1) {$t-s$};
\node at (-1,0) {0};
\node at (-1,1) {1};
\node at (-1,2) {2};
\node at (-1,3) {3};
\node at (-1,4) {4};
\node at (-1,5) {5};
\node at (-1,6) {$s$};

\draw[->] (-0.5,-0.5) -- (-0.5,6);
\draw[->] (-0.5,-0.5) -- (7,-0.5);

\draw (0,0) -- (0,5);
\draw (4,0) -- (4,5);
\draw (6,1) -- (6,5);
\end{tikzpicture}
\end{center}
\end{multicols}

\caption{The $\A_2(1)$-module $L_3$ which appears in Figure \ref{fig:A_2(1)MSO(6)MSO(4)} and its $E_2$ page.}
\label{fig:L_3}
 \end{figure}

     \begin{figure}[H]
\begin{multicols}{2}
 
\begin{center}
\begin{tikzpicture}[scale=0.5]

\draw[fill] (0,2) circle(.1);

\draw[fill] (0,3) circle(.1);

\draw (0,2) -- (0,3);

\draw[fill] (0,4) circle(.1);

\draw[fill] (1,5) circle(.1);

\draw[fill] (1,6) circle(.1);
\draw (1,5) -- (1,6);

\draw (0,2) to [out=150,in=150] (0,4);

\draw (0,3) to [out=30,in=150] (1,5);

\draw (0,4) to [out=30,in=150] (1,6);

\end{tikzpicture}
\end{center}

\begin{center}
\begin{tikzpicture}[scale=0.5]
\node at (0,-1) {0};
\node at (1,-1) {1};
\node at (2,-1) {2};
\node at (3,-1) {3};
\node at (4,-1) {4};
\node at (5,-1) {5};
\node at (6,-1) {6};
\node at (7.5,-1) {$t-s$};
\node at (-1,0) {0};
\node at (-1,1) {1};
\node at (-1,2) {2};
\node at (-1,3) {3};
\node at (-1,4) {4};
\node at (-1,5) {5};
\node at (-1,6) {$s$};

\draw[->] (-0.5,-0.5) -- (-0.5,6);
\draw[->] (-0.5,-0.5) -- (7,-0.5);

\draw[fill] (0,0) circle(.1);
\draw (2,1) -- (2,5);

\draw (6,2) -- (6,5);

\end{tikzpicture}
\end{center}
\end{multicols}

\caption{The $\A_2(1)$-module $L_4$ which appears in Figure \ref{fig:A_2(1)MG} and \ref{fig:A_2(1)SOn} and its $E_2$ page.}
\label{fig:L_4}
 \end{figure}

  \begin{figure}[H]
\begin{multicols}{2}
 
\begin{center}
\begin{tikzpicture}[scale=0.5]
\draw[fill] (0,0) circle(.1);
\draw[fill] (0,2) circle(.1);
\draw (0,0) to [out=150,in=150] (0,2);
\draw[fill] (0,3) circle(.1);
\draw (0,2) -- (0,3);
\draw[fill] (0,4) circle(.1);
\draw[fill] (0,5) circle(.1);
\draw (0,3) to [out=30,in=30] (0,5);
\draw (0,4) -- (0,5);
\draw[fill] (0,6) circle(.1);
\draw (0,4) to [out=150,in=150] (0,6);
\draw[fill] (0,7) circle(.1);
\draw (0,6) -- (0,7);
\draw[fill] (0,8) circle(.1);
\draw[fill] (0,9) circle(.1);
\draw (0,7) to [out=150,in=150] (0,9);
\draw (0,8) -- (0,9);
\draw[fill] (0,10) circle(.1);
\draw (0,8) to [out=30,in=30] (0,10);
\end{tikzpicture}
\end{center}

\begin{center}
\begin{tikzpicture}[scale=0.5]
\node at (0,-1) {0};
\node at (1,-1) {1};
\node at (2,-1) {2};
\node at (3,-1) {3};
\node at (4,-1) {4};
\node at (5,-1) {5};
\node at (6,-1) {6};
\node at (7.5,-1) {$t-s$};
\node at (-1,0) {0};
\node at (-1,1) {1};
\node at (-1,2) {2};
\node at (-1,3) {3};
\node at (-1,4) {4};
\node at (-1,5) {5};
\node at (-1,6) {$s$};

\draw[->] (-0.5,-0.5) -- (-0.5,6);
\draw[->] (-0.5,-0.5) -- (7,-0.5);

\draw (0,0) -- (0,5);
\draw (4,0) -- (4,5);

\end{tikzpicture}
\end{center}
\end{multicols}

\caption{The $\A_2(1)$-module $L_5$ which appears in Figure \ref{fig:A_2(1)MSO(6)MSO(4)} and \ref{fig:A_2(1)MSO10} and its $E_2$ page.}
\label{fig:L_5}
 \end{figure}
 
  \begin{figure}[H]
\begin{multicols}{2}
 
\begin{center}
\begin{tikzpicture}[scale=0.5]
\draw[fill] (0,0) circle(.1);
\draw[fill] (0,2) circle(.1);
\draw (0,0) to [out=150,in=150] (0,2);
\draw[fill] (0,3) circle(.1);
\draw (0,2) -- (0,3);

\end{tikzpicture}
\end{center}

\begin{center}
\begin{tikzpicture}[scale=0.5]
\node at (0,-1) {0};
\node at (1,-1) {1};
\node at (2,-1) {2};
\node at (3,-1) {3};
\node at (4,-1) {4};
\node at (5,-1) {5};
\node at (6,-1) {6};
\node at (7.5,-1) {$t-s$};
\node at (-1,0) {0};
\node at (-1,1) {1};
\node at (-1,2) {2};
\node at (-1,3) {3};
\node at (-1,4) {4};
\node at (-1,5) {5};
\node at (-1,6) {$s$};

\draw[->] (-0.5,-0.5) -- (-0.5,6);
\draw[->] (-0.5,-0.5) -- (7,-0.5);

\draw (0,0) -- (0,5);
\draw (4,1) -- (4,5);
\draw (4,1) -- (6,3);

\end{tikzpicture}
\end{center}
\end{multicols}

\caption{The $\A_2(1)$-module $L_6$ which appears in Figure \ref{fig:A_2(1)MG} and \ref{fig:A_2(1)Spinn} and its $E_2$ page.}
\label{fig:L_6}
 \end{figure}

   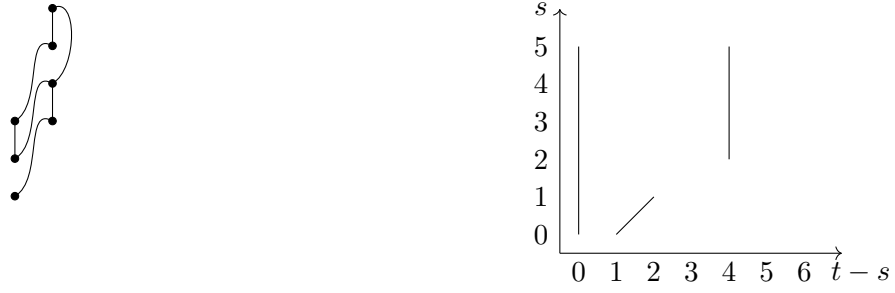
\begin{figure}[H]
\begin{multicols}{2}
 
\begin{center}
\begin{tikzpicture}[scale=0.5]
\draw[fill] (4,4) circle(.1);
\draw[fill] (4,5) circle(.1);
\draw[fill] (4,6) circle(.1);
\draw[fill] (5,6) circle(.1);
\draw (4,4) to [out=30,in=150] (5,6);
\draw (4,5) -- (4,6);
\draw[fill] (5,7) circle(.1);
\draw (5,6) -- (5,7);
\draw (4,5) to [out=30,in=150] (5,7);
\draw[fill] (5,8) circle(.1);
\draw[fill] (5,9) circle(.1);
\draw (4,6) to [out=30,in=150] (5,8);
\draw (5,7) to [out=30,in=30] (5,9);
\draw (5,8) -- (5,9);

\end{tikzpicture}
\end{center}

\begin{center}
\begin{tikzpicture}[scale=0.5]
\node at (0,-1) {0};
\node at (1,-1) {1};
\node at (2,-1) {2};
\node at (3,-1) {3};
\node at (4,-1) {4};
\node at (5,-1) {5};
\node at (6,-1) {6};
\node at (7.5,-1) {$t-s$};
\node at (-1,0) {0};
\node at (-1,1) {1};
\node at (-1,2) {2};
\node at (-1,3) {3};
\node at (-1,4) {4};
\node at (-1,5) {5};
\node at (-1,6) {$s$};

\draw[->] (-0.5,-0.5) -- (-0.5,6);
\draw[->] (-0.5,-0.5) -- (7,-0.5);

\draw (0,0) -- (0,5);
\draw (1,0) -- (2,1);
\draw (4,2) -- (4,5);

\end{tikzpicture}
\end{center}
\end{multicols}

\caption{The $\A_2(1)$-module $L_7$ which appears in Figure \ref{fig:A_2(1)MG} and its $E_2$ page.}
\label{fig:L_7}
 \end{figure}

  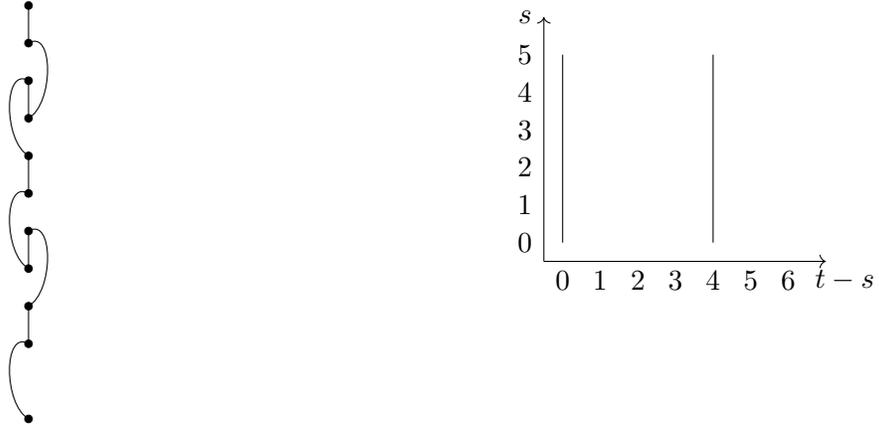
\begin{figure}[H]
\begin{multicols}{2}
 
\begin{center}
\begin{tikzpicture}[scale=0.5]
 \draw[fill] (2,4) circle(.1);
\draw[fill] (2,6) circle(.1);
\draw (2,4) to [out=150,in=150] (2,6);
\draw[fill] (2,7) circle(.1);
\draw (2,6) -- (2,7);
\draw[fill] (2,8) circle(.1);
\draw[fill] (2,9) circle(.1);
\draw (2,7) to [out=30,in=30] (2,9);
\draw (2,8) -- (2,9);
\draw[fill] (2,10) circle(.1);
\draw (2,8) to [out=150,in=150] (2,10);
\draw[fill] (2,11) circle(.1);
\draw (2,10) -- (2,11);
\draw[fill] (2,12) circle(.1);
\draw[fill] (2,13) circle(.1);
\draw (2,11) to [out=150,in=150] (2,13);
\draw (2,12) -- (2,13);
\draw[fill] (2,14) circle(.1);
\draw (2,12) to [out=30,in=30] (2,14);
\draw[fill] (2,15) circle(.1);
\draw (2,14) -- (2,15);

\end{tikzpicture}
\end{center}

\begin{center}
\begin{tikzpicture}[scale=0.5]
\node at (0,-1) {0};
\node at (1,-1) {1};
\node at (2,-1) {2};
\node at (3,-1) {3};
\node at (4,-1) {4};
\node at (5,-1) {5};
\node at (6,-1) {6};
\node at (7.5,-1) {$t-s$};
\node at (-1,0) {0};
\node at (-1,1) {1};
\node at (-1,2) {2};
\node at (-1,3) {3};
\node at (-1,4) {4};
\node at (-1,5) {5};
\node at (-1,6) {$s$};

\draw[->] (-0.5,-0.5) -- (-0.5,6);
\draw[->] (-0.5,-0.5) -- (7,-0.5);

\draw (0,0) -- (0,5);
\draw (4,0) -- (4,5);

\end{tikzpicture}
\end{center}
\end{multicols}

\caption{The $\A_2(1)$-module $L_8$ which appears in Figure \ref{fig:A_2(1)MSO(6)MSO(4)} and \ref{fig:A_2(1)MSO10} and its $E_2$ page.}
\label{fig:L_8}
 \end{figure}

 \section{Comment on the difference between Standard Models}\label{sec:comment}
 
 There is a $\Z_2$ valued bordism invariant of $\Omega_5^{\Spin\times\frac{\SU(3)\times\SU(2)\times\U(1)}{\Z_q}}$ for $q=1,3$ while not for $q=2,6$.
 In this appendix, we will comment on this difference.
 
\begin{figure}[H]
\begin{multicols}{2}
 
\begin{center}
\begin{tikzpicture}[scale=0.5]
\node[below] at (0,0) {$c_2(\SU(2))$};
\draw[fill] (0,0) circle(.1);

\end{tikzpicture}
\end{center}

\begin{center}
\begin{tikzpicture}[scale=0.5]
\node[below] at (0,0) {$c_2(\U(2))$};
\node[above] at (0,2) {$c_1(\U(2))c_2(\U(2))$};
\draw[fill] (0,0) circle(.1);
\draw[fill] (0,2) circle(.1);
\draw (0,0) to [out=150,in=150] (0,2);
\end{tikzpicture}
\end{center}
\end{multicols}

\caption{The local difference between the $\A_2(1)$-module structures.}
\label{fig:difference}
 \end{figure}
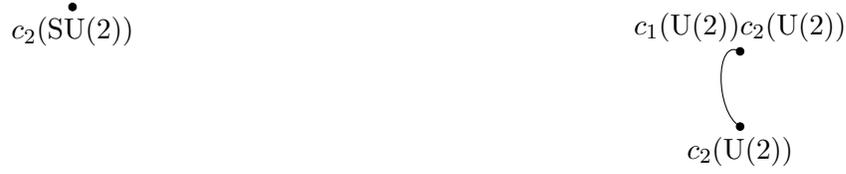

 There is no odd torsion in the bordism groups $\Omega_d^{\Spin\times\frac{\SU(3)\times\SU(2)\times\U(1)}{\Z_q}}$ (see footnote \ref{ft:no-odd-torsion}).
 So we only need to use the mod 2 Adams spectral sequence whose input is only the mod 2 cohomology. Since $\H^*(\B(\frac{\SU(3)\times\SU(2)\times\U(1)}{\Z_3}),\Z_2)=\H^*(\B(\SU(3)\times\SU(2)\times\U(1)),\Z_2)$ and $\H^*(\B(\frac{\SU(3)\times\SU(2)\times\U(1)}{\Z_6}),\Z_2)=\H^*(\B(\frac{\SU(3)\times\SU(2)\times\U(1)}{\Z_2}),\Z_2)$. So the results of bordism groups $\Omega_d^{\Spin\times\SU(3)\times\SU(2)\times\U(1)}$ and $\Omega_d^{\Spin\times\frac{\SU(3)\times\SU(2)\times\U(1)}{\Z_3}}$ are the same, while the results of bordism groups $\Omega_d^{\Spin\times\frac{\SU(3)\times\SU(2)\times\U(1)}{\Z_2}}$ and $\Omega_d^{\Spin\times\frac{\SU(3)\times\SU(2)\times\U(1)}{\Z_6}}$ are the same.
 So we need only to comment on the difference between $q=1$ and 2.
 For $q=1$, there is an $\SU(2)$ factor in $\SU(3)\times\SU(2)\times\U(1)$. While for $q=2$, there is a $\U(2)$ factor in $\frac{\SU(3)\times\SU(2)\times\U(1)}{\Z_2}=\U(2)\times\SU(3)$.
 Consider the second Chern class $c_2$ in both cases.
We have $\H^*(\B\SU(2),\Z_2)=\Z_2[c_2]$ while $\H^*(\B\U(2),\Z_2)=\Z_2[c_1,c_2]$.
In $\H^*(\B\SU(2),\Z_2)$, $\Sq^2c_2=0$, while in $\H^*(\B\U(2),\Z_2)$, $\Sq^2c_2=c_1c_2$.
The local difference between the $\A_2(1)$-module structures is shown in Figure \ref{fig:difference}.

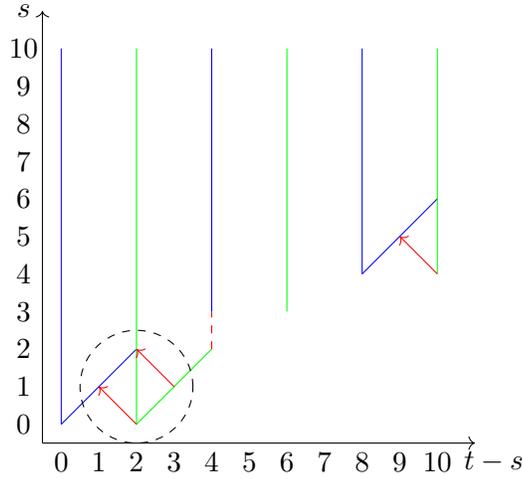
\begin{figure}[H]
\begin{center}
\begin{tikzpicture}[scale=0.5]
\node at (0,-1) {0};
\node at (1,-1) {1};
\node at (2,-1) {2};
\node at (3,-1) {3};
\node at (4,-1) {4};
\node at (5,-1) {5};
\node at (6,-1) {6};
\node at (7,-1) {7};
\node at (8,-1) {8};
\node at (9,-1) {9};
\node at (10,-1) {10};

\node at (11.5,-1) {$t-s$};
\node at (-1,0) {0};
\node at (-1,1) {1};
\node at (-1,2) {2};
\node at (-1,3) {3};
\node at (-1,4) {4};
\node at (-1,5) {5};
\node at (-1,6) {6};
\node at (-1,7) {7};
\node at (-1,8) {8};
\node at (-1,9) {9};
\node at (-1,10) {10};

\node at (-1,11) {$s$};

\draw[->] (-0.5,-0.5) -- (-0.5,11);
\draw[->] (-0.5,-0.5) -- (11,-0.5);

\draw[color=blue] (0,0) -- (0,10);
\draw[color=blue] (0,0) -- (2,2);

\draw[color=blue] (4,3) -- (4,10);

\draw[color=blue] (8,4) -- (8,10);
\draw[color=blue] (8,4) -- (10,6);

\draw[color=green] (2,0) -- (2,10);
\draw[color=green] (2,0) -- (4,2);

\draw[color=green] (6,3) -- (6,10);

\draw[color=green] (10,4) -- (10,10);

\draw[->][color=red] (2,0) -- (1,1);
\draw[->][color=red] (3,1) -- (2,2);
\draw[dashed,color=red] (4,2) -- (4,3);

\draw[->][color=red] (10,4) -- (9,5);
\draw[dashed] (2,1) circle(1.5);

\end{tikzpicture}
\end{center}
\caption{The detail of Figure \ref{fig:Ceta}. The arrows indicate the differential $d_1$, the dashed line indicates the extension. The circled part is the key part.}
\label{fig:detail}
\end{figure}

 The two parts of Figure \ref{fig:difference} are Figure \ref{fig:Z_2} and \ref{fig:Ceta} respectively. The detail of Figure \ref{fig:Ceta} is shown in Figure \ref{fig:detail} where we have used the result of Figure \ref{fig:Z_2} and Lemma \ref{principle}.
 
 Note that the $\Z_2$ valued bordism invariant of $\Omega_5^{\Spin\times\SU(3)\times\SU(2)\times\U(1)}$ is $c_2(\SU(2))\tilde\eta$, while there is a $\Z$ valued bordism invariant $c_1(\U(1))c_2(\SU(2))$ of $\Omega_6^{\Spin\times\SU(3)\times\SU(2)\times\U(1)}$.
 Also note that there is no $\Z_2$ valued bordism invariant of $\Omega_5^{\Spin\times\frac{\SU(3)\times\SU(2)\times\U(1)}{\Z_2}}$, while there is a 
 $\Z$ valued bordism invariant $\frac{c_1(\U(2))c_2(\U(2))}{2}$ of $\Omega_6^{\Spin\times\frac{\SU(3)\times\SU(2)\times\U(1)}{\Z_2}}$.
The arrow from (2,0) to (1,1) in the circled part of Figure \ref{fig:detail} kills the $\Z_2$ at (1,1), also reduces the $\Z$ starting from (2,0) to its half. This explains the difference above.

  \section{Comment on the difference between Standard Models with additional discrete symmetries}\label{sec:comment-2}

  There is a $\Z_4$ valued bordism invariant $c_2(\SU(2))\eta'$ of $\Omega_5^{\Spin\times_{\Z_2}\Z_4\times\frac{\SU(3)\times\SU(2)\times\U(1)}{\Z_q}}$ for $q=1,3$ while there is a $\Z_2$ valued bordism invariant $c_2(\U(2))a$ for $q=2,6$.
 In this appendix, we will comment on this difference.
 
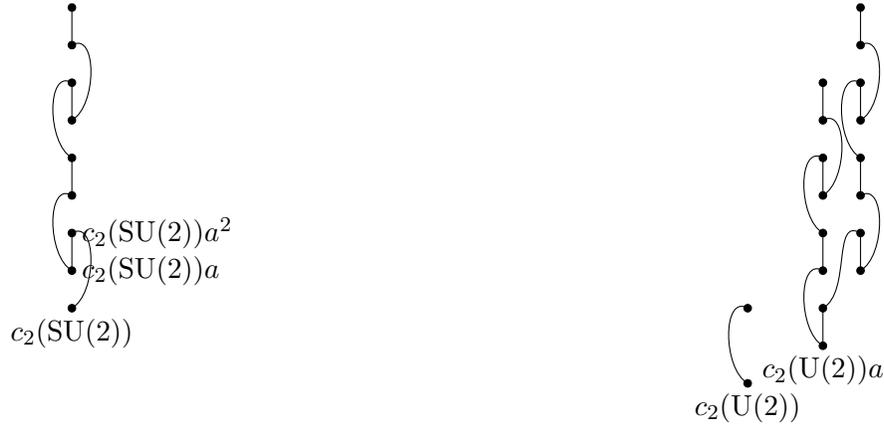
\begin{figure}[H]
\begin{multicols}{2}
 
\begin{center}
\begin{tikzpicture}[scale=0.5]
\node[below] at (0,0) {$c_2(\SU(2))$};
\draw[fill] (0,0) circle(.1);
\draw[fill] (0,1) circle(.1);
\draw[fill] (0,2) circle(.1);
\draw (0,0) to [out=30,in=30] (0,2);
\draw (0,1) -- (0,2);
\draw[fill] (0,3) circle(.1);
\draw (0,1) to [out=150,in=150] (0,3);
\draw[fill] (0,4) circle(.1);
\draw (0,3) -- (0,4);
\draw[fill] (0,5) circle(.1);
\draw[fill] (0,6) circle(.1);
\draw (0,4) to [out=150,in=150] (0,6);
\draw (0,5) -- (0,6);
\draw[fill] (0,7) circle(.1);
\draw (0,5) to [out=30,in=30] (0,7);
\draw[fill] (0,8) circle(.1);
\draw (0,7) -- (0,8);

\node[right] at (0,1) {$c_2(\SU(2))a$};
\node[right] at (0,2) {$c_2(\SU(2))a^2$};

\end{tikzpicture}
\end{center}

\begin{center}
\begin{tikzpicture}[scale=0.5]
\node[below] at (0,0) {$c_2(\U(2))$};

\def\rx{0}
\def\ry{12}

\draw[fill] (0+\rx,-12+\ry) circle(.1);
\draw[fill] (0+\rx,-10+\ry) circle(.1);
\draw (0+\rx,-12+\ry) to [out=150,in=150] (0+\rx,-10+\ry);

\node[below] at (2+\rx,-11+\ry) {$c_2(\U(2))a$};

\draw[fill] (2+\rx,-11+\ry) circle(.1);
\draw[fill] (2+\rx,-10+\ry) circle(.1);
\draw (2+\rx,-11+\ry) -- (2+\rx,-10+\ry);
\draw[fill] (2+\rx,-9+\ry) circle(.1);
\draw (2+\rx,-11+\ry) to [out=150,in=150] (2+\rx,-9+\ry);
\draw[fill] (3+\rx,-9+\ry) circle(.1);
\draw[fill] (3+\rx,-8+\ry) circle(.1);
\draw (3+\rx,-9+\ry) -- (3+\rx,-8+\ry);
\draw (2+\rx,-10+\ry) to [out=30,in=150] (3+\rx,-8+\ry);
\draw[fill] (2+\rx,-8+\ry) circle(.1);
\draw (2+\rx,-9+\ry) -- (2+\rx,-8+\ry);
\draw[fill] (2+\rx,-7+\ry) circle(.1);
\draw[fill] (2+\rx,-6+\ry) circle(.1);
\draw (2+\rx,-7+\ry) -- (2+\rx,-6+\ry);
\draw (2+\rx,-8+\ry) to [out=150,in=150] (2+\rx,-6+\ry);
\draw[fill] (3+\rx,-7+\ry) circle(.1);
\draw (3+\rx,-9+\ry) to [out=30,in=30] (3+\rx,-7+\ry);
\draw[fill] (3+\rx,-6+\ry) circle(.1);
\draw (3+\rx,-7+\ry) -- (3+\rx,-6+\ry);
\draw[fill] (2+\rx,-5+\ry) circle(.1);
\draw (2+\rx,-7+\ry) to [out=30,in=30] (2+\rx,-5+\ry);
\draw[fill] (3+\rx,-5+\ry) circle(.1);
\draw[fill] (3+\rx,-4+\ry) circle(.1);
\draw (3+\rx,-5+\ry) -- (3+\rx,-4+\ry);
\draw (3+\rx,-6+\ry) to [out=150,in=150] (3+\rx,-4+\ry);
\draw[fill] (2+\rx,-4+\ry) circle(.1);
\draw (2+\rx,-5+\ry) -- (2+\rx,-4+\ry);
\draw[fill] (3+\rx,-3+\ry) circle(.1);
\draw[fill] (3+\rx,-2+\ry) circle(.1);
\draw (3+\rx,-3+\ry) -- (3+\rx,-2+\ry);
\draw (3+\rx,-5+\ry) to [out=30,in=30] (3+\rx,-3+\ry);

\end{tikzpicture}
\end{center}
\end{multicols}

\caption{The local difference between the tensor product of the $\A_2(1)$-module structures shown in Figure \ref{fig:difference} and the $\A_2(1)$-module structure of $\H^{*+2}(\RP_2^{\infty},\Z_2)$ shown in Figure \ref{fig:A_2(1)RP_2}.}
\label{fig:difference-2}
 \end{figure}

 There is no odd torsion in the bordism groups $\Omega_d^{\Spin\times\frac{\SU(3)\times\SU(2)\times\U(1)}{\Z_q}}$ (see footnote \ref{ft:no-odd-torsion}).
 So we only need to use the mod 2 Adams spectral sequence whose input is only the mod 2 cohomology. Since $\H^*(\B(\frac{\SU(3)\times\SU(2)\times\U(1)}{\Z_3}),\Z_2)=\H^*(\B(\SU(3)\times\SU(2)\times\U(1)),\Z_2)$ and $\H^*(\B(\frac{\SU(3)\times\SU(2)\times\U(1)}{\Z_6}),\Z_2)=\H^*(\B(\frac{\SU(3)\times\SU(2)\times\U(1)}{\Z_2}),\Z_2)$. So the results of bordism groups $\Omega_d^{\Spin\times_{\Z_2}\Z_4\times\SU(3)\times\SU(2)\times\U(1)}$ and $\Omega_d^{\Spin\times_{\Z_2}\Z_4\times\frac{\SU(3)\times\SU(2)\times\U(1)}{\Z_3}}$ are the same, while the results of bordism groups $\Omega_d^{\Spin\times_{\Z_2}\Z_4\times\frac{\SU(3)\times\SU(2)\times\U(1)}{\Z_2}}$ and $\Omega_d^{\Spin\times_{\Z_2}\Z_4\times\frac{\SU(3)\times\SU(2)\times\U(1)}{\Z_6}}$ are the same.
 So we need only to comment on the difference between $q=1$ and 2.
 For $q=1$, there is an $\SU(2)$ factor in $\SU(3)\times\SU(2)\times\U(1)$. While for $q=2$, there is a $\U(2)$ factor in $\frac{\SU(3)\times\SU(2)\times\U(1)}{\Z_2}=\U(2)\times\SU(3)$.
 Consider the second Chern class $c_2$ in both cases.
We have $\H^*(\B\SU(2),\Z_2)=\Z_2[c_2]$ while $\H^*(\B\U(2),\Z_2)=\Z_2[c_1,c_2]$.
In $\H^*(\B\SU(2),\Z_2)$, $\Sq^2c_2=0$, while in $\H^*(\B\U(2),\Z_2)$, $\Sq^2c_2=c_1c_2$. While $MT(\Spin\times_{\Z_2}\Z_4)=M\Spin\wedge\Sigma^{-2}\RP_2^{\infty}$.
The local difference between the tensor product of the $\A_2(1)$-module structures shown in Figure \ref{fig:difference} and the $\A_2(1)$-module structure of $\H^{*+2}(\RP_2^{\infty},\Z_2)$ shown in Figure \ref{fig:A_2(1)RP_2} is shown in Figure \ref{fig:difference-2}.

 The two parts of Figure \ref{fig:difference-2} are Figure \ref{fig:RP_2} and Figure \ref{fig:Ceta}, \ref{fig:L_1} respectively. The detail of Figure \ref{fig:RP_2} is given in \cite{2017arXiv170804264C} and is based on Lemma \ref{principle}.

  Note that one of the $\Z_4$ valued bordism invariant of $\Omega_5^{\Spin\times_{\Z_2}\Z_4\times\SU(3)\times\SU(2)\times\U(1)}$ is $c_2(\SU(2))\eta'$. Here $\eta'$ is an extension of $a$ by $\tilde\eta$ and $\tilde\eta$ is the mod 2 index of 1d Dirac operator.
 Also note that one of the $\Z_2$ valued bordism invariant of $\Omega_5^{\Spin\times_{\Z_2}\Z_4\times\frac{\SU(3)\times\SU(2)\times\U(1)}{\Z_2} }$
 is $c_2(\U(2))a$.
Figure \ref{fig:RP_2} gives a $\Z_4$ in 1d which is shifted to 5d by $c_2(\SU(2))$, while Figure \ref{fig:L_1} gives a $\Z_2$ in 0d which is shifted to 5d by $c_2(\U(2))a$. This explains the difference above.

\section{Pati-Salam model and bordism invariants} 
 \label{sec:explanation}

The red part in Figure \ref{fig:E_2SU4SU2SU2} corresponds to the three bordism invariants in Table \ref{table:SU4SU2SU2Bordism} which are hard to describe directly, but the reduction of $\Spin\times\SU(2)\times\SU(2)\times\SU(4)$ structure to $\Spin\times_{\Z_2}(\SU(2)\times\SU(2)\times\SU(4))$ structure given by 
\bea
\xymatrix{&\B(\Spin\times\SU(2)\times\SU(2)\times\SU(4))\ar[d]\\
M\ar[ru]\ar[r]&\B(\Spin\times_{\Z_2}(\SU(2)\times\SU(2)\times\SU(4)))}
\eea
 induces a natural group homomorphism $f:\Omega_d^{\Spin\times\SU(2)\times\SU(2)\times\SU(4)}\to \Omega_d^{\Spin\times_{\Z_2}(\SU(2)\times\SU(2)\times\SU(4))}$.
 We will try to express the three bordism invariants corresponding to the red part in Figure \ref{fig:E_2SU4SU2SU2} with the help of $f$.
 
We can compute $\Omega_d^{\Spin\times\SU(2)\times\SU(2)\times\SU(4)}$ 
 using the Adams spectral sequence. 
For $G=\Spin\times\SU(2)\times\SU(2)\times\SU(4)$, the Madsen-Tillmann spectrum $MTG$ of the group $G$ is 
\bea
MTG=M\Spin\wedge (\B (\SU(2)\times\SU(2)\times\SU(4)))_+.
\eea
The $(\B(\SU(2)\times\SU(2)\times\SU(4)))_+$ is the disjoint union of the classifying space $\B(\SU(2)\times\SU(2)\times\SU(4))$ and a point, see footnote \ref{ft:X_+}.

For the dimension $d=t-s<8$, since there is no odd torsion (see footnote \ref{ft:no-odd-torsion}), by \eqref{eq:ExtA_2(1)}, we have the Adams spectral sequence
\bea
\Ext_{\A_2(1)}^{s,t}(\H^*(\B (\SU(2)\times\SU(2)\times\SU(4)),\Z_2),\Z_2)\Rightarrow\Omega_{t-s}^{\Spin \times \SU(2)\times\SU(2)\times\SU(4)}.
\eea

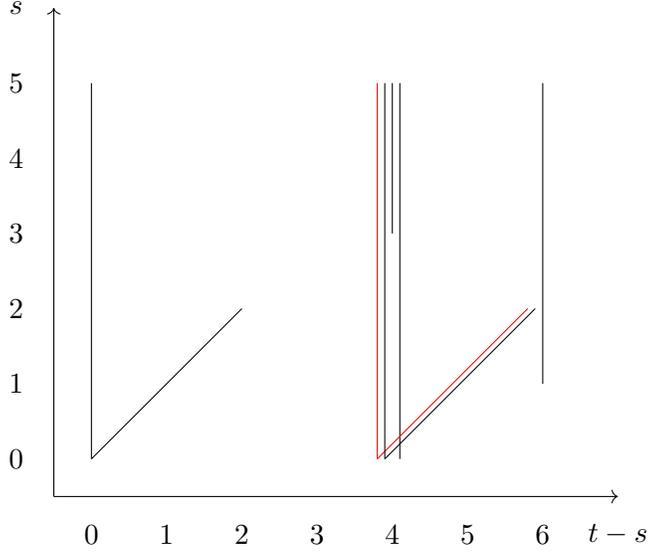
\begin{figure}[H]
\begin{center}
\begin{tikzpicture}
\node at (0,-1) {0};
\node at (1,-1) {1};
\node at (2,-1) {2};
\node at (3,-1) {3};
\node at (4,-1) {4};
\node at (5,-1) {5};
\node at (6,-1) {6};
\node at (7,-1) {$t-s$};
\node at (-1,0) {0};
\node at (-1,1) {1};
\node at (-1,2) {2};
\node at (-1,3) {3};
\node at (-1,4) {4};
\node at (-1,5) {5};
\node at (-1,6) {$s$};

\draw[->] (-0.5,-0.5) -- (-0.5,6);
\draw[->] (-0.5,-0.5) -- (7,-0.5);

\draw (0,0) -- (0,5);
\draw (0,0) -- (2,2);
\draw (4,3) -- (4,5);

\draw (4.1,0) -- (4.1,5);
\draw (6,1) -- (6,5);

\draw[color=red] (3.8,0) -- (3.8,5);
\draw[color=red] (3.8,0) -- (5.8,2);
\draw (3.9,0) -- (3.9,5);
\draw (3.9,0) -- (5.9,2);

\end{tikzpicture}
\end{center}
\caption{$\Omega_*^{\Spin \times \SU(2)\times\SU(2)\times\SU(4)}$. Here the red part is related to the red part in Figure \ref{fig:E_2SU4SU2SU2}.}
\label{fig:E_2SU2SU2SU4}
\end{figure}

The $\A_2(1)$-module structure of $\H^*(\B (\SU(2)\times\SU(2)\times\SU(4)),\Z_2)$ below degree 6 and the $E_2$ page are shown in Figure \ref{fig:A_2(1)SU2SU2SU4}, \ref{fig:E_2SU2SU2SU4}.
Here we have used the correspondence between $\A_2(1)$-module structure and the $E_2$ page shown in Figure \ref{fig:Z_2} and \ref{fig:Ceta}.

\begin{figure}[H]
\begin{center}
\begin{tikzpicture}[scale=0.5]

\node[below] at (0,0) {$1$};
\draw[fill] (0,0) circle(.1);
\node[right] at (0,4) {$c_2$};
\node[right] at (2,4) {$c_2'$};
\node[right] at (4,4) {$c_2''$};
\node[right] at (4,6) {$c_3''$};
\draw[fill] (0,4) circle(.1);
\draw[fill] (2,4) circle(.1);
\draw[fill] (4,4) circle(.1);
\draw[fill] (4,6) circle(.1);

\draw (4,4) to [out=150,in=150] (4,6);

\end{tikzpicture}
\end{center}
\caption{The $\A_2(1)$-module structure of $\H^*(\B (\SU(2)\times\SU(2)\times\SU(4)),\Z_2)$ below degree 6.}
\label{fig:A_2(1)SU2SU2SU4}
\end{figure}

Thus we obtain the bordism group $\Omega^{\Spin \times \SU(2)\times\SU(2)\times\SU(4)}_d$ shown in Table \ref{table:SU2SU2SU4Bordism}.

\begin{table}[H]
\centering
\begin{tabular}{ c c c}
\hline
\multicolumn{3}{c}{Bordism group}\\
\hline
$d$ & 
$\Omega^{\Spin \times \SU(2)\times\SU(2)\times\SU(4)}_d$
& bordism invariants \\
\hline
0& $\Z$\\
1& $\Z_2$ & $\tilde\eta$\\
2& $\Z_2$ & Arf\\
3 & $0$\\
4 & $\Z^4$ & $\frac{\sigma}{16},c_2(\SU(2)),c_2(\SU(2))',c_2(\SU(4))$\\
5 & $\Z_2^2$ & $c_2(\SU(2))\tilde\eta,c_2(\SU(2))'\tilde\eta$ \\
6 & $\Z\times\Z_2^2$ & $\frac{c_3(\SU(4))}{2},c_2(\SU(2))\text{Arf},c_2(\SU(2))'\text{Arf}$ \\
\hline
\end{tabular}
\caption{Bordism group. 
$\tilde\eta$ is a mod 2 index of 1d Dirac operator.
Arf is a 2d Arf invariant.
$\sigma$ is the signature of manifold.
Here $c_2(\SU(2)),c_2(\SU(2))'$ are the Chern classes of the two $\SU(2)$ bundles respectively.
$c_i(\SU(4))$ is the Chern class of the $\SU(4)$ bundle.
Note that $c_3=\Sq^2c_2=(w_2+w_1^2)c_2=0\mod2$ on Spin 6-manifolds.
}
\label{table:SU2SU2SU4Bordism}
\end{table}

By \eqref{eq:TPexact}, we obtain the cobordism group $\TP_d(\Spin \times \SU(2)\times\SU(2)\times\SU(4))$ shown in Table \ref{table:SU2SU2SU4TP}.

\begin{table}[H]
\centering
\begin{tabular}{ c c c}
\hline
\multicolumn{3}{c}{Cobordism group}\\
\hline
$d$ & 
$\TP_d(\Spin \times \SU(2)\times\SU(2)\times\SU(4))$
& topological terms \\
\hline
0& $0$\\
1& $\Z_2$ & $\tilde\eta$\\
2& $\Z_2$ & Arf \\
3 & $\Z^4$ & $\frac{1}{48}$CS$_3^{TM}$, CS$_3^{\SU(2)}$, CS$_3^{\SU(2)'}$, CS$_3^{\SU(4)}$\\
4 & $0$ \\
5 & $\Z\times\Z_2^2$ & $\frac{1}{2}$CS$_5^{\SU(4)}$, $c_2(\SU(2))\tilde\eta,c_2(\SU(2))'\tilde\eta$  \\
\hline
\end{tabular}
\caption{Topological phase classification ($\equiv$ TP) as a cobordism group, following Table \ref{table:SU2SU2SU4Bordism}. 
$\tilde\eta$ is a mod 2 index of 1d Dirac operator.
Arf is a 2d Arf invariant.
The $TM$ is the spacetime tangent bundle.
$\text{CS}_{2n-1}^V$ or $\text{CS}_{2n-1}^G$ is the Chern-Simons form of the vector bundle $V$ or the associated vector bundle of the principal $G$-bundle (associated to the Chern/Pontryagin class).
}
\label{table:SU2SU2SU4TP}
\end{table}

We find that the red part in Figure \ref{fig:E_2SU4SU2SU2} is related to the red part in Figure \ref{fig:E_2SU2SU2SU4}. More precisely, the image of the red part in Figure \ref{fig:E_2SU2SU2SU4} under the group homomorphism $f$ is exactly the red part in 
 Figure \ref{fig:E_2SU4SU2SU2}. So we can express the three bordism invariants  corresponding to the red part in Figure \ref{fig:E_2SU4SU2SU2} as we did in in Table \ref{table:SU4SU2SU2Bordism}. Correspondingly, we can express the topological term corresponding to the $\Z$ valued bordism invariant as we did in Table \ref{table:SU4SU2SU2TP}. Notice that a bordism invariant is a homomorphism from the bordism group to its cyclic factor, while by \eqref{eq:TPexact}, the free parts of bordism group and cobordism group are related by
 \bea
 (\TP_d(G))_{\text{free}}=\Hom((\Omega_{d+1}^G)_{\text{free}},\Z).
 \eea
 So both $f^*$ and $g$ are dual to $f$.
 
 \section{Another form of one of the bordism invariants of $\Omega_4^{\Spin\times_{\Z_2}\Spin(n)}$}\label{sec:another}

The group homomorphism $\Spin\times_{\Z_2}\Spin(n)\to\SO$ induces a reduction of 
$\Spin\times_{\Z_2}\Spin(n)$-structures to $\SO$-structures, this yields a group homomorphism
$h:\Omega_d^{\Spin\times_{\Z_2}\Spin(n)}\to\Omega_d^{\SO}$. In particular, for $d=4$, the bordism invariant of $\Omega_4^{\SO}=\Z$ is the signature $\sigma$. There is a $\Z$ valued bordism invariant of $\Omega_4^{\Spin\times_{\Z_2}\Spin(n)}$ which can be expressed as $h^*(\sigma)$. On the other hand, there is an $\SO(n)$ bundle $V_{\SO(n)}$ with $w_2(TM)=w_2(V_{\SO(n)})$. In our computation in \Sec{sec:SpinSpin10modZ2},
we find that there is a $\Z$ valued bordism invariant of $\Omega_4^{\Spin\times_{\Z_2}\Spin(n)}$ which can be expressed as $p_1(V_{\SO(n)})$. We find that $p_1(V_{\SO(n)})$ and $h^*(\sigma)$ are the same.
Since $p_1(V_{\SO(n)})=w_2(V_{\SO(n)})^2=w_2(TM)^2\mod2$ and $\sigma=3\sigma=p_1(TM)=w_2(TM)^2\mod2$, $p_1(V_{\SO(n)})$ and $h^*(\sigma)$ are the same modulo 2.
While we can find the manifold generator for this $\Z$ valued bordism invariant, it is exactly $\CP^2$ with the natural $\Spin^c=\Spin\times_{\Z_2}\Spin(2)$ structure. Since this $\Spin^c$ structure implies a $\Spin\times_{\Z_2}\Spin(n)$ structure for all $n\ge2$ (just take $V_{\SO(n)}=V_{\SO(2)}\oplus\underline{\R^{n-2}}$), $p_1(V_{\SO(n)})=p_1(V_{\SO(2)})=-c_2(V_{\SO(2)}\otimes_{\R}\C)=-c_2(V_{\SO(2)}\oplus\overline{V_{\SO(2)}})=c_1(V_{\SO(2)})^2=1$. On the other hand, $\sigma(\CP^2)=1$. So $p_1(V_{\SO(n)})$ and $h^*(\sigma)$ have the same value on the manifold generator $\CP^2$. So they are the same bordism invariant.
 
\section{Bibliography}
\bibliographystyle{Yang-Mills}
\bibliography{SU3SU2U1-cobordism.bib} 

\end{document}